%% file: thesis_printable.tex
\documentclass[12pt,a4paper]{book}

\input{packages.tex}

\input{macros.tex}

\makeatletter
\renewcommand{\cleardoublepage}{%
  \clearpage\thispagestyle{empty}%
  \if@twoside
    \ifodd\c@page
    \else
      \hbox{}\newpage
      \if@twocolumn
         \hbox{}\newpage
      \fi
    \fi
  \fi}
\makeatother

\makeatletter
\newcommand*{\cleartoleftpage}{%
  \clearpage
    \if@twoside
    \ifodd\c@page
      \hbox{}\newpage
      \if@twocolumn
        \hbox{}\newpage
      \fi
    \fi
  \fi
}
\makeatother

\newlength{\chapterrulelength}
\newlength{\chapterlabellength}
\newcommand{\mychapterlabel}{\colorbox{black}{\color{white}\huge\textsc{\chaptertitlename}~\thechapter}}

\titleformat{\chapter}
[display]
  {}
  {\settowidth{\chapterlabellength}{\mychapterlabel}
  \setlength{\chapterrulelength}{\textwidth}
  \addtolength{\chapterrulelength}{-\chapterlabellength}
  \rule[-3pt]{\chapterrulelength}{2pt}\mychapterlabel}
  {8pt}
  {\sffamily\bfseries\Huge}
  
  \titleformat{\part}
[frame]
  {\normalfont}
  {\Huge\filcenter\colorbox{black}{\color{white}\enspace\textsc{Part}~\thepart\enspace}}
  {36pt}
  {\Huge\sc\filcenter}

\title{Light propagation in inhomogeneous and anisotropic cosmologies}
\author{Pierre FLEURY}
\date\today
\makeindex

\begin{document}
\dominitoc
\frontmatter 


\input{coverpage.tex}

\cleardoublepage


\newlength{\pageboxlength}
\settowidth{\pageboxlength}{000}
\newlength{\cornerdrag}
\setlength{\cornerdrag}{0.5cm}
\addtolength{\cornerdrag}{\pageboxlength}

\fancypagestyle{plain}{%
\fancyhf{}
\fancyhead[LE]{\hspace*{-\cornerdrag}\makebox[\pageboxlength][r]{\sf\thepage}}
\fancyhead[RO]{\makebox[\pageboxlength][l]{\sf\thepage}\hspace*{-\cornerdrag}}
}

\pagestyle{fancy}
\fancyhf{}
\renewcommand{\sectionmark}[1]{\markright{\thesection \: #1}}
\renewcommand{\chaptermark}[1]{\markboth{\chaptername \ \thechapter \quad #1}{}}
\renewcommand{\headrulewidth}{0pt}
\fancyhead[LE]{\hspace*{-\cornerdrag}\makebox[\pageboxlength][r]{\sf\thepage}\hspace{0.5cm}\rule{1pt}{1.5cm}\hspace{0.5cm}\nouppercase{\sf\slshape\leftmark}}
\fancyhead[RE]{}
\fancyhead[LO]{}
\fancyhead[RO]{\nouppercase{\sf\slshape\rightmark}\hspace{0.5cm}\rule{1pt}{1.5cm}\hspace{0.5cm}\makebox[\pageboxlength][l]{\sf\thepage}\hspace*{-\cornerdrag}}
\cfoot{}


\thispagestyle{empty}
\begin{flushright}
\begin{otherlanguage}{frenchb}
\emph{\`A la mémoire d'Olivier Jean-Marie.}
\end{otherlanguage}
\end{flushright}
\newpage

%

\chapter*{Remerciements}
\phantomsection
\input{acknowledgements.tex}

\tableofcontents

\chapter*{Introduction}
\phantomsection
\addstarredchapter{Introduction}
\markboth{Introduction}{Introduction}
\input{introduction.tex}

\chapter*{Conventions, notations, and acronyms}
\phantomsection
\addstarredchapter{Conventions, notations, and acronyms}
\markboth{Conventions, notations, and acronyms}{Conventions, notations, and acronyms}
\input{notations.tex}

\mainmatter
\part{Geometric optics in curved spacetime}\label{part:geometric_optics}

\chapter[From electromagnetism to geometric optics]{From electromagnetism\\to geometric optics}
\input{chapter_1.tex}

\chapter{Light beams}\label{chapter:beams}
\input{chapter_2.tex}

\chapter{Distances}\label{chapter:distances}
\input{chapter_3.tex}

\part{Standard cosmology and observations}\label{part:standard_cosmology}

\chapter{The standard cosmological spacetimes}\label{chapter:standard_spacetimes}

\input{chapter_4.tex}

\chapter{Observations in standard cosmology}\label{chapter:observations}
\input{chapter_5.tex}

\part{Inhomogeneity beyond the fluid limit}
\label{part:Ricci-Weyl}

\input{introduction_part_III.tex}

\chapter{Swiss-cheese cosmologies}\label{chapter:SC}

\input{chapter_6.tex}

\chapter{Stochastic cosmological lensing}\label{chapter:SL}
\input{chapter_7.tex}

\part{Anisotropic cosmologies}\label{part:anisotropic_cosmologies}

\input{introduction_part_IV.tex}

\chapter{Observing an anisotropic universe}\label{chapter:optics_Bianchi_I}
\input{chapter_8.tex}

\chapter{Sources of anisotropy}\label{chapter:sources_anisotropy}
\input{chapter_9.tex}

\chapter*{Conclusion}
\phantomsection
\addstarredchapter{Conclusion}
\markboth{Conclusion}{Conclusion}
\input{conclusion.tex}

\appendix
\renewcommand{\chaptermark}[1]{\markboth{Appendix\ \thechapter \quad #1}{}}
\part*{Appendices}\label{Appendix}
\chapter{The equations of Einsteinian gravitation}\label{appendix:GR}
\input{appendix_GR.tex}

\begin{otherlanguage}{french}
\mtcselectlanguage{french}
\chapter{Compte-rendu en fran\c{c}ais}
\input{compte_rendu_francais.tex}
\end{otherlanguage}

\phantomsection
\addstarredchapter{Bibliography}
\bibliography{bibliography_thesis}
\bibliographystyle{JHEP}

\phantomsection
\addstarredchapter{Index}
\printindex

\input{backcover.tex}


\end{document}

%% file: packages.tex
\usepackage[frenchb,spanish,english]{babel}

\usepackage[utf8x]{inputenc}
\usepackage[T1]{fontenc}
\usepackage{ucs}
\usepackage{textcomp}
\usepackage{microtype}
\usepackage{ae,aecompl,aeguill}
\usepackage{lmodern}

\usepackage[small,labelfont={sf,bf},labelsep=quad]{caption}
\usepackage{geometry}
\usepackage{fancyhdr}
\usepackage{fancybox}
\usepackage{url}
\usepackage{hyperref}
\usepackage{cite}

\usepackage{xspace}
\usepackage{lastpage}
\usepackage{stmaryrd}

\usepackage{graphicx}
\usepackage{subcaption}
\usepackage[table]{xcolor}
\usepackage{rotating}
\usepackage{empheq}
\usepackage{ifthen}
\usepackage{tensor}
\usepackage{paralist}
\usepackage{wrapfig}
\usepackage{array}

\usepackage{amsmath}
\usepackage{amsfonts}
\usepackage{amssymb}
\usepackage{mathrsfs}
\usepackage{extarrows}
\usepackage{verbatim}
\usepackage{siunitx}
\usepackage{upgreek}
\usepackage{multirow}

\usepackage{makeidx}
\usepackage[sectionbib]{chapterbib}

\usepackage{lettrine}
\usepackage{minitoc}

\usepackage{pdfpages}
\usepackage[sf,bf,newlinetospace]{titlesec}
\usepackage{titletoc}

%% file: macros.tex
\newcommand\ph{\ensuremath{\varphi}}
\newcommand\eps{\ensuremath{\varepsilon}}

\newcommand{\cc}{\text{c.c.}}
\newcommand{\cst}{\mathrm{cst}}

\newcommand\define{\equiv}
\newcommand{\coordequal}{\overset{*}{=}}

\newcommand\vect[1]{\boldsymbol{#1}}
\newcommand{\mat}[1]{\boldsymbol{#1}}

\newcommand\ex[1]{\mathrm{e}^{#1}}
\renewcommand\i{\ensuremath{\mathrm{i}}}

\renewcommand\Re{\ensuremath{\mathrm{Re}}}
\renewcommand\Im{\ensuremath{\mathrm{Im}}}

\newcommand\transpose[1]{#1^{\rm T}}
\newcommand{\tr}{\mathrm{tr}}

\newcommand\e[1]{_{\text{#1}}}
\newcommand\h[1]{^{\text{#1}}}
\newcommand\U[1]{\:\mathrm{#1}}
\newcommand{\micro}{\hbox{\textmu}}

\newcommand{\dd}{\mathrm{d}}
\newcommand{\Dd}{\mathrm{D}}
\newcommand{\pd}[3][]{\frac{\partial^{#1} #2}{\partial {#3}^{#1}}}
\newcommand{\ddf}[3][]{\frac{\dd^{#1} #2}{\dd {#3}^{#1}}}
\newcommand{\Ddf}[3][]{\frac{\Dd^{#1} #2}{\dd {#3}^{#1}}}

\newcommand{\grad}{\vec{\nabla}}

\renewcommand\lim[2]{\underset{ #1 \rightarrow #2 }{ \mathrm{lim} } \,}

\newcommand{\delimiters}[4][]{
\ifthenelse{ \equal{#1}{1} }{  #2 #3 #4  }
					{ \ifthenelse{\equal{#1}{2}}{ \big#2 #3 \big#4 }
						{ \ifthenelse{\equal{#1}{3}}{ \Big#2 #3 \Big#4 }
							{ \ifthenelse{\equal{#1}{4}}{ \bigg#2 #3 \bigg#4 }
								{ \ifthenelse{\equal{#1}{5}}{ \Bigg#2 #3 \Bigg#4 }
									{ \left#2 #3 \right#4 }
								}
							}
						}
					}
													}

\newcommand{\pa}[2][]{\delimiters[#1]{(}{#2}{)}}
\newcommand{\pac}[2][]{\delimiters[#1]{[}{#2}{]}}
\newcommand{\paac}[2][]{\delimiters[#1]{\{}{#2}{\}}}
\newcommand{\abs}[2][]{\delimiters[#1]{|}{#2}{|}}
\newcommand{\ev}[2][]{\delimiters[#1]{\langle}{#2}{\rangle}}
\newcommand{\evaluate}[2][]{\delimiters[#1]{.}{#2}{|}}
\newcommand{\mean}[2][]{\delimiters[#1]{\langle}{#2}{\rangle}}

\newcommand{\source}{S}
\newcommand{\obs}{O}

\newcommand{\tidal}{\mathcal{R}}
\newcommand{\jacobi}{\mathcal{D}}
\newcommand{\wronski}{\mathcal{W}}
\newcommand{\amplification}{\mathcal{A}}
\newcommand{\deformation}{\mathcal{S}}
\newcommand{\Ricfoc}{\mathscr{R}}
\newcommand{\Weylfoc}{\mathscr{W}}

\newcommand{\Lie}{\mathcal{L}}
\newcommand{\wl}{\mathscr{L}}

\newcommand{\Hc}{\mathcal{H}}

\newcommand{\nul}[1]{
#1}

\newcommand{\zero}{\mat{0}}
\newcommand{\identity}{\mat{1}}

\newenvironment{system}
{ \left\{ \begin{aligned} }
{ \end{aligned} \right. }

\newlength{\boxtitlelength}
\newlength{\halfrulelength}
\newcommand{\boxtitle}[1]{\footnotesize\bf{\:#1\:}}

\newcommand{\hiddensection}[1]{
\refstepcounter{section}
\phantomsection
\addcontentsline{toc}{section}{\protect\numberline{\thesection}#1}
\markright{\thesection \: #1}
}

\definecolor{blue4}{RGB}{0,0,143}
\definecolor{red4}{RGB}{143,0,0}
\definecolor{orange}{RGB}{255,128,0}
\definecolor{darkcyan}{RGB}{0,128,128}
\definecolor{olive}{RGB}{0,128,0}
\definecolor{purple}{RGB}{128,0,128}
\definecolor{cyan2}{RGB}{0,255,255}
\definecolor{fushia}{RGB}{255,0,255}
\definecolor{mygray}{gray}{0.5}
\definecolor{lightgray}{gray}{0.85}


%% file: coverpage.tex
\thispagestyle{empty}
\begin{otherlanguage}{frenchb}

\noindent
\begin{minipage}{4.5cm}
\includegraphics[width=\linewidth]{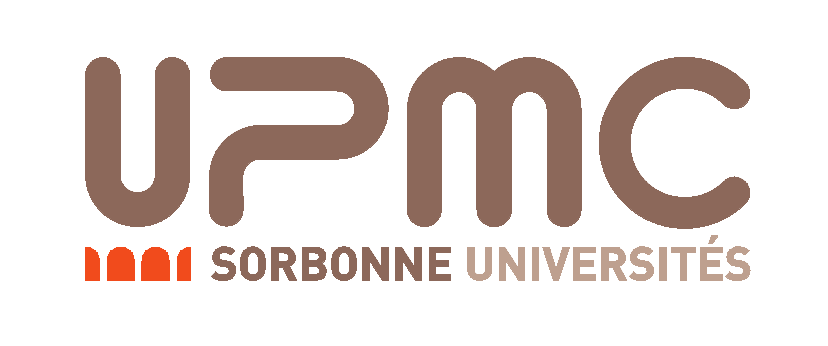}
\end{minipage}
\hspace*{1.5cm}
\begin{minipage}{4cm}
\includegraphics[width=\linewidth]{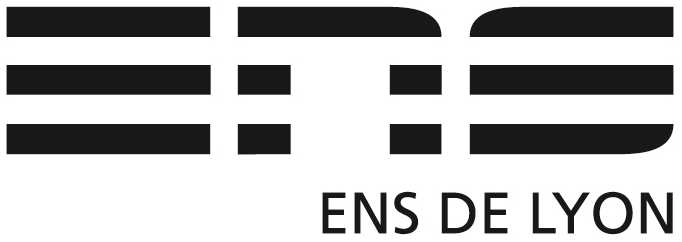}
\end{minipage}
\hfill
\begin{minipage}{4cm}
\includegraphics[width=\linewidth]{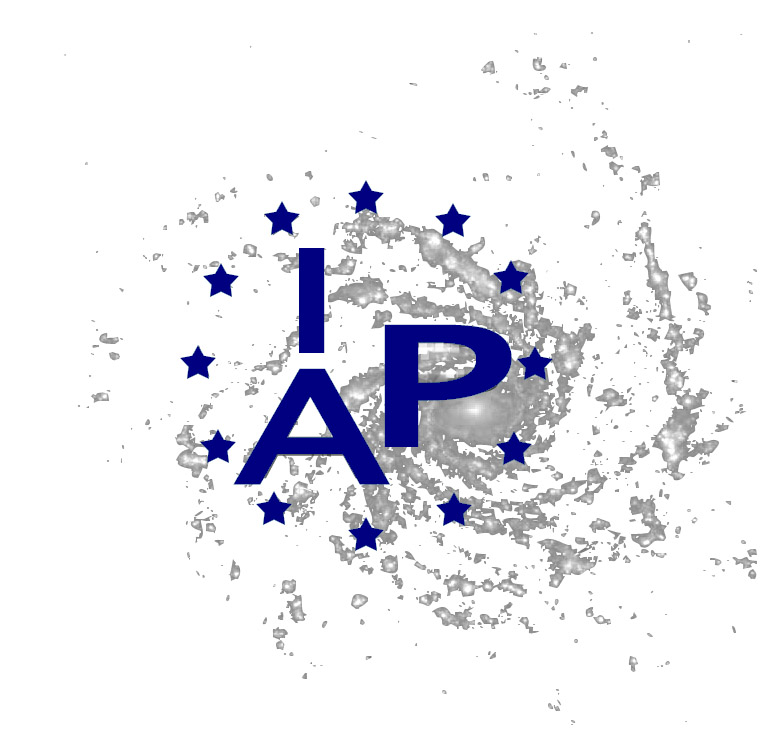}
\end{minipage}

\centering
{
\large\bfseries
TH\`{E}SE DE DOCTORAT\\
DE L'UNIVERSITÉ PIERRE ET MARIE CURIE
}

\vspace*{1.5cm}

{\large et de l'\'{E}cole Doctorale de Physique en \^{I}le-de-France}

\vspace*{1.5 cm}

réalisée à

\medskip

{\large l'Institut d'Astrophysique de Paris}

\vspace*{1.5 cm}

\fbox{\parbox{\textwidth}{
\vspace*{0.2cm}
\centering\LARGE\bfseries
Light propagation in inhomogeneous\\[0.1cm]and anisotropic cosmologies
\vspace*{0.2cm}%
}}

\vspace*{1.5cm}

présentée par

\medskip

{\LARGE Pierre \textsc{Fleury}}

%

\vfill



et soutenue publiquement le 2 novembre 2015

\vspace*{1cm}

devant le jury composé de

\bigskip

\large
\begin{tabular}{rll}
P\up{r} & Ruth \textsc{Durrer} & Rapportrice\\
P\up{r} & Pedro \textsc{Ferreira} & Rapporteur\\
P\up{r} & Michael \textsc{Joyce} & Examinateur\\
D\up{r} & David \textsc{Langlois} & Examinateur \\
D\up{r} & Reynald \textsc{Pain} & Examinateur \\
D\up{r} & \'{E}ric \textsc{Gourgoulhon} & Invité \\
D\up{r} & Jean-Philippe \textsc{Uzan} & Directeur de thèse 
\end{tabular}

\end{otherlanguage}

%% file: acknowledgements.tex
\begin{otherlanguage}{french}
\lettrine{M}{es} premiers remerciements vont à mon directeur de thèse, Jean-Philippe Uzan, d'abord pour m'avoir donné l'opportunité de travailler sur un sujet passionnant, et ensuite pour m'avoir guidé au cours de ces trois années. Si je préfère parler de guide, c'est parce que, pour mon plus grand bonheur, tu n'as jamais été directif envers moi ; je te sais particulièrement gré de cette confiance que tu m'as accordée dès les premiers mois. J'ai beaucoup appris grâce à toi, en physique bien sûr mais pas seulement, car dans ton univers la musique, la théâtre, le cinéma, la bande-dessinée et tant de choses encore gravitent autour des sciences. J'espère moi aussi atteindre, un jour, une si belle virialisation. Bien qu'il me faille à présent quitter le nid(AP), je sais que nous aurons encore beaucoup d'occasions d'échanger et de travailler ensemble. Je m'en réjouis.

Ma reconnaissance va ensuite à mon \og grand frère \fg, Cyril Pitrou, qui fut presque un second encadrant pour moi, qu'il s'agît de science pure ou de chose plus prosaïques. J'ai énormément apprécié chaque occasion de travailler avec toi, j'aime ta façon très géométrique de raisonner et j'admire la vivacité et la précision de ton esprit. Vivement que nous écrivions un livre ensemble ! Merci aussi pour tes décryptages politiques et économiques : à l'instar de Jean-Philippe, tu m'as démontré par l'exemple qu'un véritable intellectuel ne peut pas, ne doit pas, se cantonner à une seule discipline. Enfin, je n'oublie pas la méticulosité avec laquelle tu as corrigé les moindres erreurs phonétiques et déplacements d'accents lors de mes présentations en anglais. Compte sur moi pour te renvoyer l'ascenseur.

I wish to address special thanks to my other close collaborators. \`A Hélène Dupuy d'abord : cela a été un réel plaisir de travailler avec toi dès notre stage de master en binôme. Merci surtout d'avoir bien voulu dédier une partie non négligeable de ta première année de thèse pour m'aider à clore ce projet que nous avions commencé ensemble. Après tout, il le valait bien ! \begin{otherlanguage}{spanish}
A Juan Pablo Beltr\'{a}n, fue un gran placer conocerte y trabajar contigo. Muchas gracias por haber compartido un poco de tu pa\'{i}s: desde los patacones hasta Guatavita, me encant\'{o} todo.\end{otherlanguage}\xspace
\`A Julien Larena, enfin, merci pour ta sympathie si immédiate et si sincère, je suis heureux à l'idée que notre collaboration ne fait que commencer.

Je souhaite bien s\^{u}r exprimer ma gratitude envers les rapporteurs, Ruth Durrer et Pedro Ferreira, car je suis bien conscient que l'évaluation d'un texte de 300 pages représente une tâche très chronophage, ainsi qu'envers les examinateurs et invité, Michael Joyce, David Langlois, Reynald Pain et \'Eric Gourgoulhon. Je suis honoré de voir mon travail jugé par un comité d'une si grande qualité.

Je remercie également les membres de l'IAP que j'ai eu le plaisir de rencontrer et de côtoyer pendant ces trois ans: Patrick Boissé, Frédéric Daigne, Robert Mochkovitch, Pasquier Noterdaeme et Elisabeth Vangioni au troisième étage ; Luc Blanchet, Gilles Esposito-Farèse, Guillaume Faye, Patrick Peter et Sébastien Renaux-Petel au sein du $\mathcal{G}\mathbb{R}\eps\mathbb{C}\mathcal{O}$. \`A Guillaume, merci de ta gestion exemplaire du trésor, et de partager aussi spontanément tes idées ou connaissances, parfois surprenantes, souvent longues à raconter, mais toujours passionnantes ! \`A Sébastien, merci pour ton amitié, et pour ce super-pouvoir que tu as de mettre immédiatement à l'aise toute personne avec qui tu échanges ne serait-ce que quelques mots. C'est une qualité très précieuse que tu partages avec Gilles. 

Grâce à la confiance que m'ont accordé Jacques Chauveau, Nicolas Treps et Philippe Thomen, cette thèse a aussi eu une composante d'enseignement passionnante ; je les en remercie. Je salue par ailleurs Charles Antoine, Mathieu Bertin et Remi Geiger pour leur accueil chaleureux au sein de l'équipe de Quanta et Relativité, et pour avoir supporté mon perfectionnisme sur les sujets de TD.

\end{otherlanguage}

\begin{otherlanguage}{spanish}
Esta tesis no habr\'{i}a sido tan enriquecedora sin los dos viajes extraordinarios a Colombia, que fueron posibles gracias al intercambio iniciado por Yeinzon Rodr\'{i}guez. Agradezco también a todos los que me recibieron de manera tan calurosa, y me hicieron descubrir un pa\'{i}s tan fascinante como magn\'{i}fico, en particular a César Valenzuela y a Juan Pablo Beltr\'an. Y c\'{o}mo no acordarme de la salsa Cale\~{n}a!
\end{otherlanguage}

I have given a few seminars across Europe during my thesis. This has always been a nice and stimulating experience, but it would not have been so without those who kindly devoted a bit of their time to take care of me when I was visiting their institutes. For this, I thank a lot Miguel Zumalac\'{a}rregui, Andrew Jaffe, Nathalie Schwaiger, Vincent Vennin, Thibaut Louis, Johannes Noller, Thomas Buchert, Pierre Salati, Bj\"{o}rn Hermann, Giovanni Marozzi, Ruth Durrer, Sami Nurmi, Syksy R\"{a}s\"{a}nen, Viraj Sanghai, and Timothy Clifton.

\begin{otherlanguage}{french}
Ces trois années aurait été tellement plus fades, tellement moins stimulantes, et tellement moins drôles sans vous tous : les doctorants ! Alba, Alice, Caterina, Charlotte, Clément, Clotilde, Erwan, Federico, Flavien, Florent, Guillaume D. et Guillaume P., Hayley, Hélène, Jean, Jean-Baptiste, Julia, Laura, Maxime, Mélanie, Nicolas C. et Nicolas M., Rebekka, Sandrine, Sébastien, Sylvain, Thomas, Tilman, Vincent B. et Vincent V. et enfin Vivien ; nombre d'entre vous sont aujourd'hui des amis plus que des collègues, et ce n'est que pour garder une taille raisonnable à ces remerciements que je me retiens d'écrire dix lignes sur chacun. Nos YMCA autant que nos pauses café ont indéniablement compté parmi les moments les plus essentiels de mon passage à l'IAP. Mais il n'est pas question que j'oublie ceux qui n'ont pas eu la chance d'effectuer leur thèse dans le meilleur institut d'astrophysique \emph{du monde}\footnote{Dixit Joe Silk et Pedro Ferreira. Des gens très bien.}, et que j'ai (malgré tout) eu beaucoup de plaisir à rencontrer, les uns aux Houches : Agnès, Anais, Mathilde, Julian, Linda, Tico et Thibaut ; les autres à Elbereth, dans des séminaires divers et variés, ou simplement autour d'un verre à la Butte aux Cailles : Alexis, Benjamin, Jibril et Marta. Je sais que nous nous recroiserons, car après tout notre monde est homéomorphe à $\mathbb{S}_2$, et cette idée me plaît.

Une thèse marque la fin de longues études, et il serait illusoire de nier l'influence que certains professeurs de physique ont pu avoir sur mes choix d'orientation. Dès mes premiers pas dans cette discipline jusqu'à l'ENS de Lyon, j'ai eu la chance d'être accompagné par des enseignants qui m'ont sans cesse incité à donner le meilleur de moi-même, par l'admiration qu'ils m'inspiraient. Ce fut le cas d'Olivier Calvosa, Martine Brenier, \'{E}ric Brottier, Philippe Cren, François Gieres, Henning Samtleben, Pierre Salati, et surtout d'Olivier Jean-Marie. C'est en grande partie gr\^{a}ce à cet enseignant hors-pair que je suis arrivé jusqu'ici. Je lui dois ma rigueur et ma pédagogie. Cette thèse est dédiée à sa mémoire ; puisse-t-elle vivre dans chaque graine de scientifique qu'il a semée au cours de sa vie.

Les remerciements d'une thèse ne sont, à mon sens, pas le lieu pour exposer sa vie personnelle. C'est une excuse bien commode, car je ne saurais trouver de mots assez justes pour exprimer toute la gratitude et l'affection que j'éprouve envers ma famille. C'est ce même mélange de pudeur et de maladresse qui me fera taire le nom de celle avec qui j'ai partagé ma vie pendant ces quelques années, et à qui je dois tant.
\end{otherlanguage}

%% file: introduction.tex
\lettrine{2}{015} has seen the very appropriate conjunction of the international year of light and the centenary of Einstein's general theory of relativity; it would have been difficult to find a better occasion for presenting a thesis dedicated to the propagation of light in cosmology. The paradigmatic revolution that represented general relativity one hundred years ago indeed deeply changed our vision of the Universe, setting the foundations of modern cosmology. Since then, this discipline grew and lived, becoming an entire field of research. Physical phenomena that were formerly considered unmeasurable, such as gravitational lensing, or whose very existence was ignored, such as the cosmic microwave background radiation, became powerful cosmological probes, and have now reached an unprecedented level of precision.

In contrast with this increasing accuracy of observations---which had us enter into the so-called \emph{precision era} of cosmology---, the theoretical framework currently used to interpret the associated data is remarkably simple. In fact, for most practical purposes, it has barely changed since the pioneering works of Friedmann and Lema\^itre in the 1920s. There was however no reason for such a change, since to date the Friedmann-Lema\^{i}tre cosmological model, based on the assumption that the Universe is spatially homogeneous and isotropic, has shamelessly passed all observational tests. Of course, the Universe is not \emph{strictly} homogeneous, especially on small scales, but it seems that such an inhomogeneity does not need to be taken into account when interpreting cosmological observations.

The tremendous success of the homogeneous and isotropic model is particularly striking if we consider the case of distance measurements. On astronomical and cosmological scales, distances are mostly measured by comparing the apparent size or luminosity of a light source with its intrinsic size or luminous power. Consequently, such measurements rely on the good understanding of light propagation through the cosmos, in particular the way light beams are focused by matter lying between the sources and us. The point is that current observations involve beams with extremely different sizes: from less than a microarcsecond for supernova observation, to a few degrees with baryon acoustic oscillation experiments, in terms of angular aperture. Depending on the observations at stake, light is thus expected to experience a completely different Universe. Yet the Friedmann-Lema\^itre model arises as \emph{one model to fit them all.} 

Why, and to which extent, such a simplistic model can be considered a good approximation for interpreting cosmological observations? If not, how to go beyond the assumptions of perfect homogeneity and isotropy? are the fundamental questions which motivated the present thesis. I chose to divide this dissertation in four parts, the first two being dedicated to fundamentals, while the last two report the original research that I have performed during the last three years. More precisely, in Part~\ref{part:geometric_optics}, I introduce the laws governing light propagation in curved spacetime, from a relativist's point of view. The presentation is intended to be modern and pedagogical, with a few novel elements absent from textbooks. Part~\ref{part:standard_cosmology} is dedicated to standard cosmology: after having presented the Friedmann-Lema\^{i}tre model and the standard perturbation theory about it, I review the current observational status, emphasizing the precise points on which the understanding of light propagation through the Universe is required, and where the standard model is used. Part~\ref{part:Ricci-Weyl} is the heart of this thesis, it is devoted to the analysis of the effect of small-scale structures on the interpretation of observations, in particular the Hubble diagram of supernovae. I demonstrate that this effect may be non-negligible given the accuracy reached by current and future measurements, and I propose a new theoretical framework for addressing this issue. Finally, Part~\ref{part:anisotropic_cosmologies} deals with two works related to cosmology beyond perfect isotropy. Besides, elements of differential geometry and general relativity can be found in Appendix.

\bigskip

The research reported in this dissertation has been done in collaboration with several colleagues, and led to the following articles:
\begin{enumerate}
\item P. Fleury, H. Dupuy, and J.-P. Uzan
\textit{Interpretation of the Hubble diagram in a nonhomogeneous universe}.
\href{http://journals.aps.org/prd/abstract/10.1103/PhysRevD.87.123526}{Phys. Rev. D 87, 123526 (2013)},
\href{http://arxiv.org/abs/1302.5308}{\tt [arXiv:1302.5308]}.
\item P. Fleury, H. Dupuy, and J.-P. Uzan.
\textit{Can all cosmological observations be accurately interpreted with a unique geometry?}
\href{http://link.aps.org/doi/10.1103/PhysRevLett.111.091302}{Phys. Rev. Lett. 111, 091302 (2013)},
\href{http://arxiv.org/abs/1304.7791}{\tt [arXiv:1304.7791]}. This letter has been highlighted by the science popularisation website~\href{http://phys.org/news/2013-09-universe-conflicting-cosmological.html}{Phys.org}.
\item P. Fleury. \textit{Swiss-cheese models and the Dyer-Roeder approximation}.
\href{http://iopscience.iop.org/1475-7516/2014/06/054/}{JCAP06(2014)054},
\href{http://arxiv.org/abs/1402.3123}{\tt [arXiv:1402.3123]}.
\item P. Fleury, J. P. Beltr\'{a}n Almeida, C. Pitrou, and J.-P. Uzan.
\textit{On the stability and causality of scalar-vector theories}.
\href{http://iopscience.iop.org/1475-7516/2014/11/043/}{JCAP11(2014)043},
\href{http://arxiv.org/abs/1406.6254}{\tt [arXiv:1406.6254]}
\item P. Fleury, C. Pitrou, and J.-P. Uzan.
\textit{Light propagation in a homogeneous and anisotropic universe}.
\href{http://journals.aps.org/prd/abstract/10.1103/PhysRevD.91.043511}{Phys. Rev. D 91, 043511},
\href{http://arxiv.org/abs/1410.8473}{\tt [arXiv:1410.8473]}.
\item P. Fleury, J. Larena, and J.-P. Uzan.
\textit{The theory of stochastic cosmological lensing}.
Accepted for publication in JCAP.
\href{http://arxiv.org/abs/1508.07903}{\tt [arXiv:1508.07903]}.
\end{enumerate}

%% file: notations.tex
\paragraph{Units.} Numerical results are mostly given in terms of units of the International System. In abstract calculations, we adopt the usual relativistic convention of $c=1$, so that lengths and times have the same dimension.

\paragraph{Differential geometry.} We follow the conventions of Misner, Thorne, Wheeler~\cite{1973grav.book.....M}. In particular, the signature of spacetime's metric is taken to be $(-+++)$\index{signature of the metric}. We use Einstein's summation rule over repeated indices, the range of the sum being dictated by the nature of the indices (see Table~\ref{tab:notations}). As often as possible, we dedicate indices of the beginning of the alphabet to components over orthonormal bases.

\paragraph{Notations.} See Table~\ref{tab:notations} for a list of the recurrent symbols used in this dissertation. As often in the relativity literature, vectors and tensors are identified with their components over an arbitrary coordinate basis (e.g. $\vect{u}$ will be equivalently denoted $u^\mu$). Partial derivatives and covariant derivatives with respect to, e.g. coordinate $x^\mu$ are abridged as
\begin{align}
\pd{f}{x^\mu} &\define \partial_\mu f \define f_{,\mu}\\
\nabla_{\partial/\partial x^\mu} f &\define \nabla_\mu f = f_{;\mu}
\end{align}
for any function or tensor $f$.

Sets of indices are symmetrised and antisymmetrised\index{symmetrisation}\index{antisymmetrisation} according to
\begin{align}
T_{(\mu_1\ldots\mu_n)} &\define \frac{1}{n!} \sum_{\sigma\in S_n} T_{\sigma(\mu_1\ldots\mu_n)} \\
T_{[\mu_1\ldots\mu_n]} &\define \frac{1}{n!} \sum_{\sigma\in S_n} \eps(\sigma)\, T_{\sigma(\mu_1\ldots\mu_n)},
\end{align}
for any tensor or subtensor~$\vect{T}$, where $S_n$ denotes the set of all permutations~$\sigma$ of $n$ elements, and $\eps(\sigma)$ is the signature of $\sigma$, i.e. $+1$ or $-1$ depending of whether it consists of an even or odd number of transpositions, respectively. In particular,
\begin{equation}
T_{(\mu\nu)} \define \frac{1}{2} \pa{ T_{\mu\nu} + T_{\nu\mu} },
\qquad
T_{[\mu\nu]} \define \frac{1}{2} \pa{ T_{\mu\nu} - T_{\nu\mu} }.
\end{equation}

\begin{table}[h!]
\centering
\begin{tabular}{|r|l|}
\hline
\rowcolor{lightgray} {\sf\bfseries Notation} & {\sf\bfseries Description} \\
\hline
$\equiv$ & definition \\
$\coordequal$ & equality in a given coordinate system \\
bold symbols $\vect{A}$, $\vect{g}$, $\vect{\jacobi}$, \ldots & vectors, tensors, and matrices \\
greek indices $\alpha,\beta,\ldots\mu,\nu,\ldots$ & run from $0$ to $3$ \\
lowercase latin indices $a,b,\ldots,i,j,\ldots$ & run from $1$ to $3$ \\
uppercase latin indices $A,B,\ldots,I,J,\ldots$ & run from $1$ to $2$ \\
$\dd\Omega^2 \define \dd\theta^2 + \sin^2\theta \dd\ph^2$ & infinitesimal solid angle in spherical coordinates \\
$[\mu\nu\rho\sigma]$ & permutation symbol with $[0123]=1$ \\
$\eps_{\mu\nu\rho\sigma} \define \sqrt{-g}[\mu\nu\rho\sigma]$ & Levi-Civita tensor \\
$\Lie_{\vect{V}}$ & Lie derivative along vector~$\vect{V}$ \\
$\vect{\xi}$ & Killing vector, or separation vector \\
$\nabla, \Dd$ & covariant derivatives \\
$\Gamma\indices{^\rho_\mu_\nu}$ & connection coefficients \\
$R\indices{^\mu_\nu_\rho_\sigma}$ & curvature tensor \\
$R_{\mu\nu}$, $R$ & Ricci tensor, scalar \\
$E_{\mu\nu}$ & Einstein tensor \\
$\Sigma$ & hypersurface \\
$\perp_{\mu\nu}$ & induced metric on a hypersurface \\
$\wl$ & worldline \\
$\vect{u}$ & four-velocity \\
$\tau$ & proper time \\
$\vect{k}$ & wave four-vector \\
$v$ & affine parameter along null geodesics \\
$S, O$ & source, observation events \\
$(\vect{s}_A)_{A=1,2}$ & Sachs (screen) basis \\
$S_{\mu\nu}$ & screen projector \\
$\vect{\tidal}$ & optical tidal matrix \\
$\Ricfoc$, $\Weylfoc$ & Ricci, Weyl lensing scalars \\
$\vect{\wronski} $ & Wronski matrix \\
$\vect{\jacobi}$ & Jacobi matrix \\
$\vect{\amplification}$ & amplification matrix \\
$\vect{\deformation}$ & deformation rate matrix \\
$\theta, \sigma$ & null expansion rate, null shear rate \\
$t$ & cosmic time \\
$a$ & cosmic scale factor \\
$\eta$ & conformal time, with $\dd t = a\dd\eta$ \\
$H\define a^{-1} \dd a/\dd t $ & Hubble expansion rate \\
$\Hc\define a^{-1} \dd a/\dd\eta=aH$ & conformal Hubble expansion rate \\
$\Omega_x$ & cosmological parameter associated with $x$ \\
$\delta=\delta\rho/\bar{\rho}$ & density contrast\\
$\delta\e{D}$ & Dirac distribution \\
$f,\alpha,\bar{\alpha}$ & smoothness parameter \\
$M_\odot$ & Solar mass\\ 
\hline
\end{tabular}
\caption{Description of the main notations used in this thesis.}
\label{tab:notations}
\end{table}

\paragraph{Acronyms.} They will be defined when used for the first time. See also Table~\ref{tab:acronyms}.

\begin{table}[h!]
\centering
\begin{tabular}{|r|l|}
\hline
\rowcolor{lightgray} {\sf\bfseries Acronym} & {\sf\bfseries Signification} \\
\hline
BAO &  Baryon Acoustic Oscillation \\
BOSS & Baryon Oscillation Spectroscopic Survey \\
c.c. & complex conjugate \\
CFHT(LenS) & Canada-France-Hawaii-Telescope (Lensing Survey) \\
CMB & Cosmic Microwave Background \\
COSMOGRAIL & COsmological MOnitoring of GRAvItational Lenses \\
FL & Friedmann-Lema\^{i}tre \\
GR & General Relativity \\
HST & Hubble Space Telescope \\
JLA & Joint Lightcurve Analysis \\
(K)DR & (Kantowski)-Dyer-Roeder \\
$\Lambda$CDM & $\Lambda$ Cold Dark Matter \\
LRG & Luminous Red Galaxy \\
LSST & Large Synoptic Survey Telescope \\
LTB & Lema\^{i}tre-Tolman-Bondi \\
MLCS & Multicolour Light Curve Shape \\
RS & Rees-Sciama \\
SALT & Spectral Adaptive Lightcurve Template \\
SDSS & Sloan Digital Sky Survey \\
SNLS & SuperNova Legacy Survey \\
SC & Swiss Cheese \\
SL & Strong gravitational Lensing \\
SN(e)(Ia) & (Type Ia) SuperNova(e) \\
(I)SW & (Integrated) Sachs-Wolfe \\
VLBI & Very Long Baseline Interferometry \\
WFIRST & Wide-Field InfraRed Survey Telescope \\
WL & Weak gravitational Lensing \\
WMAP & Wilkinson Microwave Anisotropy Probe \\
\hline
\end{tabular}
\caption{List of acronyms and their signification.}
\label{tab:acronyms}
\end{table}

%% file: chapter_1.tex
\lettrine{I}{n} 1860, Maxwell unified electricity, magnetism, and light in a single physical theory, thus providing foundations for the phenomenological laws of geometric optics formulated earlier by Euclid, Newton, Snell, Descartes, or Fermat. Because it is quite naturally extended in the presence of gravity, described by Einstein's general relativity, Maxwell's theory not only teaches us about the nature of light, it also tells us how its propagation is affected by gravitational fields. This first chapter is a journey from standard Maxwell's electromagnetism to gravitational lensing, in which we review in details how gravity is able to modify the path of light, but also its frequency, energy, and polarisation.

\bigskip

\minitoc
\newpage

\section{Electromagnetism}

Maxwell's theory of electromagnetism is naturally formulated in Minkowski spacetime, the structure of which it historically revealed, leading to the development of special, and later general, relativity. This section aims at reminding this standard formulation and describing its extension in the presence of gravity. Because our main purpose is the analysis of electromagnetic waves, we will only consider here Maxwell's theory in electric vacuum, that is in the absence of electric charges and currents; see Ref.~\cite{1975clel.book.....J} for details about the coupling between electromagnetism and matter.

\subsection{In Minkowski spacetime}
\label{sec:electromagnetism_Minkowski}
\index{electromagnetism!in flat spacetime}

In Maxwell's theory of electromagnetism, the fundamental object is a vector field~$\vect{A}$, which represents the four-vector potential from which derives the electromagnetic field. The latter is encoded into the Faraday tensor~$\vect{F}$\index{Faraday tensor} defined as the field strength of $\vect{A}$, i.e.
\begin{equation}
F_{\mu\nu} 
\define \partial_\mu A_\nu - \partial_\nu A_\mu .
\label{eq:def_Faraday_tensor}
\end{equation}
This definition ensures that the \emph{intrinsic} Maxwell equations $\partial_{[\mu} F_{\nu\rho]}=0$ are satisfied.
The dynamics of the electromagnetic field is then described by an action principle, the Maxwell action\index{Maxwell!action in flat spacetime}\index{action!of electromagnetism|see{Maxwell}} being
\begin{equation}
S\e{M}[\vect{A}] = -\frac{1}{16\pi} \int_\mathcal{M} \dd^4 x \; F^{\mu\nu} F_{\mu\nu}
\qquad
\text{(flat spacetime)},
\label{eq:action_EM_flat}
\end{equation}
where $\mathcal{M}$ denotes the spacetime manifold. The $1/16\pi$ prefactor in $S\e{EM}$ indicates that we work in the Gaussian system of units, in which the electric and magnetic fields have the same dimension; see e.g. Ref.~\cite{1975clel.book.....J} for a detailed correspondence between Gaussian units and the units of the International System.

In classical electromagnetism, the four-vector potential~$\vect{A}$ is not observable, contrary to its field strength~$\vect{F}$, directly related to the force that the electromagnetic field applies on charged particles\footnote{In quantum mechanics, however, the vector potential influences the phase of the wavefunction of charged particles~\cite{Baez:1995sj}, in a way that depends on their path. Such an effect, named after Aharonov and Bohm, is experimentally accessible via interferometry, and is used in Superconducting Quantum Interference Devices (SQIDs) to measure magnetic fluxes.}. From the definition~\eqref{eq:def_Faraday_tensor} of $F_{\mu\nu}$, we see that the potential~$\vect{A}$ from which it derives is not unique; any \emph{gauge transformation}\index{gauge!transformation (electromagnetism)}
\begin{equation}
A_\mu \mapsto A_\mu + \partial_\mu s,
\label{eq:gauge_transformation}
\end{equation}
where $s$ is a scalar quantity, indeed leaves $F_{\mu\nu}$ unchanged. The action~\eqref{eq:action_EM_flat} is thus also unaffected by \eqref{eq:gauge_transformation}: Maxwell's theory is gauge invariant. As any symmetry in the action, gauge invariance is associated via Noether theorem~\cite{ItzyksonZuber,DeruelleUzan} with the conservation of a physical quantity, which in this case is the vector $Q^\nu\equiv\partial_\mu F^{\mu\nu}$. When matter is (minimally) coupled to the electromagnetic field, this is equivalent to the conservation of electric charge.

In the absence of any electric source, the four-vector potential~$\vect{A}$ only appears in $S\e{EM}$; stationarity of the latter with respect to variations of the former thus leads to the equations of motion
\begin{equation}
0 = 4\pi\frac{\delta S\e{M}}{\delta A_\nu} = \partial_\mu F^{\mu\nu}
\qquad \text{(flat spacetime)},
\label{eq:extrinsic_Maxwell_flat}
\end{equation}
which represent the \emph{extrinsic} Maxwell equations\index{Maxwell!equations in flat spacetime} in vacuum. When we additionally impose the Lorenz gauge condition~$\partial_\mu A^\mu=0$\index{Lorenz gauge!in flat spacetime}, the extrinsic Maxwell equations~\eqref{eq:extrinsic_Maxwell_flat} take the form of a simple wave equation
\begin{equation}
\Box A_\mu = 0 \qquad \text{(flat spacetime)},
\label{eq:EOM_potential_flat}
\end{equation}
where $\Box\equiv\partial^\nu \partial_\nu$ is the d'Alembertian operator in flat spacetime.

\subsection{In curved spacetime}
\label{sec:EM_curved_spacetime}
\index{electromagnetism!in curved spacetime}

Let us now examine how the laws of electromagnetism are affected by the presence of gravity, that we will assume to be described by Einstein's general theory of relativity. Because the dynamics of the gravitational field is not the topic of this chapter, we will not address it here, but we refer the reader to the appendix~\ref{appendix:GR} for a summary of the notations, conventions, physical quantities, and concepts of general relativity that we will use throughout this thesis.

\subsubsection{Minimally coupled electrodynamics}

As stated by Einstein's equivalence principle, the laws of non-gravitational physics must be locally the same in the presence or in the absence of gravitation, provided they are worked out in a freely falling frame. Hence, a theory of electromagnetism in curved spacetime must give back the Lagrangian density of Eq.~\eqref{eq:action_EM_flat} about any event~$E$ of $\mathcal{M}$, provided it is written using, e.g., Gaussian normal coordinates, for which $g_{\mu\nu}(E)\mapsto\eta_{\mu\nu}$ and $\Gamma\indices{^\rho_\mu_\nu}(E)\mapsto 0$.

Besides, since curvature cannot be eliminated by any coordinate transformation, we deduce that any direct coupling in the action between $A_\mu$ and the Riemann tensor (or higher derivatives of the metric) would violate Einstein's equivalence principle as stated above. Imposing this principle then results into the \emph{minimal coupling prescription}\index{minimal coupling} for making curved-spacetime laws from flat-spacetimes laws: starting from the action in Minkowski spacetime,
\begin{enumerate}
\item Replace the coordinate volume element~$\dd^4 x$ by the covariant volume element~$\sqrt{-g}\,\dd^4 x$, where $g$ denotes the determinant of spacetime's metric\index{metric!determinant of the}. This quantity is defined with respect to the \emph{covariant} components~$g_{\mu\nu}$ of the metric,
\begin{equation}
g \equiv \frac{1}{4!} \, [\alpha\beta\gamma\delta]\,[\mu\nu\rho\sigma]\,g_{\alpha\mu}\,g_{\beta\nu}\,g_{\gamma\rho}\,g_{\delta\sigma}
\qquad
\text{(sum on all indices)}
\label{eq:metric_determinant}
\end{equation}
where $[\alpha\beta\gamma\delta]$\index{permutation symbol} is the completely antisymmetric permutation symbol, with the convention $[0123]=1$.
\item Replace partial derivatives~$\partial_\mu$ by covariant derivatives~$\nabla_\mu$.
\end{enumerate}

Applying the minimal coupling prescription to the Maxwell action~\eqref{eq:action_EM_flat} leads to the standard action of electromagnetism in curved spacetime\index{Maxwell!action in curved spacetime}
\begin{empheq}[box=\fbox]{equation}
S\e{M}[\vect{A},\vect{g}] = -\frac{1}{16\pi} \int_\mathcal{M} \dd^4 x \; \sqrt{-g} \, F^{\mu\nu} F_{\mu\nu},
\label{eq:action_EM}
\end{empheq}
where the Faraday tensor now reads~$F_{\mu\nu}=\nabla_\mu A_\nu-\nabla_\nu A_\mu$, which turns out to be equal to its expression in flat spacetime (with partial derivatives), provided spacetime geometry is torsion free as assumed in GR.

The equations of motion deriving from the stationarity of $S\e{EM}$ are now\index{Maxwell!equations in curved spacetime}
\begin{equation}
0 = 4\pi\frac{\delta S\e{EM}}{\delta A_\nu} = \frac{1}{\sqrt{-g}}\,\partial_\mu\pa{\sqrt{-g}\,F^{\mu\nu}} = \nabla_\mu F^{\mu\nu},
\end{equation}
which can be rewritten in terms of the four-vector potential as
\begin{align}
0 &= \nabla_\mu F^{\mu\nu} \\
	&= \nabla_\mu \nabla^\mu A^\nu - \nabla_\mu \nabla^\nu A^\mu \\
	&= \nabla_\mu \nabla^\mu A^\nu - \nabla^\nu \nabla_\mu A^\mu - R\indices{^\mu_\rho_\mu^\nu} A^\rho,
\end{align}
where the Riemann tensor appeared due to the commutation of two covariant derivatives; imposing in addition the general-relativistic form of the Lorenz gauge~$\nabla_\mu A^\mu=0$,\index{Lorenz gauge!in curved spacetime} finally yields
\begin{empheq}[box=\fbox]{equation}
\Box A_\mu - R^\nu_\mu A_\nu = 0,
\label{eq:EOM_potential_curved}
\end{empheq}
with the covariant d'Alembertian $\Box\equiv\nabla^\nu\nabla_\nu$ and the Ricci tensor~$R_{\mu\nu}\define R\indices{^\rho_\mu_\rho_\nu}$. It is interesting here to note the importance of applying the minimal coupling prescription \emph{in the action} rather than in the equations of motion; indeed if we had replaced $\partial_\nu$ by $\nabla_\nu$ directly in Eq.~\eqref{eq:EOM_potential_flat}, then we would have missed the Ricci curvature term that appears in Eq.~\eqref{eq:EOM_potential_curved}.

The presence of covariant derivatives in $\Box$ and of the Ricci tensor in Eq.~\eqref{eq:EOM_potential_curved} clearly indicates that spacetime geometry affects the electromagnetic field. Note that the converse is also true, since the electromagnetic field has energy and momentum, encoded into its stress-energy tensor\index{stress-energy tensor!of the electromagnetic field}
\begin{equation}
T_{\mu\nu}\h{EM} \equiv \frac{-2}{\sqrt{-g}} \frac{\delta S\e{M}}{\delta g^{\mu\nu}}
								= \frac{1}{4\pi} \pac{ F_{\mu\rho}F\indices{_\nu^\rho} - \frac{1}{4} (F^{\rho\sigma}F_{\rho\sigma})g_{\mu\nu} } ,
\label{eq:electromagnetic_stress-energy_tensor}
\end{equation}
which is a source of gravity in the Einstein field equations. It is a significant difference, that it is worth emphasizing, between Newtonian gravity in which mass is the only form of energy to gravitate, and general relativity in which all forms of energy have the ability to curve spacetime; in Einstein's theory, even photons attract each others!

\subsubsection{Beyond minimal coupling?\index{Horndeski!vector-tensor theory}}

If one relaxes the assumption of minimal coupling between the vector field~$A_\mu$ and spacetime geometry, then various terms involving spacetime curvature can appear in the action, e.g. $R_{\mu\nu\rho\sigma}F^{\mu\nu}F^{\rho\sigma}$, $R_{\mu\nu}A^\mu A^\nu$, $R F^{\mu\nu} F_{\mu\nu}$, etc. Such couplings turn out to be generically unhealthy~\cite{2010PhRvD..81f3519E}, by generating Hamiltonian instabilities or, since they are already second-order derivatives of the metric, third-order derivatives in the equations of motion; the presence of a bare~$\vect{A}$ also potentially violates the gauge-invariance of the electromagnetic sector, which implies the nonconservation of electric charge if $\vect{A}$ is minimally coupled to matter.

There is however at least one exception. In a very technical article~\cite{Horndeski:1976gi}, Horndeski proved that if a theory which couples a vector field~$\vect{A}$ to spacetime geometry
\begin{inparaenum}[(i)]
\item derives from an action principle involving~$\vect{A}$ and $\vect{g}$;
\item generates second-order equations of motion;
\item conserves the electric charge; and
\item reduces to standard electromagnetism in flat spacetime,
\end{inparaenum}
then its action reads
\begin{equation}\label{eq:Horndeski_action}
S\e{H}[\vect{A},\vect{g}] = S\e{EH}[g_{\mu\nu}] + S\e{M}[A_\mu,g_{\mu\nu}] + \frac{\ell^2}{16\pi} \int \dd^4 x\sqrt{-g} \; L^{\mu\nu\rho\sigma} F_{\mu\nu} F_{\rho\sigma}
\end{equation}
where the first term is the Einstein-Hilbert action (see appendix~\ref{appendix:GR}), the second term is the Maxwell action as seen in the previous paragraph, and the last term is the only nonminimal coupling term that is allowed by the above four assumptions. The $L^{\mu\nu\rho\sigma}$ tensor is defined~by
\begin{align}
L^{\mu\nu\rho\sigma} &\define -\frac{1}{2} \eps^{\mu\nu\alpha\beta}\eps^{\rho\sigma\gamma\delta} R_{\alpha\beta\gamma\delta}\\
									&= 2 R^{\mu\nu\rho\sigma} + 4 \pa{ R^{\mu[\sigma}g^{\rho]\nu} + R^{\nu[\sigma}g^{\rho]\mu} }
										+ 2 R g^{\mu[\rho}g^{\sigma]\nu},
\end{align}
it enjoys the same symmetries as the Riemann tensor, and is divergence free (i.e. $\nabla_\mu L^{\mu\nu\rho\sigma}=0$). The coupling constant~$\ell$, which has the dimension of a length, physically represents the typical curvature radius scale below which this theory would significantly deviate from minimally coupled electromagnetism. The sign of this new term in Eq.~\eqref{eq:Horndeski_action} ensures the Hamiltonian stability of the theory~\cite{Jimenez:2013qsa}.

Due to its non-trivial coupling between electromagnetism and gravity, Horndeski's vector-tensor theory is expected to present a rich phenomenology beyond the standard Einstein-Maxwell framework, such as photon-graviton oscillations, varying speed of light, gravitational birefringence and optical activity, etc.


\subsection{Electric and magnetic fields}
\index{electric field}\index{magnetic field}
\label{sec:electric_magnetic}

Let us close this section on pure electromagnetism by indicating how to disentangle electric and magnetic fields from the Faraday tensor~$\vect{F}$. In a special-relativistic context, if $F_{\mu\nu}$ is written as a matrix whose $\mu$, $\nu$ respectively label the lines and columns, then
\begin{equation}
[F_{\mu\nu}] =
\begin{bmatrix}
0 & -E_1 & -E_2 & -E_3 \\
E_1 & 0 & B_3 & -B_2 \\
E_2 & -B_3 & 0 & B_1 \\
E_3 & B_2 & -B_1 & 0
\end{bmatrix}
\qquad \text{(flat spacetime)},
\label{eq:Faraday_electric_magnetic_flat}
\end{equation}
where $E_i$ and $B_i$ are the components of the electric and magnetic fields. This comes from an identification between
\begin{inparaenum}[(i)]
\item the definition of the Faraday tensor~$F_{\mu\nu}=\partial_\mu A_\nu - \partial_\nu A_\mu$, and
\item the relation linking fields and potentials, usually written as $\vec{E}=-\grad V-\partial_t \vec{A}$ and $\vec{B}=\grad\times\vec{A}$, with $(A^\mu)=(V,\vec{A})$.
\end{inparaenum}
As parts of an order-two tensor, $\vec{E}$ and $\vec{B}$ are not, properly speaking, vectors: they are \emph{observer dependent}, and thus transform in a non-trivial way under Lorentz boosts.

Let $\vect{u}$ be the four-velocity of an experimentalist (with $u^\mu u_\mu=-1$) who wishes to characterise the electric and magnetic parts of $\vect{F}$ in her rest frame. By definition, the components of the Faraday tensor over an orthonormal tetrad representing such a frame must take the form of Eq.~\eqref{eq:Faraday_electric_magnetic_flat}. One can therefore define the electric and magnetic fields $(\vect{E},\vect{B})$ as purely spatial vectors, i.e.
\begin{equation}
u_\mu E^\mu = u_\mu B^\mu = 0,
\end{equation}
such that
\begin{empheq}[box=\fbox]{equation}
F_{\mu\nu} = 2 u_{[\mu} E_{\nu]} + \eps_{\mu\nu\rho\sigma} u^\rho B^\sigma,
\label{eq:Faraday_electric_magnetic}
\end{empheq}
where we have introduced the \emph{Levi-Civita tensor}\index{Levi-Civita!tensor}
\begin{equation}
\eps_{\mu\nu\rho\sigma} \define \sqrt{-g} \, [\mu\nu\rho\sigma]
\label{eq:Levi_Civita}
\end{equation}
following the convention of Ref.~\cite{2004rtmb.book.....P}. Note that, given the definition of the metric determinant~$g$, the completely contravariant counterpart of Eq.~\eqref{eq:Levi_Civita} is $\eps^{\mu\nu\rho\sigma}=-[\mu\nu\rho\sigma]/\sqrt{-g}$. By inverting Eq.~\eqref{eq:Faraday_electric_magnetic} one obtains the expressions of the electric and magnetic fields:
\begin{align}
E^\mu &= u_\nu F^{\mu\nu},\\
B^\mu &= -u_\nu \tilde{F}^{\mu\nu},
\end{align}
where $\tilde{F}_{\mu\nu}=\eps_{\mu\nu\rho\sigma}F^{\rho\sigma}/2$ is the \emph{Hodge dual} of the Faraday two-form. It enjoys a decomposition similar to Eq.~\eqref{eq:Faraday_electric_magnetic}, as $\tilde{F}_{\mu\nu}=-2 u_{[\mu} B_{\nu]} + \eps_{\mu\nu\rho\sigma} u^\rho E^\sigma$. Summarising, the Hodge duality turns $\vect{E}$ into $-\vect{B}$ and $\vect{B}$ into $\vect{E}$.

These two quantities can also be used to rewrite the electromagnetic stress-energy tensor~$T\h{EM}_{\mu\nu}$\index{stress-energy tensor!of the electromagnetic field} in a way that makes its interpretation in terms of energy, energy flux, pressure, etc. easier. Introducing Eq.~\eqref{eq:Faraday_electric_magnetic} into Eq.~\eqref{eq:electromagnetic_stress-energy_tensor} one indeed obtains
%
%
\begin{equation}
T_{\mu\nu}\h{EM}=\rho \, u_\mu u_\nu + 2 u_{(\mu}\Pi_{\nu)} +  p \perp_{\mu\nu} + \Pi_{\mu\nu},
\label{eq:stress-energy_EM_electric_magnetic}
\end{equation}
where we introduced the spatial part of the metric~$\perp_{\mu\nu}=u_\mu u_\nu + g_{\mu\nu}$, while
\begin{align}
\rho &\define \frac{E^2+B^2}{8\pi} \qquad (X^2\define X^\mu X_\mu),\\
p &\define \frac{\rho}{3}, \\ 
\Pi_\nu &\define \frac{u^\mu \eps_{\mu\nu\rho\sigma}E^\rho B^\sigma}{4\pi}, \\
\Pi_{\mu\nu} &= 2p\perp_{\mu\nu} - \frac{E_\mu E_\nu + B_\mu B_\nu}{4\pi},
\end{align}
are respectively the energy density, isotropic radiation pressure\index{radiation pressure}, Poynting vector\index{Poynting vector}, and anisotropic stress of the electromagnetic field. Both the Poynting vector and the anisotropic stress are purely spatial ($\Pi_\mu u^\mu=0$, $\Pi_{\mu\nu}u^\mu=0$), and the latter is trace free ($\Pi^\mu_\mu=0$).


\section{Light rays in geometric optics}

In this section, we introduce what will be the framework for all the remainder of this thesis, namely geometric optics. After having discussed the underlying assumptions---the eikonal approximation---, we examine how the Maxwell equations in curved spacetime govern the propagation of electromagnetic waves in this regime, and analyse some features of their trajectories.

\subsection{The geometric optics regime}

\subsubsection{An ansatz for electromagnetic waves}

The geometric optics framework is conveniently discussed when one considers the following ansatz for the vector four-potential of an electromagnetic wave,
\begin{equation}
\vect{A} = \vect{a} \ex{\i \phi} + \cc,
\label{eq:ansatz_EM_wave}
\end{equation}
where $\vect{a}$, $\phi$ respectively stand for the wave's amplitude and the phase. While the latter is real by definition, the former can be complex in general; here nevertheless \emph{we assume that $a_\mu$ is real}, i.e. we restrict our study to \emph{linearly polarised} waves. This assumption can actually be made without any loss of generality: thanks to the linearity of the Maxwell equations~\eqref{eq:EOM_potential_curved}, any elliptically polarised wave is indeed the superposition of two linearly polarised ones.

To the wave~\eqref{eq:ansatz_EM_wave} can be associated a \emph{wave four-vector} defined as the gradient of the phase,
\begin{equation}
k_\mu \define \partial_\mu \phi,
\end{equation}
which thus represents the local direction of propagation of the wave through spacetime. Note that the Lorenz gauge condition already imposes conditions on $\vect{a}$ and $\vect{k}$, namely
\begin{equation}
0 = \nabla_\mu A^\mu = \pa{ \nabla_\mu a^\mu + \i k_\mu a^\mu } \ex{\i\phi} + \cc
\end{equation}
whose real and imaginary parts respectively imply
\begin{equation}
\nabla_\mu a^\mu = 0,
\qquad
k_\mu a^\mu = 0.
\end{equation}
Such a wave is therefore \emph{transverse}, in the sense that its directions of excitation and propagation are orthogonal.

\subsubsection{The eikonal approximation\index{eikonal approximation}}
\label{sec:eikonal}

Geometric optics\index{geometric optics regime|see{eikonal approximation}} corresponds to a regime for which the genuinely undulatory properties of light are irrelevant, so that it behaves as a stream of classical point particles: well-defined trajectories and no interference nor diffraction phenomena. In flat spacetime, such a situation is achieved when the phase of the electromagnetic wave varies much faster than its amplitude, i.e.
\begin{equation}
\partial\phi \gg a^{-1} \partial a,
\end{equation}
which is known as the \emph{eikonal approximation}.

In the presence of gravity, however, the approximation requires an additional assumption since another spatio-temporal scale enters into the game: spacetime curvature itself. In order for the undulatory phenomena not to be affected by this additional element, the phase of the electromagnetic wave must evolve on scales much smaller than the typical spacetime curvature radius~$\ell\e{c}$, i.e.
\begin{equation}
\partial\phi \gg \ell\e{c}^{-1},
\label{eq:eikonal_2}
\end{equation}
where $\ell\e{c}^{-1}$ must be understood as the square-root of a  typical component of the Riemann curvature tensor. Assumption~\eqref{eq:eikonal_2} is therefore the second part of the eikonal approximation in curved spacetime~\cite{1973grav.book.....M}.

Let us examine the restrictiveness of \eqref{eq:eikonal_2} in terms of the wavelength~$\lambda$---which is indeed the typical evolution scale of $\phi$---on two examples. In a cosmological context, the typical curvature radius is given by the inverse of the expansion rate $\ell\e{c}\sim c/H\sim 4\U{Gpc}$ today (that is essentially the size of the observable Universe) or $\sim 0.2\U{Mpc}$ at the epoch of recombination; hence $\lambda\ll \ell\e{c}$ is not particularly restrictive in this context. In the vicinity of a spherically symmetric massive object, $\ell\e{c}$ can be evaluated using the Kretschmann scalar\footnote{The Kretschmann scalar is defined as the contraction of the Riemann tensor with itself, $K\equiv R^{\mu\nu\rho\sigma} R_{\mu\nu\rho\sigma}$, and is useful to investigate the properties of the complete spacetime curvature, rather than restricting to the Ricci scalar which, for example, is useless in vacuum.}~$K$\index{Kretschmann scalar} as $\ell\e{c}\sim K^{-1/4}\sim \sqrt{r^3/r\e{S}}$, where $r\e{S}$ denotes the Schwarzshild radius of the object, and $r$ the coordinate distance to it. On the surface of Earth, this implies $\lambda\ll 2\U{AU}$, while on the horizon of a stellar black hole it leads to $\lambda\ll 3\U{km}$. Only in the latter---rather extreme---case, a part of the radio domain does not satisfy the eikonal approximation.

\subsubsection{The Maxwell equations in geometric optics}

Inserting the ansatz~\eqref{eq:ansatz_EM_wave} for $\vect{A}$ into the Maxwell equations~\eqref{eq:EOM_potential_curved} gives
\begin{equation}
\Box a^\mu - (k^\nu k_\nu) a^\mu - R^\mu_\nu a^\nu 
+ \i\pa{ 2 k^\nu \nabla_\nu a^\mu +  a^\mu \nabla_\nu k^\nu } = 0.
\label{eq:EOM_ansatz_potential}
\end{equation}
The real part of Eq.~\eqref{eq:EOM_ansatz_potential} implies $(k^\nu k_\nu) a^\mu = \Box a^\mu-R_\nu^\mu a^\nu$, but the eikonal approximation indicates that the left-hand side is much larger than the right-hand side; both must therefore vanish identically for the situation to be acceptable. Putting it together with the imaginary part of Eq.~\eqref{eq:EOM_ansatz_potential} results into the propagation equations of electromagnetic waves in the geometric optics regime:\index{Maxwell equations!in the geometric optics regime}
\begin{empheq}[box=\fbox]{gather}
k^\mu k_\mu = 0,
\label{eq:dispersion_relation}\\
k^\nu \nabla_\nu a^\mu + \frac{1}{2} (\nabla_\nu k^\nu) a^\mu =0.
\label{eq:conservation_photon_polarisation}
\end{empheq}
The first one~\eqref{eq:dispersion_relation} is the dispersion relation, while the second one~\eqref{eq:conservation_photon_polarisation} contains both photon conservation and the evolution equation of polarisation, as we shall see respectively in \S~\ref{sec:energetical_observables} and \S~\ref{sec:polarisation}.

The eikonal approximation also leads to a familiar structure for the trio $(k^\mu,E^\mu,B^\mu)$. The Faraday tensor indeed reads
%
$F_{\mu\nu} = 2\i k_{[\mu}A_{\nu]} + \cc$
%
at leading order, so that $E_\mu=2\i u^\nu k_{[\mu}A_{\nu]} + \cc$ and $B_\mu=\i u^\nu\eps_{\nu\mu\rho\sigma}k^\rho A^\sigma+\cc$, whence
\begin{empheq}[box=\fbox]{equation}
k^\mu E_\mu = k^\mu B_\mu = E^\mu B_\mu = 0,
\end{empheq}
which express the wave's transversality in terms of its electric and magnetic fields.

\subsection{Photons follow null geodesics}
\label{photons_follow_geodesics}

We now focus on the trajectory followed by electromagnetic waves. Start with taking the gradient of the dispersion relation $k^\nu k_\nu = 0$,
\begin{equation}
0 = \nabla_\mu (k^\nu k_\nu) = 2 k^\nu \nabla_\mu k_\nu.
\end{equation}
Because $k_\nu$ is defined itself as the gradient of the phase~$\phi$, the indices of $\nabla_\mu k_\nu$ can be inverted due to the symmetry of the Christoffel coefficients,
\begin{equation*}
\nabla_\mu k_\nu = \nabla_\mu \partial_\nu \phi = \partial_\mu \partial_\nu \phi - \Gamma\indices{^\rho_\mu_\nu} \partial_\rho \phi,
\end{equation*}
%
and the above equation takes the familiar form\index{geodesic!equation}
\begin{empheq}[box=\fbox]{equation}
k^\nu \nabla_\nu k^\mu = 0.
\label{eq:geodesic_equation}
\end{empheq}
The integral curves of the vector field $k^\mu$---the curves to which $k^\mu$ is everywhere tangent---are therefore \emph{null geodesics}, since they satisfy the geodesic equation\footnote{It is interesting to note here that the geodesic equation emerges as a consequence of the null condition $k^\mu k_\mu=0$, and it would be tempting to conclude that any null curve is a geodesic. This is actually the case \emph{only if its tangent vector is a gradient}, as this property of $k^\mu$ was a crucial ingredient in the derivation. A counterexample is easily constructed in Minkowski spacetime; consider the curve defined as $x=r\cos(ct/r)$, $y=r\sin(ct/r)$, which corresponds to a circular motion---a helicoid in spacetime---of radius~$r$ at velocity~$c$. Its tangent vector $(u^\mu)=[1,-c\sin(ct/r),c\cos(ct/r),0]$ is clearly null, but it is not a geodesic.}~\eqref{eq:geodesic_equation}
and their tangent vectors are null~\eqref{eq:dispersion_relation}. Such curves, that we shall call light rays in the following\footnote{Albeit in their classical definition, light rays are rather the spatial projections of null geodesics.}, can be interpreted as the worldlines of photons, referring to the Einstein-de Broglie duality between waves and particles, according to which the momentum of a photon is $p^\mu = \hbar k^\mu$, where $\hbar\equiv h/(2\pi)$ denotes the reduced Planck constant.

A parameter~$v$ along a given light ray is naturally defined through its tangent vector~by
\begin{equation}
k^\mu \equiv \ddf{x^\mu}{v},
\end{equation}
which tells us that a small variation~$\dd v$ of parameter~$v$ corresponds to the small displacement~$\dd x^\mu = k^\mu \dd v$ along the ray. The geodesic equation~\eqref{eq:geodesic_equation} then reads, in terms of $v$,
\begin{equation}
0 = \Ddf{k^\mu}{v} \define \ddf{k^\mu}{v} + \Gamma\indices{^\mu_\nu_\rho} k^\nu k^\rho
	= \ddf[2]{x^\mu}{v} + \Gamma\indices{^\mu_\nu_\rho} \ddf{x^\nu}{v} \ddf{x^\rho}{v},
\label{eq:geodesic_equation_affine_parameter}
\end{equation}
where $\Dd/\dd v$ denotes the \emph{covariant derivative} with respect to $v$. The particular form~\eqref{eq:geodesic_equation_affine_parameter} of the geodesic equation, where $\Dd k^\mu / \dd v$ completely vanishes instead of just being proportional to $k^\mu$, indicates that $v$ is not any parameter along the curve but an \emph{affine parameter}\index{affine parameter!definition} (see e.g. Ref.~\cite{2004rtmb.book.....P} for further details). Any affine transformation $v\mapsto av+b$ indeed preserves the form of Eq.~\eqref{eq:geodesic_equation_affine_parameter}. The physical meaning of~$v$, which has been introduced so far as a mathematical object, will be discussed in \S~\ref{sec:physical_interpretation_affine_parameter}.

\subsection{Conformal invariance}\label{sec:conformal_invariance}

Null geodesics turn out to enjoy a particular symmetry that they do not share with their timelike or spacelike equivalents, namely the invariance under conformal transformations of the metric. This property can be stated as the following theorem.

\medskip

\noindent{\sffamily\bfseries Theorem.}\index{conformal!invariance of null geodesics} Let~$\vect{g}$ and $\vect{\tilde{g}}$ be two metric tensors for a same spacetime manifold~$\mathcal{M}$ described with an arbitrary coordinate system $\{x^\mu\}_{\mu=0\ldots 3}$, related by a \emph{conformal transformation}\index{conformal!transformation}:
\begin{equation}
g_{\mu\nu}(x^\rho) = \Omega^2(x^\rho) \tilde{g}_{\mu\nu}(x^\rho),
\label{eq:conformal_transformation_metric}
\end{equation}
where $\Omega$ is an arbitrary scalar function on $\mathcal{M}$. If a curve~$\gamma$ is a null geodesic for $\vect{g}$, then it is also a null geodesic for $\vect{\tilde{g}}$. Moreover, if $v$ is an affine parameter of $\gamma$ for $\vect{g}$, then any $\tilde{v}$ so that $\dd v=\Omega^2 \dd \tilde{v}$ is an affine parameter of $\gamma$ for $\vect{\tilde{g}}$.

\medskip

The above property is practically very useful when there exists a conformal transformation such that $\tilde{g}_{\mu\nu}$ is much simpler than $g_{\mu\nu}$; the analysis of light propagation is then more easily performed in terms $\tilde{g}_{\mu\nu}$, while its counterpart in $g_{\mu\nu}$ can be recovered by a systematic procedure (see \S~\ref{sec:conformal_trick} for a complete dictionary and an example). Besides practical calculations, conformal invariance also immediately explains the failure of Nordstr\o m's theory of gravitation~\cite{2011cqvz.book..247D,2011GReGr..43.3337D}, proposed in 1912---three years before Einstein's general relativity. In this theory, as reformulated by Einstein and Fokker in 1914, spacetime's geometry is conformally flat: its metric reads $g_{\mu\nu}=\exp(2\Phi)\eta_{\mu\nu}$, where $\eta_{\mu\nu}$ denotes the Minkowski metric and $\Phi$ corresponds to the Newtonian potential in the weak-field regime. Because of the conformal invariance of null geodesics, Nordstr\o m's gravity thus predicts no gravitational deflection of light, and is therefore ruled out by observations.

\subsubsection{Proof of the theorem}

Suppose that $\gamma$ is a null geodesic affinely parametrised by $x^\mu(v)$, so that its tangent vector~$k^\mu \equiv \dd x^\mu/\dd v$ satisfies $\Dd k^\mu/\dd v=0$. Then consider the same curve~$\gamma$ but parametrised by $\tilde{v}$, with $\dd v = \Omega^2 \dd\tilde{v}$; the associated tangent vector reads $\tilde{k}^\mu \define \dd x^\mu/\dd\tilde{v}=\Omega^2 k^\mu$. The covariant derivative
\begin{equation}
\frac{\tilde{\Dd}\tilde{k}^\mu}{\dd \tilde{v}}
\define \tilde{k}^\nu\tilde{\nabla}_\nu \tilde{k}^\mu 
= \tilde{k}^\nu\partial_\nu \tilde{k}^\mu + \tilde{\Gamma}\indices{^\mu_\nu_\rho} \tilde{k}^\nu \tilde{k}^\rho
\end{equation}
can be rewritten using the following correspondence between the Christoffel coefficients of $\vect{\tilde{g}}$ and $\vect{g}$:
\begin{align}
\tilde{\Gamma}\indices{^\mu_\nu_\rho} &\define \frac{1}{2} \, \tilde{g}^{\mu\sigma} \pa{\partial_\nu \tilde{g}_{\sigma\rho} 
																																				+ \partial_\rho \tilde{g}_{\sigma\nu}
																																				- \partial_\sigma \tilde{g}_{\nu\rho}}  \\
																	&= \Gamma\indices{^\mu_\nu_\rho} - 2\delta^\mu_{(\nu}\partial_{\rho)} \ln \Omega 
																																+ g_{\nu\rho} g^{\mu\sigma} \partial_\sigma\ln\Omega,
\end{align}
so that
\begin{equation}
\frac{\tilde{\Dd}\tilde{k}^\mu}{\dd \tilde{v}} = \Omega^4 \pac{ \Ddf{k^\mu}{v} + (k^\nu k_\nu) g^{\mu\sigma} \partial_\sigma\ln\Omega } = 0,
\end{equation}
which concludes the proof.

\subsubsection{Microscopic or emergent symmetry?}

It is tempting to see in the conformal invariance of lightcones a consequence of the conformal invariance of electromagnetism\index{conformal!invariance of electromagnetism}. Indeed, the Maxwell action~\eqref{eq:action_EM} is unchanged by the transformation~\eqref{eq:conformal_transformation_metric} of the metric, without the need of any transformation of the vector field~$\vect{A}$, it was therefore expected to find such a symmetry for the propagation of light. However, this rationale only holds in four dimensions, since in dimension~$d$,
\begin{align}
S\e{M}[\vect{A},\tilde{\vect{g}}]
&\define
\int \dd^d x \; \sqrt{-\tilde{g}} \, \tilde{g}^{\mu\rho}\tilde{g}^{\nu\sigma}F_{\mu\nu} F_{\rho\sigma}\\
&=
\int \dd^d x \; \sqrt{-g} \, \Omega^{d-4} g^{\mu\rho}g^{\nu\sigma}F_{\mu\nu} F_{\rho\sigma}\label{eq:conformal_action}\\
&\not=
S\e{M}[\vect{A},\vect{g}] \qquad \text{if $d\not=4$},
\end{align}
while the conformal invariance of null geodesics is fully general, regardless of the dimension---as it was not involved in the proof of the previous paragraph. Yet photons do follow null geodesics, even for $d\not= 4$, since the dimension was not involved in the derivation of \S~\ref{photons_follow_geodesics} either. 

This paradoxical situation suggests that, in general, conformal invariance must emerge somewhere on the way from electromagnetism to geometric optics. It can be understood by considering the equation of motion for $\vect{A}$ which derives from the action~\eqref{eq:conformal_action},
\begin{equation}
\Box A_\mu - R^\nu_\mu A_\nu + (d-4)F\indices{^\nu_\mu}\partial_\nu \ln\Omega = 0,
\label{eq:EOM_potential_conformal}
\end{equation}
where last term makes Eq.~\eqref{eq:EOM_potential_conformal} differ from the Maxwell equations~\eqref{eq:EOM_potential_curved}, and thus spoils conformal invariance. This new term is however proportional to $\partial\ln\Omega\sim \ell_\Omega^{-1}$, where $\ell_\Omega$ denotes the typical evolution scale of $\Omega$. Similar terms turn out to appear in the transformation of the Ricci tensor between the metrics $\vect{g}$ and $\tilde{\vect{g}}$ (see e.g. Ref.~\cite{1984ucp..book.....W} for details), so that $\ell_\Omega^{-2}$ can be considered a contribution to spacetime curvature. Hence, by virtue of the eikonal approximation, the last term of \eqref{eq:EOM_potential_conformal} can be neglected, and conformal invariance is approximately recovered.
 
The conformal invariance of lightcones is therefore an \emph{emergent} symmetry in general, which accidentally coincides with a fundamental symmetry of electromagnetism in four dimensions.

\subsection{Conserved quantities}

In many situations of interest, spacetime geometry itself enjoys some symmetries, which implies the existence of conserved quantities for its geodesics. Mathematically speaking, such a symmetry is defined by the existence of a Killing vector~$\vect{\xi}$ along which the (Lie) derivative of the metric vanishes~\cite{2004rtmb.book.....P,DeruelleUzan}:
\begin{equation}
\Lie_{\vect{\xi}} \vect{g} = \zero.
\end{equation}
In terms of components, the above equation reads
\begin{align}
0 	&= \xi^\rho \partial_\rho g_{\mu\nu} + \partial_\mu \xi^\rho g_{\rho\nu} + \partial_\nu \xi^\rho g_{\rho\mu}\\
	&= \xi^\rho \nabla_\rho g_{\mu\nu} + \nabla_\mu \xi^\rho g_{\rho\nu} + \nabla_\nu \xi^\rho g_{\rho\mu}\\
	&= 2\nabla_{(\mu} \xi_{\nu)};
\end{align}
the tensor $\nabla_\mu \xi_\nu$ is thus antisymmetric if $\vect{\xi}$ is a Killing vector. This property implies that its scalar product with the wave-four vector is a constant of motion, since
\begin{align}
\ddf{(k^\mu \xi_\mu)}{v} &= k^\mu k^\nu\nabla_\nu \xi_\mu \\
											&= 2 k^\mu k^\nu \nabla_{(\nu} \xi_{\mu)} \\
											&= 0.
\end{align}
A spacetime with $N$ Killing vectors $(\vect{\xi}_i)_{i=1\ldots N}$ thus generates in principle $N$ constants $C_i \define \xi_i^\mu k_\mu$ along any geodesic motion. The above calculations are indeed true for any geodesic (timelike, spacelike or null).

\subsection{Fermat's principle}\index{Fermat's principle}

While both timelike and spacelike geodesics enjoy a intuitive geometrical meaning in terms of extremalisation of the associated proper time or length, null geodesics are a priori harder to interpret this way, because the line element $\dd s^2$ between two neighbouring events of any null curve vanishes, by definition. It is thus unclear what extremalising such an always zero function is supposed to mean.

\begin{figure}[h!]
\begin{flushright}
\begin{minipage}{5cm}
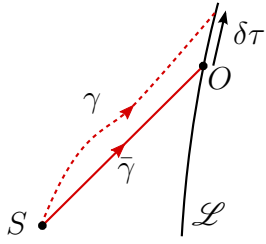
\end{minipage}
\begin{minipage}{8cm}
\caption{Relativistic Fermat's principle. The null curve~$\bar{\gamma}$ is a geodesic iff $\tau_O$ is a local extremum of arrival times, i.e. $\delta \tau=0$ for any null curve~$\gamma$ slightly differing from $\bar{\gamma}$.}\label{fig:Fermat}
\end{minipage}
\end{flushright}
\end{figure}

Yet there exists a similar characterisation of null geodesics, reminiscent of classical Fermat's principle (see e.g.~\cite{1999prop.book.....B,Perez_optique}) which essentially states that light always follows the quickest way between two points. In a general relativistic context, this can be formulated as the following theorem, illustrated in Fig.~\ref{fig:Fermat}.


\medskip

\noindent{\sffamily\bfseries Theorem.}~\cite{1990ApJ...351..114K,1990CQGra...7.1319P} Let $S$ be an arbitrary event (light emission) and consider a null curve~$\bar{\gamma}$ connecting $S$ to the worldline~$\wl$ of an observer. Call $O$ the intersection between $\bar{\gamma}$ and $\wl$ (observation event), and $\tau_O$ the proper time measured by the observer at $O$, with respect to an arbitrary origin. Then~$\bar{\gamma}$ is a null geodesic if and only if $\tau_O$ extremises the arrival times~$\tau$ of all the \emph{null} curves~$\gamma$ connecting $S$ to $\wl$ which slightly deviate from $\bar{\gamma}$.

\medskip

Let us now prove this theorem. Our demonstration is partially inspired from Ref.~\cite{1992grle.book.....S}, pp. 101-102, though we tried to propose a more intuitive formulation of the converse part. In all that follows, we respectively denote with $\bar{x}^\mu(v)$ and $x^\mu(v)\equiv \bar{x}^\mu(v)+\delta x^\mu(v)$ parametrisations of the neighbouring null curves $\bar{\gamma}$ and $\gamma$; the associated tangent vectors are defined by $\bar{k}^\mu\define \dd \bar{x}^\mu/\dd v$ and $k^\mu\define \dd x^\mu /\dd v$. As an infinitesimal quantity, $\delta x^\mu(v)$ can be considered a vector field along $\bar{\gamma}$; we assume that its values at the events $S$ and $O$ read
\begin{equation}
\delta x^\mu_S = 0, \qquad \delta x^\mu_O=u^\mu_O \delta \tau,
\end{equation}
where $u^\mu_O$ is the four-velocity of the observer at $O$. These assumptions mean that both $\bar{\gamma}$ and $\gamma$ emerge from $S$, and cross $\wl$ with a relative delay $\delta\tau$.

Finally, since the (covariant) derivative of $\delta x^\mu$ with respect to $v$ reads $\Dd \delta x^\mu/\dd v=k^\mu-\bar{k}^\mu\define\delta k^\mu$, a consequence of the nullity of $\gamma$ is then
\begin{equation}
k^\mu k_\mu = 2 \bar{k}^\mu \delta k_\mu = 0,
\label{eq:Fermat_orthogonality}
\end{equation}
at first order in $\delta k^\mu$.

\subsubsection{$\boldsymbol{\bar{\gamma}}$ geodesic $\boldsymbol{\Longrightarrow \delta\tau=0}$}

Suppose that $\bar{\gamma}$ is a null geodesic affinely parametrised by $v$ (without loss of generality), we then have
\begin{equation}
\ddf{}{v} \pa{\bar{k}_\mu \delta x^\mu} = \Ddf{\bar{k}^\mu}{v} \delta x^\mu + \bar{k}^\mu \delta k_\mu = 0,
\end{equation}
where both terms in the middle are zero because of, respectively, the geodesic equation~\eqref{eq:geodesic_equation_affine_parameter}, and Eq.~\eqref{eq:Fermat_orthogonality}. We conclude that $\bar{k}_\mu \delta x^\mu$ is a constant all along $\bar{\gamma}$, which moreover vanishes because $\delta x^\mu_S=0$, hence
\begin{equation}
0 = (\bar{k}_\mu \delta x^\mu)_O = (\bar{k}_\mu u^\mu)_O \delta\tau,
\end{equation}
that is $\delta \tau=0$ since the null vector $\bar{\vect{k}}_O$ and the timelike vector $\vect{u}_O$ cannot be orthogonal to each other.

\subsubsection{$\boldsymbol{\bar{\gamma}}$ geodesic $\boldsymbol{\Longleftarrow \delta\tau=0}$}

\begin{figure}[h!]
\centering
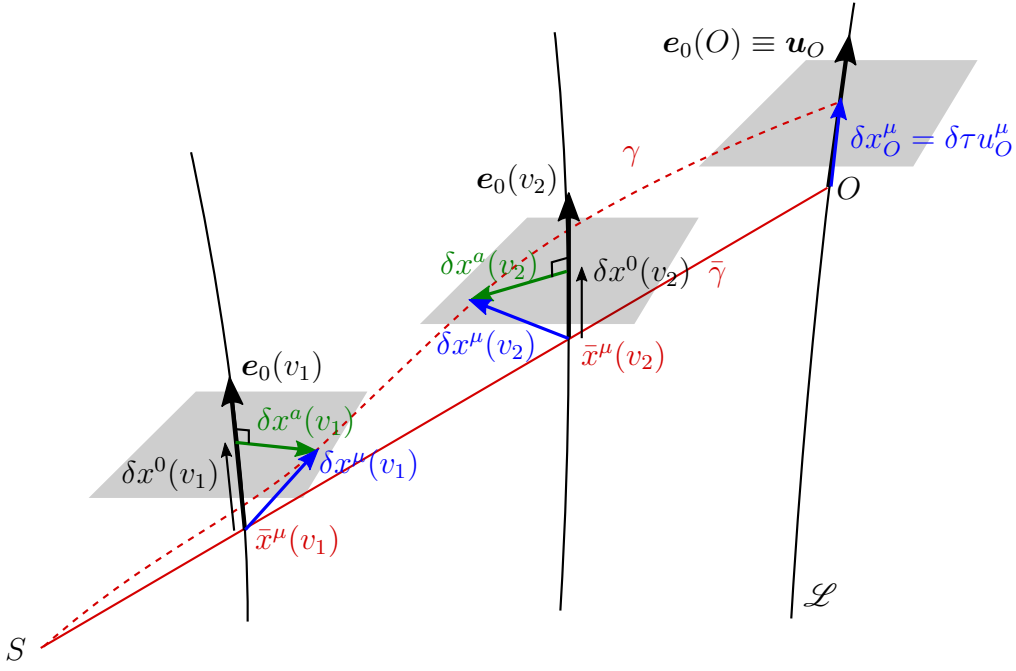
\caption{Geometry of the proof of (the converse of) Fermat's principle. The four-velocity $u^\mu_O$ of the observer at $O$ is parallely transported along $\bar{\gamma}$ to generate a vector field~$\vect{e}_0(v)$, interpreted as the four-velocity of fictive observers (two of them are represented, at $v_1$ and $v_2$). The null curve~$\gamma$ is defined as the trajectory of a particle moving at the speed of light which is detected at spatial positions $\delta x^a(v)$ by each of the fictive observers. This imposes the delay~$\delta x^0(v)$ of this particle with respect to the one following $\bar{\gamma}$, and thus the complete separation $\delta x^\mu=\delta x^0 e_0^\mu + \delta x^a e_a^\mu$ between $\gamma$ and $\bar{\gamma}$.}
\label{fig:Fermat_converse}
\end{figure}

Proving this converse assertion requires a more constructive approach. Let $\{\vect{e}_\alpha\}_{\alpha=0\ldots 3}$ be a tetrad field so that $\vect{e}_0(O)\define \vect{u}_O$, and parallely transported along $\bar{\gamma}$ from $O$ to $S$. This procedure basically generates a family of fictive observers with four-velocities~$\vect{e}_0(v)$ along $\bar{\gamma}$, as illustrated in Fig.~\ref{fig:Fermat_converse}.

\newcommand{\localdeltaxO}{\delta x^\mu_O=\delta\tau u^\mu_O}

Suppose each of these observers selects an arbitrary \emph{spatial} position~$\{\delta x^a\}_{a=1\ldots 3}$ in his rest frame---i.e., with respect to the local tetrad $\{\vect{e}_\alpha(v)\}$---so that the $\delta x^a(v)$ are smooth, and consider the worldline~$\gamma$ of a particle that would interpolate all those spatial positions at the speed of light. By construction, $\gamma$ is therefore a null curve, and its deviation with respect to $\bar{\gamma}$ is
\begin{equation}
\delta x^\mu = \delta x^0 e_0^\mu + \delta x^a e_a^\mu,
\end{equation}
where the value of $\delta x^0$ is imposed $\delta x^a$ via the speed-of-light condition, i.e. $\bar{k}^\alpha \delta k_\alpha=0$ in tetrad components. When integrated from $S$ to $O$, this condition implies
\begin{equation}
\delta \tau = \delta x^0_O = \int_S^O \dd v \; \frac{\bar{k}_a \delta k^a}{\bar{k}_0} = - \int_S^O \ddf{}{v} \pa{ \frac{\bar{k}_a}{\bar{k}_0} } \delta x^a
\label{eq:Fermat_converse}
\end{equation}
after an integration by parts.

Now if $\delta \tau=0$ for any null curve $\gamma$ in the vicinity of $\bar{\gamma}$, then Eq.~\eqref{eq:Fermat_converse} implies
\begin{equation}
\ddf{}{v} \pa{ \frac{\bar{k}_a}{\bar{k}_0} } = 0,
\label{eq:geodesic_tetrad}
\end{equation}
because the $\delta x^a(v)$ were arbitrary fields. Besides, using $\bar{k}_\alpha\define \bar{k}_\mu e_\alpha^\mu$ and the fact that the tetrad field $\{\vect{e}_\alpha\}$ is parallely transported along $\bar{\gamma}$ ($\Dd \vect{e}_\alpha/\dd v=\vect{0}$), we can rewrite Eq.~\eqref{eq:geodesic_tetrad} as
\begin{equation}
e_a^\mu \pa{ \Ddf{\bar{k}_\mu}{v} - \frac{\bar{k}_\mu}{\bar{k}_0} \ddf{\bar{k}_0}{v} } = 0,
\end{equation}
so that the expression between parentheses vanishes, since its projection over $\vect{e}_0$ is also clearly zero. The null curve $\bar{\gamma}$ is therefore a geodesic as its tangent vector remains parallel to itself. One can also recover the affinely parametrised geodesic equation using e.g. the parameter $w$ such that $\dd v=\bar{k}_0 \dd w$ instead of $v$. 

\section{Observables of an electromagnetic wave}

So far, we have investigated the propagation of light from a four-dimensional, fully covariant, point of view. In this section, on the contrary, we examine the properties of electromagnetic waves which are actually observable, and therefore observer-dependent by definition. We chose to divide them into three categories: the kinematical observables (\S~\ref{sec:kinematical_observables}), associated with light's trajectory and frequency; the energetical observables (\S~\ref{sec:energetical_observables}); and finally polarisation (\S~\ref{sec:polarisation}).

\subsection{Kinematics}
\label{sec:kinematical_observables}

\subsubsection{The 3+1 decomposition of the wave vector}

In the rest frame of an observer with four-velocity $\vect{u}$, the electric field associated with a wave~$\vect{A}=\vect{a}\ex{\i\phi}+\cc$ oscillates with an angular frequency~$\omega$\index{angular frequency} defined as the (absolute value of the) rate of change of the phase~$\phi$. If $\tau$ denotes the proper time of this observer, we thus have
\begin{equation}
\omega\equiv \abs{ \ddf{\phi}{\tau} } = \abs{ u^\mu\partial_\mu\phi } = -u^\mu k_\mu,
\end{equation}
where the minus sign comes from the fact that we consider $\vect{k}$ as being future oriented, i.e. $k^0>0$ in any coordinate system. The frequency~$\nu$\index{frequency} is related to $\omega$ via the usual relation~$\omega=2\pi\nu$. 

Besides frequency, the wave is also characterised, in the frame of the observer, by its direction of propagation\index{direction of propagation}, that is the opposite of the direction in which the observer must look to actually detect the wave. It can be defined as being parallel to the spatial gradient of the phase,
\begin{equation}
d^\mu \propto \perp^{\mu\nu} \partial_\nu\phi \define \pa{g^{\mu\nu}+u^\mu u^\nu}k_\nu,
\label{eq:direction_propagation}
\end{equation}
where $\vect{\perp}$ denotes the projector on the observer's local space. In the above equation, we used a proportionality symbol~$\propto$ instead of an equality, because we also want $\vect{d}$ to be a unit vector;
\begin{wrapfigure}{l}{5cm}
\centering
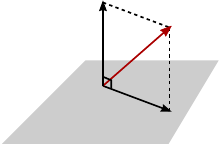
\caption{Decomposition of the wave four-vector.}
\label{fig:3+1_wavevector}
\end{wrapfigure}
it is actually easy to check that the norm of $\perp^\mu_\nu k^\nu$ is $\omega$, so that the right-hand side of Eq.~\eqref{eq:direction_propagation} must be divided by $\omega$.

The above definitions finally lead to the so-called 3+1 decomposition of the wave four-vector, illustrated in Fig.~\ref{fig:3+1_wavevector},
\begin{empheq}[box=\fbox]{equation}
\vect{k} = \omega \pa{ \vect{u} + \vect{d} } ,
\label{eq:3+1_wavevector}
\end{empheq}
with the orthonormality relations
\begin{equation}
u^\mu u_\mu =-1,
\qquad
d^\mu d_\mu = 1,
\qquad
u^\mu d_\mu = 0.
\end{equation}

\subsubsection{Physical meaning of the affine parameter\index{affine parameter!physical interpretation}}\label{sec:physical_interpretation_affine_parameter}

The 3+1 decomposition~\eqref{eq:3+1_wavevector} of the wave four-vector with respect to a given observer provides a simple physical interpretation for the affine parameter~$v$. Between $v$ and $v+\dd v$, the displacement of a photon through spacetime is given by $\dd x^\mu=k^\mu \dd v$. Hence, in the frame of the observer, the photon travels over a distance
\begin{equation}
\dd\ell \define d_\mu \dd x^\mu = d_\mu k^\mu \dd v = \omega \dd v,
\end{equation}
which gives to $\dd v$ its meaning. Equivalently, $\omega \dd v$ can be interpreted as the time lapse~$\dd\tau$ measured by the observer between the events~$x^\mu(v)$ and $x^\mu(v+\dd v)$, but it is somehow harder to visualise than the corresponding travel distance.

\subsubsection{The redshift and its interpretation}\index{redshift!definition}

The relative difference between the frequency emitted by a source, $\omega_S$, and the one actually measured by an observer, $\omega_O$, is quantified by the redshift~$z$ as
\begin{empheq}[box=\fbox]{equation}
1+z \equiv \frac{\omega_S}{\omega_O} = \frac{(u^\mu k_\mu)_S}{(u^\mu k_\mu)_O}.
\end{empheq}

Because the expression of $z$ involves both the four-velocities and the wave four-vectors at the source and observation events (respectively $S$ and $O$), there is a fundamental degeneracy in its interpretation---is it a Doppler effect due to their relative velocity, or an Einstein effect due to, e.g., gravitational dilation of time?\index{redshift!interpretation}

\begin{figure}[h!]
\centering
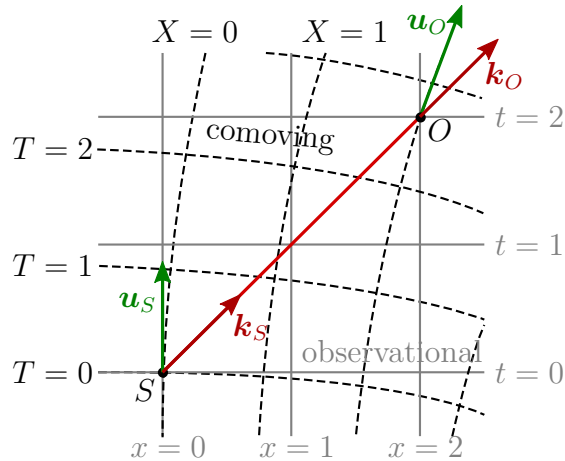
\caption{Illustration of the coordinate dependence of the interpretation of the redshift. With comoving coordinates $(T,X)$---black, dashed---only the components of the wave four-vector differ between $S$ and $O$: $z$ is interpreted as a gravitational redshift. With observational coordinates $(t,x)$---grey, solid---only the components of the four-velocity of the source and the observer differ: $z$ is interpreted as a Doppler effect.}
\label{fig:interpretation_redshift}
\end{figure}

This turns out to be a bad question in general, for its answer is \emph{coordinate dependent}. We are indeed always free to pick a \emph{comoving} coordinate system (see Fig.~\ref{fig:interpretation_redshift}), with respect to which both the source and the observer are at rest, $u^i_S \coordequal u^i_O \coordequal 0$, so that any difference between $\omega_S$ and $\omega_O$ would be due to $k_\mu^O\not\coordequal k_\mu^S$. In such a case, $z$ shall be interpreted as a gravitational redshift. Besides, we could also make the opposite choice and pick an \emph{observational} coordinate system, such that now $k^O_\alpha \coordequal k^S_\alpha$; any redshift would then be attributed to $u^\alpha_O\not\coordequal u^\alpha_S$, and interpreted as a Doppler effect.

The point I try to make here is that there is fundamentally no preferred interpretation for the redshift of an observed signal\footnote{See however Ref.~\cite{2014MNRAS.438.2456K} for a case against this conclusion.}. In a cosmological context~\cite{2009AmJPh..77..688B} for instance, it is \emph{not} more sensible to attribute the redshift of remote galaxies to a stretching of wavelengths, due to the expansion of the Universe, than to a Doppler effect due to their recession velocities. They are just two different points of view on a same physical phenomenon\footnote{This does not mean that the expansion of the Universe can be modelled by the motion of test particles in a Minkowski spacetime. See e.g. Ref.~\cite{2007AcA....57..139A} for a thought experiment illustrating the difference between both situations.}.

\subsubsection{Observational notion of relative velocity}

Pushing further the above reasoning, the redshift allows us to define a very general notion of velocity of a light source as seen by an observer. It is essentially the velocity~$\vec{V}$ that the observer would associate to the source by applying the special-relativistic Doppler formula
\begin{equation}
1+z =\frac{1-\vec{d}\cdot\vec{V}}{1-V^2},
\label{eq:SR_Doppler}
\end{equation}
where $-\vec{d}$ is an Euclidean unit vector materialising the observer's line of sight.

\begin{figure}[h!]
\centering
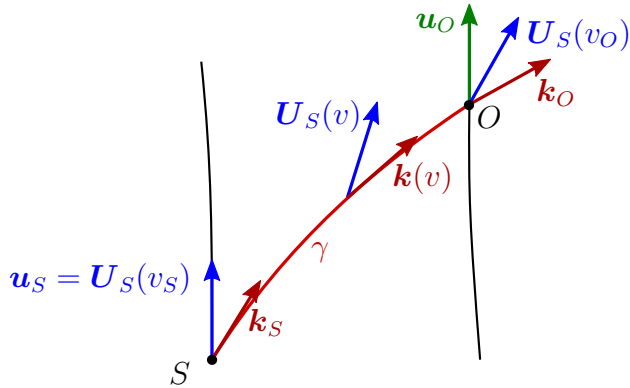
\caption{Defining an observational notion of relative velocity. The four-vector $\vect{U}_S(v)$ is constructed by parallel-transporting $\vect{u}_S$ from $S$ to $O$ along the null geodesic $\gamma$ connecting them. The resulting vector~$\vect{U}_S(v_O)$ at $O$ represents the four--velocity of the source as seen by the observer: the observed redshift is computed from it by applying the special-relativistic formula~\eqref{eq:SR_Doppler}.}
\label{fig:Doppler_velocity}
\end{figure}

It can seem very artificial, but there is actually a fully covariant way of constructing such this velocity (see Fig.~\ref{fig:Doppler_velocity}). Let $\gamma$ be a light ray connecting a source $S$ to an observer $O$, affinely parametrised by $v$, and define the four-vector~$\vect{U}_S(v)$ as the parallel transportation of the source's four-velocity~$\vect{u}_S$ from $S$ to $O$ along $\gamma$,
\begin{equation}
\Ddf{U^\mu_S}{v} = 0 \qquad \text{with} \qquad U_S^\mu(v_S)=u^\mu_S,
\label{eq:relative_four-velocity}
\end{equation}
Since both $\vect{U}_S^\mu$ and $\vect{k}$ are parallel transported along $\gamma$, their scalar product $U^\mu k_\mu$ is a constant along this geodesic; in particular $U^\mu_S(v_O) k_\mu^O=U^\mu_S(v_S) k_\mu^S=u^\mu_S k_\mu^S=-\omega_S$, the redshift can therefore be written as
\begin{equation}
1+z = \frac{U^\mu_S(v_O) k_\mu^O}{u^\mu_O k_\mu^O}.
\label{eq:SR_redshift_2}
\end{equation}
If we work at $O$ in the observer's frame, for which $u^\mu_O\coordequal(1,\vec{0})$, $k^\mu_O\coordequal \omega_O(1,\vec{d})$, then $U^\mu_S(v_O)\coordequal(1,\vec{V})/\sqrt{1-V^2}$ in general, so that Eq.~\eqref{eq:SR_redshift_2} is equivalent to Eq.~\eqref{eq:SR_Doppler}. This construction is orginally due to Synge~\cite{SyngeGRBook}.

Let us give two concrete examples of this \emph{Synge velocity}\index{Synge!velocity}. In homogeneous cosmology, where spacetime is described by the Friedmann-Lemaître geometry (see Chap.~\ref{chapter:standard_spacetimes}), the recession velocity of a comoving source with respect to a comoving observer reads $V=(a_O^2-a_S^2)/(a_O^2+a_S^2)$, where $a_E\equiv a(t_E)$ denotes the scale factor at the cosmic time~$t_E$ of the event $E$. Note that for $t_O\gtrsim t_S$ we recover the Hubble law $V\approx H_0(t_O-t_S)$.

In the vicinity of non-rotating body of mass $M$, where spacetime is described by the Schwarzschild geometry, the recession Doppler velocity between a source at $r_S$ and an observer at $r_O>r_S$ with the same angular coordinates is~$V=(A_S-A_O)/(A_S+A_O)$, where $A(r)\equiv1-2GM/r$. If $r_S\rightarrow 2GM$, i.e. if the massive body is a black hole whose horizon is approached by the source, then $V$ approaches the speed of light.

\subsection{Energetics and photon conservation}
\label{sec:energetical_observables}

As already mentioned in \S~\ref{sec:EM_curved_spacetime} and \S~\ref{sec:electric_magnetic}, any electromagnetic field possesses energy and momentum, whose various components are encoded in the decomposition~\eqref{eq:stress-energy_EM_electric_magnetic} of its stress-energy tensor. In the special case of an electromagnetic wave of the form~$\vect{A}=\vect{a}\ex{i\phi}+\cc$, the energy density~$\rho$ and the Poynting vector~$\vect{\Pi}$---which represents the energy flux density, or momentum density---read
\begin{align}
\rho &\define \frac{E^2+B^2}{8\pi} = \frac{\omega^2a^2\sin^2\phi}{\pi}, 
\label{eq:energy_density_EM_wave} \\
\Pi_\mu &\define \frac{u^\nu \eps_{\nu\mu\rho\sigma} E^\rho B^\sigma}{4\pi} = \rho \, d_\mu,
\end{align}
with $a^2\define a^\mu a_\mu\geq 0$, as ensured by the Lorenz gauge $k^\mu a_\mu=0$ together with the dispersion relation~$k^\mu k_\mu=0$, and $\vect{d}$ is the unit vector in the direction of propagation introduced in Eq.~\eqref{eq:3+1_wavevector}.

In practice, experiments are not sensitive to the detailed time evolution of such quantities, but rather to their average over a certain integration time $\Delta\tau\gg \omega^{-1}$. For instance, the luminous power~$P$ associated with an electromagnetic wave is the time average of the flux of the Poynting vector over the detector surface
\begin{equation}
P(\tau)
\define \iint\e{detect.} \dd^2\Sigma_\mu \int_\tau^{\tau+\Delta\tau}\dd t \; \Pi^\mu  
= \iint\e{detect.} \dd^2\Sigma_\mu d^\mu \mean{\rho},
\label{eq:luminous_power}
\end{equation}
where $\mean{\ldots}$ denotes (proper) time averaging. From Eq.~\eqref{eq:energy_density_EM_wave} we get $\mean{\rho}=\omega^2 a^2/(2\pi)$ for the mean energy density of the electromagnetic wave. Note that, by virtue of Eq.~\eqref{eq:luminous_power}, $\mean{\rho}$ also represent the power per unit area, that is, the luminous intensity~$I$ detected by the observer.\index{luminous intensity}

If we now adopt a corpuscular description of light, i.e. if we consider the electromagnetic wave as a photon stream, then the mean energy density can be written as $\mean{\rho}=n\times\hbar\omega$, where $n$ denotes the number of photons per unit volume within the wave. Similarly, because the Poynting vector is the momentum density of the wave, it reads~$\mean{\vect{\Pi}}=n\times\hbar\omega \vect{d}$ in terms of photons. These considerations naturally drive us to define the \emph{photon flux density}\index{photon!flux density} associated with the wave as
\begin{equation}
\vect{j} \equiv \frac{\mean{\rho} \vect{u} + \mean{\vect{\Pi}}}{\hbar\omega} = \frac{a^2}{2\pi \hbar} \, \vect{k} .
\end{equation}
Note that thanks to the above second equality, $\vect{j}$ is a real four-vector in the sense that it is \emph{observer-independent}, contrary to $\rho$, $\vect{\Pi}$, or the photon number density given by $n=-u^\mu j_\mu$.

Going back to the propagation equations for electromagnetic waves, the contraction of Eq.~\eqref{eq:conservation_photon_polarisation} with $\vect{a}$ immediately leads to a continuity equation
\begin{empheq}[box=\fbox]{equation}
\nabla_\mu j^\mu = 0,
\label{eq:photon_conservation}
\end{empheq}
which translates\index{photon!conservation} photon number conservation. Of course, this law is only valid in vacuum, or at least in the absence of coupling between matter and the electromagnetic field, since in the latter case photons can potentially be emitted of absorbed, which would generate a source term in Eq.~\eqref{eq:photon_conservation}.

\subsection{Polarisation}
\label{sec:polarisation}

\subsubsection{In terms of the vector potential}\index{polarisation!from the vector potential}

So far we have only exploited the information that Eq.~\eqref{eq:conservation_photon_polarisation} provides about the norm of the wave amplitude $\vect{a}$, but it actually also drives the evolution of its direction. Let us assume that $a^2 \not=0$ (else the wave has no energy, i.e. does not exist at all), and define the unit potential polarisation vector as
\begin{equation}
\vect{\alpha} \equiv \frac{\vect{a}}{\sqrt{a^2}}.
\end{equation}

Inserting $a^\mu = \sqrt{a^2}\alpha^\mu$ in Eq.~\eqref{eq:conservation_photon_polarisation}, we get
\begin{align}
0 
= \sqrt{a^2} \, k^\nu\nabla_\nu\alpha^\mu + \frac{\alpha^\mu}{2\sqrt{a^2}} \, \nabla_\nu (a^2k^\nu).
\end{align}
In the second term we recognise the divergence of the photon flux density $2\pi\hbar\nabla_\nu j^\nu$, which vanishes by virtue of photon conservation as seen in the previous paragraph, so that finally
\begin{empheq}[box=\fbox]{equation}
\Ddf{\vect{\alpha}}{v} = 0.
\label{eq:propagation_polarisation_potential}
\end{empheq}
The potential polarisation vector is thus parallely transported along the light ray.

\subsubsection{In terms of the electric field}\index{polarisation!from the electric field}

Albeit very convenient for a covariant description of polarisation, the vector potential is not directly measurable. In practice, polarisation is rather characterised by the behaviour of the electric field, but since the latter is observer-dependent, we can already suspect that parallel transportation, satisfied for $\vect{\alpha}$, will not hold here.
Defining the electric field all along the wave's worldline~$\gamma$ indeed requires to define a family of observers, i.e. a field of four-velocities~$\vect{u}$, which has no reason to be parallel-transported along $\gamma$. This issue becomes evident when one calculates the covariant derivative of $E^\mu=2\i u_\nu k^{[\mu}A^{\nu]}+\cc$ along $\gamma$,
\begin{equation}
\Ddf{E^\mu}{v} = -\frac{1}{2}(\nabla_\nu k^\nu) E^\mu + \pac{ 2\i k^{[\mu}A^{\nu]} \Ddf{u_\nu}{v} +\cc },
\label{eq:covariant_derivative_electric_field}
\end{equation}
where we see that the second term, which prevents $\Dd \vect{E}/\dd v$ from being proportional to $\vect{E}$, and thus prevents the electric polarisation vector
\begin{equation}
\vect{\epsilon}\define\frac{\vect{E}}{\sqrt{E^2}}
\end{equation}
from being parallely transported, precisely encodes the deviation from parallel transportation of $\vect{u}$ along $\gamma$.

Yet the electric polarisation partially inherits the propagation properties of $\vect{\alpha}$; it is actually parallel-transported \emph{as much as possible}, while keeping orthogonal to both $\vect{u}$ and~$\vect{d}$. Such a notion can be formalised by defining the projector on the two-plane orthogonal to those vectors,
\begin{equation}
S^\mu_\nu \define \delta^\mu_\nu + u^\mu u_\nu - d^\mu d_\nu,
\end{equation}
that we shall call \emph{screen projector}, since the plane it projects on materialises a spatial ($\perp \vect{u}$) screen, orthogonal to the observer's line of sight~$\vect{d}$. The electric polarisation then reads
\begin{empheq}[box=\fbox]{equation}
S^\mu_\nu \Ddf{\epsilon^\nu}{v} = 0.
\label{eq:transport_electric_polarisation}
\end{empheq}

Let us now prove Eq.~\eqref{eq:transport_electric_polarisation}. First, because $u^\mu E_\mu=0=k^\mu E_\mu$, we also have $d^\mu E_\mu=0$, so that $E^\mu=S^\mu_\nu E^\nu$ (the electric field belongs to the screen plane). Plugging-in the relation between $E^\mu$ and $A^\mu$ yields
\begin{equation}
E^\rho = S^\rho_\mu \, 2\i u_\nu k^{[\mu}A^{\nu]}+\cc = \i \omega S^\rho_\mu A^\mu +\cc,
\end{equation}
where we used $S^\mu_\nu k^\nu=0$. The screen projection Eq.~\eqref{eq:covariant_derivative_electric_field} thus reads
\begin{equation}
S^\rho_\mu \Ddf{E^\mu}{v} = \pa{ - \frac{1}{2}\nabla_\nu k^\nu + \frac{1}{\omega} \ddf{\omega}{v} } E^\rho.
\end{equation}
The above equation can then be contracted with $E_\rho$ in order to get an equation governing~$E^2$; combining the latter with the former finally gives Eq.~\eqref{eq:transport_electric_polarisation}. Note that the same calculations could have been done with the magnetic field instead of the electric field.

%% file: Fermat.pdf_tex
\begingroup%
  \makeatletter%
  \providecommand\color[2][]{%
    \errmessage{(Inkscape) Color is used for the text in Inkscape, but the package 'color.sty' is not loaded}%
    \renewcommand\color[2][]{}%
  }%
  \providecommand\transparent[1]{%
    \errmessage{(Inkscape) Transparency is used (non-zero) for the text in Inkscape, but the package 'transparent.sty' is not loaded}%
    \renewcommand\transparent[1]{}%
  }%
  \providecommand\rotatebox[2]{#2}%
  \ifx\svgwidth\undefined%
    \setlength{\unitlength}{128.95866699bp}%
    \ifx\svgscale\undefined%
      \relax%
    \else%
      \setlength{\unitlength}{\unitlength * \real{\svgscale}}%
    \fi%
  \else%
    \setlength{\unitlength}{\svgwidth}%
  \fi%
  \global\let\svgwidth\undefined%
  \global\let\svgscale\undefined%
  \makeatother%
  \begin{picture}(1,0.68122602)%
    \put(0,0){\includegraphics[width=\unitlength]{Fermat.pdf}}%
    \put(0.57830216,0.42787412){\color[rgb]{0,0,0}\makebox(0,0)[lb]{\smash{$O$}}}%
    \put(-0.00392568,0.00658558){\color[rgb]{0,0,0}\makebox(0,0)[lb]{\smash{$S$}}}%
    \put(0.53299501,0.03190862){\color[rgb]{0,0,0}\makebox(0,0)[lb]{\smash{$\wl$}}}%
    \put(0.65261398,0.56082536){\color[rgb]{0,0,0}\makebox(0,0)[lb]{\smash{$\delta\tau$}}}%
    \put(0.3182579,0.17170404){\color[rgb]{0,0,0}\makebox(0,0)[lb]{\smash{$\bar{\gamma}$}}}%
    \put(0.21881576,0.37635302){\color[rgb]{0,0,0}\makebox(0,0)[lb]{\smash{$\gamma$}}}%
  \end{picture}%
\endgroup%

%% file: Fermat_converse.pdf_tex
\begingroup%
  \makeatletter%
  \providecommand\color[2][]{%
    \errmessage{(Inkscape) Color is used for the text in Inkscape, but the package 'color.sty' is not loaded}%
    \renewcommand\color[2][]{}%
  }%
  \providecommand\transparent[1]{%
    \errmessage{(Inkscape) Transparency is used (non-zero) for the text in Inkscape, but the package 'transparent.sty' is not loaded}%
    \renewcommand\transparent[1]{}%
  }%
  \providecommand\rotatebox[2]{#2}%
  \ifx\svgwidth\undefined%
    \setlength{\unitlength}{382.31365967bp}%
    \ifx\svgscale\undefined%
      \relax%
    \else%
      \setlength{\unitlength}{\unitlength * \real{\svgscale}}%
    \fi%
  \else%
    \setlength{\unitlength}{\svgwidth}%
  \fi%
  \global\let\svgwidth\undefined%
  \global\let\svgscale\undefined%
  \makeatother%
  \begin{picture}(1,0.64857436)%
    \put(0,0){\includegraphics[width=\unitlength]{Fermat_converse.pdf}}%
    \put(-0.00132417,0.0019127){\color[rgb]{0,0,0}\makebox(0,0)[lb]{\smash{$S$}}}%
    \put(0.7775056,0.05575607){\color[rgb]{0,0,0}\makebox(0,0)[lb]{\smash{$\wl$}}}%
    \put(0.30515131,0.18591573){\color[rgb]{0,0,1}\makebox(0,0)[lb]{\smash{$\delta x^\mu(v_1)$}}}%
    \put(0.41958608,0.30315844){\color[rgb]{0,0,1}\makebox(0,0)[lb]{\smash{$\delta x^\mu(v_2)$}}}%
    \put(0.42319697,0.38174745){\color[rgb]{0,0.50196078,0}\makebox(0,0)[lb]{\smash{$\delta x^a(v_2)$}}}%
    \put(0.24416099,0.22824283){\color[rgb]{0,0.50196078,0}\makebox(0,0)[lb]{\smash{$\delta x^a(v_1)$}}}%
    \put(0.82307826,0.50556166){\color[rgb]{0,0,1}\makebox(0,0)[lb]{\smash{$\localdeltaxO$}}}%
    \put(0.68474404,0.37517588){\color[rgb]{0.83137255,0,0}\makebox(0,0)[lb]{\smash{$\bar{\gamma}$}}}%
    \put(0.60087052,0.4934635){\color[rgb]{0.83137255,0,0}\makebox(0,0)[lb]{\smash{$\gamma$}}}%
    \put(0.64005749,0.59992238){\color[rgb]{0,0,0}\makebox(0,0)[lb]{\smash{$\vect{e}_0(O)\define \vect{u}_O$}}}%
    \put(0.45810994,0.46613248){\color[rgb]{0,0,0}\makebox(0,0)[lb]{\smash{$\vect{e}_0(v_2)$}}}%
    \put(0.22894378,0.28212963){\color[rgb]{0,0,0}\makebox(0,0)[lb]{\smash{$\vect{e}_0(v_1)$}}}%
    \put(0.24255724,0.11126961){\color[rgb]{0.83137255,0,0}\makebox(0,0)[lb]{\smash{$\bar{x}^\mu(v_1)$}}}%
    \put(0.56061954,0.29264397){\color[rgb]{0.83137255,0,0}\makebox(0,0)[lb]{\smash{$\bar{x}^\mu(v_2)$}}}%
    \put(0.80960317,0.45534866){\color[rgb]{0,0,0}\makebox(0,0)[lb]{\smash{$O$}}}%
    \put(0.11237223,0.17632786){\color[rgb]{0,0,0}\makebox(0,0)[lb]{\smash{$\delta x^0(v_1)$}}}%
    \put(0.57250282,0.374788){\color[rgb]{0,0,0}\makebox(0,0)[lb]{\smash{$\delta x^0(v_2)$}}}%
  \end{picture}%
\endgroup%

%% file: 3+1_wavevector.pdf_tex
\begingroup%
  \makeatletter%
  \providecommand\color[2][]{%
    \errmessage{(Inkscape) Color is used for the text in Inkscape, but the package 'color.sty' is not loaded}%
    \renewcommand\color[2][]{}%
  }%
  \providecommand\transparent[1]{%
    \errmessage{(Inkscape) Transparency is used (non-zero) for the text in Inkscape, but the package 'transparent.sty' is not loaded}%
    \renewcommand\transparent[1]{}%
  }%
  \providecommand\rotatebox[2]{#2}%
  \ifx\svgwidth\undefined%
    \setlength{\unitlength}{105.675bp}%
    \ifx\svgscale\undefined%
      \relax%
    \else%
      \setlength{\unitlength}{\unitlength * \real{\svgscale}}%
    \fi%
  \else%
    \setlength{\unitlength}{\svgwidth}%
  \fi%
  \global\let\svgwidth\undefined%
  \global\let\svgscale\undefined%
  \makeatother%
  \begin{picture}(1,0.6566657)%
    \put(0,0){\includegraphics[width=\unitlength]{3+1_wavevector.pdf}}%
    \put(0.03849481,0.10753928){\color[rgb]{0,0,0}\rotatebox{42.87250047}{\makebox(0,0)[lb]{\smash{space}}}}%
    \put(0.79400307,0.53599119){\color[rgb]{0.66666667,0,0}\makebox(0,0)[lb]{\smash{$k^\mu$}}}%
    \put(0.23387644,0.56999784){\color[rgb]{0,0,0}\makebox(0,0)[lb]{\smash{$\omega u^\mu$}}}%
    \put(0.5530056,0.05640451){\color[rgb]{0,0,0}\makebox(0,0)[lb]{\smash{$\omega d^\mu$}}}%
    \put(0.37341514,0.16982022){\color[rgb]{0,0,0}\makebox(0,0)[lb]{\smash{$O$}}}%
  \end{picture}%
\endgroup%

%% file: interpretation_redshift.pdf_tex
\begingroup%
  \makeatletter%
  \providecommand\color[2][]{%
    \errmessage{(Inkscape) Color is used for the text in Inkscape, but the package 'color.sty' is not loaded}%
    \renewcommand\color[2][]{}%
  }%
  \providecommand\transparent[1]{%
    \errmessage{(Inkscape) Transparency is used (non-zero) for the text in Inkscape, but the package 'transparent.sty' is not loaded}%
    \renewcommand\transparent[1]{}%
  }%
  \providecommand\rotatebox[2]{#2}%
  \ifx\svgwidth\undefined%
    \setlength{\unitlength}{229.16826172bp}%
    \ifx\svgscale\undefined%
      \relax%
    \else%
      \setlength{\unitlength}{\unitlength * \real{\svgscale}}%
    \fi%
  \else%
    \setlength{\unitlength}{\svgwidth}%
  \fi%
  \global\let\svgwidth\undefined%
  \global\let\svgscale\undefined%
  \makeatother%
  \begin{picture}(1,0.74516013)%
    \put(0,0){\includegraphics[width=\unitlength]{interpretation_redshift.pdf}}%
    \put(0.19601156,0.09645636){\color[rgb]{0,0,0}\makebox(0,0)[lb]{\smash{$S$}}}%
    \put(0.67315952,0.52226185){\color[rgb]{0,0,0}\makebox(0,0)[lb]{\smash{$O$}}}%
    \put(0.63941828,0.70555418){\color[rgb]{0,0.50196078,0}\makebox(0,0)[lb]{\smash{$\vect{u}_O$}}}%
    \put(0.17020737,0.24996303){\color[rgb]{0,0.50196078,0}\makebox(0,0)[lb]{\smash{$\vect{u}_S$}}}%
    \put(0.31197053,0.52682363){\color[rgb]{0,0,0}\rotatebox{-5.68419439}{\makebox(0,0)[lb]{\smash{comoving}}}}%
    \put(0.46906037,0.1602807){\color[rgb]{0.50196078,0.50196078,0.50196078}\makebox(0,0)[lb]{\smash{observational}}}%
    \put(-0.00220908,0.12537185){\color[rgb]{0,0,0}\makebox(0,0)[lb]{\smash{$T=0$}}}%
    \put(-0.00220908,0.49191476){\color[rgb]{0,0,0}\makebox(0,0)[lb]{\smash{$T=2$}}}%
    \put(-0.00220908,0.29991609){\color[rgb]{0,0,0}\makebox(0,0)[lb]{\smash{$T=1$}}}%
    \put(0.22469844,0.68391342){\color[rgb]{0,0,0}\makebox(0,0)[lb]{\smash{$X=0$}}}%
    \put(0.46906037,0.68391342){\color[rgb]{0,0,0}\makebox(0,0)[lb]{\smash{$X=1$}}}%
    \put(0.18978959,0.00319089){\color[rgb]{0.50196078,0.50196078,0.50196078}\makebox(0,0)[lb]{\smash{$x=0$}}}%
    \put(0.39924268,0.00319089){\color[rgb]{0.50196078,0.50196078,0.50196078}\makebox(0,0)[lb]{\smash{$x=1$}}}%
    \put(0.60869576,0.00319089){\color[rgb]{0.50196078,0.50196078,0.50196078}\makebox(0,0)[lb]{\smash{$x=2$}}}%
    \put(0.78324,0.12537185){\color[rgb]{0.50196078,0.50196078,0.50196078}\makebox(0,0)[lb]{\smash{$t=0$}}}%
    \put(0.78324,0.33482494){\color[rgb]{0.50196078,0.50196078,0.50196078}\makebox(0,0)[lb]{\smash{$t=1$}}}%
    \put(0.78324,0.54427803){\color[rgb]{0.50196078,0.50196078,0.50196078}\makebox(0,0)[lb]{\smash{$t=2$}}}%
    \put(0.35664755,0.20287572){\color[rgb]{0.66666667,0,0}\makebox(0,0)[lb]{\smash{$\vect{k}_S$}}}%
    \put(0.76647208,0.61549121){\color[rgb]{0.66666667,0,0}\makebox(0,0)[lb]{\smash{$\vect{k}_O$}}}%
  \end{picture}%
\endgroup%

%% file: Doppler_velocity.pdf_tex
\begingroup%
  \makeatletter%
  \providecommand\color[2][]{%
    \errmessage{(Inkscape) Color is used for the text in Inkscape, but the package 'color.sty' is not loaded}%
    \renewcommand\color[2][]{}%
  }%
  \providecommand\transparent[1]{%
    \errmessage{(Inkscape) Transparency is used (non-zero) for the text in Inkscape, but the package 'transparent.sty' is not loaded}%
    \renewcommand\transparent[1]{}%
  }%
  \providecommand\rotatebox[2]{#2}%
  \ifx\svgwidth\undefined%
    \setlength{\unitlength}{269.53786621bp}%
    \ifx\svgscale\undefined%
      \relax%
    \else%
      \setlength{\unitlength}{\unitlength * \real{\svgscale}}%
    \fi%
  \else%
    \setlength{\unitlength}{\svgwidth}%
  \fi%
  \global\let\svgwidth\undefined%
  \global\let\svgscale\undefined%
  \makeatother%
  \begin{picture}(1,0.53266654)%
    \put(0,0){\includegraphics[width=\unitlength]{Doppler_velocity.pdf}}%
    \put(0.21920721,0.00271298){\color[rgb]{0,0,0}\makebox(0,0)[lb]{\smash{$S$}}}%
    \put(0.64502139,0.35882362){\color[rgb]{0,0,0}\makebox(0,0)[lb]{\smash{$O$}}}%
    \put(0.56011904,0.49918584){\color[rgb]{0,0.50196078,0}\makebox(0,0)[lb]{\smash{$\vect{u}_O$}}}%
    \put(-0.00187821,0.13743974){\color[rgb]{0,0,1}\makebox(0,0)[lb]{\smash{$\vect{u}_S=\vect{U}_S(v_S)$}}}%
    \put(0.41593119,0.18017782){\color[rgb]{0.83137255,0,0}\makebox(0,0)[lb]{\smash{$\gamma$}}}%
    \put(0.71428527,0.47826516){\color[rgb]{0,0,1}\makebox(0,0)[lb]{\smash{$\vect{U}_S(v_O)$}}}%
    \put(0.3281933,0.075911){\color[rgb]{0.66666667,0,0}\makebox(0,0)[lb]{\smash{$\vect{k}_S$}}}%
    \put(0.72766339,0.39681091){\color[rgb]{0.66666667,0,0}\makebox(0,0)[lb]{\smash{$\vect{k}_O$}}}%
    \put(0.52827761,0.2787129){\color[rgb]{0.66666667,0,0}\makebox(0,0)[lb]{\smash{$\vect{k}(v)$}}}%
    \put(0.36630931,0.36549701){\color[rgb]{0,0,1}\makebox(0,0)[lb]{\smash{$\vect{U}_S(v)$}}}%
  \end{picture}%
\endgroup%

%% file: chapter_2.tex
\lettrine{O}{bservations} in astronomy and cosmology often rely on the measurement of the apparent size, shape, and luminosity of distant light sources. Such notions cannot be described from a single light ray, a single geodesic, but rather require a collection of rays---a light beam---which connect each point of the extended source to the observer. If the rays remain close enough to each other, then their collective behaviour can be studied as a whole, and the beam considered an object in itself. This chapter is dedicated to the laws governing the propagation of such narrow light beams, and their connection with observables. In particular, we introduce the two fundamental tools of the gravitational lensing theory, namely the Jacobi matrix and the optical scalars, which will be crucial for discussing the various observable notions of distance in Chap.~\ref{chapter:distances}.

\bigskip

\minitoc
\newpage

\section{Description of a light beam}

In this section, we show how the covariant description of light beams (\S~\ref{sec:covariant_description_light_beam}), which is indeed the most natural way they are defined, can be turned to a more observation-oriented description. This requires to introduce a notion of screen (\S~\ref{sec:screen_space}) on which observers can project the beam and characterise its shape and extension. We will show in particular that such morphological properties are actually observer independent. Their evolution with light propagation will finally be discussed in \S~\ref{sec:propagation_screen_space}.

\subsection{Covariant approach}\label{sec:covariant_description_light_beam}

We define a light beam\index{light beam!definition} as a set of light rays which \emph{all intersect at one event}, the lightcone of which they belong to. Depending on the physical situation that one wishes to address, this event can either represent light emission~$S$ from which the rays emerge, or light reception~$O$ towards which they converge. In this chapter, without loss of generality, we will consider the second option.

\subsubsection{Intrinsic coordinate system and basis}

\begin{figure}[h!]
\centering
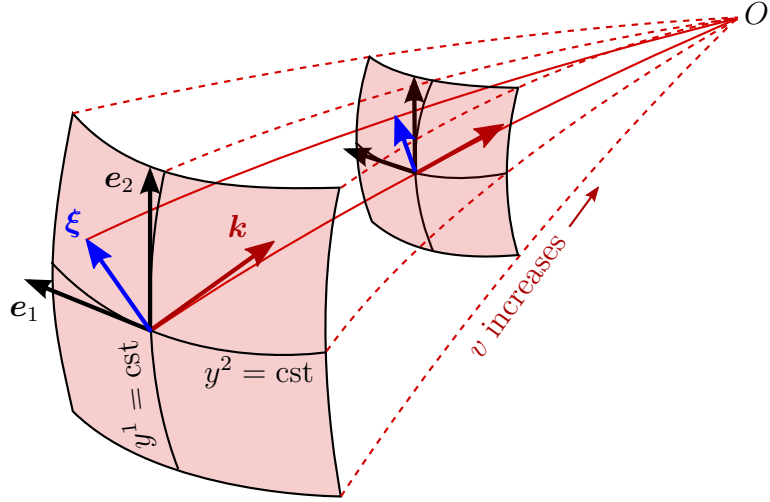
\caption{Schematic representation of a light beam converging at $O$, with its coordinate system~$(v,y^1,y^2)$ and the associated basis $(\vect{k},\vect{e}_1,\vect{e}_2)$. Red lines (solid and dashed) represent light rays within the beam; the separation~$\vect{\xi}$ between two of them is indicated in blue. Red surfaces are iso-$v$; horizontal or vertical black lines are respectively the intersection between iso-$y^1$ or iso-$y^2$ surfaces and iso-$v$ surfaces.}
\label{fig:beam}
\end{figure}

Geometrically speaking, a light beam is thus modelled by a bundle of null geodesic with a vertex point. Its structure therefore only requires three coordinates to be explored (see Fig.~\ref{fig:beam}). One is naturally chosen to be an affine parameter~$v$ along each individual geodesic within the bundle. The other two $(y^I)_{I=1,2}$ can be thought of as labels for the geodesics, such as angular coordinates on the observer's celestial sphere~\cite{1985PhR...124..315E}\index{light beam!intrinsic coordinates}. To this coordinate system is canonically associated a vector basis formed by
\begin{equation}
\vect{k} \define \pd{}{v}, \qquad \vect{e}_I \define \pd{}{y^I}.
\label{eq:beam_basis}
\end{equation}

Since all the geodesics of the bundle intersect at $O$, it is natural to affect to this event a common value~$v_\obs$ (often set to zero) of their affine parametrisation: $v=v_\obs\Leftrightarrow O$. This choice implies that the coordinate system~$(v,y^I)$ is singular at $O$---just like spherical coordinates are singular at their origin---in particular the vectors~$\vect{e}_I$ vanish at $O$. Consider indeed any two neighbouring rays $v\mapsto x^\mu(v,y^I)$ and $v\mapsto x^\mu(v,y^I+\delta y^I)$; because they intersect at $v_\obs$, we have
\begin{equation}
0 = x^\mu(v_\obs,y^I+\delta y^I) - x^\mu(v_\obs,y^I) = e^\mu_I(O) \delta y^I,
\end{equation}
which implies $e^\mu_I(O) = 0$ since $\delta y^I$ is arbitrary.

A second consequence of the existence of a vertex point in the geodesic bundle is the orthogonality of $\vect{k}$ and $\vect{e}_I$. Proving this property requires to note that $(\vect{k},\vect{e}_1,\vect{e}_2)$ is not any vector set but a \emph{coordinate basis} (or \emph{holonomous basis}), which means that the Lie brackets~$[\vect{k},\vect{e}_I]$ and $[\vect{e}_I,\vect{e}_J]$ must vanish~\cite{Baez:1995sj}. In terms of components, $[\vect{k},\vect{e}_I]=\vect{0}$ reads
\begin{equation}
0 = k^\nu \partial_\nu e^\mu_I - e_I^\nu \partial_\nu k^\mu = k^\nu \nabla_\nu e^\mu_I - e_I^\nu \nabla_\nu k^\mu;
\label{eq:commutation_k_e}
\end{equation}
it follows that
\begin{align}
\pd{}{v}(e_I^\mu k_\mu) 	&= k_\mu k^\nu \nabla_\nu e_I^\mu \qquad \text{since }k^\nu \nabla_\nu k_\mu=0 \\
											&= k_\mu  e_I^\nu \nabla_\nu k^\mu \qquad \text{because of Eq.~\eqref{eq:commutation_k_e}}\\
											&= \frac{1}{2} e_I^\nu \nabla_\nu (k_\mu k^\mu) \\
											&= 0 \qquad \text{as $\vect{k}$ is null.}
\end{align}
So the scalar product $e_I^\mu k_\mu$ does not depend on $v$, whence
\begin{equation}
e_I^\mu k_\mu = (e_I^\mu k_\mu)_O = 0.
\end{equation}
Physically speaking, this means that the phase~$\phi$ of the electromagnetic wave represented by the beam is the same for two events separated by $\dd x^\mu=e_I^\mu \dd y^I$, since for such a displacement $\dd\phi = k_\mu \dd x^\mu = 0$. The wave four-vector~$\vect{k}$ being also orthogonal to itself, we conclude that any displacement $\dd x^\mu = k^\mu\dd v + e^\mu_I \dd y^I$ within the beam keeps $\phi$ unchanged: the beam entirely belongs to a $\phi=\text{cst}$-hypersurface, which defines the lightcone of $O$.

\subsubsection{Separation vector}

The relative behaviour of two neighbouring rays, respectively labelled by $y^I$ and $y^I+\delta y^I$, is usually described by their \emph{separation vector} (or connecting vector)\index{separation vector}\index{connecting vector|see{separation vector}}
\begin{equation}\label{eq:separation_vector_definition}
\xi^\mu(v,y^I,y^I+\delta y^I) \define x^\mu(v,y^I+\delta y^I) - x^\mu(v,y^I) = e^\mu_I \delta y^I.
\end{equation}
As a linear combination of the $\vect{e}_I$, $\vect{\xi}$ inherits all their properties, namely
\begin{empheq}[box=\fbox]{align}
\xi^\mu_O &= 0, \\
k^\nu \nabla_\nu \xi^\mu &= \xi^\nu\nabla_\nu k^\mu, \label{eq:commutation_xi_k}\\
\xi^\mu k_\mu &= 0,\label{eq:orthogonality_xi_k}
\end{empheq}
which will turn out to be particularly useful for the remainder of this chapter.

\subsubsection{Geodesic deviation equation}

The fact that $\vect{\xi}$ is the separation between two geodesics imposes its evolution with~$v$. Taking the gradient of the geodesic equation in the direction of $\vect{\xi}$, we indeed have
\begin{align}
0 &= \xi^\rho \nabla_\rho (k^\nu \nabla_\nu k^\mu) \\
	&= (\xi^\rho \nabla_\rho k^\nu)(\nabla_\nu k^\mu) + \xi^\rho k^\nu \nabla_\rho \nabla_\nu k^\mu \\
	&= (k^\rho \nabla_\rho \xi^\nu)(\nabla_\nu k^\mu) + \xi^\rho k^\nu \nabla_\nu \nabla_\rho k^\mu 
																							- \xi^\rho k^\nu R\indices{^\mu_\sigma_\nu_\rho} k^\sigma 
			\label{eq:GDE_derivation_1}\\
	&= k^\rho \nabla_\rho(\xi^\nu\nabla_\nu k^\mu) - R\indices{^\mu_\sigma_\nu_\rho} k^\sigma k^\nu  \xi^\rho 
			\label{eq:GDE_derivation_2}
\end{align}
where, in Eq.~\eqref{eq:GDE_derivation_1}, we used Eq.~\eqref{eq:commutation_xi_k} to exchange the roles of $\vect{\xi}$ and $\vect{k}$ in the first term, and the definition of the Riemann tensor to exchange $\nabla_\rho$ and $\nabla_\nu$ in the second term. Using again Eq.~\eqref{eq:commutation_k_e}, we can rewrite the first term of Eq.~\eqref{eq:GDE_derivation_2} as $k^\rho \nabla_\rho k^\nu\nabla_\nu \xi^\mu$, which is nothing but the second covariant derivative of $\vect{\xi}$ with respect to $v$; the result is known as the \emph{geodesic deviation equation}\index{geodesic!deviation equation}
\begin{empheq}[box=\fbox]{equation}
\Ddf[2]{\xi^\mu}{v} = R\indices{^\mu_\sigma_\nu_\rho} k^\sigma k^\nu  \xi^\rho  .
\label{eq:GDE}
\end{empheq}
This equation surely provides the best geometrical interpretation of curvature, as the quantity which rules the relative acceleration between neighbouring geodesic motions. Note that Eq.~\eqref{eq:GDE} also applies to timelike and spacelike geodesics.

\subsection{Screen space}\label{sec:screen_space}

From now on, we restrict to \emph{infinitesimal light beams}, within which any two rays are close enough for their relative behaviour to be well-described by their separation vector~$\vect{\xi}$. This vector is then the key tool for the analysis of a beam's morphology, but its relation to actually observable quantities requires to introduce a notion of screen. 

\subsubsection{Defining a screen}\index{screen space}

Consider an observer with four-velocity $\vect{u}$ whose worldline intersects the beam at an event~$E$ different from $O$. Since in general $e^\mu_I(E)\not=0$, the beam has a nonzero extension around $E$. In order to characterise its morphology, the observer shall project it on a screen, defined as a two-dimensional spatial ($\perp \vect{u}$) plane, and chosen to be orthogonal to the local line-of-sight ($\perp \vect{d}$).

The relative position, on the screen, of the two light spots associated with two rays separated by $\xi^\mu$, is then
\begin{equation}
\xi^\mu_\perp \define S^\mu_\nu \xi^\nu,
\label{eq:separation_projected}
\end{equation}
where the screen projector is defined by
\begin{equation}
S^{\mu\nu} \define g^{\mu\nu} + u^\mu u^\nu - d^\mu d^\nu,
\end{equation}
and indeed coincides with the one introduced in \S~\ref{sec:polarisation}.

\subsubsection{The beam's morphology is frame independent}
\label{sec:beam_morphology_frame_independent}

\begin{figure}[h!]
\centering
\begin{minipage}{6cm}
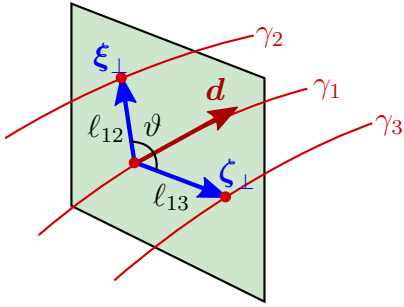
\end{minipage}
\hspace*{1cm}
\begin{minipage}{5cm}
\caption{Relative position, on an observer's screen, of the three light spots associated with the three rays $\gamma_{1,2,3}$.}
\label{fig:frame_independence}
\end{minipage}
\end{figure}

Consider a set of any three rays $\gamma_{1,2,3}$ within the beam, the last two being respectively separated from the first one by $\vect{\xi}$ and $\vect{\zeta}$, as depicted in Fig.~\ref{fig:frame_independence}. The scalar product of their screen projection is then
\begin{equation}
\xi^\mu_\perp \zeta_\mu^\perp = \ell_{12} \ell_{13} \cos\vartheta,
\end{equation}
where $\ell_{ij}$ stands for the (proper) distance between the spots $i$ and $j$, while $\vartheta$ is the angle between the segments relating $1$ to $2$ and $1$ to $3$.


The orthogonality between $\vect{\xi}$ and $\vect{k}$~\eqref{eq:orthogonality_xi_k} allows us to rewrite Eq.~\eqref{eq:separation_projected} as
\begin{equation}
\xi^\mu_\perp = \xi^\mu + \frac{\xi^\nu u_\nu}{\omega} k^\mu .
\end{equation}
and the analog relation for $\zeta_\perp^\mu$, so that
\begin{empheq}[box=\fbox]{equation}
\xi^\mu_\perp \zeta_\mu^\perp = \xi^\mu \zeta_\mu .
\end{empheq}
Therefore, the scalar product~$\xi^\mu_\perp \zeta_\mu^\perp$ does not depend on $\vect{u}$, and so do the lengths $\ell_{12}$, $\ell_{13}$ (the above rationale can be done with $\vect{\xi}=\vect{\zeta}$), and the angle $\vartheta$. We conclude that \emph{the morphology of an infinitesimal light beam is an intrinsic property, neither its size nor its shape depend on the frame in which they are measured.} Let us emphasize that this property is valid for measurements of lengths and angles on a screen by an experimentalist located at a point where the beam has a nonzero extension. It does \emph{not} apply to the observer situated at $O$, where the beam converges. In particular, the relativistic aberration effects that will be discussed in \S~\ref{sec:angular_diameter_distance} are precisely due to the fact that angles measured by an observer at $O$ depend on its four-velocity.

\subsubsection{The Sachs basis}\index{Sachs!basis}

For the purpose of characterising the morphology of a beam, it is convenient to introduce an orthonormal basis~$(\vect{s}_A)_{A=1,2}$ for the screen. By definition, its vectors must satisfy
\begin{equation}
s_A^\mu u_\mu = s_A^\mu d_\mu = 0,
\qquad
s_A^\mu s_{B\mu} = \delta_{AB},
\label{eq:Sachs_orthogonality}
\end{equation}
which also implies $s_A^\mu k_\mu=0$ since $k^\mu=\omega(u^\mu+d^\mu)$. The screen projector is naturally decomposed in terms of this basis as
\begin{equation}
S^{\mu\nu} = \delta^{AB} s_A^\mu s_B^\nu.
\end{equation}

Suppose that a family of observers, lying all along the beam and thus defining a four-velocity field $\vect{u}(v)$, want to compare the patterns they observe on their respective screen. Each of them can define a basis $(\vect{s}_A)_{A=1,2}(v)$ according to the requirements~\eqref{eq:Sachs_orthogonality}, but these requirements do not specify any orientation---the bases of two different observers could be arbitrarily rotated with respect to each other.

In order for the orientation of the patterns on the screens to be tractable, the screen vectors~$\vect{s}_A(v)$ should be parallely transported along the beam. However, just like for polarisation (\S~\ref{sec:polarisation}), a complete parallel transportation is forbidden if $\vect{u}(v)$ is not, because $\vect{s}_A$ and $\vect{u}$ must keep orthogonal to each other. The solution is again a partial parallel transportation, in the sense of
\begin{equation}
S^\mu_\nu \Ddf{s^\nu_A}{v} = 0.
\label{eq:Sachs_transport}
\end{equation}
Vector~$\dot{\vect{s}}_A$, where a dot denotes the covariant derivative $\Dd/\dd v$, can thus be decomposed as $\dot{\vect{s}}_A=\alpha \vect{u}+\beta\vect{d}$. Furthermore, since $k_\mu s^\mu_A=0$ and $\dot{\vect{k}}=\vect{0}$ (geodesic equation), we also have $k_\mu \dot{s}_A^\mu=0$, which implies $\alpha=\beta$; in other words,
\begin{equation}
\dot{\vect{s}}_A \propto \vect{k}.
\end{equation}
The above relation, together with $\dot{\vect{k}}=\vect{0}$, then implies that any $n$th-order covariant derivative of $\vect{s}_A$ along $\vect{k}$ is parallel to $\vect{k}$.

In many references on gravitational lensing theory (e.g. Refs.~\cite{1992grle.book.....S,2004LRR.....7....9P}), the authors take advantage of observer independence of the beam's morphology and assume, without loss of generality, that $\vect{u}(v)$ is actually parallely transported along the beam, hence so are the $\vect{s}_A(v)$. With this additional requirement, $(\vect{s}_1,\vect{s}_2)$ is called the \emph{Sachs basis}. We choose here not to make this assumption and work with a generic $\vect{u}(v)$ field. We will however use the same terminology for simplicity.

An advantage of our choice is that our Sachs basis enjoys a simple physical interpretation. The transportation law~\eqref{eq:Sachs_transport} is indeed reminiscent of the electric polarisation vector~$\vect{\epsilon}$ defined in \S~\ref{sec:polarisation}, and we can make the following identification:
\begin{equation}
\vect{s}_1 = \vect{\epsilon}, \qquad \vect{s}_2 = \vect{\beta},
\label{eq:Sachs=polarisation}
\end{equation}
where $\vect{\beta}$ is the magnetic analog of $\vect{\epsilon}$; the Sachs vectors then indicate the directions of oscillation of the electric and magnetic fields of a given photon within the beam. Of course, any global isometry between $(\vect{s}_1,\vect{s}_2)$ and $(\vect{\epsilon},\vect{\beta})$ is allowed as well. Note also that the identification~\eqref{eq:Sachs=polarisation} does not hold if other physical phenomena besides gravitation affect polarisation, such as optically active matter or Faraday rotation in the presence of strong background magnetic fields.

\subsection{Propagation in screen space}\label{sec:propagation_screen_space}

We now want to investigate the consequences of the geodesic deviation equation~\eqref{eq:GDE} on the evolution, with $v$, of the beam's morphology. For that purpose, it is natural to focus on the components of $\vect{\xi}$ on the Sachs basis, defined by
\begin{equation}
\xi_A \define s_A^\mu \xi_\mu = s_A^\mu \xi^\perp_\mu,
\end{equation}
and therefore such that
\begin{equation}
\vect{\xi}_\perp = \xi^A \vect{s}_A.
\end{equation}
Note that the position of indices $A,B,C\ldots$ does not matter (here $\xi^A=\xi_A$) as they are raised and lowered by $\delta_{AB}$.

\subsubsection{The Sachs vector equation}\index{Sachs!equation (vector)}

An evolution equation for $\xi_A$, directly inherited from the geodesic deviation equation, can be derived the following way. We first decompose its second derivative with respect to $v$ thanks to the Leibnitz rule,
\begin{equation}
\ddot{\xi}_A = \ddot{\xi}_\mu s_A^\mu + 2 \dot{\xi}_\mu \dot{s}_A^\mu + \xi_\mu \ddot{s}_A^\mu,
\label{eq:Sachs_equation_derivation}
\end{equation}
where, again, a dot denotes a covariant derivative $\Dd/\dd v$. We have seen in the previous paragraph that both $\dot{\vect{s}}_A$ and $\ddot{\vect{s}}_A$ are parallel to $\vect{k}$; the orthogonality of $\vect{k}$ and $\vect{\xi}$ then implies that the last two terms in the right-hand side of Eq.~\eqref{eq:Sachs_equation_derivation} vanish identically. Inserting Eq.~\eqref{eq:GDE} in the first one therefore leads to
\begin{equation}
\ddot{\xi}_A = R_{\mu\nu\rho\sigma} s_A^\mu k^\nu k^\rho \xi^\sigma.
\end{equation}
Finally, using $\xi^\sigma=\xi_\perp^\sigma + (\xi^\nu u_\nu/\omega) k^\sigma$, and the antisymmetry of the Riemann tensor under $\rho\leftrightarrow\sigma$, we obtain the \emph{Sachs vector equation}
\begin{empheq}[box=\fbox]{equation}
\ddf[2]{\xi^A}{v} = \tidal{^A_B} \xi^B,
\label{eq:Sachs_vector_equation}
\end{empheq}
in which the left-hand side involves a simple (not covariant) derivative, because $\xi_A$ is actually a scalar, and where we introduced the \emph{optical tidal matrix}\index{optical tidal matrix}
\begin{equation}
\tidal_{AB} \define R_{\mu\nu\rho\sigma} s_A^\mu k^\nu k^\rho s_B^\sigma.
\label{eq:optical_tidal_matrix_def}
\end{equation}
Because the Riemann tensor is invariant under the exchange of the first pair of indices with second one, the optical tidal matrix is \emph{symmetric} $\tidal_{AB}=\tidal_{BA}$, which justifies a posteriori the notation $\tidal^A_B$ in Eq.~\eqref{eq:Sachs_vector_equation}.

\subsubsection{Ricci and Weyl lensing}\index{Ricci!lensing}\index{Weyl!lensing}\label{sec:Ricci_Weyl_lensing}

As any $2\times 2$ matrix, $\vect{\tidal}$ can be decomposed into a pure-trace part and a trace-free part. This decomposition turns out to fit very well with the decomposition of the Riemann tensor as\index{Weyl!tensor}
\begin{equation}
R_{\mu\nu\rho\sigma} = R_{\mu[\rho} g_{\sigma]\nu} 
										- R_{\nu[\rho} g_{\sigma]\mu} 
										- \frac{1}{3} R \, g_{\mu[\rho} g_{\sigma]\nu}
										+ C_{\mu\nu\rho\sigma},
\end{equation}
where $C_{\mu\nu\rho\sigma}$ denotes the Weyl (or conformal curvature) tensor; it has the same symmetries as the Riemann tensor, and it is trace free, in the sense that $C\indices{^\mu_\nu_\mu_\sigma}=0$. Inserting this decomposition in the definition~\eqref{eq:optical_tidal_matrix_def}, and using the orthogonality relations involving $k^\mu$, $s_A^\mu$, we get
\begin{equation}
\tidal_{AB} = - \frac{1}{2} R_{\mu\nu} k^\mu k^\nu \, \delta_{AB} + C_{\mu\nu\rho\sigma} s_A^\mu k^\nu k^\rho s_B^\sigma ,
\label{eq:optical_tidal_matrix_Ricci-Weyl_1}
\end{equation}
whose last term is a trace-free matrix, indeed
\begin{align}
\delta^{AB} C_{\mu\nu\rho\sigma} s_A^\mu k^\nu k^\rho s_B^\sigma 
&= C_{\mu\nu\rho\sigma} k^\nu k^\rho S^{\mu\sigma}  \\
&= C_{\mu\nu\rho\sigma} k^\nu k^\rho (u^\mu u^\sigma - d^\mu d^\sigma) \\
&= \omega^2 ( C_{\mu\nu\rho\sigma} d^\nu d^\rho u^\mu u^\sigma - C_{\mu\nu\rho\sigma} u^\nu u^\rho d^\mu d^\sigma) \\
&= 0,
\end{align}
where we used that the Weyl tensor is trace free in the second line, and its symmetries in the third and fourth lines.

The optical tidal matrix therefore takes the simple form
\begin{equation}\label{eq:optical_tidal_matrix_Ricci-Weyl_2}
\vect{\tidal} =
\begin{pmatrix}
\Ricfoc & 0 \\ 0 & \Ricfoc
\end{pmatrix}
+
\begin{pmatrix}
- {\rm Re}\,\Weylfoc & {\rm Im}\,\Weylfoc \\
  {\rm Im}\,\Weylfoc & {\rm Re}\,\Weylfoc
\end{pmatrix}
\end{equation}
which involves the Ricci lensing and Weyl lensing scalars, respectively defined by
\begin{align}\label{eq:Ricci-Weyl}
\Ricfoc &\define -\frac12 R_{\mu\nu} k^\mu k^\nu, \\
\Weylfoc &\define -\frac{1}{2} C_{\mu\nu\rho\sigma}
(s_1^{\mu} - {\rm i} s_2^{\mu}) k^\nu k^\rho (s_1^{\sigma} - {\rm i} s_2^{\sigma}).
\end{align}

\subsubsection{Physical interpretation}\label{sec:interpretation_Ricci_Weyl}

The decomposition~\eqref{eq:optical_tidal_matrix_Ricci-Weyl_2} of $\vect{\tidal}$ is useful to qualitatively understand the physics of gravitational lensing. It is indeed clear that Ricci and Weyl curvatures affect the beam's geometry in different ways:
\begin{itemize}
\item Suppose that only $\Ricfoc$ is at work, then the separation~$\xi_A$ between any two light spots on a screen evolves as $\ddot{\xi}_A=\Ricfoc \xi_A$ as the beam propagates. Ricci lensing thus induces a \emph{homothetic transformation} of the beam's pattern.
\item The Weyl lensing matrix, on the contrary, has two different eigenvalues $\mp |\Weylfoc|$, whose eigendirections are respectively rotated by $\beta$ and $\beta+\pi/2$, with $\Weylfoc=|\Weylfoc|\ex{-2\i\beta}$, with respect to the Sachs basis. The separation vectors $\xi_A^\pm$ aligned with those directions thus evolve as $\ddot{\xi}^\pm_A = \pm |\Weylfoc| \xi^\pm_A$, so that the beam's shape gets elongated in the first direction and contracted in the second one. Weyl lensing thus tends to \emph{shear} light beams.
\end{itemize}

Ricci and Weyl curvatures have distinct physical origins. On the one hand, the Ricci tensor is directly to matter's \emph{local} density of energy and momentum via the Einstein equation $R_{\mu\nu}-R g_{\mu\nu}/2 + \Lambda g_{\mu\nu} = 8\pi G T_{\mu\nu}$, which implies
\begin{equation}
\Ricfoc = -4\pi G T_{\mu\nu} k^\mu k^\nu \leq 0
\end{equation}
under the null energy condition. In the case of a perfect fluid with rest-frame energy density~$\rho$ and pressure~$p$, the stress-energy tensor reads $T_{\mu\nu}=(\rho+p) u_\mu u_\nu + p g_{\mu\nu}$, and the above relation becomes simply $\Ricfoc=-4\pi G (\rho+p)\omega^2$. \emph{Ricci lensing thus tells us how a light beam is focused by the matter it encloses}. Note by the way that the cosmological constant does not have any focusing effect, like any other form of matter equivalent to a perfect fluid with $p=-\rho$.

The Weyl tensor, on the other hand, describes the \emph{nonlocal} effects of gravity. The simplest examples are the tidal forces created around a massive body, but gravitational waves or frame dragging are also gravitational phenomena encoded in the Weyl tensor. Hence, unlike Ricci lensing, \emph{Weyl lensing is mostly due to matter lying outside the beam}. This fits quite well with our Newtonian intuition: a mass inside the beam attracts all the rays towards it, generating convergence, while a mass outside shears it by tidal effects.


\subsubsection{Self focusing}\index{self focusing of a light beam}\label{sec:self_focusing}

Let us close this section by a remark on the potential ability of light beams to focus themselves. As mentioned in \S~\ref{sec:EM_curved_spacetime}, light indeed possesses energy and momentum, whose contribution to Ricci focusing reads
\begin{equation}
\Ricfoc\e{self} \define -4\pi G \, T_{\mu\nu}\h{beam} k^\mu k^\nu = -16\pi G I \omega^2,
\end{equation}
where $I$ is the luminous intensity (power per unit area) of the beam, also equal to its energy density~$\rho\e{beam}$. This quantity is not a constant during light propagation, the energy being diluted over the growing wavefront and redshifted. If the light source has an absolute luminosity (power)~$L$, the observed intensity goes like $I=L/(4\pi D\e{L}^2)$, where $D\e{L}$ denotes the \emph{luminosity distance}\index{distance!luminosity} from the source to the observer---see Chap.~\ref{chapter:distances} for more details about distances in curved spacetime.

The amplitude of self focusing can be evaluated by comparing it to the cosmic Ricci focusing~$\Ricfoc\e{cosm} \define -4\pi G \rho\e{m} \omega^2$, associated with the mean matter density~$\rho\e{m}=2.8\times 10^{-27}(1+z)^3\U{kg/m^3}$ in the Universe. We get
\begin{equation}
\frac{\Ricfoc\e{self}}{\Ricfoc\e{cosm}} = \frac{1.7\times 10^{-6}}{(1+z)^3} \, \frac{L}{L_\odot} \pa{\frac{1\U{pc}}{D\e{L}}}^2,
\end{equation}
where $L_\odot=3.8\times 10^{26}\U{W}$ is the solar luminosity. For stellar sources, this is therefore a very small number; but for much brighter sources such as quasars, with a typical luminosity of $L\sim 10^{40}\U{W} \sim 10^{14} L_\odot$, self focusing turns out to dominate over cosmic focusing as far as $D\e{L} < 10\U{kpc}$. This effect makes very luminous sources appear even brighter than they already are.

\section{The Jacobi matrix}

The comparison between the physical morphology of a source and how it appears to an observer is the heart of gravitational lensing experiments. The Jacobi matrix, which we introduce in this section, precisely contains this information.

\subsection{Definition and interpretation}
\label{sec:Jacobi_definition}

\subsubsection{Wronski matrix}\index{Wronski matrix}\label{sec:Wronski_matrix}

The Sachs vector equation~\eqref{eq:Sachs_vector_equation} is a second-order \emph{linear} differential equation, which satisfies the Cauchy-Lipshitz conditions if we assume that $\tidal_{AB}(v)$ is smooth. Hence there is a linear one-to-one and onto relation between any solution of Eq.~\eqref{eq:Sachs_vector_equation} and its initial conditions; in other words, there exists a $4\times 4$ invertible matrix $\mat{\wronski}$ such that
\begin{equation}
\begin{pmatrix}
\xi_1\\
\xi_2 \\
\dot{\xi}_1 \\
\dot{\xi}_2
\end{pmatrix}(v_2)
=
\mat{\wronski}(v_2 \leftarrow v_1)
\begin{pmatrix}
\xi_1\\
\xi_2 \\
\dot{\xi}_1 \\
\dot{\xi}_2
\end{pmatrix}(v_1) .
\label{eq:Wronski_def}
\end{equation}
We shall call $\mat{\wronski}$ the \emph{Wronski matrix} of the Sachs equation\footnote{In a mathematically rigorous way, $\mat{\wronski}(v\leftarrow v_0)$ is the Wronski matrix associated with a fundamental system of solutions $(\mat{M}_1,\mat{M}_2)$ of the matrix equation $\ddot{\mat{M}}=\mat{\tidal}\mat{M}$, with initial conditions $\mat{M}_1(v_0)=\dot{\mat{M}}_2(v_0)=\identity_2$ and $\dot{\mat{M}}_1(v_0)=\mat{M}_2(v_0)=\zero_2$}.
The notation $v_2\leftarrow v_1$, which indicates that Eq.~\eqref{eq:Sachs_vector_equation} is integrated from $v_1$ to $v_2$, is useful for expressing the elementary properties of~$\mat{\wronski}$ implied by its very definition, namely
\begin{align}
\mat{\wronski}(v_3 \leftarrow v_1) &= \mat{\wronski}(v_3 \leftarrow v_2) \mat{\wronski}(v_2 \leftarrow v_1),
\label{eq:Chasles}\\
\mat{\wronski}(v_2 \leftarrow v_1) &= [\mat{\wronski}(v_1 \leftarrow v_2)]^{-1}.
\label{eq:Wronski_reciprocity}
\end{align}

Its propagation equation, inherited from Eq.~\eqref{eq:Sachs_vector_equation}, reads
\begin{equation}\label{eq:master_equation_Wronski}
\pd{}{v_2}\mat{\wronski}(v_2 \leftarrow v_1) =
\begin{pmatrix}
\zero_2 & \identity_2 \\
\mat{\tidal}(v_2) & \zero_2
\end{pmatrix}
\mat{\wronski}(v_2 \leftarrow v_1),
\end{equation}
with initial condition~$\mat{\wronski}(v_1 \leftarrow v_1) = \identity_4$, and where $\zero_n$, $\identity_n$ respectively denote the $n\times n$ zero and unity matrices. This Cauchy problem is formally solved by
\begin{equation}\label{eq:formal_solution_Wronski}
\mat{\wronski}(v_2 \leftarrow v_1)
= \mathrm{Vexp} \int_{v_1}^{v_2}
\begin{pmatrix}
\zero_2 & \identity_2 \\
\mat{\tidal}(v) & \zero_2
\end{pmatrix} \dd v,
\end{equation}
where $\mathrm{Vexp}$ is the affine-parameter ordered exponential defined, for any matrix-valued function $\mat{M}(v)$, by
\begin{equation}
\mathrm{Vexp}\int_{v_1}^{v_2} \mat{M}(v)\dd v
\define
\sum_{n=0}^{\infty} \int_{v_1}^{v_2} \dd w_1 \int_{v_0}^{w_1} \dd w_2 \ldots \int_{v_0}^{w_{n-1}} \dd w_n \;
					\mat{M}(w_1) \mat{M}(w_2) \ldots \mat{M}(w_n) .
\end{equation}
This expression reduces to a regular exponential if, for all $v$, $v'$, $\mat{M}(v)$ commutes with $\mat{M}(v')$. In the case of Eq.~\eqref{eq:formal_solution_Wronski}, this occurs if, and only if, $\mat{\tidal}(v)$ is a constant. 

Because of its ``Chasles relation''~\eqref{eq:Chasles}, the Wronski matrix is a very convenient tool for solving the Sachs equation~\eqref{eq:Sachs_vector_equation} piecewise, as the junction between various pieces of solution is simply achieved by matrix multiplications. This is indeed the reason why the Wronski matrix was introduced independently by Ref.~\cite{2012PhRvD..85b3510F} (not under this name) and by the author of this thesis in Ref.~\cite{2013PhRvD..87l3526F}, in order to deal with light propagation in Swiss-cheese models (see Chap.~\ref{chapter:SC}). In most cases, however, only a $2\times 2$ part of $\vect{\wronski}$ is really useful, as we will see below.

\subsubsection{Jacobi matrix}\index{Jacobi matrix!definition}\label{sec:Jacobi_matrix_definition}

If we choose the initial conditions for the integration of the Sachs equation to be a vertex point of the light beam (here the observation event~$O$), then by definition $\xi_A(v_\obs)=0$, and Eq.~\eqref{eq:Wronski_def} indicates that $\xi_A(v\not= v_\obs)$ is related to $\dot{\xi}_A(v_\obs)$ as
\begin{empheq}[box=\fbox]{equation}
\xi^A(v) = \jacobi\indices{^A_B}(v\leftarrow v_\obs) \ddf{\xi^B}{v}(v_\obs),
\label{eq:Jacobi_def}
\end{empheq}
where $\vect{\jacobi}(v\leftarrow v_\obs)$ is the $2\times 2$ top-right block of $\mat{\wronski}(v\leftarrow v_\obs)$; it is called \emph{Jacobi matrix} for reasons that shall become clearer below.

\begin{figure}[h!]
\centering
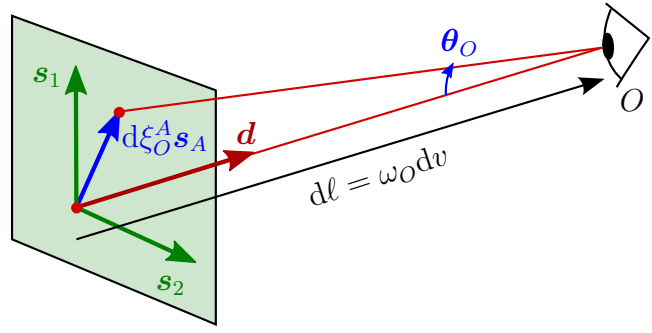
\caption{The vector $\dot{\xi}^A_\obs \vect{s}_A$ at $O$ is directly proportional to the observed angular separation $\vect{\theta}_\obs=\theta^A_\obs \vect{s}_A$ between the rays, since $\theta^A=\dd\xi^A/\dd\ell$.}
\label{fig:angular_separation}
\end{figure}

In \S~\ref{sec:physical_interpretation_affine_parameter} of the previous chapter, we have seen that an increase $\dd v$ of the affine parameter corresponds, in the rest frame of arbitrary observer, to a displacement $\dd\ell/\omega$ of the photon. It follows, as depicted in Fig.~\ref{fig:angular_separation}, that the derivative~$\dot{\xi}^B (v_\obs)$ involved in the definition~\eqref{eq:Jacobi_def} of $\mat{\jacobi}$ reads
\begin{equation}
\evaluate{\ddf{\xi^B}{v}}_\obs = \omega_O \evaluate{\ddf{\xi^B}{\ell}}_\obs = -\omega_\obs \theta^B_\obs,
\label{eq:derivative=angle}
\end{equation}
where $\theta^B_\obs \ll 1$ denotes the angular separation, on the observer's celestial sphere, between the two images associated with two rays separated by $\vect{\xi}$. The minus sign on the right-hand side is due to the fact that $k^\mu$ is future oriented, so that $\dd\ell$ is positive for a displacement towards the future, whereas $\theta^B_\obs$ is more naturally defined from a past-oriented displacement.

In light of the identification~\eqref{eq:derivative=angle}, the matrix~$\mat{\jacobi}(\source\leftarrow\obs)$ corresponding to the integration of the Sachs equation from the observation event~$O$ to the source event~$S$ can be written as\index{Jacobi matrix!geometrical interpretation}
\begin{equation}
\jacobi\indices{^A_B}(\source\leftarrow\obs) = -\frac{1}{\omega_\obs}\pd{\xi^A_\source}{\theta^B_\obs}.
\end{equation}
We conclude that, modulo the normalisation factor $-1/\omega_\obs$, $\mat{\jacobi}(\source\leftarrow\obs)$ indeed represents the Jacobi matrix of the map $\{\theta^A_\obs\}\mapsto\{\xi^A_\source\}$, which relates the positions of images on the observer's celestial sphere to the physical separations of the associated sources. As such, $\mat{\jacobi}(\source\leftarrow\obs)$ encodes all the gravitational distortions of the image of a very small light source.

%
%
%

\subsection{Decompositions}\label{sec:decomposition_Jacobi}

\subsubsection{General case}\index{Jacobi matrix!decompositions}

The Jacobi matrix~$\mat{\jacobi}(\source\leftarrow\obs)$ can be decomposed in a way that emphasize the geometrical transformations between the observed image and the actual source---whose intrinsic morphology cannot be observed. First note that its determinant is, still modulo a frequency factor, the Jacobian of $\{\theta^A_\obs\}\mapsto\{\xi^A_\source\}$, that is, the ratio between the physical area of the source~$A_\source$ and its apparent angular size (observed solid angle)~$\Omega_\obs$, both assumed to be infinitesimal quantities. In other words,\index{distance!angular diameter}
\begin{equation}
\det(\omega_\obs\mat{\jacobi}) = \frac{A_\source}{\Omega_\obs}  \define D\e{A}^2,
\end{equation}
where we anticipated on Chap.~\ref{chapter:distances}, \S~\ref{sec:angular_diameter_distance} identifying the above ratio with the square of the angular diameter distance $D\e{A}$ between the source and the observer.

As any $2\times 2$ matrix, $\mat{\jacobi}/\sqrt{\det\mat{\jacobi}}$ can be written as the product $\mat{R}\mat{S}$ of a rotation matrix~$\mat{R}$ and a symmetric matrix~$\mat{S}$. Moreover, since its determinant is unity, we can write $\mat{S}$ as the exponential of traceless symmetric matrix. The resulting decomposition of the Jacobi matrix thus reads
\begin{equation}
\mat{\jacobi} 
= 
-\frac{D\e{A}}{\omega_\obs}
\begin{pmatrix}
\cos\psi & -\sin\psi \\
\sin\psi & \cos\psi
\end{pmatrix}
\exp
\begin{pmatrix}
-\gamma_1 & \gamma_2 \\
\gamma_2 & \gamma_1
\end{pmatrix} .
\label{eq:Jacobi_decomposition}
\end{equation}
A further step consists in diagonalising the exponential matrix; it is convenient for that purpose to define $\gamma\geq 0$ and $\varphi$ such that $\gamma_1 + \i\gamma_2 = \gamma \ex{-2\i\varphi}$, leading to
\begin{equation}
\exp
\begin{pmatrix}
-\gamma_1 & \gamma_2 \\
\gamma_2 & \gamma_1
\end{pmatrix}
= 
\begin{pmatrix}
\cos\varphi & -\sin\varphi \\
\sin\varphi & \cos\varphi
\end{pmatrix}
\begin{pmatrix}
\ex{-\gamma} & 0 \\
0 & \ex{\gamma}
\end{pmatrix}
\begin{pmatrix}
\cos\varphi & \sin\varphi \\
-\sin\varphi & \cos\varphi
\end{pmatrix}.
\end{equation}

\begin{figure}[h!]
\centering
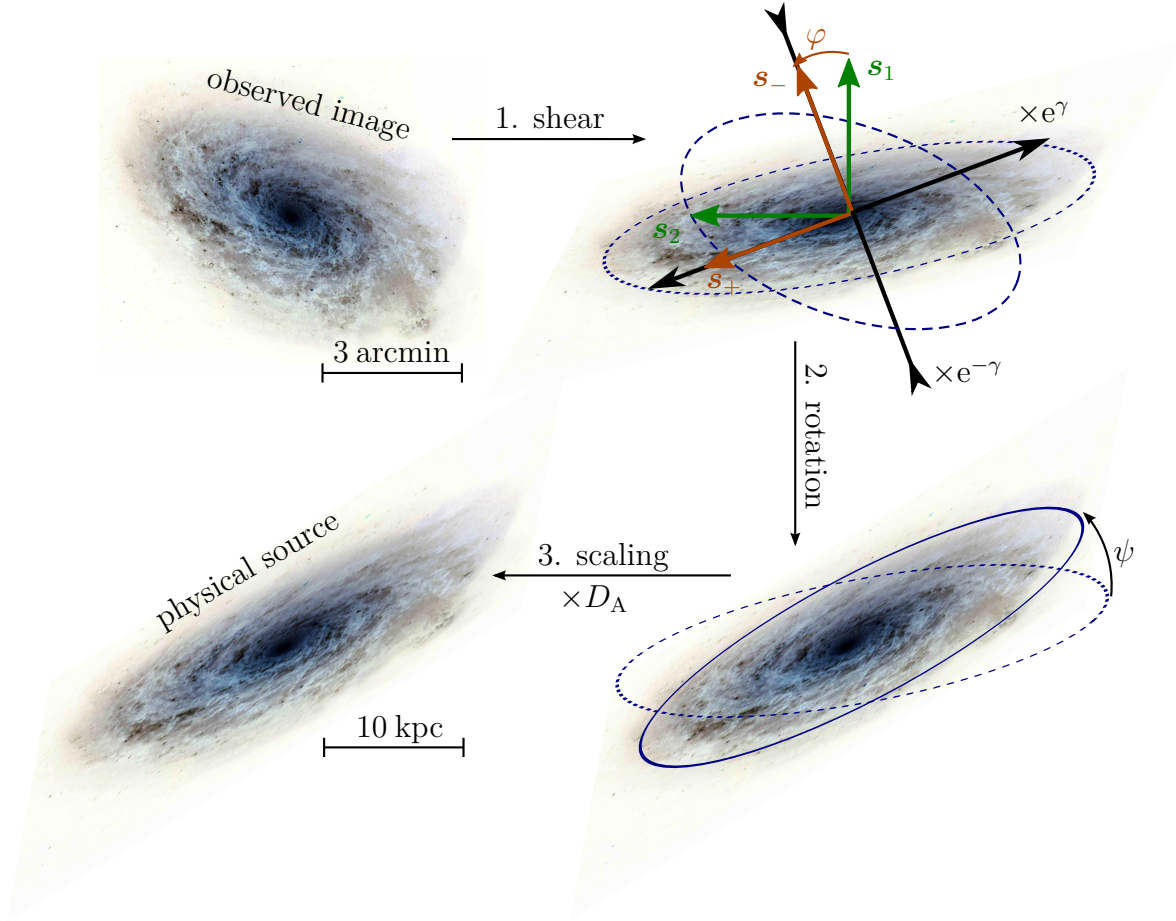
\caption{Transformations of an image due to gravitational lensing, as encoded in the decomposition~\eqref{eq:Jacobi_decomposition} of the Jacobi matrix. In this example, the angular distance between the source galaxy and the observer is $D\e{A}=10\U{kpc}/3\U{arcmin}\approx 1\U{Mpc}$.}
\label{fig:decomposition_Jacobi}
\end{figure}

The various quantities involved in the above decomposition must be interpreted the following way (see Fig.~\ref{fig:decomposition_Jacobi}). Starting from an observed image, the intrinsic properties of the source are reconstructed by successively:
\begin{enumerate}
\item contracting and expanding the image by factors $\ex{-\gamma}$ and $\ex{\gamma}$, respectively, in the directions
\begin{align}
\vect{s}_-  &\define \cos\varphi \vect{s}_1 + \sin\varphi\vect{s}_2, \\
\vect{s}_+ &\define -\sin\varphi \vect{s}_1 + \cos\varphi\vect{s}_2,
\end{align}
this \emph{shear}\index{shear!net effect on images} operation preserves the area of the image;
\item rotating anticlockwise the result by the angle $\psi$;
\item translating angles into physical distances by multiplying them with $D\e{A}$.
\end{enumerate}


The changes of orientation of the image with respect to its source---induced by shear~(2.) and rotation~(3.)---are a priori not measurable, because it is a priori impossible to know what is the intrinsic orientation of the source. Nevertheless, for a source of polarised light, whose shape is aligned with the direction of polarisation (e.g. in quasar jets), such effects can become observable because they generically break the alignment between polarisation---materialised by the Sachs basis---and the image's shape.

\subsubsection{Perturbative case}\label{sec:amplification_matrix}

In a number of practical situations,
the lensing effects encoded in the Jacobi matrix result from small perturbations of the spacetime geometry, with respect to a background which generates no shear nor rotation of images (Friedmann-Lema\^{i}tre or Minkowski), i.e. for which $\overline{\mat{\jacobi}}=- \bar{D}\e{A}/\omega_\obs \identity_2$. In such cases, the general decomposition~\eqref{eq:Jacobi_decomposition} can be expanded at order $1$ in the deformation scalars $\gamma_1,\gamma_2,\psi \ll 1$, and in the \emph{convergence}\index{convergence!definition}
\begin{equation}
\kappa \define \frac{\bar{D}\e{A}-D\e{A}}{\bar{D}\e{A}} \ll 1,
\label{eq:convergence}
\end{equation}
under the form
\begin{equation}
\mat{\jacobi} = \mat{\amplification}\overline{\mat{\jacobi}} + \ldots,
\label{eq:Jacobi_perturbative_decomposition}
\end{equation}
where the \emph{amplification} (or \emph{magnification})\index{amplification matrix!definition}\index{magnification matrix|see{amplification matrix}} matrix reads
\begin{equation}
\mat{\amplification} 
\define
\begin{pmatrix}
1 - \kappa - \gamma_1 & \gamma_2 - \psi \\
\gamma_2 + \psi & 1-\kappa + \gamma_1
\end{pmatrix}.
\label{eq:amplification_matrix}
\end{equation}
In fact, we will see in \S~\ref{sec:geometrical_meaning_optical_scalars} that if $\gamma\ll 1$, then $\psi\sim\gamma^2$ and can thus be neglected in the above first-order expansion. This is the reason why the amplification matrix is usually considered symmetric in the weak-lensing literature~\cite{1992grle.book.....S,2001PhR...340..291B}.

In principle, it is possible to use a decomposition of the form~\eqref{eq:amplification_matrix} even for finite deformations, but the quantities $\kappa,\gamma,\psi \sim 1$ then differ from the ones defined in Eqs.~\eqref{eq:Jacobi_decomposition}, \eqref{eq:convergence}, and thus lose their geometrical meaning. For instance, the relative correction to the angular distance is, in that case, no longer equal to the convergence, but rather to
\begin{equation}
\frac{\bar{D}\e{A}-D\e{A}}{\bar{D}\e{A}} = 1 - \frac{1}{\sqrt{\mu}},
\end{equation}
where the \emph{magnification}~$\mu$ reads
\begin{equation}
\mu \define \frac{1}{\det\mat{\amplification}} = \frac{1}{(1-\kappa)^2 - \gamma^2 + \psi^2}.
\end{equation} 
Note by the way that the conventional name ``magnification matrix'' for $\mat{\amplification}$ is somewhat misleading, and would be more adapted to $\mat{\amplification}^{-1}$. It is precisely this loss of geometrical interpretation for the commonly used $\kappa,\gamma,\psi$ which led the author of this thesis to propose the more generally sensitive decomposition~\eqref{eq:Jacobi_decomposition}.

Let us finally mention that a perturbative expansion of the form~\eqref{eq:Jacobi_perturbative_decomposition} can also be performed with respect to a background with nonzero shear and rotation---e.g. for weak lensing in a perturbed Bianchi~I universe \cite{2015arXiv150301125P,2015arXiv150301127P}---but the expression~\eqref{eq:amplification_matrix} of $\mat{\amplification}$ needs to be slightly adapted to account for it.

\subsection{Evolution}

Formally, the evolution of the Jacobi matrix with $v$ is entirely determined by simply extracting the $2\times 2$ top-right block of the Wronski matrix given by Eq.~\eqref{eq:formal_solution_Wronski}. However, affine-parameter ordered exponentials are not particularly easy to handle, and thus of limited interest for practical calculations.

\subsubsection{Jacobi matrix equation}

An evolution equation for $\mat{\jacobi}$ only is directly obtained by taking the second derivative of its definition~\eqref{eq:Jacobi_def} and inserting the Sachs equation~\eqref{eq:Sachs_vector_equation}, which yields
\begin{empheq}[box=\fbox]{equation}
\pd[2]{}{v} \mat{\jacobi}(v\leftarrow v_\obs) = \mat{\tidal}(v) \mat{\jacobi}(v\leftarrow v_\obs),
\label{eq:Jacobi_matrix_equation}
\end{empheq}
that we will refer to as the \emph{Jacobi matrix equation}\index{Jacobi matrix!equation}, and abridge as $\ddot{\mat{\jacobi}}=\mat{\tidal}\mat{\jacobi}$, where it is understood that a dot stands for a derivative with respect to the \emph{final} affine parameter~$v$ of $\mat{\jacobi}(v\leftarrow v_\obs)$.

The initial conditions ($v=v_\obs$) for the differential equation~\eqref{eq:Jacobi_matrix_equation} derive from the very definition~\eqref{eq:Jacobi_def} of the Jacobi matrix, and read
\begin{align}
\mat{\jacobi}(v_\obs\leftarrow v_\obs) &= \zero_2,
\label{eq:Jacobi_initial_condition_1}\\
\dot{\mat{\jacobi}}(v_\obs\leftarrow v_\obs) &= \identity_2.
\label{eq:Jacobi_initial_condition_2}
\end{align}
The set of Eqs.~\eqref{eq:Jacobi_matrix_equation}, \eqref{eq:Jacobi_initial_condition_1}, and \eqref{eq:Jacobi_initial_condition_2} thus completely characterises the evolution of the Jacobi matrix with $v$.

%

\subsubsection{Etherington's reciprocity law}\index{Etherington's reciprocity law}

We have seen in Eq.~\eqref{eq:Wronski_reciprocity} that the Wronski matrix enjoys a simple reciprocity law: inverting the initial and final conditions simply turns $\mat{\wronski}$ into its inverse. There exist a similar relation for the Jacobi matrix, known as Etherington's reciprocity law~\cite{1933PMag...15..761E}, but contrary to the former, the latter does not trivially follow from a definition: it relies on the specific equation~\eqref{eq:Jacobi_matrix_equation} governing the evolution of $\mat{\jacobi}$.

Following Refs.~\cite{1971grc..conf..104E,2004LRR.....7....9P}, we consider two solutions~$v\mapsto\mat{\jacobi}(v\leftarrow v_1)$ and $v\mapsto\mat{\jacobi}(v\leftarrow v_2)$ of the Jacobi matrix equation, and define
\begin{equation}
\mat{C}(v) \define 
\transpose{\dot{\mat{\jacobi}}}(v\leftarrow v_1) \mat{\jacobi}(v\leftarrow v_2)
- \transpose{\mat{\jacobi}}(v\leftarrow v_1) \dot{\mat{\jacobi}}(v\leftarrow v_2) ,
\label{eq:conservation_law_Jacobi}
\end{equation}
where $\transpose{}$ denotes matrix transposition. The symmetry of the optical tidal matrix~$\transpose{\mat{\tidal}}=\mat{\tidal}$ implies that \emph{$\mat{C}$ is a constant}; in particular, $\mat{C}(v_1)=\mat{C}(v_2)$ implies
\begin{empheq}[box=\fbox]{equation}
\mat{\jacobi}(v_1\leftarrow v_2) = -\transpose{\mat{\jacobi}}(v_2\leftarrow v_1)
\end{empheq}
by virtue of Eqs.~\eqref{eq:Jacobi_initial_condition_1}, \eqref{eq:Jacobi_initial_condition_2}. A particular consequence of the above relation is that inverting $v_1$ and $v_2$ leaves the determinant of the Jacobi matrix unchanged:
\begin{equation}
\det\mat{\jacobi}(v_1\leftarrow v_2) = \det\mat{\jacobi}(v_2\leftarrow v_1),
\label{eq:symmetry_determinant_Jacobi}
\end{equation}
which will allow us to derive the duality relation between the angular and luminosity distances in Chap.~\ref{chapter:distances}.

\section{The optical scalars}

The propagation equations for light beams can be reformulated in terms of their deformation rates, also known as the optical scalars. This formulation has the advantage of exhibiting even more clearly the respective roles of Ricci or Weyl curvatures, and allows one to derive the so-called focusing theorem.

\subsection{Definitions}

\subsubsection{From the Jacobi matrix}

Since the deformations of a light beam are described by the Jacobi matrix, the associated deformation \emph{rates} are naturally defined by its logarithmic derivative with respect to the affine parameter; we thus introduce the \emph{deformation rate matrix}\index{deformation rate!matrix}
\begin{equation}
\boxed{
\mat{\deformation}\define \dot{\mat{\jacobi}} \mat{\jacobi}^{-1}
}
\qquad \text{i.e.} \qquad
\deformation\indices{^A_B} \define \pd{\dot{\xi}^A}{\xi^B}.
\label{eq:deformation_rate_matrix}
\end{equation}
The conservation law~\eqref{eq:conservation_law_Jacobi} applied for $v_1=v_2$ (hence $\mat{C}=\mat{0}_2$) reads
\begin{equation}
\zero_2 = \transpose{\mat{\jacobi}} \transpose{\mat{\deformation}}\mat{\jacobi} 
- \transpose{\mat{\jacobi}} \mat{\deformation} \mat{\jacobi}
\end{equation}
which, if we assume $\det\mat{\jacobi}\not=0$, implies that $\mat{\deformation}$ \emph{is a symmetric matrix}.

Decomposing $\mat{\deformation}$ into its pure-trace and trace-free parts then yields
\begin{equation}
\mat{\deformation}
=
\begin{pmatrix}
\nul{\theta} & 0 \\
0 & \nul{\theta}
\end{pmatrix}
+
\begin{pmatrix}
-\nul{\sigma}_1 & \nul{\sigma}_2 \\
\nul{\sigma}_2 & \nul{\sigma}_1
\end{pmatrix},
\label{eq:decomposition_deformation}
\end{equation}
where $\nul{\theta}$ and $\nul{\sigma}\define\nul{\sigma}_1+\i\nul{\sigma}_2$ are the \emph{optical scalars}\index{optical scalars!definition}\index{deformation rate!scalars|see{optical scalars}}; they are called respectively the beam's expansion rate\index{expansion rate (of a light beam)} and shear rate\index{shear!rate}, for reasons that shall become clearer in \S~\ref{sec:geometrical_meaning_optical_scalars}. Note that, despite its notation, this $\theta$ must not be confused with an angle, both $\theta$ and $\sigma$ have the dimension of $[v]^{-1}$.

\subsubsection{From the gradient of the wave four-vector}

Alternatively, one can define the deformation rate matrix~$\mat{\deformation}$, and thus the optical scalars $\theta,\sigma$, via
\begin{empheq}[box=\fbox]{equation}
\deformation_{AB} = s_A^\mu s_B^\nu \nabla_\mu k_\nu .
\label{eq:deformation_rate_matrix_alternative}
\end{empheq}
Let us show that this definition is indeed equivalent to the previous one~\eqref{eq:deformation_rate_matrix},
\begin{align}
\dot{\xi}_A &\equiv k^\mu \nabla_\mu (s_A^\nu \xi_\nu) \\
					&= s_A^\nu k^\mu \nabla_\mu \xi_\nu \qquad \text{since }\dot{s}_A^\mu \propto k^\mu \perp s_A^\mu\\
					&= s_A^\nu \xi^\mu \nabla_\mu k_\nu \qquad \text{because of Eq.~\eqref{eq:commutation_xi_k}} \\
					&= s^\nu_A \xi_\perp^\mu \nabla_\mu k_\nu \qquad \text{as }k^\mu\nabla_\mu k_\nu=0 \\
					&= s^\nu_A \xi^B s_B^\mu \nabla_\mu k_\nu,
\end{align}
whose derivative with respect to $\xi^B$ indeed leads to the expression~\eqref{eq:deformation_rate_matrix_alternative}, modulo an inversion of the indices $\mu\leftrightarrow\nu$, which is allowed since $\nabla_\mu k_\nu$ is a symmetric tensor. This property, which comes from the fact that $k_\mu$ is the gradient of the wave's phase, is an alternative proof for the symmetry of $\mat{\deformation}$.

More generally, the tensor $\nabla_\mu k_\nu$ can be decomposed over the four-dimensional orthonormal basis $(u^\mu,d^\mu,s_1^\mu,s_2^\mu)$ according to
\begin{equation}
\nabla_\mu k_\nu = \deformation_{AB} s^A_\mu s^B_\nu
								 - 2\omega^{-1} S_{(\mu}^\rho k_{\nu)} u^\sigma \nabla_{\rho} k_\sigma
								 + \omega^{-2} k_\mu k_\nu u^\rho u^\sigma \nabla_{\rho} k_\sigma,
\label{eq:decomposition_gradient_k}
\end{equation}
which can be derived starting from $\nabla_\mu k_\nu=\delta_\mu^\rho \delta_\nu^\sigma \nabla_\rho k_\sigma$, with $\delta_\mu^\rho=S_\mu^\rho + d_\mu d^\rho - u_\mu u^\rho$, and using $k^\mu k_\mu=0=k^\nu \nabla_\nu k_\mu$. Taking the trace of Eq.~\eqref{eq:decomposition_gradient_k} then yields
\begin{empheq}[box=\fbox]{equation}
\nabla_\mu k^\mu = \tr \mat{\deformation} = 2\theta,
\end{empheq}
while the trace of its square gives
\begin{empheq}[box=\fbox]{equation}
(\nabla_\mu k_\nu)(\nabla^\mu k^\nu) = \tr(\mat{\deformation}^2) = 2(\theta^2 + \abs{\sigma^2}).
\end{empheq}
The quantities $\theta$ and $\abs{\sigma}^2$ are thus fully covariant quantities, which reflects the frame independence of the beam's morphology.

\subsection{Geometrical interpretation}
\label{sec:geometrical_meaning_optical_scalars}
\index{optical scalars!geometrical interpretation}

\subsubsection{Expansion rate}

The physical cross-sectional area of a light beam is defined as
\begin{equation}
A \define \int\e{beam} \dd\xi^1 \dd\xi^2 = \int\e{beam} \det{\mat{\jacobi}} \; \dd\dot{\xi}^1_\obs \dd\dot{\xi}^2_\obs .
\end{equation}
For an infinitesimal light beam, $\mat{\jacobi}$ can be considered constant in the above integral, and the evolution rate of $A$ with the affine parameter reads
\begin{equation}
\frac{\dot A}{A} = \frac{1}{\det\mat{\jacobi}}\ddf{(\det\mat{\jacobi})}{v}
						 = \tr (\dot{\mat{\jacobi}} \mat{\jacobi}^{-1}) \define \tr\mat{\deformation},
\end{equation}
whence
\begin{empheq}[box=\fbox]{equation}
\theta = \frac{1}{2 A} \ddf{A}{v} = \frac{1}{D\e{A}} \ddf{D\e{A}}{v},
\label{eq:interpretation_expansion}
\end{empheq}
where we reintroduced the angular diameter distance $D\e{A}\propto\sqrt{A}$. The quantity $2\theta$ therefore represents the evolution rate of the beam's area.

\subsubsection{Shear rate}

Consider two light rays separated by $\vect{\xi}$. The distance~$\ell$ between the associated light spots on the local screen reads
\begin{equation}
\ell^2 \define \xi^\mu_\perp \xi_\mu^\perp = \xi^A \xi_A = \jacobi\indices{^A_B} \jacobi\indices{_A_C} \, \dot{\xi}^B_\obs \dot{\xi}^C_\obs,
\end{equation}
and since $(\dd \transpose{\mat{\jacobi}} \mat{\jacobi})/\dd v = 2 \transpose{\mat{\jacobi}} \mat{\deformation} \mat{\jacobi}$, we conclude that the evolution rate of $\ell$ reads
\begin{equation}
\frac{1}{\ell} \ddf{\ell}{v} = \frac{\xi^A \deformation_{AB} \xi^B}{\xi^A \xi_A} = \theta - \abs{\sigma} \cos 2\iota,
\end{equation}
where $\iota$ denotes here the angle between $(\xi^A)$ and the eigendirection of $\mat{\deformation}$ associated with the eigenvalue $\theta-\abs{\sigma}$. The quantity $\dot{\ell}/\ell$ thus belongs to the interval $[\theta-\abs{\sigma},\theta+\abs{\sigma}]$, which gives $\abs{\sigma}$ its geometrical meaning:
\begin{empheq}[box=\fbox]{equation}
2\abs{\sigma} = \pa{\frac{1}{\ell} \ddf{\ell}{v}}\e{max} - \pa{\frac{1}{\ell} \ddf{\ell}{v}}\e{min}
\end{empheq}
is the rate of stretching of the light beam. Note by the way the alternative expression for the expansion rate, $2\theta=(\dot{\ell}/\ell)\e{max}+(\dot{\ell}/\ell)\e{min}$.

\subsubsection{Relation with the deformation scalars}

We have just seen that the expansion rate $\theta$ is related to $D\e{A}$ via Eq.~\eqref{eq:interpretation_expansion}. Similarly, there exist relations between the shear rate~$\sigma$, the net shear~$\gamma$, its direction~$\varphi$, and the rotation angle~$\psi$, which can be derived by inserting the decomposition~\eqref{eq:Jacobi_decomposition} of~$\mat{\jacobi}$ in the definition~\eqref{eq:deformation_rate_matrix} of~$\mat{\deformation}$. The result is explicitly
%
%
%
%
\begin{multline}
\mat{\deformation}
=
\frac{\dot{D}\e{A}}{D\e{A}}
\begin{pmatrix}
1 & 0 \\
0 & 1
\end{pmatrix}
+
(\dot{\psi}+\dot{\varphi})
\begin{pmatrix}
0 & -1 \\
1 & 0
\end{pmatrix}
+
\dot{\gamma}
\begin{pmatrix}
-\cos2(\psi+\varphi) & -\sin2(\psi+\varphi) \\
-\sin2(\psi+\varphi) & \cos2(\psi+\varphi)
\end{pmatrix}\\
+
\dot{\varphi}
\begin{pmatrix}
\sin2(\psi+\varphi)\sinh2\gamma & \cosh2\gamma - \cos2(\psi+\varphi) \sinh2\gamma \\
-\cosh2\gamma - \cos2(\psi+\varphi) \sinh2\gamma & -\sin2(\psi+\varphi)\sinh2\gamma
\end{pmatrix},
\end{multline}
which, after regrouping the trace, trace-free symmetric, and antisymmetric parts, and identifying with Eq.~\eqref{eq:decomposition_deformation}, indeed confirms Eq.~\eqref{eq:interpretation_expansion} and yields
\begin{empheq}[box=\fbox]{align}
\nul{\sigma} &= \pa{\dot{\gamma}-\i\dot{\varphi}\sinh2\gamma} \ex{-2\i(\psi+\varphi)},
\label{eq:sigma_gamma}\\
0 &= \dot{\psi} - 2 \dot{\varphi} \sinh^2\gamma.
\label{eq:zero_rotation}
\end{empheq}
An alternative expression for $\sigma$ can also be derived by combining Eqs.~\eqref{eq:sigma_gamma}, \eqref{eq:zero_rotation}, and reads
\begin{equation}
\nul{\sigma} = \frac{1}{2\cosh2\gamma}\ddf{}{v}\pac{\ex{ -2\i(\psi+\varphi) }\sinh 2\gamma}.
\label{eq:sigma_alternative}
\end{equation}
In the weak-lensing case ($\gamma\ll 1$), Eq.~\eqref{eq:zero_rotation} implies that $\psi\sim \gamma^2$ is a second-order quantity and can be neglected. Besides, Eq.~\eqref{eq:sigma_alternative} becomes
\begin{equation}
\sigma \approx \ddf{}{v} \pa{ \gamma\ex{-2\i\ph} }.
\end{equation}

\begin{figure}[h!]
\centering
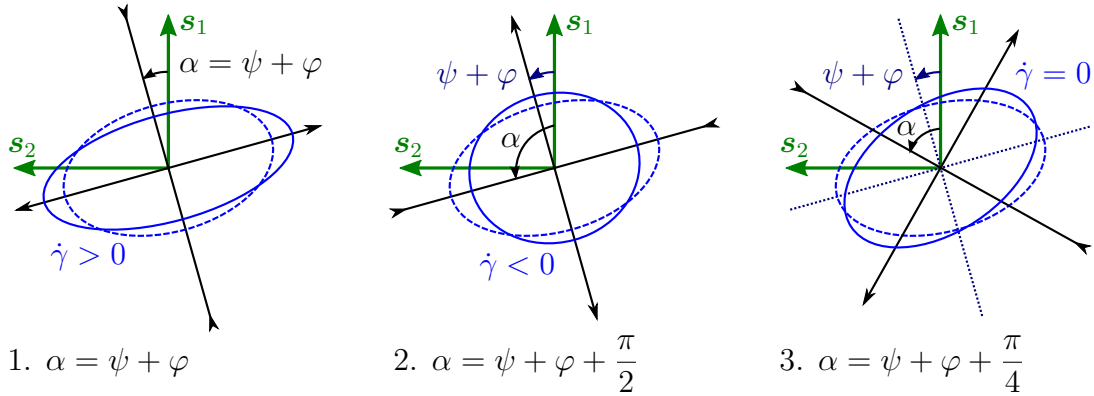
\caption{Illustration of the effect of the shear rate~$\sigma$ for the three configurations discussed in the text. Blue ellipses represent the cross-sectional shape of an initially circular light beam, for two successive values of the affine parameter: $v$ (dashed) and $v+\delta v$ (solid). Black arrows indicate the eigendirection of the shear rate matrix.}
\label{fig:deformation_rate}
\end{figure}

The geometrical interpretation of Eq.~\eqref{eq:sigma_gamma} is more easily discussed if we introduce the angle $\alpha$ such that $\sigma=\abs{\sigma}\ex{-2\i\alpha}$; moving the phase term~$\ex{2\i(\psi+\varphi)}$ to the left-hand side of Eq.~\eqref{eq:sigma_gamma} and taking the real and imaginary parts then gives
\begin{align}
\abs{\sigma} \cos 2(\varphi+\psi-\alpha) &= \dot{\gamma}, \\
\abs{\sigma} \sin 2(\varphi+\psi-\alpha) &= -\dot{\varphi}\sinh 2\gamma,
\end{align}
which tell us how the shear rate works on an already sheared image, depending on their relative orientation. On the one hand, as already illustrated in Fig.~\ref{fig:decomposition_Jacobi}, $\varphi+\psi$ is the angle between $\vect{s}_1$ (resp. $\vect{s}_2$) and the direction in which the beam has been effectively contracted (resp. expanded). On the other hand, $\alpha$ is the angle between $\vect{s}_1$ (resp.~$\vect{s}_2$) along which the beam undergoes minimum (resp. maximum) elongation \emph{rate}~$\dot{\ell}/\ell$. The following three situations, depicted in Fig.~\ref{fig:deformation_rate} are then easily understood:
\begin{enumerate}
\item If $\alpha=\psi+\varphi$, then the stretching described by $\sigma$ occurs precisely in the same direction along which the beam is already stretched. Its deformation is then amplified, without any change of its orientation: $\dot{\gamma}=\abs{\sigma}$, $\dot{\varphi}=\dot{\psi}=0$.
\item If $\alpha=\psi+\varphi+\pi/2$, it is the contrary of the above, elongation occurs in the direction for which the beam is contracted, and vice versa. The deformation thus tends to attenuate, $\dot{\gamma}=-\abs{\sigma}$, $\dot{\varphi}=\dot{\psi}=0$.
\item If $\alpha=\psi+\varphi+\pi/4$, we have an interesting situation for which the shear rate actually generates no distortion of the beam ($\dot{\gamma}=0$). The pattern displayed on the screen thus conserves its degree of deformation, but the direction in which the latter occurs changes according to $\dot{\varphi}=\abs{\sigma}/\sinh 2\gamma$. It also undergoes a global rotation according to $\dot{\psi}=2\dot{\varphi}\sinh^2\gamma = \abs{\sigma}\tanh\gamma$.
\end{enumerate}

\subsection{Evolution}

\subsubsection{Sachs scalar equations}
\index{optical scalars!transport equations}

Besides the geometrical meaning of its components, the deformation rate matrix~$\mat{\deformation}$ can be considered a \emph{Riccati variable} associated with $\mat{\jacobi}$, since its allows one to trade the second-order linear Jacobi matrix equation~\eqref{eq:Jacobi_matrix_equation} for a first-order nonlinear Riccati equation
\begin{equation}
\dot{\mat{\deformation}} + \mat{\deformation}^2 = \mat{\tidal},
\label{eq:evolution_deformation}
\end{equation}
that one easily obtain by taking the derivative of $\mat{\deformation}$ and replacing $\ddot{\mat{\jacobi}}$ by $\mat{\tidal}\mat{\jacobi}$. Inserting the decomposition~\eqref{eq:decomposition_deformation} of $\mat{\deformation}$ then leads to the \emph{Sachs scalar equations}\index{Sachs!equations (scalar)}
\begin{empheq}[box=\fbox]{align}
\dot{\nul{\theta}} + \nul{\theta}^2 + \abs{\nul{\sigma}}^2 &= \Ricfoc ,
\label{eq:evolution_expansion}\\
\dot{\nul{\sigma}} + 2\nul{\theta} \nul{\sigma} &= \Weylfoc .
\label{eq:evolution_shear_rate}
\end{empheq}
They confirm the discussion of \S~\ref{sec:interpretation_Ricci_Weyl} regarding the respective role of Ricci and Weyl tensors in gravitational lensing, as $\Ricfoc$ stands for the source of convergence (antiexpansion), while $\Weylfoc$ is a source of shear. Note however the presence of $\abs{\sigma}^2$ in Eq.~\eqref{eq:evolution_expansion}, which adds to the focusing effect of $\Ricfoc$ and therefore makes Weyl lensing an \emph{indirect} source of convergence. This property is far from being obvious if one works within the Jacobi matrix formalism only.

\subsubsection{Initial conditions}

Due to the initial condition~$\mat{\jacobi}(v_\obs\leftarrow v_\obs)=\zero_2$ for the Jacobi matrix, the deformation matrix $\mat{\deformation}\define \dot{\mat{\jacobi}}\mat{\jacobi}^{-1}$ diverges at $O$, and so do in principle the optical scalars $\theta,\sigma$. This divergence can be analysed from the Taylor series of the Jacobi matrix in the vicinity of $v_\obs$, that is
\begin{equation}
\mat{\jacobi}(v\leftarrow v_\obs) = (v-v_\obs) \identity_2 + \frac{(v-v_\obs)^3}{3!}\,\mat{\tidal}_\obs +\mathcal{O}(v-v_\obs)^4,
\label{eq:initial_expansion_Jacobi}
\end{equation}
hence
\begin{align}
\mat{\deformation}
&= \pac{ \identity_2 + \mathcal{O}(v-v_\obs)^2 } \pac{ (v-v_\obs) \identity_2 + \mathcal{O}(v-v_\obs)^3 }^{-1} \\
&= (v-v_\obs)^{-1}\identity_2+ \mathcal{O}(v-v_\obs).
\label{eq:initial_deformation}
\end{align}
The deformation matrix therefore has a simple pole at $O$, which moreover only concerns its trace part (expansion rate). In other words, the initial conditions for the optical scalars are
\begin{align}
\nul{\theta} &= (v-v_\obs)^{-1} + \mathcal{O}(v-v_\obs),\\
\nul{\sigma} &= \mathcal{O}(v-v_\obs).
\end{align}
The initial divergence of $\theta$ makes it inconvenient for numerical calculations. It is preferable, in practice, to use directly the area~$A$ of the beam, or the angular distance~$D\e{A}$, as in the equations exhibited hereafter.

\subsubsection{Focusing theorem}\label{sec:focusing_theorem}

Replacing $\theta$ by $(\dd\sqrt{A}/\dd v)/\sqrt{A}$ in the first Sachs scalar equation~\eqref{eq:evolution_expansion} leads to the following evolution equation for the beam's area $A$, known as the \emph{focusing theorem}\index{focusing theorem},
\begin{empheq}[box=\fbox]{equation}\label{eq:focusing_theorem}
\ddf[2]{\sqrt{A}}{v} = \pa{ \Ricfoc - \abs{\sigma}^2 } \sqrt{A} \leq 0,
\end{empheq}
where we used $\Ricfoc\leq 0$, as ensured by the null energy condition (see \S~\ref{sec:interpretation_Ricci_Weyl}). Note that, by virtue of Eq.~\eqref{eq:interpretation_expansion}, $\sqrt{A}$ could also have been replaced by the angular distance $D\e{A}$ in the above. A similar introduction of $D\e{A}$ in Eq.~\eqref{eq:evolution_shear_rate} then leads to the following reformulation of the Sachs scalar equations\index{Sachs!equations (scalar) reformulated}
\begin{align}
\ddf[2]{D\e{A}}{v} &= \pa{\Ricfoc-\abs{\sigma}^2} D\e{A}, \\
\ddf{D\e{A}^2 \sigma}{v} &= D\e{A}^2 \Weylfoc,
\end{align}
which enjoy a better behaviour at $O$ than the original ones, and are therefore more adapted to numerical calculations of $D\e{A}(v)$.

Physically speaking, the focusing theorem tells us that $\sqrt{A}$ cannot increase more than linearly with $v$, and that \emph{any} gravitational effect tends to focus the beam. This seems to imply that there exist no divergent gravitational lenses. Yet that is wrong in general: for example, a beam going through the interior of a matter circle \emph{is} defocused. The point is that the focusing theorem is true for \emph{infinitesimal} light beams only. If this assumption is relaxed, if the beam has a finite extension, then its rays no longer have the same direction of propagation, and their separations no longer belong to the same screen space, etc. It is precisely the existence of a collection of different screen spaces within a finite beam that can make it locally focused and globally defocused.




%% file: beam.pdf_tex
\begingroup%
  \makeatletter%
  \providecommand\color[2][]{%
    \errmessage{(Inkscape) Color is used for the text in Inkscape, but the package 'color.sty' is not loaded}%
    \renewcommand\color[2][]{}%
  }%
  \providecommand\transparent[1]{%
    \errmessage{(Inkscape) Transparency is used (non-zero) for the text in Inkscape, but the package 'transparent.sty' is not loaded}%
    \renewcommand\transparent[1]{}%
  }%
  \providecommand\rotatebox[2]{#2}%
  \ifx\svgwidth\undefined%
    \setlength{\unitlength}{289.64615479bp}%
    \ifx\svgscale\undefined%
      \relax%
    \else%
      \setlength{\unitlength}{\unitlength * \real{\svgscale}}%
    \fi%
  \else%
    \setlength{\unitlength}{\svgwidth}%
  \fi%
  \global\let\svgwidth\undefined%
  \global\let\svgscale\undefined%
  \makeatother%
  \begin{picture}(1,0.6378437)%
    \put(0,0){\includegraphics[width=\unitlength]{beam.pdf}}%
    \put(0.27955341,0.33754211){\color[rgb]{0.66666667,0,0}\makebox(0,0)[lb]{\smash{$\vect{k}$}}}%
    \put(0.94457138,0.6140539){\color[rgb]{0,0,0}\makebox(0,0)[lb]{\smash{$O$}}}%
    \put(-0.00174782,0.23587037){\color[rgb]{0,0,0}\makebox(0,0)[lb]{\smash{$\vect{e}_1$}}}%
    \put(0.12025023,0.40258392){\color[rgb]{0,0,0}\makebox(0,0)[lb]{\smash{$\vect{e}_2$}}}%
    \put(0.1798075,0.06483697){\color[rgb]{0,0,0}\rotatebox{97.7448872}{\makebox(0,0)[lb]{\smash{$y^1=\text{cst}$}}}}%
    \put(0.24912901,0.1503908){\color[rgb]{0,0,0}\rotatebox{-0.82619252}{\makebox(0,0)[lb]{\smash{$y^2=\text{cst}$}}}}%
    \put(0.60533733,0.17820361){\color[rgb]{0.66666667,0,0}\rotatebox{54.3750334}{\makebox(0,0)[lb]{\smash{$v$ increases}}}}%
    \put(0.06984845,0.34583887){\color[rgb]{0,0,1}\makebox(0,0)[lb]{\smash{$\vect{\xi}$}}}%
  \end{picture}%
\endgroup%

%% file: frame_independence.pdf_tex
\begingroup%
  \makeatletter%
  \providecommand\color[2][]{%
    \errmessage{(Inkscape) Color is used for the text in Inkscape, but the package 'color.sty' is not loaded}%
    \renewcommand\color[2][]{}%
  }%
  \providecommand\transparent[1]{%
    \errmessage{(Inkscape) Transparency is used (non-zero) for the text in Inkscape, but the package 'transparent.sty' is not loaded}%
    \renewcommand\transparent[1]{}%
  }%
  \providecommand\rotatebox[2]{#2}%
  \ifx\svgwidth\undefined%
    \setlength{\unitlength}{187.85895996bp}%
    \ifx\svgscale\undefined%
      \relax%
    \else%
      \setlength{\unitlength}{\unitlength * \real{\svgscale}}%
    \fi%
  \else%
    \setlength{\unitlength}{\svgwidth}%
  \fi%
  \global\let\svgwidth\undefined%
  \global\let\svgscale\undefined%
  \makeatother%
  \begin{picture}(1,0.6027128)%
    \put(0,0){\includegraphics[width=\unitlength]{frame_independence.pdf}}%
    \put(0.3988846,0.40685859){\color[rgb]{0.66666667,0,0}\makebox(0,0)[lb]{\smash{$\vect{d}$}}}%
    \put(0.17590737,0.47612806){\color[rgb]{0,0,1}\makebox(0,0)[lb]{\smash{$\vect{\xi}_\perp$}}}%
    \put(0.42512388,0.24417629){\color[rgb]{0,0,1}\makebox(0,0)[lb]{\smash{$\vect{\zeta}_\perp$}}}%
    \put(0.16491117,0.3330998){\color[rgb]{0,0,0}\makebox(0,0)[lb]{\smash{$\ell_{12}$}}}%
    \put(0.6128512,0.42017008){\color[rgb]{0.83137255,0,0}\makebox(0,0)[lb]{\smash{$\gamma_1$}}}%
    \put(0.73568103,0.34814833){\color[rgb]{0.83137255,0,0}\makebox(0,0)[lb]{\smash{$\gamma_3$}}}%
    \put(0.49859806,0.52438697){\color[rgb]{0.83137255,0,0}\makebox(0,0)[lb]{\smash{$\gamma_2$}}}%
    \put(0.294551,0.19543886){\color[rgb]{0,0,0}\makebox(0,0)[lb]{\smash{$\ell_{13}$}}}%
    \put(0.27566827,0.33496108){\color[rgb]{0,0,0}\makebox(0,0)[lb]{\smash{$\vartheta$}}}%
  \end{picture}%
\endgroup%

%% file: interpretation_Jacobi.pdf_tex
\begingroup%
  \makeatletter%
  \providecommand\color[2][]{%
    \errmessage{(Inkscape) Color is used for the text in Inkscape, but the package 'color.sty' is not loaded}%
    \renewcommand\color[2][]{}%
  }%
  \providecommand\transparent[1]{%
    \errmessage{(Inkscape) Transparency is used (non-zero) for the text in Inkscape, but the package 'transparent.sty' is not loaded}%
    \renewcommand\transparent[1]{}%
  }%
  \providecommand\rotatebox[2]{#2}%
  \ifx\svgwidth\undefined%
    \setlength{\unitlength}{243.20488281bp}%
    \ifx\svgscale\undefined%
      \relax%
    \else%
      \setlength{\unitlength}{\unitlength * \real{\svgscale}}%
    \fi%
  \else%
    \setlength{\unitlength}{\svgwidth}%
  \fi%
  \global\let\svgwidth\undefined%
  \global\let\svgscale\undefined%
  \makeatother%
  \begin{picture}(1,0.49998996)%
    \put(0,0){\includegraphics[width=\unitlength]{interpretation_Jacobi.pdf}}%
    \put(0.34558757,0.28105764){\color[rgb]{0.66666667,0,0}\makebox(0,0)[lb]{\smash{$\vect{d}$}}}%
    \put(0.03267735,0.37730895){\color[rgb]{0,0.50196078,0}\makebox(0,0)[lb]{\smash{$\vect{s}_1$}}}%
    \put(0.2219387,0.05614278){\color[rgb]{0,0.50196078,0}\makebox(0,0)[lb]{\smash{$\vect{s}_2$}}}%
    \put(0.65878389,0.42518244){\color[rgb]{0,0,1}\makebox(0,0)[lb]{\smash{$\vect{\theta}_O$}}}%
    \put(0.4621618,0.18342656){\color[rgb]{0,0,0}\rotatebox{18.75952186}{\makebox(0,0)[lb]{\smash{$\dd\ell=\omega_\obs \dd v$}}}}%
    \put(0.93398699,0.33207116){\color[rgb]{0,0,0}\makebox(0,0)[lb]{\smash{$O$}}}%
    \put(0.17016661,0.27724703){\color[rgb]{0,0,1}\makebox(0,0)[lb]{\smash{$\dd\xi^A_\obs \vect{s}_A$}}}%
  \end{picture}%
\endgroup%

%% file: Jacobi_decomposition.pdf_tex
\begingroup%
  \makeatletter%
  \providecommand\color[2][]{%
    \errmessage{(Inkscape) Color is used for the text in Inkscape, but the package 'color.sty' is not loaded}%
    \renewcommand\color[2][]{}%
  }%
  \providecommand\transparent[1]{%
    \errmessage{(Inkscape) Transparency is used (non-zero) for the text in Inkscape, but the package 'transparent.sty' is not loaded}%
    \renewcommand\transparent[1]{}%
  }%
  \providecommand\rotatebox[2]{#2}%
  \ifx\svgwidth\undefined%
    \setlength{\unitlength}{463.4321068bp}%
    \ifx\svgscale\undefined%
      \relax%
    \else%
      \setlength{\unitlength}{\unitlength * \real{\svgscale}}%
    \fi%
  \else%
    \setlength{\unitlength}{\svgwidth}%
  \fi%
  \global\let\svgwidth\undefined%
  \global\let\svgscale\undefined%
  \makeatother%
  \begin{picture}(1,0.74616391)%
    \put(0,0){\includegraphics[width=\unitlength]{Jacobi_decomposition.pdf}}%
    \put(0.69163361,0.68642487){\color[rgb]{0,0.50196078,0}\makebox(0,0)[lb]{\smash{$\vect{s}_1$}}}%
    \put(0.15965917,0.67779359){\color[rgb]{0,0,0}\rotatebox{-15.00000022}{\makebox(0,0)[lb]{\smash{observed image}}}}%
    \put(0.12513416,0.2375997){\color[rgb]{0,0,0}\rotatebox{30.00000011}{\makebox(0,0)[lb]{\smash{physical source}}}}%
    \put(0.39270302,0.6432686){\color[rgb]{0,0,0}\makebox(0,0)[lb]{\smash{1. shear}}}%
    \put(0.64300937,0.45338102){\color[rgb]{0,0,0}\rotatebox{-90}{\makebox(0,0)[lb]{\smash{2. rotation}}}}%
    \put(0.42722804,0.28938721){\color[rgb]{0,0,0}\makebox(0,0)[lb]{\smash{3. scaling}}}%
    \put(0.44449054,0.25486219){\color[rgb]{0,0,0}\makebox(0,0)[lb]{\smash{$\times D\e{A}$}}}%
    \put(0.59985311,0.67779362){\color[rgb]{0.66666667,0.26666667,0}\makebox(0,0)[lb]{\smash{$\vect{s}_-$}}}%
    \put(0.56259411,0.51362479){\color[rgb]{0.66666667,0.26666667,0}\makebox(0,0)[lb]{\smash{$\vect{s}_+$}}}%
    \put(0.74436181,0.43737801){\color[rgb]{0,0,0}\makebox(0,0)[lb]{\smash{$\times \ex{-\gamma}$}}}%
    \put(0.81258379,0.6484266){\color[rgb]{0,0,0}\makebox(0,0)[lb]{\smash{$\times \ex{\gamma}$}}}%
    \put(0.89051573,0.29279367){\color[rgb]{0,0,0}\makebox(0,0)[lb]{\smash{$\psi$}}}%
    \put(0.64300937,0.71231863){\color[rgb]{0.66666667,0.26666667,0}\makebox(0,0)[lb]{\smash{$\varphi$}}}%
    \put(0.51892936,0.55457215){\color[rgb]{0,0.50196078,0}\makebox(0,0)[lb]{\smash{$\vect{s}_2$}}}%
    \put(0.28110103,0.14992018){\color[rgb]{0,0,0}\makebox(0,0)[lb]{\smash{$10\U{kpc}$}}}%
    \put(0.26256364,0.45358132){\color[rgb]{0,0,0}\makebox(0,0)[lb]{\smash{$3\U{arcmin}$}}}%
  \end{picture}%
\endgroup%

%% file: deformation_rate.pdf_tex
\begingroup%
  \makeatletter%
  \providecommand\color[2][]{%
    \errmessage{(Inkscape) Color is used for the text in Inkscape, but the package 'color.sty' is not loaded}%
    \renewcommand\color[2][]{}%
  }%
  \providecommand\transparent[1]{%
    \errmessage{(Inkscape) Transparency is used (non-zero) for the text in Inkscape, but the package 'transparent.sty' is not loaded}%
    \renewcommand\transparent[1]{}%
  }%
  \providecommand\rotatebox[2]{#2}%
  \ifx\svgwidth\undefined%
    \setlength{\unitlength}{449.9171875bp}%
    \ifx\svgscale\undefined%
      \relax%
    \else%
      \setlength{\unitlength}{\unitlength * \real{\svgscale}}%
    \fi%
  \else%
    \setlength{\unitlength}{\svgwidth}%
  \fi%
  \global\let\svgwidth\undefined%
  \global\let\svgscale\undefined%
  \makeatother%
  \begin{picture}(1,0.3111458)%
    \put(0,0){\includegraphics[width=\unitlength]{deformation_rate.pdf}}%
    \put(0.14112319,0.28910181){\color[rgb]{0,0.50196078,0}\makebox(0,0)[lb]{\smash{$\vect{s}_1$}}}%
    \put(-0.00112521,0.18706334){\color[rgb]{0,0.50196078,0}\makebox(0,0)[lb]{\smash{$\vect{s}_2$}}}%
    \put(0.46118209,0.28910181){\color[rgb]{0,0.50196078,0}\makebox(0,0)[lb]{\smash{$\vect{s}_1$}}}%
    \put(0.31893369,0.18706334){\color[rgb]{0,0.50196078,0}\makebox(0,0)[lb]{\smash{$\vect{s}_2$}}}%
    \put(0.63899259,0.1868051){\color[rgb]{0,0.50196078,0}\makebox(0,0)[lb]{\smash{$\vect{s}_2$}}}%
    \put(0.78124099,0.28910181){\color[rgb]{0,0.50196078,0}\makebox(0,0)[lb]{\smash{$\vect{s}_1$}}}%
    \put(0.14112319,0.25353971){\color[rgb]{0,0,0}\makebox(0,0)[lb]{\smash{$\alpha=\psi+\varphi$}}}%
    \put(0.40783894,0.19130604){\color[rgb]{0,0,0}\makebox(0,0)[lb]{\smash{$\alpha$}}}%
    \put(0.35449579,0.24464919){\color[rgb]{0,0,0.50196078}\makebox(0,0)[lb]{\smash{$\psi+\varphi$}}}%
    \put(0.67455469,0.24464919){\color[rgb]{0,0,0.50196078}\makebox(0,0)[lb]{\smash{$\psi+\varphi$}}}%
    \put(0.73597463,0.20129773){\color[rgb]{0,0,0}\makebox(0,0)[lb]{\smash{$\alpha$}}}%
    \put(-0.00112521,0.00460501){\color[rgb]{0,0,0}\makebox(0,0)[lb]{\smash{1. $\alpha=\psi+\varphi$}}}%
    \put(0.31893369,0.00460501){\color[rgb]{0,0,0}\makebox(0,0)[lb]{\smash{2. $\alpha=\psi+\varphi+\dfrac{\pi}{2}$}}}%
    \put(0.63899259,0.00460501){\color[rgb]{0,0,0}\makebox(0,0)[lb]{\smash{3. $\alpha=\psi+\varphi+\dfrac{\pi}{4}$}}}%
    \put(0.39005789,0.08461974){\color[rgb]{0,0,1}\makebox(0,0)[lb]{\smash{$\dot{\gamma}<0$}}}%
    \put(0.03443689,0.09351026){\color[rgb]{0,0,1}\makebox(0,0)[lb]{\smash{$\dot{\gamma}>0$}}}%
    \put(0.83458414,0.24464919){\color[rgb]{0,0,1}\makebox(0,0)[lb]{\smash{$\dot{\gamma}=0$}}}%
  \end{picture}%
\endgroup%

%% file: chapter_3.tex
\lettrine{T}{he} greatest achievement of the theory of relativity is certainly the unification of the concepts of space and time, which implies in particular that the notion of spatial distance is fundamentally ambiguous. Yet astronomy---hence, to some extent, cosmology---is all about distance measurements, which encourages us to try to generalise the notion of distance in a relativistic context, rather than abandoning it. We here review a number of attempts to address this issue, both from purely theoretical and observational points of view. Throughout this chapter, at least ten different well-defined and well-motivated notions of distance are presented.

\bigskip

\minitoc

\newpage

\section{Defining distances}

In this section, we expose several theoretical constructions for characterising spatial distances in relativity. The first difficulty being to artificially disentangle space from time, we start by investigating the issue in the context of special relativity (\S~\ref{sec:distances_special_relativity}), where we show that distances can be univocally defined. The conclusion is nevertheless drastically different if we allow observers to be noninertial, or spacetime to be curved (\S~\ref{sec:distances_GR}).

\subsection{In special relativity}\label{sec:distances_special_relativity}

We suppose in all this subsection that spacetime geometry is described by the Minkowski metric, and we use an \emph{inertial} coordinate system $\{x^\alpha\}$ for which its components read $\vect{g}(\vect{\partial}_\alpha,\vect{\partial}_\beta)=\eta_{\alpha\beta}\define[\mathrm{diag}(-1,1,1,1)]_{\alpha\beta}$.

\subsubsection{Distance between two events}

Let $A$ and $B$ be two events with coordinates $x^\alpha_{A,B}$. Their \emph{spatio-temporal} separation is defined as the norm of the four-vector\footnote{This notion of four-vector connecting two arbitrary events is meaningless in general, but Minkowski spacetime is an exception.}~$\vect{AB}$ connecting them, that is
\begin{equation}
\Delta s^2(A,B) \define \vect{g}(\vect{AB},\vect{AB}) = \eta_{\alpha\beta} \Delta x^\alpha \Delta x^\beta,
\label{eq:distance_A_B}
\end{equation}
with $\Delta x^\alpha \define x^\alpha_B-x^\alpha_A$. This separation is timelike, null, or spacelike respectively for $\Delta s^2<0$, $\Delta s^2=0$, or $\Delta s^2>0$ respectively. In the third case, it can be interpreted as the square of the spatial distance between $A$ and $B$, 
\begin{empheq}[box=\fbox]{equation}
D(A,B) \define \sqrt{\Delta s^2(A,B)}.
\end{empheq}

\begin{figure}[t]
\centering
	\begin{subfigure}[t]{0.26\linewidth}
	\centering
	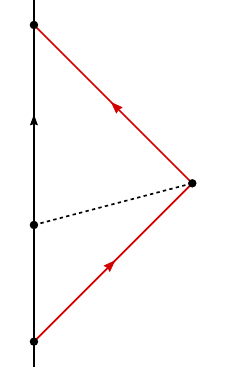
	\caption{Distance~$D(A,B)$ between two events $A$, $B$.}
	\label{fig:Synge_distance}
	\end{subfigure}
\hspace{1cm}
	\begin{subfigure}[t]{0.28\linewidth}
	\centering
	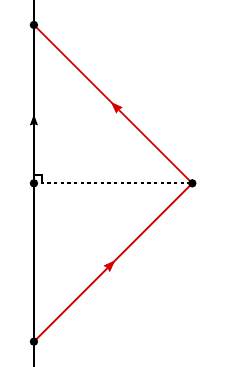
	\caption{Distance~$D(\wl,B)$ between an inertial worldline~$\wl$ and an event $B$.}
	\end{subfigure}
\hspace{1cm}
	\begin{subfigure}[t]{0.3\linewidth}
	\centering
	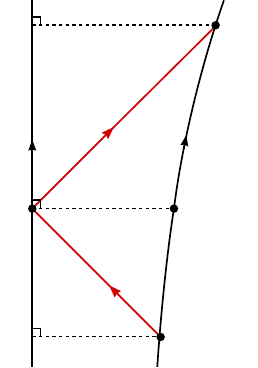
	\caption{Distances between an inertial worldline~$\wl$ and a (possible noninertial) particle~$\wl'$.}
	\label{fig:distance_observer-particle_Minkowski}
	\end{subfigure}	
\caption{Defining distances in special relativity.}
\label{fig:theoretical distances}
\end{figure}

As originally shown in Ref.~\cite{Robb1911}, this notion of distance can also be expressed in terms of time measurements only. Consider an arbitrary inertial observer whose worldline~$\wl$ passes through $A$. Without loss of generality, we assume the coordinate system~$\{x^\alpha\}$ to be adapted to this observer, in particular $x^0=t$ is her proper time, and $t_A$ is the date of $A$ in her rest frame. We define two events~$E,R\in\wl$  from the construction depicted in Fig.~\ref{fig:Synge_distance}: the observer emits a photon at $E$, which is reflected at $B$ back to the observer, who finally receives it at $R$. We call $t_E,t_R$ the corresponding dates.

This construction implies that both the four-vectors $\vect{EB}$ and $\vect{BR}$ are null, because proportional to the emitted and reflected wave four-vectors. From $\vect{EB}=\vect{EA}+\vect{AB}$ we deduce
\begin{align}
0 
&= \vect{g}(\vect{EB},\vect{EB}) \\
&= \vect{g}(\vect{EA},\vect{EA}) + \vect{g}(\vect{EA},\vect{AB}) + \vect{g}(\vect{AB},\vect{AB}) \\
&= -(t_A-t_E)^2 + (t_A-t_E) \, \vect{g}(\vect{\partial}_t,\vect{AB}) + \vect{g}(\vect{AB},\vect{AB}),
\label{eq:Synge_int_1}
\end{align}
%
%
where we used $\vect{EA}=(t_A-t_E)\vect{\partial}_t$ and $\vect{g}(\vect{\partial}_t,\vect{\partial}_t)=-1$; the same calculation with $\vect{BR}=\vect{BA}+\vect{AR}$ then yields
\begin{equation}
0 = -(t_R-t_A)^2 - (t_R-t_A) \, \vect{g}(\vect{\partial}_t,\vect{AB}) + \vect{g}(\vect{AB},\vect{AB}).
\label{eq:Synge_int_2}
\end{equation}
Combining Eqs.~\eqref{eq:Synge_int_1}, \eqref{eq:Synge_int_2} to eliminate the scalar product between $\vect{\partial}_t$ and $\vect{AB}$, we finally obtain
\begin{equation}
\vect{g}(\vect{AB},\vect{AB}) = (t_A-t_E)(t_R-t_A),
\end{equation}
which is positive iff $t_A\in[t_E,t_R]$, that is, as expected, iff $A$ lies outside the lightcone of $B$. When this condition is fulfilled, the above relation yields the \emph{Synge formula}~\cite{GourgoulhonRR}\index{distance!Synge formula}\index{Synge!formula|see{distance}} for the distance between two events
\begin{empheq}[box=\fbox]{equation}
D(A,B) = \sqrt{(t_A-t_E)(t_R-t_A)}.
\label{eq:Synge_distance}
\end{empheq}
Although it involves here the proper time of a particular observer, $D(A,B)$ is by definition a Lorentz-invariant quantity, thus any inertial observer can calculate $D(A,B)$ using the Synge formula.

Observer independence is, however, the reason why $D(A,B)$ actually fails in describing what we usually call a spatial distance. For example, the couple of events corresponding to ($A$) the emission of a photon by this text, and ($B$) its reception by your eye, has $D(A,B)=0$ by definition. Yet the text does not touch your eye, hopefully. The reason for such a failure is that the natural questions associated with spatial distances concern the distance between an event and an worldline (how far is this supernova explosion from us?) or between two worldlines (how far is this text from your eye?), rather than the distance between two events.

\subsubsection{Distance between an event and an inertial observer}

Let $\wl$ be the worldline of an inertial observer and $B$ an arbitrary event. The distance~$D(\wl,B)$ between them is naturally defined as $D(A^*,B)$, such that $A^*\in\wl$ and $B$ are \emph{simultaneous} in the observer's frame. The notion of simultaneity invoked here admits three equivalent definitions in special relativity:\index{simultaneity!in special relativity}
\begin{enumerate}
\item $t_{A^*}=t_B$ in an inertial coordinate system associated with $\wl$.
\item $t_R-t_{A^*}=t_{A^*}-t_E$, where $E,R\in\wl$ are, as in the previous paragraph, the emission and reception by the observer of a photon reflected at $B$. This definition is known as the Einstein-Poincar\'{e} simultaneity criterion~\cite{1905AnP...322..891E,PoincareOeuvres}.\index{simultaneity!Einstein-Poincaré criterion}
\item $\vect{g}(\vect{A^* B},\vect{\partial}_t)=0$.
\end{enumerate}
Therefore
\begin{equation}
D(\wl,B) \define D(A^*,B) = \delta_{ab} x_B^a x_B^b = \frac{t_R-t_E}{2},
\end{equation}
where we have chosen $\wl$ as the origin of spatial coordinates $\{x^a\}_{a=1\ldots3}$ in the penultimate expression, while the last one have been obtained from the Synge formula~\eqref{eq:Synge_distance} by replacing $t_A$ by $t_{A^*}=(t_E+t_R)/2$. By virtue of the same Synge formula, it is easy to check that $D(\wl,B)$ is also the maximal distance between $B$ and any event of the observer's worldline,
\begin{equation}
D(\wl,B)=\underset{A\in\wl}{\mathrm{max}} \, D(A,B),
\end{equation}
provided this quantity exists.

Another possible characterisation of $D(\wl,B)$ relies on the affine parametrisation of null geodesics connecting $B$ with $\wl$. We have seen in \S~\ref{sec:physical_interpretation_affine_parameter} that the affine parameter~$v$ indeed represents the distance travelled by light, modulo a frequency factor. Here the geodesic equation is easily solved, e.g. from $B$ to $R$, leading in particular to $k^t=\cst=\omega_R$, where $\omega_R$ is the cyclic frequency measured by the observer at $R$. In other words, $\dd t/\dd v=\omega_R$, which we immediately integrate as $\omega_R(v_R-v_B)=t_R-t_B=D(\wl,B)$. We could also have chosen to integrate along the past ray (from $E$ to $B$), and get $\omega_E(v_B-v_E)=t_B-t_E=D(\wl,B)$ as well.

Summarising, the special-relativistic distance between a worldline and an event is univocally defined, because
\begin{empheq}[box=\fbox]{equation}
D(\wl,B) = \underset{A\in\wl}{\mathrm{max}} \, D(A,B) = \omega_R(v_R-v_B) = \frac{t_R-t_E}{2},
\end{empheq}
though each expression actually corresponds to a different geometrical construction.

\newcommand{\tAstar}{\frac{t_E+t_R}{2}}

\subsubsection{Distance between a particle and an inertial observer}

Consider a (possibly noninertial) particle following a worldline~$\wl'$. The distance between this particle and the inertial observer following $\wl$ is, in general, a time-dependent quantity, because of their relative motion. There are three possible definitions for this distance represented in Fig.~\ref{fig:distance_observer-particle_Minkowski}: let $P$ be an event on $\wl$ and $t$ the associated date in the observer's frame,
\begin{itemize}
\item the \emph{instantaneous distance} is $D\e{inst}(t)\define D(\wl,I)$, where $I\in\wl'$ and $P$ are simultaneous in the observer's frame;
\item the \emph{retarded distance} reads $D\e{ret}(t)\define D(\wl,R)$, where $R\in\wl'$ is the emission of a photon received at $P$---it is the notion of distance involved, e.g., in special-relativistic electrodynamics~\cite{1975clel.book.....J}; and finally
\item the \emph{advanced distance} is $D\e{adv}(t)\define D(\wl,A)$, where $A\in\wl'$ is the reception of a photon emitted at $P$.
\end{itemize}
They all differ, except when they are constant.

\subsection{In general relativity}\label{sec:distances_GR}

In curved spacetime, or for non-inertial observers in Minkowski spacetime, the previous reasonings still apply locally, i.e. for small distances compared to
\begin{inparaenum}[(i)]
\item spacetime's curvature radii, and
\item the inverse of the observer's acceleration.
\end{inparaenum}
In the previous chapters, we tacitly took advantage of this property to univocally invoke physical distances---e.g. between two light spots on a screen---when they were infinitesimal. When they are not, many constructions which coincide locally turn out to differ globally. 

\subsubsection{Distance between two events}

\begin{figure}[h!]
\centering
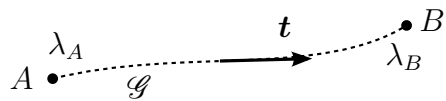
\caption{Geodesic~$\mathscr{G}$ linking two events $A$, $B$, and its tangent vector $t^\mu=\dd x^\mu/\dd\lambda$.}
\label{fig:distance_events_curved}
\end{figure}

Let $A,B$ be two events, and assume that they are connected by a unique geodesic~$\mathscr{G}$ affinely parametrised by $\lambda$, as represented in Fig.~\ref{fig:distance_events_curved}. In mathematical terms, $B$ is said to lie in a normal neighbourhood of $A$, and conversely. A natural extension of the spatio-temporal separation~$\Delta s^2$ defined in \S~\ref{sec:distances_special_relativity} to the general-relativistic case is then given by (twice) \emph{Synge's worldfunction}\index{Synge!worldfunction}
\begin{equation}
\sigma(A,B) \define \frac{1}{2} (\lambda_B-\lambda_A) \int_A^B t^\mu t_\mu \; \dd\lambda,
\end{equation}
where $t^\mu \define \dd x^\mu/\dd\lambda$ is the tangent vector to $\mathscr{G}$ associated with $\lambda$. The geodesic equation $\Dd \vect{t}/\dd\lambda=\vect{0}$ implies that $t^\mu t_\mu$ is a constant along $\mathscr{G}$, so $\sigma(A,B)=(\lambda_B-\lambda_A)^2 t^\mu t_\mu/2$. Just like $\Delta s^2(A,B)$ in special relativity, the sign of $\sigma(A,B)$ dictates the nature of $\mathscr{G}$:
\begin{itemize}
\item $\sigma(A,B)<0 \Leftrightarrow \mathscr{G}$ is timelike. In this case, the affine parameter~$\lambda$ can be chosen as the proper time~$\tau$ along $\mathscr{G}$, so that $t^\mu t_\mu = -1$, and $\sigma(A,B)=-(\tau_B-\tau_A)^2/2$.
\item $\sigma(A,B)=0 \Leftrightarrow \mathscr{G}$ is null. 
\item $\sigma(A,B)>0 \Leftrightarrow \mathscr{G}$ is spacelike. If we choose an affine parameter~$\lambda=s$  such that $t^\mu t_\mu = 1$, then $\sigma(A,B)=(s_B-s_A)^2/2$.
\end{itemize}

In the last case, we conclude that
\begin{empheq}[box=\fbox]{equation}
D(A,B) \define \sqrt{2\sigma(A,B)} =  |s_B-s_A|
\label{eq:distance_A_B_curved}
\end{empheq}
generalises the special-relativistic notion of spatial distance between two events. There is however no equivalent of the Synge formula~\eqref{eq:Synge_distance} in general relativity, in the sense that a quantity of the form $\sqrt{(\tau_A-\tau_E)(\tau_R-\tau_A)}$---where $\tau$ is the proper time of an observer whose worldline~$\wl$ contains $A$, and $E,R\in\wl$ are the emission and the reception of a photon reflected at $B$---is observer dependent and does not coincide with Eq.~\eqref{eq:distance_A_B_curved}.

\subsubsection{Distances between an event and an observer}\index{simultaneity!in general relativity}

\begin{figure}[t]
\centering
	\begin{subfigure}[b]{0.4\linewidth}
	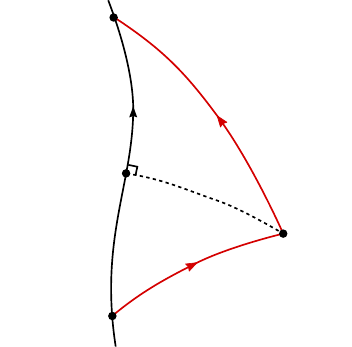
	\caption{Distances based on geodesics.}
	\label{fig:distances_worldline_event_curved}
	\end{subfigure}
\hspace{1cm}
	\begin{subfigure}[b]{0.4\linewidth}
	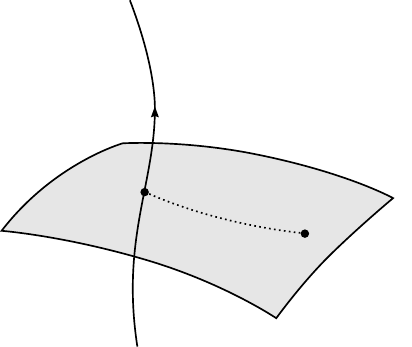
	\caption{Distance based on a foliation.}
	\label{fig:foliation-based_distance}
	\end{subfigure}
\caption{Defining distances between a worldline~$\wl$ and an event $B$ in general relativity.}
\end{figure}

Let $\wl$ be the worldline of a possibly noninertial observer and $B$ an event. A straightforward generalisation of the special-relativistic distance $D(\wl,B)$, that we shall call \emph{spatial-geodesic distance}\index{distance!spatial-geodesic}, is
\begin{empheq}[box=\fbox]{equation}
D\e{S}(\wl,B) \define \underset{A\in\wl}{\mathrm{max}} \, D(A,B),
\end{empheq}
where $D(A,B)$ is now given by Eq.~\eqref{eq:distance_A_B_curved}. An equivalent definition for the same quantity, whose construction is depicted in Fig.~\ref{fig:distances_worldline_event_curved}, is the following:
find the assumed-to-be unique spacelike geodesic~$\mathscr{G}^*$ starting from $B$ and intersecting $\wl$ orthogonally;
call $A^*$ their intersection;
then $D\e{S}(\wl,B)=D(A^*,B)$. Note the similarity with the construction of Fermi normal coordinates in the vicinity of a timelike geodesic~\cite{2004rtmb.book.....P}.

The equivalence between both definitions follows from the properties of Synge's worldfunction. One shows~\cite{2004LRR.....7....6P} that if the event $A$ is displaced by $\delta x^\mu_A=u^\mu_A \delta \tau$ along $\wl$, where $\vect{u}_A$ denotes the four-velocity of the observer at $A$, then $\sigma(A,B)$ changes by
\begin{equation}
\delta \sigma = (\lambda_B-\lambda_A) \, (t^\mu u_\mu)_A \delta \tau.
\end{equation}
We conclude that $\sigma(A,B)$ is stationary with respect to displacements of $A$ along $\wl$ iff the geodesic $\mathscr{G}$ along which it is computed is orthogonal to $\wl$. Moreover, this stationary point is a maximum if we suppose that it is unique, i.e., if we suppose that only one geodesic $\mathscr{G}^*$ connects $B$ and $\wl$ orthogonally. Indeed, $\sigma(A,B)$ is positive if $A$ lies between the intersections $E$ and $R$ of $\wl$ with the lightcone of $B$ [see Fig.~\ref{fig:distances_worldline_event_curved}] where it vanishes. Hence a unique stationary point between them must be a maximum.

Besides distances, the above construction also defines a notion of simultaneity: $A^*\in\wl$ and $B$ are \emph{spatial-geodesic simultaneous} for the observer iff the geodesic~$\mathscr{G}^*$ connecting them is orthogonal to $\wl$ at $A^*$. Contrary to the special-relativistic case, this prescription for simultaneity does not necessarily coincide with the Einstein-Poincar\'{e} criterion: $\tau_R-\tau_{A^*}$ and $\tau_{A^*}-\tau_E$, where $E$ and $R$ are defined as before, are different in general. However, the equality can be shown to approximately hold, up to second order in $D\e{S}(A,B)$, if $\wl$ is a timelike geodesic.

We thus expect from constructions based on light rays to define distinct notions of distance between $\wl$ and $B$. For example, the \emph{radar distance}\index{distance!radar} defined as half the duration of light's round trip from $\wl$ to $B$,
\begin{empheq}[box=\fbox]{equation}
D\e{R}(\wl,B) \define \frac{\tau_R-\tau_E}{2},
\label{eq:radar_distance}
\end{empheq}
has no reason to be equal to $D\e{S}$. Similarly, a \emph{null-geodesic distance}, relying on the affine parametrisation of a null geodesic connecting $B$ to $\wl$, e.g.,
\begin{empheq}[box=\fbox]{equation}
D\e{N}(\wl,B) \define \omega_R (v_R-v_B),
\end{empheq}
generically differs from both $D\e{S}$ and $D\e{R}$. It could also have been defined as $\omega_E(v_B-v_E)$ leading to a fourth distinct distance.

A last option for connecting an event to a worldline consists in relying on a particular foliation of spacetime by spacelike hypersurfaces. Suppose $\mathcal{M}= \bigcup \Sigma_t$, where $t$ is a label for the hypersurfaces $\Sigma_t$. Spacetime's metric $\vect{g}$ induces on each of them an intrinsic metric $\vect{h}(t)=\vect{g}|_{\Sigma_t}$. The associated Levi-Civita connection then allows us to define $\vect{h}$-geodesics on $\Sigma_t$ which are not, in general, $\vect{g}$-geodesics of $\mathcal{M}$. A \emph{foliation-based distance}\index{distance!based on a foliation}~$D\e{F}(\wl,B)$ between $\wl$ and $B$ can be constructed the following way, illustrated in Fig.~\ref{fig:foliation-based_distance}:
\begin{inparaenum}[(i)] \item identify the hypersurface $\Sigma_{t}$ such that $B\in\Sigma_{t}$; \item call $A_t\define \wl\cap\Sigma_t$ and $\mathscr{G}_{\vect{h}}$ the $\vect{h}$-geodesic of $\Sigma_t$ connecting $A_t$ with $B$; \item define 
\begin{empheq}[box=\fbox]{equation}
D\e{F}(\wl,B) \equiv D_{\vect{h}}(A_t,B) = \sqrt{2\sigma_{\vect{h}}(A_t,B)},
\end{empheq}
\end{inparaenum}
where $\sigma_{\vect{h}}$ is Synge's worldfunction for the manifold~$\Sigma_t$ equipped with $\vect{h}(t)$.

This procedure can seem quite natural for spacetimes which admit a preferred foliation, for example to exhibit their stationarity (Schwarzschild, Reissner-Nordstr\o m, Kerr, Majumdar-Papapetrou, etc.), or their homogeneity\footnote{In cosmology, the $D\e{F}$ associated with the foliation of spacetime by homogeneous hypersurfaces is usually referred to as the physical distance, not because it is more physical than $D\e{S}$, $D\e{N}$, or $D\e{R}$, but rather to distinguish it from the comoving (or conformal) distance. It is also the notion of distance which is implicitly chosen in, e.g., Buchert's approach to the backreaction issue~\cite{2011CQGra..28p4007B}.} (Friedmann-Lema\^itre, Bianchi); however, for $D\e{F}$ to locally coincide with the other notions of distance, the foliation must be orthogonal to the observer's worldline~$\wl$. A way to construct such a foliation in \emph{any} spacetime consists in using the Einstein-Poincar\'{e} criterion: for any event~$A\in\wl$, the set of all $B$ which are Einstein-Poincar\'{e} simultaneous with $A$ indeed form a hypersurface which is orthogonal to $\wl$.


\subsubsection{Distance between a particle and an observer}

Like in special relativity, the instantaneous, retarded, and advanced distances between an observer following $\wl$ and a particle following $\wl'$ can be constructed from each of the event-worldline distances exposed in the previous paragraph. The associated prescriptions for simultaneity can be used for defining the event $I\in\wl'$ simultaneous to a given $P\in\wl$, in the case of the instantaneous distance.

\section{Measuring distances}\label{sec:measuring_distances}

In the previous section, we have demonstrated the ambiguity of the notion of spatial distance in general relativity by proposing half a dozen theoretically well-motivated definitions for it. However, none of them is actually observable---except the radar distance, but it is practically very limited. In the present section, we thus adopt a complementary approach, reviewing the main observables used to measure distances in astronomy and cosmology, and the different notions of distance they define. Table~\ref{tab:distances} provides a summary of what follows, together with some orders of magnitude.

\renewcommand{\arraystretch}{1.2}
\begin{table}[h!]
\centering
\begin{tabular}{|ccccc|}
\hline 
\rowcolor{lightgray}
\sf\textbf{observable} & \sf\textbf{distance}	& \sf\textbf{applicability} & \sf\textbf{range}	& \sf\textbf{precision} \\ 
\hline 
		time 						& radar $D\e{R}$ 		& Solar system & AU 		& $10^{-11}$ \\
\hline 
angle								& parallax $D\e{P}$ 	& Milky way 		&  kpc 		& 1-10\% \\
 										& angular $D\e{A}$		& extragalactic	& Gpc 		&  10-50\% (clusters) \\
 										& 									&						&				&	5\% (BAO)\\ 
\hline 
intensity								& luminosity $D\e{L}$ & extragalactic 	&  10 Mpc	& 5\% (Cepheids) \\
											&									&							&	Gpc			& 20\% (SNe) \\
\hline 
\end{tabular} 
\caption{Summary of the observable notions of distance in astronomy and cosmology, with their domains of applicability, and orders of magnitude for their maximum range and current level of precision. The precision on $D\e{R}$ refers to the Viking Earth-Mars distance measurement~\cite{1979ApJ...234L.219R}; on $D\e{P}$ to the objectives of the Gaia mission~\cite{Gaia}; on $D\e{A}$ to galaxy-cluster distances measured from the X-ray emission/SZ effect~\cite{2006ApJ...647...25B}, or to the BAO scale with BOSS~\cite{2015A&A...574A..59D}; finally the precisions on $D\e{L}$ for Cepheids and SNe are based respectively on Refs.~\cite{2005ApJ...627..579R} and \cite{2014A&A...568A..22B}.}
\label{tab:distances}
\end{table}

\subsection{Radar distance}

Already defined in Eq.~\eqref{eq:radar_distance} of the previous section, the radar distance~$D\e{R}$ corresponds to half the duration of a light signal's round trip between the observer and its target, as measured in the observer's frame.

This method is of daily use for short-distance measurements on the Earth, but it is not adapted to astronomy as it requires high-reflexivity objects. As such, it is limited to distance measurements within the Solar system. A notable example is the Earth-Moon distance, thanks to the five retroreflector arrays installed by the US missions Apollo 11, 14, 15, and Soviet missions Luna 17, 21. The associated Lunar Laser Ranging experiments have determined the radar distance to the Moon with a precision on the order of the centimetre, that is, enough to carry tests of the equivalence principle and Lorentz invariance~\cite{2006LRR.....9....3W,155,10891,18966}. The radar distances to Venus, Mars, and Cassini have also been used to measure the Shapiro time delay, which is a standard test of GR in the Solar system~\cite{1973grav.book.....M}.

The radar distance is also involved in gravitational wave detection experiments, such as the ground interferometers LIGO~\cite{LIGO}, VIRGO~\cite{VIRGO}, and the future space mission eLISA~\cite{eLISA}. All three are based on the same principle: as a gravitational wave propagates through the interferometer, the \emph{radar length} of each arm is affected differently, inducing a phase difference between light signals propagating inside, and therefore a characteristic interference signal.

\subsection{Parallax distance}

\subsubsection{Definition}

\begin{figure}[h!]
\centering
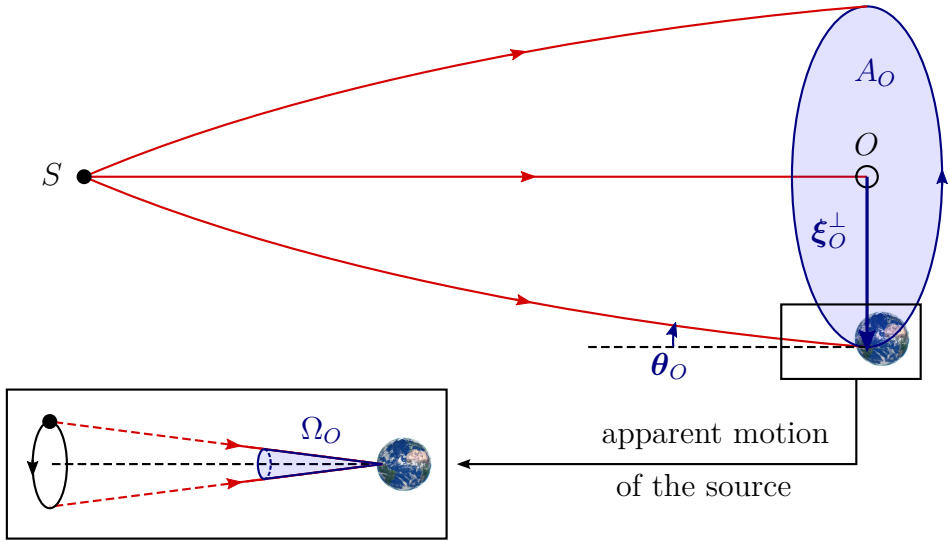
\caption{Solar parallax. The revolution of the Earth around the Sun induces an apparent motion of the source on the observer's celestial sphere. The solid angle~$\Omega_\obs$ corresponding to this apparent motion is smaller as the source is farther. The parallax distance is obtained by comparing $\Omega_\obs$ with the area~$A_\obs$ of the observer's motion---here $A_\obs=\pi(\mathrm{AU})^2$---according to $D\e{P}^2\define A_\obs/\Omega_\obs$. We chose to depict a beam with some focusing, which tends to \emph{increase} $D\e{P}$.}
\label{fig:parallax}
\end{figure}

Parallax is the apparent displacement of a light source caused by a displacement of its observer. The larger the distance between them, the smaller the parallax, so that a measurement of the apparent motion of the source by the observer, together with the knowledge of its own actual motion, is a method for measuring distances. For simplicity, we restrict here to the case where the observer's motion is orthogonal to the line of sight (see Fig.~\ref{fig:parallax}); the \emph{parallax distance}\index{distance!parallax}\index{parallax} is then
\begin{empheq}[box=\fbox]{equation}
D\e{P} \define \sqrt{\frac{A_\obs}{\Omega_\obs}},
\label{eq:parallax_definition}
\end{empheq}
where $A_\obs$ is the physical area of the observer's trajectory and $\Omega_\obs\ll 1$ is the solid angle occupied by the apparent trajectory of the source on the observer's celestial sphere.

In astronomy, solar parallax---due to the revolution of the Earth around the Sun---is the most common technique for determining the distance of stars within the Milky Way. It gave birth to the \emph{parsec}\index{parsec} unit, defined as the distance such that a source has a solar parallax of one arcsecond. The European astrometry satellite Hipparcos~\cite{Hipparcos}, launched in 1989, measured the parallax of 2.5 millions stars. It has been replaced in 2013 by Gaia~\cite{Gaia}, which is expected to deliver a catalogue of 1 billion stars, with a precision of $20$--$200\U{\micro as}$ on their parallax. This level of precision is the best that we can achieve today, even with Very Long Baseline Interferometry (VLBI) whose resolution is on the order $1\U{mas}$, setting the current limit on parallax distance measurements to $D\e{P}<10\U{kpc}$.

\subsubsection{Theoretical expression}

Let us relate the definition~\eqref{eq:parallax_definition} of the parallax distance to the properties of a light beam connecting the source to the observer. For that purpose, consider the beam delimited by all the rays emerging from $S$ and reaching a possible position of the Earth on its trajectory around the Sun, as depicted in Fig.~\ref{fig:parallax}. Let $O$ be, for example, the intersection between the beam with the Sun's worldline. Note that, contrary to the convention adopted in Chap.~\ref{chapter:beams}, this beam has a vertex at $S$ and a nonzero extension around $O$. The ecliptic plane then plays the role of screen space at $O$.

On the one hand, the area~$A_\obs$ of the Earth's trajectory is clearly identified with the area~$\dd^2 \xi^A_\obs$ of the beam at $O$, with $\xi^A_O$ the components of the separation vector~$\vect{\xi}$ over the Sachs basis at $O$. On the other hand, $\Omega_O=\omega_\obs^{-2}\dd^2\dot{\xi}^A_O$, where $\omega_\obs$ is the observed frequency since, as discussed in \S~\ref{sec:Jacobi_definition}, $\omega^{-1}|\dot{\xi}^A|$ represents the angle between two rays separated by~$\vect{\xi}$. We conclude that
\begin{equation}
D\e{P}^2 = \frac{\dd^2 \xi^A_\obs}{\omega_O^{-2}\dd^2 \dot{\xi}^A_\obs}
			= \frac{\omega_\obs^2}{\det\mat{\deformation}(\obs\leftarrow\source)},
\end{equation}
by definition~\eqref{eq:deformation_rate_matrix} of the deformation rate matrix~$\mat{\deformation}$. The parallax distance can therefore be expressed in terms of the optical scalars as
\begin{empheq}[box=\fbox]{equation}
D\e{P} = \frac{\omega_\obs}{\sqrt{\theta^2-\abs{\sigma}^2}},
\label{eq:parallax_optical_scalars}
\end{empheq}
where $\theta$ and $\sigma$ correspond to $\mat{\deformation}(O\leftarrow S)$, i.e. to an integration of the Sachs equation from the source to the observer. Note that our expression~\eqref{eq:parallax_optical_scalars} differs from the one given in Ref.~\cite{2013GReGr..45.2691J} and used in Ref.~\cite{1988ApJ...331..648R}, in which the shear rate~$\abs{\sigma}^2$ has been neglected.

Since $D\e{P}$ is an decreasing function of $\theta^2$ and an increasing function of $\abs{\sigma}^2$, the Sachs scalar equations~\eqref{eq:evolution_expansion}, \eqref{eq:evolution_shear_rate} imply that any gravitational lensing effect tends to \emph{increase} the parallax distance.

\subsection{Angular diameter distance}
\label{sec:angular_diameter_distance}

\subsubsection{Definition}

\begin{figure}[h!]
\centering
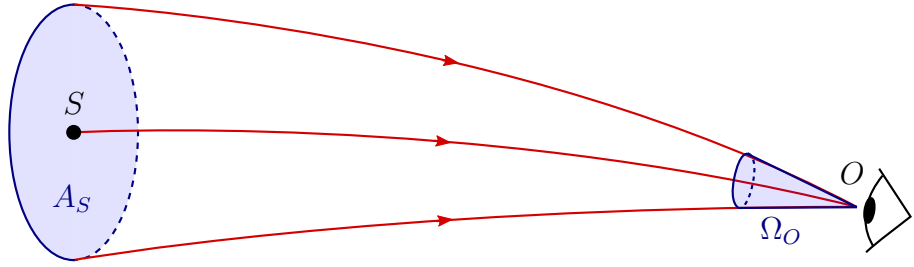
\caption{The angular diameter distance, or area distance, is obtained by comparing the actual and apparent sizes of a light source, according to $D^2\e{A}=A_\source/\Omega_\obs$.}
\label{fig:angular_distance}
\end{figure}

The notion of angular diameter distance, or area distance\index{distance!angular diameter}\index{distance!area}\footnote{Angular diameter distance and area distance can actually be considered two slightly different notion~\cite{2004LRR.....7....9P}. Strictly speaking, the former is a comparison between the proper and apparent diameter of the source, which thus depends on its orientation if some shear is at work. The area distance do not suffer from this ambiguity.} is based on the fact that a given object appears smaller as it lies farther form us. It is defined by
\begin{empheq}[box=\fbox]{equation}
D\e{A} \define \sqrt{\frac{A_\source}{\Omega_\obs}},
\label{eq:angular_distance_definition}
\end{empheq}
where $A_\source$ is the physical area of the light source, and $\Omega_\obs\ll 1$ its apparent angular size for the observer, as depicted in Fig.~\ref{fig:angular_distance}.

The difficulty of measuring angular distances in astronomy is that it requires standard rulers, i.e. sources whose size is known, or at least can be calibrated by independent experiments. An important example in cosmology is the Baryon Acoustic Oscillation (BAO) scale, which corresponds to the maximum distance travelled by a sound wave in the primordial Universe. It can be extracted from the analysis of the anisotropies of the Cosmic Microwave Background (CMB), and in the distribution of galaxies. The angular diameter distance is also naturally involved in strong gravitational lensing and time delays experiments.

\subsubsection{Theoretical expression}

The angular diameter distance is directly related to the determinant of the Jacobi matrix (see \S~\ref{sec:Jacobi_definition}) via
\begin{empheq}[box=\fbox]{equation}
D\e{A} = \omega_\obs\sqrt{\det\mat{\jacobi}(v_\source \leftarrow v_\obs)}.
\label{eq:angular_distance_theoretical}
\end{empheq}
The notation $v_\source\leftarrow v_\obs$ indicates that the Jacobi matrix equation~\eqref{eq:Jacobi_matrix_equation} must be solved from the observation event~$O$, where all the rays converge, to the source event $S$.

By comparing Eqs.~\eqref{eq:parallax_optical_scalars} and Eq.~\eqref{eq:angular_distance_theoretical}, we immediately deduce that $D\e{A}\not=D\e{P}$ in general. If an observer performs two experiments to measure his distance to a light source, the first one using the solar parallax and the second one using the angular-diameter method, the results will generically disagree. In particular, the focusing theorem derived in \S~\ref{sec:focusing_theorem} implies that any gravitational effect tends to \emph{reduce} $D\e{A}$, contrary to $D\e{P}$.

In the gravitational lensing literature, it is customary to set the observed frequency~$\omega_\obs$ to one---which can be considered a particular choice of units for frequencies---in order to simplify Eq.~\eqref{eq:angular_distance_theoretical}. However, keeping explicitly $\omega_\obs$ in the expression of $D\e{A}$ has a significant pedagogical advantage: it allows one to easily understand (i) aberration effects, and (ii) the relation between the angular diameter distance and the luminosity distance, defined in \S~\ref{sec:luminosity_distance}.


\subsubsection{Aberration effects}\index{aberration}\label{sec:aberration}

Because the cross-sectional area of a light beam is frame independent (see \S~\ref{sec:beam_morphology_frame_independent}), the source's area~$A_\source$ involved in the definition~\eqref{eq:angular_distance_definition} of the angular distance does not depend on the source's four-velocity~$\vect{u}_\source$. On the contrary, the observer angular size~$\Omega_\obs$ does depend on $\vect{u}_\obs$ in general. We thus expect $D\e{A}$ to be independent from the source's velocity, but to be affected by the observer's velocity.

These dependences are evident in Eq.~\eqref{eq:angular_distance_theoretical}. The Jacobi matrix is indeed independent from any four-velocity, as it is driven by the optical tidal matrix $\vect{\tidal}$, independent of the frame in which the screen space is defined. Besides, the $\omega_\obs$ term exhibits a $\vect{u}_\obs$-dependence of $D\e{A}$, which is responsible for \emph{aberration} effects. If the observer moves towards the source, $\omega_\obs$ increases and the source thus appears smaller (i.e. farther, $D\e{A}$ is larger) to her, than if she were receding from it. The same phenomenon potentially occurs if the observer lies within a stronger gravitational field---hence increasing $\omega_\obs$---though it also potentially affects the Jacobi matrix.

\subsection{Luminosity distance}
\label{sec:luminosity_distance}

\subsubsection{Definition}

\begin{figure}[h!]
\centering
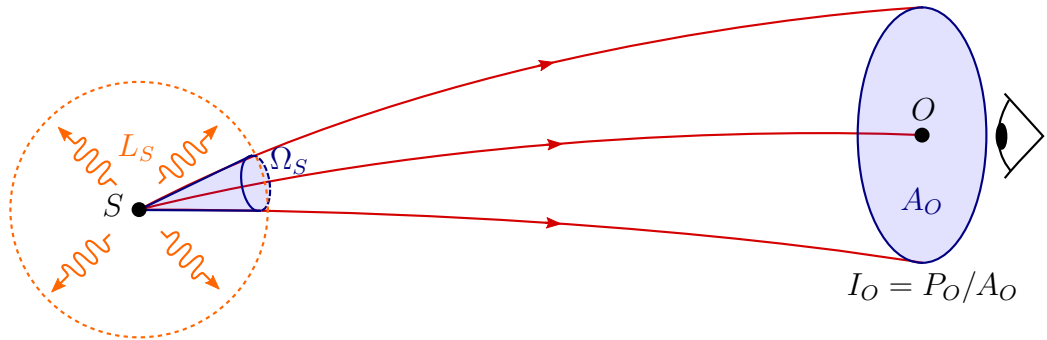
\caption{The luminosity distance is defined from the ratio between the intrinsic luminosity (total emission power)~$L_\source$ of the source and the observed luminous intensity $I_\obs\define P_\obs/A_\obs$, where $P_\obs$ is the power detected at $O$ by a small detector of area~$A_\obs$. $\Omega_\source$ denotes the angular aperture, in the source's frame, of the beam intercepted by the observer. The relevant picture is thus reversed compared to the angular distance's, since the vertex  of the beam is here $S$ rather than $O$, like for the parallax.}
\label{fig:luminosity_distance}
\end{figure}

A light source not only appears smaller but also fainter as it lies farther from the observer. In a nonrelativistic picture, if the source has isotropic light emission, then the energy~$\delta E$ it emits during a short period of time~$\delta t$ is homogeneously distributed on a spherical shell (the photosphere) with a surface density $\delta E/(4\pi r^2)$, where $r$ is the shell's radius which increases as light propagates. An observer located at $r_\obs$ thus receives an energy per unit time and area equal to $\delta E/(4\pi r_\obs^2)/\delta t$. This motivates the following definition for the \emph{luminosity distance}:\index{distance!luminosity}
\begin{empheq}[box=\fbox]{equation}
D\e{L} = \sqrt{\frac{L_\source}{4\pi I_\obs}},
\end{empheq}
where $L_\source$ denotes the intrinsic luminosity of the source, that is, the total luminous power it emits in all directions; $I_\obs$ is the observed luminous intensity, defined as $I_\obs\define P_\obs/A_\obs$, where $A_\obs$ is the area of the observer's detector, and $P_\obs$ the luminous power measured by this detector. The relevant geometry is represented in Fig.~\ref{fig:luminosity_distance}.

In astronomy, the luminosity distance is often used under a logarithmic form called \emph{distance modulus}\index{distance!modulus}, and defined as
\begin{equation}
\mu\e{L} = 5 \log\pa{\frac{D\e{L}}{10\U{pc}}}.
\end{equation}
It is the most used notion of distance in practice, as it applies to extra-galactic sources contrary to the parallax, and to unresolved sources contrary to the angular diameter distance. Nevertheless, similarly to the latter, any direct measurement of $D\e{L}$ requires sources whose intrinsic luminosity~$L_\source$ is known, or can be inferred from other observations: the so-called \emph{standard(-isable) candles}\index{standard!-isable candles}.

A good example is provided by \emph{Cepheid variables}\index{Cepheid variable}, which are pulsating stars whose luminosity oscillates with a period correlated with its mean~\cite{1912HarCi.173....1L,1917Obs....40..290E}. Once calibrated, typically with auxiliary parallax distance measurements, this relation can then be used to deduce the intrinsic luminosity of any Cepheid variable from a measurement of its period. Type Ia supernovae (SNeIa), which are mostly thought to originate from binary systems where a white dwarf reaches its Chandrasekhar mass by accreting matter from its companion---though the nature of progenitors is still debated~\cite{2012NewAR..56..122W}---, are also good standard-isable candles, as there exists a relation between the duration of the explosion and its peak luminosity~\cite{1968AJ.....73.1021K}. In cosmology, they are used to plot the \emph{Hubble diagram}\index{Hubble!diagram} from which key information on the expansion history of the Universe can be extracted (see Chap.~\ref{chapter:observations}). Recently, a Hubble diagram has been constructed from quasars~\cite{2015arXiv150507118R}, by exploiting a relation between their X-ray and UV luminosities. Because they are orders-of-magnitude brighter than supernovae, quasars have the advantage of providing a much deeper Hubble diagram, up to $z=6$.

Future detections of gravitational waves are also expected to provide excellent measurements of the luminosity distance to their sources~\cite{1986Natur.323..310S,2005ApJ...629...15H}. Indeed, gravitational waves follow null geodesics just like electromagnetic waves, hence all the notions defined for the latter apply to the former as well. In addition, theoretical analyses of the signal generated by an inspiraling binary system of compact objects (neutrons stars, black holes, etc.) show that its phase gives access to the mass of the objects, i.e. to the gravitational luminosity of the system. For that reason, binary systems of compact objects have been nicknamed \emph{standard sirens}\index{standard!sirens}---the gravitational analog of standard candles. They have, compared to supernovae, the significant advantage of relying on well-controlled theoretical predictions, and should therefore be less plagued by systematics.

\subsubsection{Distance duality relation}\index{distance!duality relation}

The similarity of the pictures that we used to define angular and luminosity distances (compare Figs.~\ref{fig:angular_distance}, \ref{fig:luminosity_distance}) strongly suggests that they are not independent notions. This expectation can be made more explicit by reexpressing~$D\e{L}$ as a function of the geometrical quantities $\Omega_\source$, $A_\obs$. Consider the fraction of all the photons emitted during a short time interval~$\delta\tau_\source$ in the source's frame which can be received by the observer's detector. By definition, there are
\begin{equation}
\delta N = \frac{L_\source}{\hbar \omega_\source} \times \delta \tau_\source \times \frac{\Omega_\source}{4\pi}
\end{equation}
such photons. \emph{Assuming that none of them is absorbed} by some interaction with matter on its way to the observer, and according to the photon conservation law derived in \S~\ref{sec:energetical_observables}, we also have
\begin{equation}
\delta N = \frac{I_\obs}{\hbar\omega_\obs} \times \delta\tau_\obs \times A_\obs,
\end{equation}
where $\delta\tau_\obs$ is the time interval, in the observer's frame, corresponding to the reception of all these photons. Note that the ratio between  $\delta\tau_\obs/\delta\tau_\source$ is the same as the ratio between the observed and emitted periods of a light signal, in other words
\begin{equation}
\frac{\delta\tau_\obs}{\delta\tau_\source} = \frac{\omega_\source}{\omega_\obs} = 1+z,
\end{equation}
so that
\begin{equation}
D\e{L} \define \sqrt{ \frac{L_\source}{4\pi I_\obs} } 
			= (1+z) \sqrt{\frac{A_\obs}{\Omega_\source}} .
\label{eq:luminosity_distance_geometric}
\end{equation}

It is then tempting to recognise the angular distance in the right-hand side of Eq.~\eqref{eq:luminosity_distance_geometric}, except that the roles of $S$ and $O$ are inverted compared to the definition~\eqref{eq:angular_distance_definition} of $D\e{A}$. This can nevertheless be solved by
\begin{equation}
\sqrt{\frac{A_\obs}{\Omega_\source}} \sqrt{\frac{\Omega_\obs}{A_\source}}
=
\frac{\omega_\source \sqrt{\det \mat{\jacobi}(v_\obs\leftarrow v_\source)}}
		{\omega_\obs\sqrt{\det\mat{\jacobi}(v_\source\leftarrow v_\obs)}}
=
1+z,
\label{eq:consequence_Etherington_reciprocity}
\end{equation}
where we have used the consequence~\eqref{eq:symmetry_determinant_Jacobi} of Etherington's reciprocity law, which states that the determinant of the Jacobi matrix is invariant under the exchange of its arguments. The \emph{distance duality relation} finally reads
\begin{empheq}[box=\fbox]{equation}
D\e{L} = (1+z)^2 D\e{A} .
\label{eq:distance_duality_relation}
\end{empheq}
The origin of the redshift factor $(1+z)^2$ can be summarised as $2=1/2 + 1/2 + 1$. The first $1/2$ comes from the fact that the energies of emitted and received photons differ; the second $1/2$ is due time dilation between the source and the observer; and finally the~$1$ originates from the exchange of the roles of $S$ and $O$ in Figs.~\ref{fig:angular_distance}, \ref{fig:luminosity_distance}. The latter can be seen as a comparative aberration effect, which for the angular distance occurs at $O$, whereas for the luminosity distance it occurs at $S$.

Equation~\eqref{eq:distance_duality_relation} holds for any spacetime, as long as any physical process capable of violating photon conservation is negligible. Note also that strict GR is not even necessary, in the sense that the dynamics of the metric has no impact on the distance duality law. It therefore remains valid for Nordstr\o m's gravity, $f(R)$, \ldots, provided electromagnetism is still minimally coupled to the metric. A counterexample is the Horndeski vector-tensor theory mentioned in \S~\ref{sec:EM_curved_spacetime}.
 
Observational tests of the distance duality relation thus potentially provide constraints on the transparency of the Universe (photon conservation) and on some potential departures from GR. A method based on the X-ray emission and the Sunyaev-Zel'dovich effect in galaxy clusters was proposed in Ref.~\cite{2004PhRvD..70h3533U}, and concluded that $(1+z)^2 D\e{A}/D\e{L}=0.93^{+0.05}_{-0.04}$, suggesting no significant violation of the distance duality relation. Besides, it has recently been shown by Ref.~\cite{2013PhRvD..87j3530E} that any violation of Etherington's reciprocity law---in particular of its consequence~\eqref{eq:consequence_Etherington_reciprocity}---would induce spectral distortions in the observed CMB. Such violations cannot exceed $0.01\%$.

%% file: distance_Synge.pdf_tex
\begingroup%
  \makeatletter%
  \providecommand\color[2][]{%
    \errmessage{(Inkscape) Color is used for the text in Inkscape, but the package 'color.sty' is not loaded}%
    \renewcommand\color[2][]{}%
  }%
  \providecommand\transparent[1]{%
    \errmessage{(Inkscape) Transparency is used (non-zero) for the text in Inkscape, but the package 'transparent.sty' is not loaded}%
    \renewcommand\transparent[1]{}%
  }%
  \providecommand\rotatebox[2]{#2}%
  \ifx\svgwidth\undefined%
    \setlength{\unitlength}{111.0234375bp}%
    \ifx\svgscale\undefined%
      \relax%
    \else%
      \setlength{\unitlength}{\unitlength * \real{\svgscale}}%
    \fi%
  \else%
    \setlength{\unitlength}{\svgwidth}%
  \fi%
  \global\let\svgwidth\undefined%
  \global\let\svgscale\undefined%
  \makeatother%
  \begin{picture}(1,1.58525086)%
    \put(0,0){\includegraphics[width=\unitlength]{distance_Synge.pdf}}%
    \put(0.86012244,0.76389505){\color[rgb]{0,0,0}\makebox(0,0)[lb]{\smash{$B$}}}%
    \put(0.00264584,0.57654722){\color[rgb]{0,0,0}\makebox(0,0)[lb]{\smash{$A$}}}%
    \put(-0.00455985,1.44843519){\color[rgb]{0,0,0}\makebox(0,0)[lb]{\smash{$R$}}}%
    \put(0.00264584,1.04491679){\color[rgb]{0,0,0}\makebox(0,0)[lb]{\smash{$\wl$}}}%
    \put(0.25484485,0.68463254){\color[rgb]{0,0,0}\rotatebox{15.00000022}{\makebox(0,0)[lb]{\smash{$D(A,B)$}}}}%
    \put(0.00264584,0.07214922){\color[rgb]{0,0,0}\makebox(0,0)[lb]{\smash{$E$}}}%
    \put(0.18278798,0.05053216){\color[rgb]{0,0,0}\makebox(0,0)[lb]{\smash{$t_E$}}}%
    \put(0.18278798,1.47725793){\color[rgb]{0,0,0}\makebox(0,0)[lb]{\smash{$t_R$}}}%
    \put(0.18278798,0.52610742){\color[rgb]{0,0,0}\makebox(0,0)[lb]{\smash{$t_A$}}}%
  \end{picture}%
\endgroup%

%% file: distance_worldline_event_Minkowski.pdf_tex
\begingroup%
  \makeatletter%
  \providecommand\color[2][]{%
    \errmessage{(Inkscape) Color is used for the text in Inkscape, but the package 'color.sty' is not loaded}%
    \renewcommand\color[2][]{}%
  }%
  \providecommand\transparent[1]{%
    \errmessage{(Inkscape) Transparency is used (non-zero) for the text in Inkscape, but the package 'transparent.sty' is not loaded}%
    \renewcommand\transparent[1]{}%
  }%
  \providecommand\rotatebox[2]{#2}%
  \ifx\svgwidth\undefined%
    \setlength{\unitlength}{111.0234375bp}%
    \ifx\svgscale\undefined%
      \relax%
    \else%
      \setlength{\unitlength}{\unitlength * \real{\svgscale}}%
    \fi%
  \else%
    \setlength{\unitlength}{\svgwidth}%
  \fi%
  \global\let\svgwidth\undefined%
  \global\let\svgscale\undefined%
  \makeatother%
  \begin{picture}(1,1.58525086)%
    \put(0,0){\includegraphics[width=\unitlength]{distance_worldline_event_Minkowski.pdf}}%
    \put(0.86012244,0.76389505){\color[rgb]{0,0,0}\makebox(0,0)[lb]{\smash{$B$}}}%
    \put(0.00264584,0.77110073){\color[rgb]{0,0,0}\makebox(0,0)[lb]{\smash{$A^*$}}}%
    \put(-0.00455985,1.44843519){\color[rgb]{0,0,0}\makebox(0,0)[lb]{\smash{$R$}}}%
    \put(0.00264584,1.04491679){\color[rgb]{0,0,0}\makebox(0,0)[lb]{\smash{$\wl$}}}%
    \put(0.25484484,0.82874622){\color[rgb]{0,0,0}\makebox(0,0)[lb]{\smash{$D(\ell,B)$}}}%
    \put(0.00264584,0.07214922){\color[rgb]{0,0,0}\makebox(0,0)[lb]{\smash{$E$}}}%
    \put(0.18278798,0.05053216){\color[rgb]{0,0,0}\makebox(0,0)[lb]{\smash{$t_E$}}}%
    \put(0.18278798,1.47725793){\color[rgb]{0,0,0}\makebox(0,0)[lb]{\smash{$t_R$}}}%
    \put(0.18278798,0.64860407){\color[rgb]{0,0,0}\makebox(0,0)[lb]{\smash{$t_{A^*}=\tAstar$}}}%
    \put(0.39895851,1.33314427){\color[rgb]{0.83137255,0,0}\rotatebox{-45}{\makebox(0,0)[lb]{\smash{$v_R-v_B$}}}}%
    \put(0.43498706,0.25229141){\color[rgb]{0.83137255,0,0}\rotatebox{45}{\makebox(0,0)[lb]{\smash{$v_B-v_E$}}}}%
  \end{picture}%
\endgroup%

%% file: distance_worldline_particle_Minkowski.pdf_tex
\begingroup%
  \makeatletter%
  \providecommand\color[2][]{%
    \errmessage{(Inkscape) Color is used for the text in Inkscape, but the package 'color.sty' is not loaded}%
    \renewcommand\color[2][]{}%
  }%
  \providecommand\transparent[1]{%
    \errmessage{(Inkscape) Transparency is used (non-zero) for the text in Inkscape, but the package 'transparent.sty' is not loaded}%
    \renewcommand\transparent[1]{}%
  }%
  \providecommand\rotatebox[2]{#2}%
  \ifx\svgwidth\undefined%
    \setlength{\unitlength}{123.5484375bp}%
    \ifx\svgscale\undefined%
      \relax%
    \else%
      \setlength{\unitlength}{\unitlength * \real{\svgscale}}%
    \fi%
  \else%
    \setlength{\unitlength}{\svgwidth}%
  \fi%
  \global\let\svgwidth\undefined%
  \global\let\svgscale\undefined%
  \makeatother%
  \begin{picture}(1,1.42595895)%
    \put(0,0){\includegraphics[width=\unitlength]{distance_worldline_particle_Minkowski.pdf}}%
    \put(0.70817367,0.58305273){\color[rgb]{0,0,0}\makebox(0,0)[lb]{\smash{$I$}}}%
    \put(-0.00409758,0.84206046){\color[rgb]{0,0,0}\makebox(0,0)[lb]{\smash{$\wl$}}}%
    \put(0.28728611,0.64780466){\color[rgb]{0,0,0}\makebox(0,0)[lb]{\smash{$D\e{inst}(t)$}}}%
    \put(0.8700535,1.29532398){\color[rgb]{0,0,0}\makebox(0,0)[lb]{\smash{$A$}}}%
    \put(0.65601603,0.08954176){\color[rgb]{0,0,0}\makebox(0,0)[lb]{\smash{$R$}}}%
    \put(0.74054963,0.84206046){\color[rgb]{0,0,0}\makebox(0,0)[lb]{\smash{$\wl'$}}}%
    \put(-0.00409758,0.58305273){\color[rgb]{0,0,0}\makebox(0,0)[lb]{\smash{$P$}}}%
    \put(0.25491014,1.23057205){\color[rgb]{0,0,0}\makebox(0,0)[lb]{\smash{$D\e{adv}(t)$}}}%
    \put(0.22253418,0.03266131){\color[rgb]{0,0,0}\makebox(0,0)[lb]{\smash{$D\e{ret}(t)$}}}%
    \put(0.15778225,0.5183008){\color[rgb]{0,0,0}\makebox(0,0)[lb]{\smash{$t$}}}%
  \end{picture}%
\endgroup%

%% file: distance_events_curved.pdf_tex
\begingroup%
  \makeatletter%
  \providecommand\color[2][]{%
    \errmessage{(Inkscape) Color is used for the text in Inkscape, but the package 'color.sty' is not loaded}%
    \renewcommand\color[2][]{}%
  }%
  \providecommand\transparent[1]{%
    \errmessage{(Inkscape) Transparency is used (non-zero) for the text in Inkscape, but the package 'transparent.sty' is not loaded}%
    \renewcommand\transparent[1]{}%
  }%
  \providecommand\rotatebox[2]{#2}%
  \ifx\svgwidth\undefined%
    \setlength{\unitlength}{190.86650391bp}%
    \ifx\svgscale\undefined%
      \relax%
    \else%
      \setlength{\unitlength}{\unitlength * \real{\svgscale}}%
    \fi%
  \else%
    \setlength{\unitlength}{\svgwidth}%
  \fi%
  \global\let\svgwidth\undefined%
  \global\let\svgscale\undefined%
  \makeatother%
  \begin{picture}(1,0.17892035)%
    \put(0,0){\includegraphics[width=\unitlength]{distance_events_curved.pdf}}%
    \put(-0.00265238,0.0288666){\color[rgb]{0,0,0}\makebox(0,0)[lb]{\smash{$A$}}}%
    \put(0.79830087,0.14281855){\color[rgb]{0,0,0}\makebox(0,0)[lb]{\smash{$B$}}}%
    \put(0.22672637,0.0108551){\color[rgb]{0,0,0}\makebox(0,0)[lb]{\smash{$\mathscr{G}$}}}%
    \put(0.52204974,0.13215455){\color[rgb]{0,0,0}\makebox(0,0)[lb]{\smash{$\vect{t}$}}}%
    \put(0.73149591,0.07667233){\color[rgb]{0,0,0}\makebox(0,0)[lb]{\smash{$\lambda_B$}}}%
    \put(0.07200554,0.09828813){\color[rgb]{0,0,0}\makebox(0,0)[lb]{\smash{$\lambda_A$}}}%
  \end{picture}%
\endgroup%

%% file: distance_worldline_event_curved.pdf_tex
\begingroup%
  \makeatletter%
  \providecommand\color[2][]{%
    \errmessage{(Inkscape) Color is used for the text in Inkscape, but the package 'color.sty' is not loaded}%
    \renewcommand\color[2][]{}%
  }%
  \providecommand\transparent[1]{%
    \errmessage{(Inkscape) Transparency is used (non-zero) for the text in Inkscape, but the package 'transparent.sty' is not loaded}%
    \renewcommand\transparent[1]{}%
  }%
  \providecommand\rotatebox[2]{#2}%
  \ifx\svgwidth\undefined%
    \setlength{\unitlength}{168.95bp}%
    \ifx\svgscale\undefined%
      \relax%
    \else%
      \setlength{\unitlength}{\unitlength * \real{\svgscale}}%
    \fi%
  \else%
    \setlength{\unitlength}{\svgwidth}%
  \fi%
  \global\let\svgwidth\undefined%
  \global\let\svgscale\undefined%
  \makeatother%
  \begin{picture}(1,0.98549867)%
    \put(0,0){\includegraphics[width=\unitlength]{distance_worldline_event_curved.pdf}}%
    \put(0.25952105,0.47789079){\color[rgb]{0,0,0}\makebox(0,0)[lb]{\smash{$A^*$}}}%
    \put(0.23816463,0.91684306){\color[rgb]{0,0,0}\makebox(0,0)[lb]{\smash{$R$}}}%
    \put(0.26523527,0.64306626){\color[rgb]{0,0,0}\makebox(0,0)[lb]{\smash{$\wl$}}}%
    \put(0.45387006,0.50446603){\color[rgb]{0,0,0}\rotatebox{-20.135512}{\makebox(0,0)[lb]{\smash{$D\e{S}(\ell,B)$}}}}%
    \put(0.23797999,0.06901109){\color[rgb]{0,0,0}\makebox(0,0)[lb]{\smash{$E$}}}%
    \put(0.56365277,0.78731628){\color[rgb]{0.83137255,0,0}\rotatebox{-49.98056037}{\makebox(0,0)[lb]{\smash{$v_R-v_B$}}}}%
    \put(0.34440974,0.93771589){\color[rgb]{0,0,0}\makebox(0,0)[lb]{\smash{$\tau_R$}}}%
    \put(0.34350174,0.05388174){\color[rgb]{0,0,0}\makebox(0,0)[lb]{\smash{$\tau_E$}}}%
    \put(0.82865367,0.29978966){\color[rgb]{0,0,0}\makebox(0,0)[lb]{\smash{$B$}}}%
    \put(0.47373888,0.35684964){\color[rgb]{0,0,0}\makebox(0,0)[lb]{\smash{$\mathscr{G}^*$}}}%
    \put(0.48116139,0.11244674){\color[rgb]{0.83137255,0,0}\rotatebox{26.0840877}{\makebox(0,0)[lb]{\smash{$v_B-v_E$}}}}%
  \end{picture}%
\endgroup%

%% file: foliation-based_distance.pdf_tex
\begingroup%
  \makeatletter%
  \providecommand\color[2][]{%
    \errmessage{(Inkscape) Color is used for the text in Inkscape, but the package 'color.sty' is not loaded}%
    \renewcommand\color[2][]{}%
  }%
  \providecommand\transparent[1]{%
    \errmessage{(Inkscape) Transparency is used (non-zero) for the text in Inkscape, but the package 'transparent.sty' is not loaded}%
    \renewcommand\transparent[1]{}%
  }%
  \providecommand\rotatebox[2]{#2}%
  \ifx\svgwidth\undefined%
    \setlength{\unitlength}{197.50825195bp}%
    \ifx\svgscale\undefined%
      \relax%
    \else%
      \setlength{\unitlength}{\unitlength * \real{\svgscale}}%
    \fi%
  \else%
    \setlength{\unitlength}{\svgwidth}%
  \fi%
  \global\let\svgwidth\undefined%
  \global\let\svgscale\undefined%
  \makeatother%
  \begin{picture}(1,0.84262302)%
    \put(0,0){\includegraphics[width=\unitlength]{foliation-based_distance.pdf}}%
    \put(0.26983495,0.36516052){\color[rgb]{0,0,0}\makebox(0,0)[lb]{\smash{$A_t$}}}%
    \put(0.28898207,0.55152091){\color[rgb]{0,0,0}\makebox(0,0)[lb]{\smash{$\wl$}}}%
    \put(0.4534685,0.36994907){\color[rgb]{0,0,0}\rotatebox{-14.79385657}{\makebox(0,0)[lb]{\smash{$D\e{F}(\ell,B)$}}}}%
    \put(0.76149242,0.25606253){\color[rgb]{0,0,0}\makebox(0,0)[lb]{\smash{$B$}}}%
    \put(0.07219539,0.29839981){\color[rgb]{0,0,0}\makebox(0,0)[lb]{\smash{$\Sigma_t$}}}%
    \put(0.46825798,0.24584366){\color[rgb]{0,0,0}\makebox(0,0)[lb]{\smash{$\mathscr{G}_{\vect{h}}$}}}%
  \end{picture}%
\endgroup%

%% file: parallax.pdf_tex
\begingroup%
  \makeatletter%
  \providecommand\color[2][]{%
    \errmessage{(Inkscape) Color is used for the text in Inkscape, but the package 'color.sty' is not loaded}%
    \renewcommand\color[2][]{}%
  }%
  \providecommand\transparent[1]{%
    \errmessage{(Inkscape) Transparency is used (non-zero) for the text in Inkscape, but the package 'transparent.sty' is not loaded}%
    \renewcommand\transparent[1]{}%
  }%
  \providecommand\rotatebox[2]{#2}%
  \ifx\svgwidth\undefined%
    \setlength{\unitlength}{392.5234375bp}%
    \ifx\svgscale\undefined%
      \relax%
    \else%
      \setlength{\unitlength}{\unitlength * \real{\svgscale}}%
    \fi%
  \else%
    \setlength{\unitlength}{\svgwidth}%
  \fi%
  \global\let\svgwidth\undefined%
  \global\let\svgscale\undefined%
  \makeatother%
  \begin{picture}(1,0.51156181)%
    \put(0,0){\includegraphics[width=\unitlength]{parallax.pdf}}%
    \put(0.03362857,0.33936892){\color[rgb]{0,0,0}\makebox(0,0)[lb]{\smash{$S$}}}%
    \put(0.80606652,0.36790225){\color[rgb]{0,0,0}\makebox(0,0)[lb]{\smash{$O$}}}%
    \put(0.80606652,0.43923557){\color[rgb]{0,0,0.50196078}\makebox(0,0)[lb]{\smash{$A_\obs$}}}%
    \put(0.76530462,0.28637845){\color[rgb]{0,0,0.50196078}\makebox(0,0)[lb]{\smash{$\vect{\xi}^\perp_\obs$}}}%
    \put(0.61244751,0.15797847){\color[rgb]{0,0,0.50196078}\makebox(0,0)[lb]{\smash{$\vect{\theta}_\obs$}}}%
    \put(0.56625202,0.09275944){\color[rgb]{0,0,0}\makebox(0,0)[lb]{\smash{apparent motion}}}%
    \put(0.28023804,0.09683563){\color[rgb]{0,0,0.50196078}\makebox(0,0)[lb]{\smash{$\Omega_\obs$}}}%
    \put(0.57983401,0.04180707){\color[rgb]{0,0,0}\makebox(0,0)[lb]{\smash{of the source}}}%
  \end{picture}%
\endgroup%

%% file: angular_distance.pdf_tex
\begingroup%
  \makeatletter%
  \providecommand\color[2][]{%
    \errmessage{(Inkscape) Color is used for the text in Inkscape, but the package 'color.sty' is not loaded}%
    \renewcommand\color[2][]{}%
  }%
  \providecommand\transparent[1]{%
    \errmessage{(Inkscape) Transparency is used (non-zero) for the text in Inkscape, but the package 'transparent.sty' is not loaded}%
    \renewcommand\transparent[1]{}%
  }%
  \providecommand\rotatebox[2]{#2}%
  \ifx\svgwidth\undefined%
    \setlength{\unitlength}{340.7046875bp}%
    \ifx\svgscale\undefined%
      \relax%
    \else%
      \setlength{\unitlength}{\unitlength * \real{\svgscale}}%
    \fi%
  \else%
    \setlength{\unitlength}{\svgwidth}%
  \fi%
  \global\let\svgwidth\undefined%
  \global\let\svgscale\undefined%
  \makeatother%
  \begin{picture}(1,0.2841904)%
    \put(0,0){\includegraphics[width=\unitlength]{angular_distance.pdf}}%
    \put(0.91026225,0.08492388){\color[rgb]{0,0,0}\makebox(0,0)[lb]{\smash{$O$}}}%
    \put(0.04813553,0.05991926){\color[rgb]{0,0,0.50196078}\makebox(0,0)[lb]{\smash{$A_\source$}}}%
    \put(0.82300012,0.02469815){\color[rgb]{0,0,0.50196078}\makebox(0,0)[lb]{\smash{$\Omega_\obs$}}}%
    \put(0.06038385,0.1619695){\color[rgb]{0,0,0}\makebox(0,0)[lb]{\smash{$S$}}}%
  \end{picture}%
\endgroup%

%% file: luminosity_distance.pdf_tex
\begingroup%
  \makeatletter%
  \providecommand\color[2][]{%
    \errmessage{(Inkscape) Color is used for the text in Inkscape, but the package 'color.sty' is not loaded}%
    \renewcommand\color[2][]{}%
  }%
  \providecommand\transparent[1]{%
    \errmessage{(Inkscape) Transparency is used (non-zero) for the text in Inkscape, but the package 'transparent.sty' is not loaded}%
    \renewcommand\transparent[1]{}%
  }%
  \providecommand\rotatebox[2]{#2}%
  \ifx\svgwidth\undefined%
    \setlength{\unitlength}{412.2671875bp}%
    \ifx\svgscale\undefined%
      \relax%
    \else%
      \setlength{\unitlength}{\unitlength * \real{\svgscale}}%
    \fi%
  \else%
    \setlength{\unitlength}{\svgwidth}%
  \fi%
  \global\let\svgwidth\undefined%
  \global\let\svgscale\undefined%
  \makeatother%
  \begin{picture}(1,0.30270534)%
    \put(0,0){\includegraphics[width=\unitlength]{luminosity_distance.pdf}}%
    \put(0.81597568,0.19886437){\color[rgb]{0,0,0}\makebox(0,0)[lb]{\smash{$O$}}}%
    \put(0.80627324,0.11736382){\color[rgb]{0,0,0.50196078}\makebox(0,0)[lb]{\smash{$A_\obs$}}}%
    \put(0.08441128,0.11154236){\color[rgb]{0,0,0}\makebox(0,0)[lb]{\smash{$S$}}}%
    \put(0.0979947,0.16587605){\color[rgb]{1,0.4,0}\makebox(0,0)[lb]{\smash{$L_\source$}}}%
    \put(0.75776101,0.03974426){\color[rgb]{0,0,0}\makebox(0,0)[lb]{\smash{$I_\obs = P_\obs/A_\obs$}}}%
    \put(0.23576943,0.15423312){\color[rgb]{0,0,0.50196078}\makebox(0,0)[lb]{\smash{$\Omega_\source$}}}%
  \end{picture}%
\endgroup%

%% file: chapter_4.tex
\lettrine{C}{osmology} has today an impressively successful standard model. Discussing the reasons of such a success is central to this thesis, which cannot be done without presenting the model itself. In this first chapter dedicated to standard cosmology, we introduce the spacetime geometries used to describe the Universe on large scales: first the purely homogeneous and isotropic Friedmann-Lema\^{i}tre model, with the cosmic history reconstructed from it; then the perturbation theory relaxing the strict assumptions of homogeneity and isotropy, whose limitations will also be discussed.

\bigskip

\minitoc

\newpage

\section{Homogeneous and isotropic cosmologies}

Observing our Universe, in particular through the cosmic microwave background, reveals that its properties are almost identical whatever the direction we look at. This fact, together with the \emph{Copernican principle}\index{Copernican principle}, according to which we do not occupy a special place in the cosmos, led cosmologists to model the Universe as statistically spatially homogeneous and isotropic, a hypothesis known as the \emph{cosmological principle}\index{cosmological!principle}. This section examines the consequences of the simplest application of this principle: strictly homogeneous and isotropic cosmologies.

\subsection{The Friedmann-Lema\^{i}tre geometry}

The first proposal of a spacetime geometry satisfying the cosmological principle was formulated in 1917 with Einstein's static Universe~\cite{1917SPAW.......142E}. Einstein's approach was then generalised, allowing for the possibility of an evolving cosmos, independently by Friedmann in 1922~\cite{1922ZPhy...10..377F}, and Lema\^{i}tre in 1927~\cite{1927ASSB...47...49L} who also predicted the redshift of receding galaxies \emph{before} its observation by Hubble~\cite{1929PNAS...15..168H} in 1929. The work of Friedmann was noticed by Robertson in 1929~\cite{1929PNAS...15..822R}, followed by Walker. In the 1930s, both of them analysed in great details~\cite{1935ApJ....82..284R,1936ApJ....83..187R,1936ApJ....83..257R,Walker01011937} the properties of the metric discovered by Friedmann and Lema\^{i}tre, which we shall call FL metric throughout this thesis.\index{Friedmann!-Lema\^{i}tre spacetime}

\subsubsection{Coordinate systems and metric}

\begin{figure}[h!]
\centering
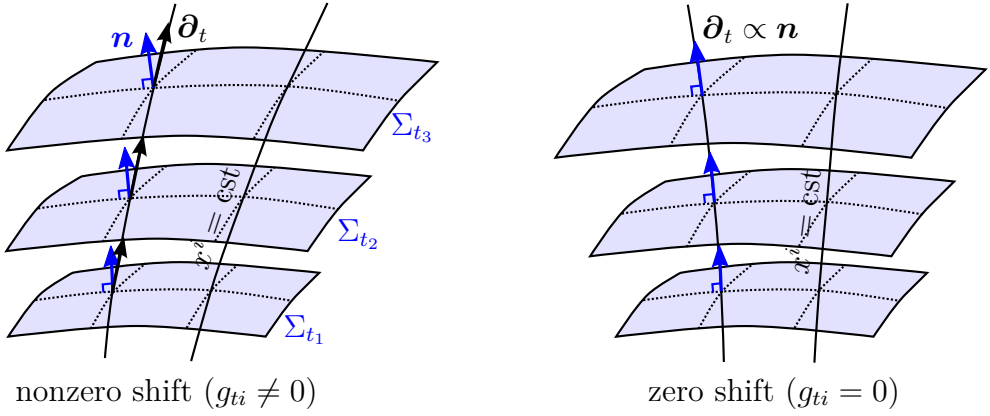
\caption{Given a foliation~$\{\Sigma_t\}$ of spacetime, the coordinate systems $\{x^i\}$ of the hypersurfaces are generally shifted with respect to each other: the $x^i=\cst$ curves are not orthogonal to the hypersurfaces. In this case $g_{ti}=\vect{g}(\vect{\partial}_t,\vect{\partial}_i)\not=0$.}
\label{fig:shift}
\end{figure}

Suppose spacetime can be foliated by a family of spacelike hypersurfaces~$\{\Sigma_t\}$ whose intrinsic geometry is homogeneous and isotropic. Their label, $t$, is naturally used as a time coordinate; each hypersurface~$\Sigma_{t_0}$ of the foliation is thus characterised by the simple equation $t=t_0$, and the one-form associated with its normal vector~$\vect{n}$ satisfies
\begin{equation}
n_\mu \dd x^\mu \propto \dd t,
\end{equation}
which implies $0=n_i=\vect{g}(\vect{n},\vect{\partial}_i)$. In other words, whatever the choice of the other three coordinates $\{x^i\}_{i=1\ldots 3}$, the associated vector fields~$\vect{\partial}_i$ are tangent to $\Sigma_t$. A convenient setting then consists in imposing that the $x^i=\cst$ curves are orthogonal to the foliation, i.e. $\vect{\partial}_t\propto\vect{n}$. This means that the metric has no shift: $g_{ti}=0$ (see Fig.~\ref{fig:shift}). Finally, the assumed homogeneity of the geometry of $\Sigma_t$ allows us to rescale the coordinate~$t$ so that $g_{tt}=\vect{g}(\vect{\partial}_t,\vect{\partial}_t)=-1$ everywhere. The metric, expressed in terms of the resulting coordinate system $(t,x^i)$---called \emph{synchronous}\index{synchronous coordinates}---, reads 
\begin{equation}
\dd s^2 = - \dd t^2 + g_{ij} \dd x^i \dd x^j.
\end{equation} 
In the context of cosmology, the coordinate $t$ is called \emph{cosmic time}\index{cosmic!time}.

The homogeneity and isotropy assumptions for $\Sigma_t$ can be shown~\cite{1984ucp..book.....W} to impose the following form of the spatial metric:
\begin{equation}
g_{ij} = a^2(t) \gamma_{ij}(x^k) \dd x^i \dd x^j,
\end{equation}
where $a(t)\geq 0$ is called the \emph{scale factor}\index{scale factor}, and $\gamma_{ij}$ denotes the intrinsic metric of $\Sigma_t$ (as within any $\Sigma_t$, the scale factor can be absorbed by a coordinate transformation). The high level of symmetry of $\gamma_{ij}$ imply that its Riemann tensor reads
\begin{equation}
\tensor[^3]{R}{_i_j_k_\ell} = 2 K \gamma_{k[i}\gamma_{j]\ell},
\end{equation}
where $K$ is a constant of dimension $L^{-2}$ called spatial curvature parameter. The cases $K<0$, $K=0$, and $K>0$ respectively correspond to hyperbolic, Euclidean, and spherical foliations. Explicit forms of the metric~$\gamma_{ij}$ can be written using various spherical-like coordinate systems, such as
\begin{align}
\gamma_{ij} \dd x^i \dd x^j 	&= \pa{\frac{1}{1+K\varrho^2/4}}^2\pa{ \dd\varrho^2 + \varrho^2 \dd\Omega^2 }\\
												&=\frac{\dd r^2}{1-K r^2} + r^2 \dd \Omega^2, \label{eq:spatial_FL_areal}\\
												&= \dd\chi^2 + f_K^2(\chi) \, \dd \Omega^2 \label{eq:spatial_FL},											
\end{align}
with $\dd\Omega^2\define\dd\theta^2+\sin^2\theta\,\dd\ph^2$, and where the radial coordinates $\varrho,r,\chi$ are related by
\begin{equation}
\frac{\varrho}{1+K\varrho^2/4} =r = f_K(\chi) 
\define
\begin{cases}
\dfrac{\sinh(\sqrt{-K}\chi)}{\sqrt{-K}} & \text{if } K<0,\\
\chi & \text{if } K=0, \\
\dfrac{\sin(\sqrt{K}\chi)}{\sqrt{K}} & \text{if } K>0.
\end{cases}
\end{equation}
The foliation-based distance~$D\e{F}$ (see \S~\ref{sec:distances_GR}) between the worldline $\chi=0$ and the event $(t,x^i)$ is easily checked to be $a(t)\chi$ which leads us to call $\chi$ the \emph{conformal radial distance}\index{conformal!radial distance}\index{comoving!distance|see{conformal distance}}. Besides, within a given spatial section~$\Sigma_t$ the proper area of a $\chi=\cst$ sphere is $4\pi a^2(t)f_K^2(\chi)=4\pi a^2(t) r^2$, hence $r$ represents the \emph{conformal areal radius}\index{conformal!areal radius} of the sphere. The last radial coordinate~$\varrho$ has no specific interpretation, but it is adapted to the introduction of Cartesian-like coordinates $x\define\varrho\sin\theta\cos\ph$, $y\define\varrho\sin\theta\sin\ph$, $z\define\varrho\cos\theta$, since $\dd\varrho^2+\varrho^2\dd\Omega^2=\dd x^2 + \dd y^2 + \dd z^2$.

It is often convenient to introduce the \emph{conformal time}~$\eta$\index{conformal!time} defined from cosmic time by $\dd t = a\,\dd\eta$, in order to completely factorise the scale factor in the expression of the metric,
\begin{equation}
\dd s^2 = a^2(\eta) \pa{ -\dd\eta^2 + \gamma_{ij} \dd x^i \dd x^j }.
\end{equation}

\subsubsection{Geometrical properties}

Table~\ref{tab:FL_geometry} summarises all the geometrical quantities (Christoffel symbols, curvatures, etc.) associated with the FL spacetime, for two different choices for the $x^0$-coordinate: cosmic time~$t$ or conformal time~$\eta$. We introduced the evolution rates of the scale factor
\begin{equation}
H \define \frac{1}{a} \ddf{a}{t} \define \frac{\dot{a}}{a},
\qquad
\Hc\define \frac{1}{a} \ddf{a}{\eta} \define \frac{a'}{a} = aH,
\end{equation}
respectively called the \emph{Hubble expansion rate}\index{Hubble!expansion rate} and conformal Hubble expansion rate\index{conformal!expansion rate}, for reasons that shall become clearer below. We see from Table~\ref{tab:FL_geometry} that the FL spacetime admits three typical curvature scales, namely $\Hc^2$, $\Hc'$, and $K$.

Interestingly, since its Weyl tensor vanishes, the FL geometry is \emph{conformally flat}. This property is obvious if $K=0$, as the metric then reads $\dd s^2= a^2 \eta_{\mu\nu} \dd x^\mu \dd x^\nu$, so that the conformal factor is simply $a^2$; for $K\not=0$ the underlying conformal transformation is more complicated, in particular it is no longer homogeneous~\cite{1973lsss.book.....H}.

\renewcommand{\arraystretch}{1.5}
\begin{table}[h]
\centering
\begin{tabular}{|>{\columncolor{lightgray}}m{3cm}|c|c|}
\cline{2-3}
\multicolumn{1}{c|}{}		& \cellcolor{lightgray} $\boldsymbol{x^0=t}$ 	& 	\cellcolor{lightgray}	$\boldsymbol{x^0=\eta}$ \\
\hline
\sf\textbf{Metric} & $\dd s^2 = -\dd t^2 + a^2(t) \gamma_{ij} \dd x^i \dd x^j$ 
				& $\dd s^2 = a^2(\eta)\pa{-\dd\eta^2 + \gamma_{ij} \dd x^i \dd x^j}$ \\
\hline
					&  $\Gamma\indices{^0_i_j}=H g_{ij} $	
							&  $\Gamma\indices{^0_0_0}=\Hc	\qquad 	\Gamma\indices{^0_i_j}=\Hc\gamma_{ij}$\\
					&	$\Gamma\indices{^i_0_j}=H \delta^i_j 	\qquad 	\Gamma\indices{^i_j_k}= \tensor[^3]{\Gamma}{^i_j_k}$	
							&	$\Gamma\indices{^i_0_j}=\Hc	\delta^i_j 	\qquad 	\Gamma\indices{^i_j_k}= \tensor[^3]{\Gamma}{^i_j_k}$ \\[2mm]
\hline 
	&  $R\indices{^0_i_0_j}=\dfrac{\ddot{a}}{a} \, g_{ij} \qquad R\indices{^i_0_j_0}=-\dfrac{\ddot{a}}{a}\,\delta^i_j$ 															& $R\indices{^0_i_0_j}=\Hc'\gamma_{ij} \qquad R\indices{^i_0_j_0}=-\Hc' \delta^i_j$ \\ 
											 &	 $R\indices{^i_j_k_\ell}=2\pa{H^2+\dfrac{K}{a^2}} \delta^i_{[k}g_{\ell]j}$	
													 	& $R\indices{^i_j_k_\ell}=2(\Hc^2+K)\delta^i_{[k}\gamma_{\ell]j}$ \\[4mm]
\hline
\vspace*{1mm}
\sf\textbf{Kretschmann scalar} & \multicolumn{2}{c|}{$\mathcal{K}=12\pac{\pa{\dfrac{\ddot{a}}{a}}^2 + \pa{ H^2 + \dfrac{K}{a^2} }^2}
																						= \dfrac{12}{a^4} \pac{ \pa{\mathcal{H}'}^2 + \pa{\mathcal{H}^2+K}^2 }$}\\[4mm]
\hline
 	& $R_{00}=-\dfrac{3\ddot{a}}{a}$  				& $R_{00}=-3\Hc'$ \\ 
									&	$R_{ij}=\pa{ \dfrac{\ddot{a}}{a} +2H^2 + \dfrac{2K}{a^2}} g_{ij}$ 	& $R_{ij}=(\Hc'+2\Hc^2+2K)\gamma_{ij}$ \\[4mm]
\hline 
\sf\textbf{Ricci scalar} & \multicolumn{2}{c|}{$R= 6\pa{ \dfrac{\ddot{a}}{a} + H^2 + \dfrac{K}{a^2} } = \dfrac{6}{a^2}(\Hc'+\Hc^2+K)$} \\ 
\hline 
\sf\textbf{Weyl tensor} & \multicolumn{2}{c|}{$C_{\mu\nu\rho\sigma}=0$} \\ 
\hline
\multirow{-24}{3cm}{\sf\bfseries Christoffel symbols}
\multirow{-20}{3cm}{\sf\bfseries Riemann tensor}
\multirow{-12}{3cm}{\sf\bfseries Ricci tensor}
\vspace*{-4mm}
\multirow{0}{3cm}{\sf\bfseries Einstein tensor} & $E_{00}=3\pa{H^2+\dfrac{K}{a^2}}$ 		& $E_{00}=3(\Hc^2+K)$ \\[4mm] 
										&	$E_{ij}=-\pa{ \dfrac{2\ddot{a}}{a} + H^2 + \dfrac{K}{a^2}} g_{ij}$	& $E_{ij}=-(2\Hc'+\Hc^2+K)\gamma_{ij}$\\[4mm]
\hline 
\end{tabular} 
\caption{Geometry of the FL spacetime for two different choices of the time coordinate: $t$ or $\eta$. A dot and a prime respectively denote a derivative with respect to $t$ and $\eta$, with the relation $X'=a \dot{X}$. The associated evolution rates for $a$ are $H\define\dot{a}/a$ and $\mathcal{H}\define a'/a=aH$. The notation $\tensor[^3]{\Gamma}{^i_j_k}$ stands for the Christoffel symbols associated with the three-dimensional metric $\gamma_{ij}$. We recall that the Kretshmann scalar is defined by $\mathcal{K}\define R^{\mu\nu\rho\sigma} R_{\mu\nu\rho\sigma}$.\index{Christoffel coefficients!of the FL metric}}
\label{tab:FL_geometry}
\end{table}

\subsubsection{Fundamental observers, cosmic expansion}\index{expansion of the Universe}\index{cosmic!expansion|see{expansion of the Universe}}

Let us now turn to the physical interpretation of the FL geometry. First note that the curves defined by $x^i=\cst$ are timelike geodesics, as easily shown by checking that the associated four-velocity $u^\mu = \delta^\mu_t$ satisfies $\dot{u}^\mu+\Gamma\indices{^\mu_\nu_\rho}u^\nu u^\rho=0$. These curves thus correspond to the worldlines of free-falling observers, called \emph{fundamental observers}\index{fundamental observers}, whose proper time is cosmic time~$t$. The spatial coordinates $\{x^i\}$ can also be viewed as Lagrangian, or \emph{comoving}, coordinates\index{comoving!coordinates} for the fundamental observers.

The kinematics of the fundamental geodesic flow can be described using Fermi normal coordinates around, e.g., the worldline~$\wl_0$ defined by $x^i=0$. A possible choice is
\begin{align}
\tau &\define t + \frac{H}{2}\,[a(t)r]^2, \\
D &\define a(t) r,
\end{align}
in terms of which the metric is indeed Minkowskian, up to terms on the order of $\text{curv.}\times D^2$,
\begin{equation}
\dd s^2 = \pa[1]{-1 + 3 H^2 D^2 + \dot{H} D^2 +\ldots} \dd \tau^2 + (1+H^2 D^2+\ldots) \dd D^2 + D^2\dd\Omega^2 + \ldots
\end{equation}
In the vicinity of $r=0$, the quantity $D=a(t) r$ is thus a good notion of physical distance\footnote{All the worldline-event distances defined in the previous chapter coincide with $D$ up to curvature terms. Note that we could have replaced $r$ by $\chi$ in the expression of $D$, since both quantities differ by terms on the order of $K \chi^2$.}. We conclude that a particle following the neighbouring fundamental geodesic $\wl_r$, with comoving coordinate $r$, moves radially with respect to $\wl_0$, with a velocity $v=\dd D/\dd\tau=\dot{a}r=H D$ as measured in the observer's frame. Because the origin of the spatial coordinate system is arbitrary, the previous reasoning equally applies to the vicinity of any fundamental observer. The cases $H\leq 0$ or $H\geq0$ therefore correspond respectively to a contracting or expanding universe, in which fundamental observers are all approaching or receding from each other. In 1929, the observation of nearby galaxies by Hubble~\cite{1929PNAS...15..168H} revealed that our Universe lies in the second case. It also established empirically the relation $v=H D$, called the \emph{Hubble law}\index{Hubble!law} in the honour of its discoverer.

\subsection{Dynamics of cosmic expansion}

The dynamics of the scale factor, hence of cosmic expansion, is governed by the matter content of the Universe via the Einstein equation. Before we analyse their specific consequences, let us first discuss the description of matter in cosmology.

\subsubsection{Description of matter}\index{matter!in homogeneous cosmology}

For matter to respect the assumptions of homogeneity and isotropy, its stress-energy tensor must read\index{perfect fluid}
\begin{empheq}[box=]{equation}
T_{\mu\nu} = \rho(t) \, u_\mu u_\nu + p(t) \perp_{\mu\nu},
\label{eq:stress-energy_tensor_perfect_fluid}
\end{empheq}
identified with a homogeneous perfect fluid\footnote{A perfect fluid is an idealised fluid model with no viscosity, anisotropic stress, or diffusive heat transport.} following the fundamental geodesic flow, with $\vect{u}=\vect{\partial}_t$, and whose energy density~$\rho$ and isotropic pressure~$p$ do not depend on spatial coordinates; $\perp_{\mu\nu}\define g_{\mu\nu}+u_\mu u_\nu$ denotes the projector on spatial sections~$t=\cst$, hence $\perp_{\mu0}=0$ and $\perp_{ij}=g_{ij}$ here. The form~\eqref{eq:stress-energy_tensor_perfect_fluid} of $T_{\mu\nu}$ forbids, for instance, the presence of a cosmic electromagnetic field which, as we will see in Chap.~\ref{chapter:sources_anisotropy},  generically creates anisotropy. Nevertheless, the superposition of a large number of electromagnetic fields with random polarisations, i.e. a gas of photons, is allowed.

The cosmological fluid described by the stress-energy tensor~\eqref{eq:stress-energy_tensor_perfect_fluid} is in principle made of several species~$s$, so that we must write
\begin{equation}
\rho = \sum_s \rho_s, \qquad p = \sum_s p_s.
\end{equation}
Each species is characterised by its equation-of-state parameter~$w$, defined as the ratio between its pressure and its energy density
\begin{equation}
w_s \define \frac{p_s}{\rho_s},
\end{equation}
or in other words, the ratio between (two thirds of) its microscopic kinetic energy and its total energy. For example, a nonrelativistic perfect gas of point particles with mass~$m$ at a temperature~$T$ has
\begin{equation}
w\e{gas} = \frac{k\e{B}T}{m c^2} = 9.20 \times 10^{-11} \pa{\frac{T}{100\U{K}}}\pa{\frac{m\e{p}}{m}},
\end{equation}
where $k\e{B}=1.38\times 10^{-23}\U{J/K}$ is the Boltzmann constant, and $m\e{p}=938\U{MeV}/c^2$ is the proton mass; $w\e{gas}$ is thus a very small quantity, except at very high temperatures, for which the gas becomes relativistic. The other end of the spectrum is ultrarelativistic matter, i.e. radiation, for which we have already seen in \S~\ref{sec:electric_magnetic} that $w\e{rad}=1/3$. The effective equation-of-state parameter for the cosmological fluid can be written as
\begin{equation}
w \define \frac{p}{\rho} = \sum_s f_s w_s, \qquad \text{with}\quad f\e{s} \define \frac{\rho_s}{\rho};
\end{equation}
it approaches $0$ when the total energy is dominated by nonrelativistic matter (baryons, dark matter), and $1/3$ when dominated by radiation (photons, neutrinos).

The conservation of total matter's energy $\nabla_{\mu}T^{\mu 0}=0$ leads to the following constraint for the time evolution of $\rho$
\begin{empheq}[box=\fbox]{equation}
\dot{\rho} + 3H(1+w)\rho = 0,
\label{eq:conservation_energy_momentum_cosmo}
\end{empheq}
while the conservation of total momentum $\nabla_\mu T^{\mu i}=0$ is here trivially satisfied.

\subsubsection{The Friedmann equations}

The Einstein equation, including a cosmological constant and using the stress-energy tensor~\eqref{eq:stress-energy_tensor_perfect_fluid} on the right-hand side leads to the two \emph{Friedmann equations}\index{Friedmann!equations} governing the dynamics of cosmic expansion
\begin{empheq}[box=\fbox]{align}
H^2 &= \frac{8\pi G}{3} \rho - \frac{K}{a^2} + \frac{\Lambda}{3}, 
\label{eq:Friedmann_1}\\
\frac{\ddot{a}}{a} &= -\frac{4\pi G}{3} (1+3 w) \rho + \frac{\Lambda}{3}.
\label{eq:Friedmann_2}
\end{empheq}
Because of the link between the conservation of energy-momentum and the Bianchi identity, Eq.~\eqref{eq:conservation_energy_momentum_cosmo} is not independent from the Friedmann equations, more precisely $\dd\eqref{eq:Friedmann_1}/\dd t-2\times\eqref{eq:Friedmann_2}\Leftrightarrow \eqref{eq:conservation_energy_momentum_cosmo}$.

The first Friedmann equation~\eqref{eq:Friedmann_1} shows that, at any time, the cosmic expansion rate has three distinct contributions: matter's energy density, spatial curvature, and cosmological constant. Their relative weight is usually quantified by introducing the associated \emph{cosmological parameters}\index{cosmological!parameters}
\begin{equation}
\Omega\e{m}\define\frac{8\pi G \rho}{3 H^2},
\qquad
\Omega_K \define \frac{-K}{a^2 H^2},
\qquad
\Omega_\Lambda \define \frac{\Lambda}{3 H^2} ,
\end{equation}
whose sum is $1$ by virtue of Eq.~\eqref{eq:Friedmann_1}. In the case where $\Omega\e{m}\gg \Omega_K,\Omega_\Lambda$, and assuming that matter's energy density is dominated by its nonrelativistic ($w=0$) or ultrarelativistic ($w=1/3$) component, we find from Eq.~\eqref{eq:conservation_energy_momentum_cosmo}
\begin{equation}
\rho \propto
\begin{cases}
a^{-3} & \text{if } w=0\\
a^{-4} & \text{if } w=1/3
\end{cases}.
\label{eq:energy_matter_radiation}
\end{equation}
Equation~\eqref{eq:Friedmann_1} is then integrated as
\begin{equation}
a \propto
\begin{cases}
t^{2/3}\propto\eta^2 &\text{if } w=0\\
t^{1/2}\propto\eta &\text{if } w=1/3
\end{cases}.
\label{eq:expansion_matter_radiation}
\end{equation}
Another interesting case is $\Omega_{K}\ll \Omega\e{m},\Omega_\Lambda>0$, with $w=0$, which yields
\begin{equation}
a(t)\propto \sinh^{2/3}\pa{ \frac{\sqrt{3\Lambda}}{2}\,t }.
\label{eq:expansion_LCDM}
\end{equation}
This solution coincides with $a\propto t^{2/3}$ in the limit $\Lambda\rightarrow 0$; on the contrary, when the cosmological constant dominates over the contribution of matter we get $a(t)\propto \exp(t\sqrt{\Lambda/3}) =\exp (Ht)$, the expansion rate being a constant in this case.

\subsubsection{Possible sources of accelerated expansion}\index{accelerated expansion!possible sources of}

The last case is an example of accelerated expansion ($\ddot{a}>0$), driven by a positive cosmological constant. As clearly shown by the second Friedmann equation~\eqref{eq:Friedmann_2}, standard matter is generally unable to produce such an acceleration---which fits with our Newtonian intuition that gravity is an attractive force---unless its energy density is negative, or $w<-1/3$. The first possibility is excluded by stability requirements; the second one would mean that matter has negative pressure, which cannot happen with standard matter since pressure is proportional to kinetic energy, hence always positive. However, such a behaviour can be \emph{mimicked} by a quantum field\footnote{Quantisation is actually not required here, but an action of the form~\eqref{eq:action_scalar_field} is motivated by quantum field theory.}, the simplest example being a scalar.

Consider a scalar field~$\phi$ minimally coupled with spacetime geometry, and self interacting through a potential~$V(\phi)$. The associated action reads\index{action!of a scalar field}\index{scalar!field}
\begin{equation}
S\e{sf}[\phi,\vect{g}] = \int_\mathcal{M} \dd^4 x \; \sqrt{-g} \pac{ -\frac12 \partial_\mu \phi \, \partial^\mu \phi - V(\phi) }, 
\label{eq:action_scalar_field}
\end{equation}
and the corresponding stress-energy tensor is\index{stress-energy tensor!of a scalar field}
\begin{equation}
T_{\mu\nu}\h{sf} 
\define \frac{-2}{\sqrt{-g}} \frac{\delta S\e{sf}}{\delta g^{\mu\nu}}
=
\partial_\mu \phi \, \partial_\nu \phi - \frac12 (\partial^\rho\phi \,\partial_\rho\phi)\,g_{\mu\nu} - V(\phi) g_{\mu\nu}.
\end{equation}
Suppose this field is homogeneous on spatial sections $t=\cst$, then $\partial_\mu\phi=-\dot{\phi} \, u_\mu$ and the above stress-energy tensor takes the form
\begin{equation}
T\h{sf}_{\mu\nu} =
\pa{ \frac{\dot{\phi}^2}{2} + V } u_\mu u_\nu
+ \pa{ \frac{\dot{\phi}^2}{2} - V } \perp_{\mu\nu},
\end{equation}
readily identified with Eq.~\eqref{eq:stress-energy_tensor_perfect_fluid}. A homogeneous scalar field thus behaves similarly to a perfect fluid with equation-of-state parameter
\begin{equation}
w\e{sf}=\frac{\dot{\phi}^2-2V(\phi)}{\dot{\phi}^2+2V(\phi)},
\end{equation}
which can be negative, and even reach $-1$ if the field evolves very slowly compared to the amplitude of its potential, $\dot{\phi}^2\ll V(\phi)$. In this extreme case, it is similar to the cosmological constant, $T_{\mu\nu}\h{hsf}=-V g_{\mu\nu}$.

As will be discussed in \S~\ref{sec:history_Universe}, our Universe seems to have experienced two eras of accelerated expansion: an early-time one known as \emph{inflation}\index{inflation}, and a recent one. Both can be driven by a scalar field---respectively called \emph{inflaton}\index{inflaton} and \emph{quintessence}\index{quintessence}~\cite{2013CQGra..30u4003T}---but there are in fact many other mechanisms capable of producing acceleration, gathered in the denomination of \emph{dark energy}\index{dark!energy}~\cite{2010deot.book.....R} in the late-time case. A noncomprehensive list includes:
\begin{itemize}
\item The presence of \textbf{exotic matter}, either minimally coupled to gravity, such as quintessence and Chaplygin gas~\cite{2006tmgm.meet..840G}\index{Chaplygin gas}, or nonminimally coupled such as chameleons~\cite{2013arXiv1312.2006K}\index{chameleon}, Galileons\index{Galileon}, Horndeski theories \cite{2013CQGra..30u4006D}\index{Horndeski!scalar theory} and beyond~\cite{2014PhRvD..89f4046Z,2015JCAP...02..018G}. Models involving nonminimally coupled fields can also be viewed as 
\item \textbf{modified theories of gravity}\index{modified theories of gravity}~\cite{2012PhR...513....1C}, more precisely scalar-tensor theories, which also contain $f(R)$ actions~\cite{2010LRR....13....3D}. Other theories beyond GR include bimetric and massive gravities~\cite{2014LRR....17....7D}, Lorentz-violating models like Einstein-\ae ther gravity~\cite{2008arXiv0801.1547J}, etc.
\item In a more conservative way, acceleration could be due to the \textbf{backreaction}\index{backreaction} of inhomogeneities on the average cosmic expansion~\cite{2011arXiv1109.2314C,2012ARNPS..62...57B,2013arXiv1311.3787W} (see also \S~\ref{sec:limits_perturbations}).
\end{itemize}

Another possibility is that acceleration is apparent, due to a misinterpretation of our observations. Two scenarios---though now ruled out as viable explanations of dark energy---can be cited:
\begin{itemize}
\item A strong absorption of the photons emitted by SNe, or their oscillations with axions~\cite{2002PhRvL..88p1302C}, which would explain their overdimming (see \S~\ref{sec:Hubble_diagram}) by violating the distance-duality relation. The photon-axion oscillation model has been eliminated by taking into account the effects intergalactic plasma~\cite{2002PhRvD..66d3517D}. Besides, the violations of the distance duality relation compatible with observations~\cite{2004PhRvD..70h3533U} are not sufficient to explain SN data without the need of dark energy.
\item A second option consists in violating the Copernical principle, by assuming that we lie at the centre of a huge void~\cite{2008PhRvL.101m1302C} (modelled e.g. by the Lema\^{i}tre-Tolman-Bondi metric) expanding faster than the homogeneous Universe. This possibility is nevertheless highly constrained by observations related to CMB scattering (see Ref.~\cite{2012CRPhy..13..682C} and references therein), and would therefore require a unnatural level of fine tuning.
\end{itemize}

\subsection{Content and history of our Universe}
\label{sec:history_Universe}

The cosmological parameters of our Universe, whose geometry is assumed to be well modelled by the FL metric, have been measured with an increasing precision over the last decades. Most observations agree~\cite{2014MNRAS.440.1138E} on the following value for today's expansion rate~\cite{2015arXiv150201589P}
\begin{equation}
H_0 = 67.74\pm 0.46\U{km/s/Mpc},
\end{equation}
often written as $H_0=h\times 100\U{km/s/Mpc}$, and with the \emph{concordance set}\index{cosmological!parameters}\index{concordance model|see{cosmological parameters}} of cosmological parameters~\cite{2015arXiv150201589P}
\begin{equation}
\Omega\e{m0} = 0.3089 \pm 0.0062,
\qquad
\Omega_{K0} = 0.0008^{+0.0040}_{-0.0039},
\qquad
\Omega_{\Lambda 0} = 0.6911 \pm 0.0062,
\end{equation}
where a subscript zero conventionally denotes today's value of a quantity. The mean matter density in today's Universe is therefore $\rho\e{m0}\sim 3\times10^{-27}\U{kg/m^3}$. Among this matter content, only one sixth is made of quarks and nonrelativistic leptons---abusively called \emph{baryonic matter}\index{baryonic matter} in cosmology, as leptons and mesons do not contribute significantly to the total amount---with $\Omega\e{b0}h^2=0.02230\pm 0.00014$~\cite{2015arXiv150201589P}, while radiation---photons and neutrinos---represent much less, $\Omega\e{r0}\sim 10^{-4}$. The actual nature of the remainder is unknown, except that it is nonrelativistic and does not seem to interact with normal matter; it has in particular no electromagnetic signature, and therefore has been denominated (cold) \emph{dark matter}\index{dark!matter}~\cite{2005PhR...405..279B}. The resulting cosmological model, in which spacetime is described by the FL geometry, whose dynamics is dictated by General Relativity with a cosmological constant, and where five sixth of the material content today is made of noninteracting and nonrelativistic dark matter, is known as $\Lambda$-Cold-Dark-Matter ($\Lambda$CDM).

\begin{figure}[p]
\centering
\includegraphics[angle=90,origin=c,width=\textwidth]{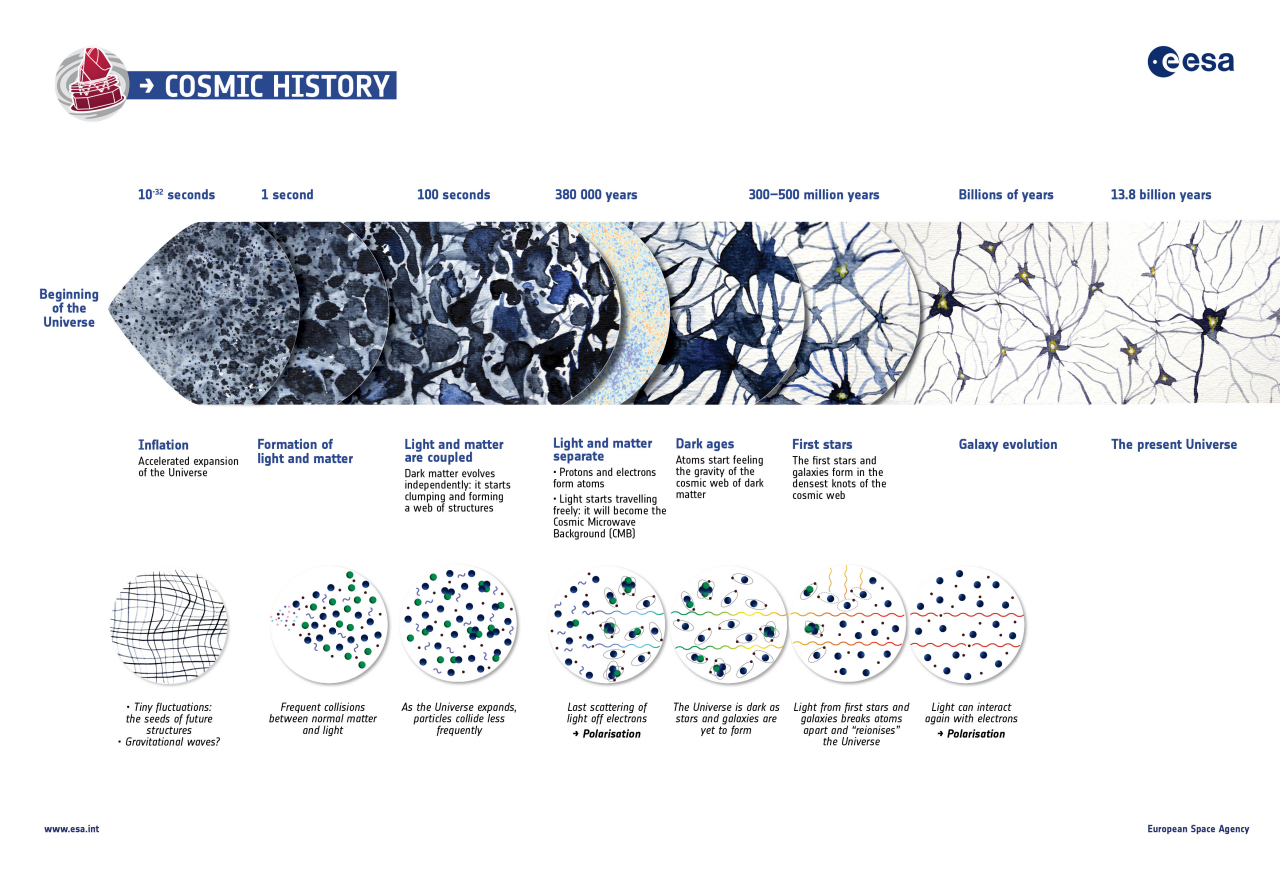}
\caption{A brief history of our Universe, in terms of cosmic time~$t$. Image courtesy of ESA.}
\label{fig:history_universe}
\end{figure}

Let us close this section with a brief history of our Universe, as inferred in the $\Lambda$CDM framework and summarised in Fig.~\ref{fig:history_universe}. First of all, following the Friedmannian dynamics backwards in time\footnote{and extrapolating it to energy scales where current physical theories are expected to break down} shows that a singularity, namely $a=0$, occurs in a finite-time past. If this so-called \emph{Big Bang}\index{Big Bang} singularity is taken as the origin of cosmic time, then today corresponds to $t_0 = 13.81\U{Gyr}$. During the first $\sim 10^{-32}\U{s}$ after the Planck era ($t\sim 10^{-35}\U{s}$), our Universe is thought to have experienced a first period of accelerated expansion, \emph{cosmic inflation}, during which distances have increased by a factor $\ex{N}$, where $N>60$ is the number of $e$-folds\index{$e$-fold number} characterising the duration of inflation. Originally introduced as a solution to the flatness and horizon problems of the Hot-Big-Bang model~\cite{1978AnPhy.115...78B,1979JETPL..30..682S,1981PhRvD..23..347G}, inflation now fully belongs to standard cosmology; see however Refs.~\cite{2001PhRvD..64l3522K,2015PhR...571....1B} for alterinflationarist theories. Although hundreds of inflationary models have been proposed and tested against observations~\cite{2014PDU.....5...75M}, the most popular ones involve a single scalar field (the inflaton)\index{inflaton}, whose slight inhomogeneities, due quantum fluctuations, have been the seeds of the structures that we observe today. The end of inflation is followed by a \emph{reheating}\index{reheating} phase, during which the inflaton decays into particles of the standard model of particle physics. Nevertheless, the underlying mechanisms are still poorly constrained by observations~\cite{2015PhRvL.114h1303M}.

After reheating, the Universe is made of a dense quark and lepton plasma. As the temperature drops due to expansion, hadrons are formed, followed by light atomic nuclei (essentially deuterium, helium, and lithium): this is \emph{primordial nucleosynthesis}\index{nucleosynthesis} (from $k\e{B}T\sim 150\U{MeV}$ to $\sim 50\U{keV}$). At the end of this period, the scale factor reads $a_0/a\e{n}=5\times 10^7$, and the expansion dynamics is still by far dominated by radiation. However, as seen in Eq.~\eqref{eq:energy_matter_radiation}, the energy density of radiation decreases faster than the one of nonrelativistic matter, and both contributions become comparable ($\Omega\e{r}\sim\Omega\e{m}$) for $a_0/a\e{eq}=\Omega\e{m0}/\Omega\e{r0}\approx 3400$. This \emph{equality}\index{equality (between matter and radiation)} represents the transition between the radiation and matter eras, i.e. between the two expansion laws of Eq.~\eqref{eq:expansion_matter_radiation}. At this stage, however, the photon fluid is still dense and energetic enough to maintain matter ionised. For $a_0/a_*=1100$ ($k\e{B}T_*\sim 0.3\U{eV}$), this condition is no longer fulfilled, and the \emph{recombination}\index{recombination} of atomic nuclei and electrons into atoms occurs, followed by the \emph{decoupling}\index{decoupling} between photons and electrons: the Universe becomes neutral and transparent. The light released at that epoch is observed today under the form of a cosmic microwave background (CMB).

After recombination, the Universe remains neutral for a few hundreds of millions of years (the \emph{dark ages}\index{dark!ages}), during which structures form via gravitational accretion. On small scales, some matter clumps collapse and get hot enough to activate the fusion of hydrogen into helium, giving birth to the first stars. The light they emit induces a \emph{reionisation}\index{reionisation} of the Universe ($a_0/a\e{re}\approx 12$). The next billions of years are then characterised by the formation and evolution of galaxies on small scales; and by the apparition of a large-scale cosmic web~\cite{1986ApJ...302L...1D}, where voids (from $10$ to $150\U{Mpc}$) are separated by walls and filaments. The cosmological constant starts to dominate over matter's energy density for $a_0/a\e{ded}\approx 1.3$, leading to an acceleration of cosmic expansion well described by Eq.~\eqref{eq:expansion_LCDM}.

\section{Linear perturbation theory}\label{sec:linear_perturbation_theory}

Our own existence, on a planet orbiting around a star, within a galaxy belonging to a supercluster, indicates that the Universe is not perfectly homogeneous, but rather presents structures over a wide variety of scales, the largest one being the cosmic web, whose typical inhomogeneity scale is $\sim 100\U{Mpc}$ today. A possible strategy for modelling this departure from strict homogeneity and isotropy consists in introducing small perturbations to FL geometry. In this section---essentially based on textbook~\cite{PeterUzan}, though much less complete---we summarise the main features of the cosmological perturbation theory at linear order.

\subsection{Perturbed quantities}

\subsubsection{The scalar-vector-tensor decomposition}\index{scalar-vector-tensor (SVT) decomposition}

Before discussing the standard perturbation scheme, let us introduce the scalar-vector-tensor (SVT) decomposition\index{scalar-vector-tensor (SVT) decomposition} of spatial vectors and tensors. Let $\vect{V}$ be a vector tangent to a spatial hypersurface $\Sigma_t$ equipped with the metric $\vect{\gamma}(t)$. It can be uniquely decomposed into a gradient part and a curl (divergent-free) part as
\begin{equation}
V_i = \partial_i V + \hat{V}_i,
\qquad \text{with} \quad
\Dd_i \hat{V}^i=0,
\end{equation}
where $\vect{\Dd}$ denotes the covariant derivative associated with the Levi-Civita connection of $\vect{\gamma}(t)$. The 3 degrees of freedom (dofs) of $\vect{V}$ are thus split into 1 scalar dof plus $3-1=2$ vector dofs. Note that, in all this section, \emph{the indices $i,j,\ldots$ are raised and lowered by $\gamma^{ij}$ and $\gamma_{ij}$, and a hat will always denote a divergence-free quantity}.

Similarly, any rank-two symmetric tensor $\vect{T}\in \mathrm{T}\Sigma_t\otimes \mathrm{T}\Sigma_t$ can be decomposed as
\begin{equation}
T_{ij} = \Dd_i \Dd_j T_1 + T_2 \, \gamma_{ij} + \Dd_{(i} \hat{T}_{j)} + \hat{T}_{ij},
\qquad \text{with} \quad
\Dd_i \hat{T}^i=\Dd_i \hat{T}^{ij}=\hat{T}^i_i=0,
\end{equation}
which spreads the 6 dofs of $\vect{T}$ into 2 scalar dofs, $3-1=2$ vector dofs, and $6-3-1=2$ tensor dofs. Such decompositions will be particularly convenient for the cosmological perturbation theory, because the scalar, vector, and tensor components will turn out to decouple from each other at linear order.

\subsubsection{Perturbed metric}

The cosmological perturbation theory consists in modelling the full spacetime metric $\vect{g}$ as approximately equal to the FL metric $\bar{\vect{g}}$---hereafter, a bar denotes an unperturbed, background quantity---from which it differs by a small perturbation~$\delta \vect{g}$. The general expression for this perturbed metric is thus\index{perturbed!metric (general form)}
\begin{align}
\dd s^2 &= \pa{ \bar{g}_{\mu\nu} + \delta g_{\mu\nu} } \dd x^\mu \dd x^\nu \\
				&= a^2(\eta) \pac{ -(1+2A) \dd\eta^2 + 2 B_i \dd x^i \dd\eta + (\gamma_{ij} + 2 C_{ij}) \dd x^i \dd x^j},
\end{align}
where $A, B_i, C_{ij}\ll 1$. All three are functions of the spacetime coordinates $(\eta,x^k)$. Mathematically speaking, they are respectively a scalar, a vector and a tensor with respect to the spatial hypersurfaces $\Sigma_t$ equipped with the background spatial metric $\vect{\gamma}$. As such, $B_i$ and $C_{ij}$ can be decomposed according to the SVT scheme as
\begin{align}
B_i &= \Dd_i B + \hat{B}_i \\
C_{ij} &= \Dd_i \Dd_j C_1  + C_2 \gamma_{ij} + \Dd_{(i} \hat{C}_{j)} + \hat{C}_{ij}
\end{align}
with the conventional requirements on hatted quantities. Among the 10 dofs of the set $\{A,B_i,C_{ij}\}$, 4 are spurious, because associated with the detailed coordinate mapping between the background and perturbed spacetimes (see e.g. Chap.~5 of Ref.~\cite{PeterUzan} for details). This issue is known as the \emph{gauge freedom}\index{gauge!freedom (cosmological perturbations)} of the cosmological perturbation theory. The remaining $10-4=6$ dofs can be encoded into the following gauge-invariant quantities
\begin{align}
\Phi &\define A + \Hc (B-C_1') + (B-C_1')',\\
\Psi &\define -C_2 - \Hc(B - C_1'), \\
\hat{\Omega}_i &\define \hat{B}_i - (\hat{C}_i)',\\
&\hat{C}_{ij}.
\end{align}

As in classical electrodynamics, the gauge can be fixed by imposing some conditions on the gauge fields (here $A,B_i,C_{ij}$). A few interesting choices exist, we here choose to work in the so-called Poisson (or Newtonian or longitudinal) gauge, for which\index{Poisson gauge}\index{Newtonian gauge|see{Poisson gauge}}\index{Longitudinal gauge|see{Poisson gauge}}
\begin{equation}
B=C_1=\hat{C}_i=0,
\end{equation}
so that the metric can be completely written in terms of gauge-invariant quantities as\index{perturbed!metric (in Poisson gauge)}
\begin{empheq}[box=\fbox]{equation}
\dd s^2 
= a^2(\eta) \pac[3]{ -(1+2\Phi) \dd\eta^2 + 2 \hat{\Omega}_i\dd x^i \dd\eta + (1-2\Psi) \gamma_{ij} \dd x^i \dd x^j + \hat{C}_{ij} \dd x^i \dd x^j }.
\label{eq:perturbed_metric_Poisson_gauge}
\end{empheq}
Note that there are some variations in what is called Newtonian gauge in the literature; for instance, in Ref.~\cite{PeterUzan} $\hat{B}_i$ is set to zero, while $\hat{C}_i$ is nonzero in general. The perturbations of geometrical quantities (Christoffel symbols, curvatures,\ldots) associated with the metric~\eqref{eq:perturbed_metric_Poisson_gauge} can be found in the Appendix C of Ref.~\cite{PeterUzan}.

Physically speaking, \begin{inparaenum}[(i)] \item the scalars $\Phi,\Psi$, called the Bardeen potentials\index{Bardeen potentials}, are analogous to the Newtonian gravitational potential; \item the vector perturbation $\Omega_i$ is a gravitomagnetic term producing inertial-frame dragging, analogously to the $g_{t\phi}$ component of the Kerr geometry~\cite{1963PhRvL..11..237K}; while \item the tensor perturbation $\hat{C}_{ij}$ represents gravitational waves.\end{inparaenum}

\subsubsection{Perturbed stress-energy tensor}\label{sec:perturbed_stress-energy_tensor}\index{perturbed!stress-energy tensor of matter}

The perturbations of spacetime's geometry are sourced by inhomogeneities of the distribution of matter. We here make the simplifying assumptions that the perturbed cosmological fluid \begin{inparaenum}[(i)] \item consists of one species; and \item can still be modelled by a perfect fluid\end{inparaenum}. The associated perturbed stress-energy tensor thus reads
\begin{align}
T_{\mu\nu} &= \bar{T}_{\mu\nu} + \delta T_{\mu\nu} \\
					&= (\rho+p) u_\mu u_\nu + p \, g_{\mu\nu}.
						\label{eq:perturbed_stress-energy-tensor}
\end{align}
While these assumptions are wrong in general, they are sufficient to model matter during the (dark-)matter-dominated and dark-energy-dominated eras, where matter's anisotropic stress is negligible.

Decomposing each quantity~$q$ of Eq.~\eqref{eq:perturbed_stress-energy-tensor} as $\bar{q}+\delta q$ yields
\begin{equation}
\delta T_{\mu\nu} = (\delta\rho + \delta p) \bar{u}_\mu \bar{u}_\nu 
								+ 2 (\bar{\rho}+\bar{p}) \bar{u}_{(\mu} \delta u_{\nu)} 
								+ \bar{p}\,\delta g_{\mu\nu}.
\end{equation}
The conventional normalisations of the background and perturbed flows $\bar{u}^\mu$ and $u^\mu$, namely
\begin{equation}
\bar{g}_{\mu\nu} \bar{u}^\mu \bar{u}^\nu = g_{\mu\nu} u^\mu u^\nu = -1
\end{equation}
imply at first order
\begin{equation}
2 g_{\mu\nu} \bar{u}^\mu \delta u^\nu + \delta g_{\mu\nu} \bar{u}^\mu \bar{u}^\nu = 0
\qquad \text{i.e.} \qquad
\delta u^0 = -\frac{\Phi}{a},
\end{equation}
where it is understood that a $0$th component refers to a component with respect to $\vect{\partial}_\eta$. As for the spatial components of the perturbation of the four-velocity, they can be written as $\delta u^i = a^{-1}\upsilon^i$, where the vector $\upsilon^i\vect{\partial}_i$ belongs to $\mathrm{T}\Sigma_t$---like the vector perturbation of the metric---and can thus be decomposed as
\begin{equation}
\upsilon_i = \partial_i \upsilon + \hat{\upsilon}_i.
\end{equation}
These perturbation of matter's stress-energy tensor is also subject to gauge freedom, and it can be shown that the following combinations
\begin{align}
\delta\rho\e{P} &\define \delta\rho + \bar{\rho}'(B-C_1') ,\\
\delta p\e{P} &\define \delta p + \bar{p}'(B-C_1'), \\
\Upsilon &= \upsilon + C_1, \\
\hat{\Upsilon}_i &= \hat{\upsilon}_i + \hat{C}_i,
\end{align}
which coincide with $\delta\rho, \delta p, \upsilon, \hat{\upsilon}_i$ respectively in the Poisson gauge ($B=C_1=\hat{C}_i=0$), are gauge invariant.

Because the cosmological fluid is assumed to contain a single matter species, its pressure and energy density are univocally related, through $p=w\rho$. As a consequence, their perturbations are related as well; in particular, the ratio
\begin{equation}
\frac{\delta p}{\delta \rho} = \ddf{p}{\rho} = w + \rho\ddf{w}{\rho} \define c\e{s}^2
\end{equation}
defines the \emph{sound velocity}~$c\e{s}$\index{sound!velocity}, i.e. the velocity of adiabatic pressure waves within the fluid. For $w=\cst$, we thus have $c\e{s}^2=w$. Note that $c\e{s}^2$ is gauge invariant, since
\begin{equation}
\frac{\delta p\e{N}}{\delta \rho\e{N}} 
= \frac{\delta p + \bar{p}'(B-C_1')}{\delta \rho + \bar{\rho}'(B-C_1')} 
= \frac{c\e{s}^2\delta \rho + c\e{s}^2\bar{\rho}'(B-C_1')}{\delta \rho + \bar{\rho}'(B-C_1')}
= c\e{s}^2
\end{equation}
at first order in perturbations.

\subsection{Evolution of perturbations}

Let us now focus on the evolution equations for the metric and matter perturbations. As previously mentioned, the equations of motion naturally separate into a decoupled set of scalar, vector, and tensor modes.

\subsubsection{Einstein's equation}\index{perturbed!Einstein's equation}

The \emph{tensor modes}\index{tensor!modes} of the Einstein's equation yields
\begin{equation}
\hat{C}_{ij}'' + 2\Hc \, \hat{C}_{ij}' + (2K-\Delta) \hat{C}_{ij} = 0,
\label{eq:evolution_tensor_modes}
\end{equation}
with $\Delta\define \gamma^{ij}\Dd_i \Dd_j$; it is analogous to a wave equation with friction ($2\Hc \hat{C}'_{ij}$) in a harmonic potential with stiffness $2K$. During the matter- and radiation-dominated eras, Eq.~\eqref{eq:evolution_tensor_modes} can be solved in Fourier space thanks to Bessel functions. The amplitude of $\hat{C}_{ij}$ turns out to decrease with $k\eta$, and is therefore negligible on small scales and at late times.

The \emph{vector modes}\index{vector!modes} satisfy the following constraint and evolution equations:
\begin{align}
(\Delta + 2 K) \hat{\Omega}_i &= -16\pi G \bar{\rho} a^2 (1+w) \hat{\Upsilon}_i 
\label{eq:metric_vector_constraint}\\
\hat{\Omega}_i' + 2\Hc \hat{\Omega}_i &= 0.
\label{eq:metric_vector_evolution}
\end{align}
From the second one, we deduce that $\hat{\Omega}_i\propto a^{-2}$, the vector perturbation of the metric is damped in an expanding Universe, and can therefore be neglected in late-time cosmology.

Finally, the \emph{scalar modes}\index{scalar!modes} of Einstein's equation read
\begin{gather}
(\Delta + 3K) \Psi = 4\pi G a^2 \pa{\delta\rho\e{P} + \bar{\rho}' \Upsilon }
\label{eq:Poisson-like_equation}\\
\Psi - \Phi = 0 
\label{eq:Phi_is_Psi}\\
\Psi' + \Hc \Phi = -4\pi G \bar{\rho} (1+w) \Upsilon \\
\Psi'' + 3\Hc (1+c\e{s}^2) \Psi' + \pac{ 2\Hc'+(\Hc^2-K)(1+3 c\e{s}^2)} \Psi - c\e{s}^2 \Delta \Psi = 0
\end{gather}
Note the similarity between Eq.~\eqref{eq:Poisson-like_equation} and the Poisson equation of Newtonian gravitation. The second term on its right-hand side, $\bar{\rho}'\Upsilon$, is related to the perturbation of the fluid's velocity, it can be understood as a kinetic energy term which, contrary to the Newtonian case, gravitates just as mass does.

\subsubsection{Conservation of energy and momentum}\index{perturbed!conservation of energy and momentum}

The conservation of the fluid's energy and momentum, encoded in the equation $\nabla_\mu T^{\mu\nu}=0$, also leads to a set of equations for the vector and scalar modes of matter perturbations. The \emph{vector mode}, on the one hand, satisfies
\begin{equation}
\hat{\Upsilon}_i' + \Hc(1-3c\e{s}^2) \hat{\Upsilon}_i = 0.
\label{eq:matter_vector_evolution}
\end{equation}
When the sound velocity~$c\e{s}$ is nonrelativistic ($c\e{s}^2\ll 1$), then $\hat{\Upsilon}_i\propto a^{-1}$. On the other hand, the \emph{scalar modes} are governed by
\begin{align}
\delta'\e{P} + 3\Hc (c\e{s}^2-w) \delta\e{P} &= -(1+w) (\Delta \Upsilon - 3\Psi') 
\label{eq:conservation_mass} \\
\Upsilon'+\Hc(1-3c\e{s}^2) \Upsilon &= -\Phi - \frac{c\e{s}^2}{1+w} \, \delta\e{P}
\label{eq:Euler}
\end{align}
where we introduced the \emph{density contrast}\index{density contrast}
\begin{equation}
\delta \define \frac{\delta \rho}{\bar{\rho}},
\end{equation}
which, here, is worked out in the Poisson gauge and thus denoted $\delta\e{P}$. Equation~\eqref{eq:conservation_mass} is analogous to the conservation of mass in fluid dynamics, while Eq.~\eqref{eq:Euler} is similar to a integral of the Euler equation. Note that, just like in the unperturbed FL case, the evolution equations for matter and spacetime's metric are not independent from each other. For instance, inserting \eqref{eq:metric_vector_evolution} into $\dd\eqref{eq:metric_vector_constraint}/\dd\eta$ yields \eqref{eq:matter_vector_evolution}. In the set of all the scalar equations, only four are independent (three if we directly replace $\Psi$ by $\Phi$).

\subsubsection{Newtonian regime}\label{sec:Newtonian_regime}

Let us focus on the late-time Universe, where the cosmological fluid essentially consists of a pressureless matter, so that we can take $w=c\e{s}^2=0$ in all the above equations. Besides, as previously mentioned, vector and tensor perturbations decrease with cosmic expansion, so that we can neglect them at late times for any reasonable initial condition, the metric therefore reads
\begin{equation}\label{eq:perturbed_metric_Newtonian_form}
\dd s^2 = a^2(\eta) \pac{ -(1+2\Phi)\dd\eta^2 + (1-2\Phi) \gamma_{ij} \dd x^i \dd x^j }
\end{equation}
in that regime.

If we consider relatively small scales (compared to $\Hc^{-1}$), then the spatial derivatives of the perturbations completely dominate over the background quantities, in particular, in Eq.~\eqref{eq:Poisson-like_equation} $\Delta\Phi\gg 3K\Phi$, and
\begin{equation}
\frac{\delta\rho\e{P}}{\bar{\rho}'\Upsilon} 
\sim \frac{\delta\e{P}}{\Hc\Upsilon} 
\sim \frac{\delta\e{P}}{\Phi} 
\sim \frac{\Delta\Phi}{\Hc^2 \Phi}\ll 1,
\end{equation}
where we used successively Eqs.~\eqref{eq:conservation_energy_momentum_cosmo}, \eqref{eq:Euler}, and \eqref{eq:Poisson-like_equation}. The resulting system of evolution equations reads
\begin{empheq}[box=\fbox]{gather}
\Delta \Phi = 4\pi G a^2 \bar{\rho} \delta, \label{eq:Newtonian_1}\\
\delta' = -\Delta \Upsilon, \label{eq:Newtonian_2}\\
\Upsilon' + \Hc \Upsilon = -\Phi, \label{eq:Newtonian_3}
\end{empheq}
which is identical to the Euler-Poisson system of Newtonian cosmology. Note that we have dropped the index P of the density contrast~$\delta$, because in this regime the discrepancies due to a gauge choice vanish (for $\delta$, but not for the metric perturbations). This is no longer true when Hubble-scale perturbations are at stake---see e.g. Fig.~5.6 of Ref.~\cite{PeterUzan}.

Equations~\eqref{eq:Newtonian_2} and \eqref{eq:Newtonian_3} can be combined to get
\begin{equation}
\delta'' + \Hc \delta' = 4\pi G \bar{\rho} \delta,
\label{eq:evolution_density_Newtonian}
\end{equation}
which rules the evolution of $\delta$ only. It is straightforward to check that, during the matter-dominated era ($a\propto\eta^2$), Eq.~\eqref{eq:evolution_density_Newtonian} admits a growing solution~$D_+$ and a damping solution~$D_-$ such that
\begin{equation}
D_+ \propto \eta^2 \propto a, \qquad D_- \propto \eta^{-3} \propto a^{-3/2},
\label{eq:growing_damping_modes}
\end{equation}
and over which the general solution can be decomposed as
\begin{equation}
\delta(\eta,x^k) = D_+(\eta\leftarrow\eta\e{ini}) \delta_+(\eta\e{ini},x^k)  +  D_-(\eta\leftarrow\eta\e{ini}) \delta_-(\eta\e{ini},x^k).
\end{equation}
If the initial time is much later than matter-radiation equality, then the decaying mode can be neglected, and the density contrast becomes proportional to its initial condition $\delta(\eta,x^k)\approx D_+(\eta\leftarrow\eta\e{ini})\delta(\eta\e{ini},x^k)$. The major part of structure formation occurs during the matter era, since the cosmological constant (or dark energy) starts to dominate 
at very late times, which justifies the relevance of the expressions~\eqref{eq:growing_damping_modes} for $D_\pm$. The growing mode can however be modified to allow for the effect of $\Lambda$ as~\cite{Bernardeau}
\begin{align}
D_+ &\propto \tensor[_2]{F}{_1}\pac{ 1,\frac{1}{3};\frac{11}{6};-\sinh^2\pa{\frac{\sqrt{3\Lambda}}{2}\,t} } \sinh^{2/3}			
							\pa{\frac{\sqrt{3\Lambda}}{2}\,t} \\
&\overset{\text{fit}}{\approx}
\frac{5}{2} \frac{a\Omega\e{m}}{\Omega\e{m}^{4/7}-\Omega_{\Lambda}+(1-\Omega\e{m}+2)(1+\Omega_{\Lambda}/70)}.
\end{align}

\subsubsection{Transfer function}

The evolution equations being (by construction) linear, they are conveniently worked out in Fourier space. From now on, we suppose for simplicity that $K=0$---a choice also motivated by observations---, so that the spatial metric is Euclidean, $\gamma_{ij}=\delta_{ij}$. Any quantity $Q(\eta,x^i)$ can then be decomposed into Fourier modes according to the convention
\begin{align}
Q(\eta,x^i) &= \int \frac{\dd^3 k}{(2\pi)^3} \; \ex{\i k_j x^j} \tilde{Q}(\eta,k_i), \\
\tilde{Q}(\eta,k_i) &\define \int \dd^3 x \; \ex{-\i k_j x^j} Q(\eta,x^i).
\end{align}
The partial differential equations governing the evolution of perturbations thus become systems of independent ordinary differential equations (with respect to $\eta$) for each mode $\vec{k}$ of each quantity, which is therefore linearly related to its initial conditions. Regarding the scalar potential~$\Phi$ in particular, it is customary to introduce the transfer function
\begin{equation}
T(\eta\leftarrow\eta\e{ini},k_i) \define \frac{\tilde{\Phi}(\eta,k_i)}{\tilde{\Phi}(\eta\e{ini},k_i)},
\label{eq:transfer_function}
\end{equation}
which depends on the cosmological parameters.

\subsection{Limits of the linear perturbation theory}\label{sec:limits_perturbations}

\subsubsection{Correlation function and power spectrum}

Because the inhomogeneity of the Universe is believed to origin from the primordial quantum fluctuations of the inflaton, any comparison between theoretical predictions and observations must rely on statistics---in particular, the statistics of matter's density contrast~$\delta$. A central object, for that purpose, is the \emph{correlation function}\index{correlation function}
\begin{equation}
\xi(\eta;x^i,y^i) \define \ev{ \delta(\eta,x^i) \delta(\eta,y^i) },
\end{equation}
which quantifies the statistical similarity between $\delta(\eta,x^i)$ and $\delta(\eta,y^i)$. A positive correlation indicates that if the region around $x^i$ is overdense (resp. underdense), then the region around $y^i$ is likely to be overdense (resp. underdense). A negative correlation (anticorrelation) indicates the opposite situation.

The assumptions of statistical homogeneity and isotropy of our Universe, based on the cosmological principle, imply that $\xi$ can only depend on the distance\footnote{We here refer to distances within a spatial hypersurface, i.e. Euclidean distances as we set $K=0$.} between $x^i$ and $y^i$,
\begin{equation}
\xi(\eta;x^i,y^i)=\xi(\eta;|x^i-y^i|).
\end{equation}
As a consequence, the Fourier transform of $\xi$ with respect to both $x$ and $y$ reads
\begin{align}
\tilde{\xi}(\eta;k_i,l_i) &\define \ev[2]{ \tilde{\delta}(\eta,k_i) \, \tilde{\delta}(\eta,l_i) } \\
									&= (2\pi)^3 \delta\e{D}(k_i+l_i) \, P_\delta(\eta;|k_i|),
\end{align}
where $\delta\e{D}$ denotes the Dirac distribution, and where we introduced the matter \emph{power spectrum}~$P$\index{power spectrum!definition}, defined as
\begin{equation}
P (\eta;k) = \int \dd^3 x \; \ex{\i k_j x^j} \xi(\eta;|x^i|) = 4\pi \int_0^\infty r^2 \dd r \; \frac{\sin kr}{kr} \, \xi(\eta;r).
\end{equation}
Analogous quantities can be defined for the other quantities of interest, in particular for the scalar potential~$\Phi$, whose evolution is described by the transfer function defined in Eq.~\eqref{eq:transfer_function}, so that
\begin{equation}
P_\Phi(\eta;k) = T^2(\eta\leftarrow\eta\e{ini},k) P_\Phi(\eta\e{ini};k).
\end{equation}

\begin{figure}[h!]
\centering
\includegraphics[width=0.7\textwidth]{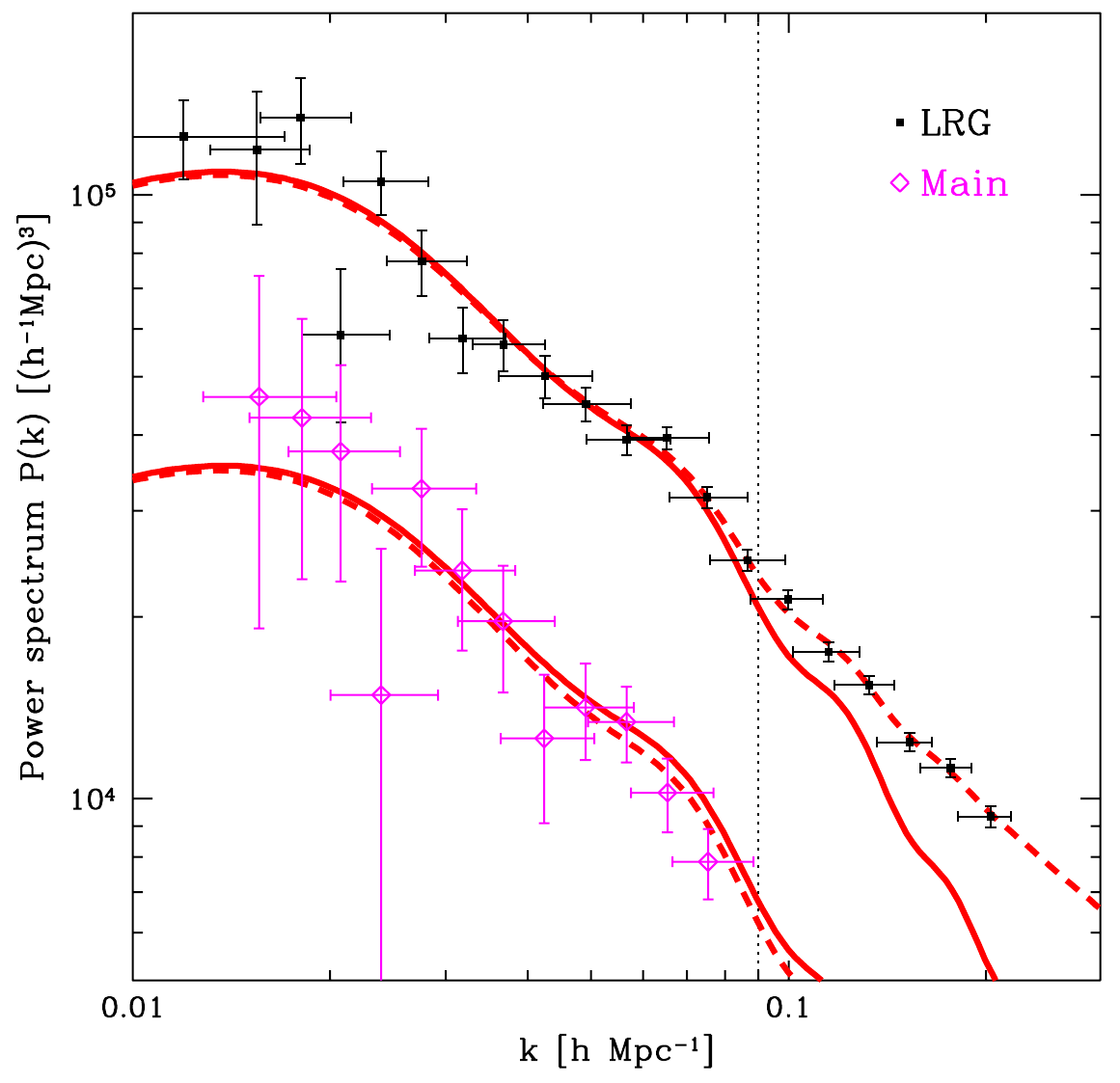}
\caption{Matter power spectra measured from the luminous red galaxy (LRG) sample and the main galaxy sample of the Sloan Digital Sky Survey (SDSS). Red solid lines indicate the predictions of the linear perturbation theory, while red dashed lines include nonlinear corrections. From Ref.~\cite{2006PhRvD..74l3507T}.}
\label{fig:power_spectrum}
\end{figure}

Figure~\ref{fig:power_spectrum} compares the observed power spectrum with theoretical predictions, in particular the linear perturbation theory (dashed line). We see that the latter fails at reproducing the actual behaviour of $P(k)$ on scales smaller than $10\U{Mpc}/h$ ($k>0.1h\U{Mpc}$), where the nonlinearities of self-gravitating fluid dynamics become significant. On such scales, theoretical models must rely on advanced perturbative techniques~\cite{2002PhR...367....1B} or $N$-body simulations (e.g.~Ref.~\cite{GADGET}).\index{power spectrum!observed versus linear}

\subsubsection{Refinements}

Besides nonlinearities at small scales, let us mention a few possible refinements for standard perturbation theory in general. Among the simplifying assumptions that we made in the present section, the crudest is the single-fluid approximation, which may be valid at late time but certainly not, e.g., at the epoch of recombination. A more precise approach considers the cosmological fluid as made of several species---see e.g. Ref.~\cite{PeterUzan} for the detailed treatment of two fluids (dark matter and radiation). Nevertheless, a fluid description is not capable of modelling precisely the matter-matter and radiation-matter interactions, which rather require kinetic theory and  the Boltzmann equation~\cite{TheseCyril}. The presence of massive neutrinos also potentially affect the formation of the large-scale structure; this issue has been investigated, e.g., in Refs.~\cite{2009JCAP...06..017L,2014JCAP...11..039B,2015JCAP...03..030D}.

\subsubsection{Backreaction and related issues}\label{sec:backreaction}

By definition, the whole formalism developed in this section assumes that spacetime geometry is well approximated by the FL metric, i.e., that perturbations are small. While it should be valid on very large scales---i.e. typically for the description of the cosmic web, where substructures are somehow smeared out---such a perturbative approach is nevertheless highly questionable on small scales. For instance, the density contrast~$\delta$ corresponding to a galaxy is $\rho\e{gal}/\rho_0 \sim 10^4 \gg 1$, which cannot be considered a small perturbation; the associated formalism thus should not be extrapolated to those scales.

While the perturbative dynamics clearly breaks down on small scales, it can be argued that the form~\eqref{eq:perturbed_metric_Newtonian_form} of the metric still holds, except in the vicinity of very compact object such as neutron stars or black holes, because the density contrast~$\delta$ is a second derivative of the metric perturbation~$\Phi$, therefore it can be very large with~$\Phi$ remaining small. This argument is reinforced by the fact that Eq.~\eqref{eq:perturbed_metric_Newtonian_form} is essentially a metric formulation of Newtonian gravity, which turns out to be very successful at describing gravitating systems on small scales, far from compact objects. This question of how well is the Universe modelled by the FL metric is still an open question, and was recently the subject of a lively debate in Refs.~\cite{2014CQGra..31w4003G,2015arXiv150507800B,2015arXiv150606452G}.

It could be argued that cosmology does not a priori aim at describing the Universe on small scales. Just like geometric optics in dielectric media does not require to describe the interactions between each photon of the ray with each quark of each atom of each molecule of the medium, but rather relies on a continuous approximation, cosmology should not have to care about each star, or each dark matter halo to get a satisfactory description of the cosmos. There are however two significant differences between cosmology and this example of optics in media.

First, cosmology is mainly based on gravitation, which, if described by the general theory of relativity, is \emph{nonlinear} contrary to electromagnetism. Linearity indeed facilitates smoothing procedures: suppose $\ev{\ldots}$ denotes a coarse-graining operator, then, in electromagnetism, the coarse-grained four-vector potential~$\ev{\vect{A}}$ is governed by the Maxwell equation with a coarse-grained four-current~$\ev{\vect{J}}$, since
\begin{equation}
\vect{\nabla}^2\ev{\vect{A}} = \ev{\vect{\nabla}^2 \vect{A}} = 4\pi\ev{\vect{J}}.
\end{equation}
On the contrary, in gravitation the Riemann curvature, and thus the Einstein tensor, is highly nonlinear with respect to the metric~$\vect{g}$, as it involves its inverse. It follows~\cite{1971grc..conf..104E} that
\begin{equation}
\vect{E}[\ev{\vect{g}}]\not=\ev{\vect{E}[\vect{g}]}=8\pi G \ev{\vect{T}};
\end{equation}
in other words, the procedures corresponding to (i) coarse graining a solution~$\vect{g}$ of the Einstein equation driven by~$\vect{T}$, or (ii) solving the Einstein equation driven by a coarse-grained $\ev{\vect{T}}$, yield different results. In cosmology, we ideally would like to perform (i), but the FL approach consists in (ii), the resulting Friedmann equations thus potentially predict a wrong dynamics for cosmic expansion. Contrary to electromagnetism, the small scales which are smeared out in cosmology can effectively reemerge in the large-scale dynamics. This issue, known as \emph{backreaction}\index{backreaction}, has been proposed in the late 1990s as an explanation of the recent acceleration of cosmic expansion without the need of dark energy~\cite{2000GReGr..32..105B}.

A second---though related---difference between gravitation and any other physical theory concerns the coarse-graining procedure itself. In a theory of macroscopic electromagnetism, for instance, the coarse-graining operator and the target of this operator (the electromagnetic field) are independent. In macroscopic gravitation, on the contrary, the metric is involved in the coarse-graining procedure (since it defines physical lengths and times) while being its target. This issue, together with the mathematical question of defining covariant averages in curved spacetime, has been addressed in Ref.~\cite{1993GReGr..25..673Z}, and summarised in Ref.~\cite{2009PhDT.......229P} in a more cosmology-oriented way.

Relativistic effects in cosmology, including backreaction, have recently stimulated various works, both from the numerical~\cite{2014CQGra..31w4006A,2015PhRvL.114e1302A} and analytical points of view, such as the timescape scenario~\cite{2013arXiv1311.3787W}; traceless backreaction~\cite{2011PhRvD..83h4020G}, a Lagrangian perturbation theory based on a relativistic Zel'dovich approximation~\cite{2012PhRvD..86b3520B,2013PhRvD..87l3503B,2015PhRvD..92b3512A}, an effective-cosmological-fluid theory~\cite{2012JCAP...07..051B},\ldots As emphasized by Ref.~\cite{2015arXiv150507800B}, no definite conclusions on the amplitude of backreaction phenomena can be drawn so far.

\bigskip

The issues raised here concern the way we model the \emph{dynamics} of cosmic expansion: How, given a realistic inhomogeneous distribution of matter in the Universe, does the latter evolve on large scales? Such a question however only represents half the way towards a fully relativistic description of the cosmos. The other half concerns \emph{kinematics}---in particular optics: How does light propagates through a realistic model of the Universe? How can we allow for a realistic distribution of matter when interpreting cosmological observations? These questions are the main concerns of the present thesis. In the next chapter, we will present the answers proposed by standard cosmology, and then focus on alternatives in Part~\ref{part:Ricci-Weyl}.

%% file: shift.pdf_tex
\begingroup%
  \makeatletter%
  \providecommand\color[2][]{%
    \errmessage{(Inkscape) Color is used for the text in Inkscape, but the package 'color.sty' is not loaded}%
    \renewcommand\color[2][]{}%
  }%
  \providecommand\transparent[1]{%
    \errmessage{(Inkscape) Transparency is used (non-zero) for the text in Inkscape, but the package 'transparent.sty' is not loaded}%
    \renewcommand\transparent[1]{}%
  }%
  \providecommand\rotatebox[2]{#2}%
  \ifx\svgwidth\undefined%
    \setlength{\unitlength}{392.19041335bp}%
    \ifx\svgscale\undefined%
      \relax%
    \else%
      \setlength{\unitlength}{\unitlength * \real{\svgscale}}%
    \fi%
  \else%
    \setlength{\unitlength}{\svgwidth}%
  \fi%
  \global\let\svgwidth\undefined%
  \global\let\svgscale\undefined%
  \makeatother%
  \begin{picture}(1,0.38670405)%
    \put(0,0){\includegraphics[width=\unitlength]{shift.pdf}}%
    \put(0.16400926,0.35368574){\color[rgb]{0,0,0}\makebox(0,0)[lb]{\smash{$\vect{\partial}_t$}}}%
    \put(0.10122655,0.34417324){\color[rgb]{0,0,1}\makebox(0,0)[lb]{\smash{$\vect{n}$}}}%
    \put(0.66341684,0.35273456){\color[rgb]{0,0,0}\makebox(0,0)[lb]{\smash{$\vect{\partial}_t\propto\vect{n}$}}}%
    \put(0.19336831,0.12983952){\color[rgb]{0,0,0}\rotatebox{72.3373352}{\makebox(0,0)[lb]{\smash{$x^i=\cst$}}}}%
    \put(0.7646518,0.12599978){\color[rgb]{0,0,0}\rotatebox{83.81856239}{\makebox(0,0)[lb]{\smash{$x^i=\cst$}}}}%
    \put(0.01073531,0.00528283){\color[rgb]{0,0,0}\makebox(0,0)[lb]{\smash{nonzero shift ($g_{ti}\not=0$)}}}%
    \put(0.61248383,0.00528283){\color[rgb]{0,0,0}\makebox(0,0)[lb]{\smash{zero shift ($g_{ti}=0$)}}}%
    \put(0.2657135,0.06647759){\color[rgb]{0,0,1}\makebox(0,0)[lb]{\smash{$\Sigma_{t_1}$}}}%
    \put(0.31670913,0.15826974){\color[rgb]{0,0,1}\makebox(0,0)[lb]{\smash{$\Sigma_{t_2}$}}}%
    \put(0.36770477,0.26026101){\color[rgb]{0,0,1}\makebox(0,0)[lb]{\smash{$\Sigma_{t_3}$}}}%
  \end{picture}%
\endgroup%

%% file: chapter_5.tex
\lettrine{A}{lmost} all cosmological measurements rely, so far, on the observation of distant light sources, such as galaxies, supernovae, or quasars, in a wide range of wavelengths. Interpreting these observations, that is, extracting from them information about the structure and dynamics of the Universe, thus requires to know its optical properties. In this second chapter dedicated to standard cosmology, we present the answers provided by the Friedmann-Lema\^{i}tre model and by the standard perturbation theory. We then review a number of cosmological observations with the constraints they impose on some cosmological parameters. Their surprising level of agreement, despite the fact that they involve totally different scales, finally leads us to discuss the motivations of this thesis. 

\bigskip

\minitoc

\newpage

\section{Optics in homogeneous and isotropic cosmologies}\label{sec:optics_FL}

This first section is dedicated to light propagation in Friedmann-Lema\^itre cosmologies. We first introduce a method for simplifying the underlying calculations, based on the invariance of lightcones under conformal transformations, and then use it to solve explicitly the null geodesic equations and calculate the lensing Jacobi matrix.

\subsection{The conformal trick}\label{sec:conformal_trick}

As demonstrated in \S~\ref{sec:conformal_invariance}, null geodesics are invariant under conformal transformations\index{conformal!invariance of null geodesics} of spacetime's metric. This property can be conveniently\index{conformal!trick} exploited to simplify the analysis of light propagation in cosmology, since both the unperturbed and perturbed FL metrics take the form
\begin{equation}
\vect{g} = a^2(\eta) \, \tilde{\vect{g}},
\end{equation}
with, in the unperturbed case\footnote{This conformal transformation is somehow incomplete for $K\not= 0$, in the sense that it does not fully take advantage of the conformal flatness of the FL geometry: there indeed exists a conformal factor $\Omega(x^\mu)$ such that $\vect{g}=\Omega^2 \vect{f}$ where $\vect{f}$ is the Minkowski metric~\cite{1973lsss.book.....H}. However, as we will see below, factorising $a^2$ out already simplifies enough the calculations.},
\begin{equation}\label{eq:FL_metric_static}
\tilde{g}_{\mu\nu}\dd x^\mu \dd x^\nu = -\dd\eta^2 + \dd\chi^2 + f_K^2(\chi) \pa{\dd\theta^2 + \sin^2\theta \, \dd\ph^2 },
\end{equation}
which does not depend on $\eta$. From a purely technical point of view, it is much simpler to analyse light propagation in terms of $\tilde{\vect{g}}$, getting rid of the geometrical terms due to the time evolution of $a$, and then recover all the lensing quantities for $\vect{g}$ thanks to the dictionary given in Table~\ref{tab:conformal_dictionary}. \emph{This dictionary is completely general}, in the sense that it applies for any metric and any conformal factor~$a$; in particular, the latter could depend on all spacetime coordinates. It was used for instance in Ref.~\cite{2015PhRvD..91d3511F}---which belongs to the present thesis, see Chap.~\ref{chapter:optics_Bianchi_I}---in order to simplify the analysis of light propagation in anisotropic cosmologies of the Bianchi~I kind.

Before applying it to the FL spacetime, let us comment a few entries of the conformal dictionary. Most relations are actually direct consequences of the definitions of the quantities at stake, in particular $\vect{\xi}=\tilde{\vect{\xi}}$ is due to the fact that the definition~\eqref{eq:separation_vector_definition} of the separation vector does not involve the metric, but only the coordinates of null geodesics. On the contrary, $\vect{u}=a^{-1}\tilde{\vect{u}}$ is a \emph{choice}, made here for simplicity, because the four-velocities of sources and observers are independent from the laws of light propagation. The correspondence between the Sachs bases follows from this choice, and from their normalisation conditions. One can also check that, with this correspondence, the partial parallel transport requirement~\eqref{eq:Sachs_transport} for $\vect{s}_A$ is satisfied iff it is for $\tilde{\vect{s}}_A$,
\begin{equation}\label{eq:equivalence_transport}
S^\mu_\nu \Ddf{s^\nu_A}{v} = 0 \Longleftrightarrow \tilde{S}^\mu_\nu \frac{\widetilde{\Dd}\tilde{s}^\nu_A}{\dd\tilde{v}} = 0.
\end{equation}

\renewcommand{\arraystretch}{1.2}
\begin{table}[t]
\centering
\begin{tabular}{|r|ll|}
\hline
\rowcolor{lightgray}
\textsf{\bfseries Quantity} & \multicolumn{2}{c|}{\textsf{\bfseries Correspondence}} \\
\hline
metric & $g_{\mu\nu}=a^2 \tilde{g}_{\mu\nu}$ & $g^{\mu\nu}=a^{-2} \tilde{g}^{\mu\nu}$ \\
affine parameter & \multicolumn{2}{l|}{$\dd v = a^2 \dd\tilde{v}$} \\
wave four-vector & $k^\mu = a^{-2} \tilde{k}^\mu$ & $k_\mu = \tilde{k}_\mu$ \\
four-velocity & $u^\mu = a^{-1} \tilde{u}^\mu$ & $u_\mu = a\tilde{u}_\mu$  \\
frequency & \multicolumn{2}{l|}{$\omega = a^{-1}\tilde{\omega}$} \\
redshift & \multicolumn{2}{l|}{$1+z= a_O a_S^{-1}(1+\tilde{z})$} \\
propagation direction & $d^\mu = a^{-1}\tilde{d}^\mu$ & $d_\mu = a \tilde{d}_\mu$  \\
screen projector & $S_{\mu\nu}=a^2 \tilde{S}_{\mu\nu}$ & $S^{\mu\nu}=a^{-2} \tilde{S}^{\mu\nu}$\\
Sachs basis & $s_A^\mu=a^{-1}\tilde{s}_A^\mu $ & $s^A_\mu=a\tilde{s}^A_\mu $ \\
separation four-vector & $\xi^\mu = \tilde{\xi}^\mu$ & $\xi_\mu = a^2 \tilde{\xi}_\mu$ \\
separation in screen space & \multicolumn{2}{l|}{$\xi_A = a \tilde{\xi}_A$} \\
Jacobi matrix & \multicolumn{2}{l|}{$\vect{\jacobi}(S\leftarrow O)=a_S a_O \widetilde{\vect{\jacobi}}(S\leftarrow O)$} \\
angular distance & \multicolumn{2}{l|}{$D\e{A}=a_S \widetilde{D}\e{A}$} \\
luminosity distance & \multicolumn{2}{l|}{$D\e{L}=a_O^2 a_S^{-1} \widetilde{D}\e{L}$} \\
deformation scalars & \multicolumn{2}{l|}{$\gamma,\varphi,\psi = \tilde{\gamma}, \tilde{\varphi},\tilde{\psi}$} \\[2mm]
deformation rate matrix & \multicolumn{2}{l|}{$\vect{\deformation}=\dfrac{1}{a}\dfrac{\dd a}{\dd v}\identity_2 + a^{-2}\widetilde{\vect{\deformation}}$} \\[3mm]
expansion rate & \multicolumn{2}{l|}{$\theta = \dfrac{1}{a}\dfrac{\dd a}{\dd v} + a^{-2}\tilde{\theta} $}\\[2mm]
shear rate & \multicolumn{2}{l|}{$\sigma=a^{-2}\tilde{\sigma}$} \\
\hline
\end{tabular} 
\caption{The conformal dictionary of geometric optics in curved spacetime. All quantities are defined in Chaps.~\ref{chapter:beams}, \ref{chapter:distances}. Tilded and untilded four-dimensional vectors and tensors are defined on the same manifold, but not with respect to the same metric: (un)tilded indices $\mu,\nu$ are raised and lowered, respectively, by the (un)tilded metric.}
\index{conformal!dictionary}
\label{tab:conformal_dictionary}
\end{table}

\subsection{Light rays}\label{sec:light_rays_FL}

Without loss of generality, we choose the observation event~$O$ (here and now) to be the centre $\chi=0$ of the spatial coordinate system. As mentioned in the previous chapter, in cosmology, quantities referring to $O$ are conventionally denoted with a zero subscript $_0$, this standard notation will here be equivalent to the $_O$ subscript.

It is straightforward to check that the radial null curves such that
\begin{equation}\label{eq:null_geodesics_FL}
\chi = \eta_0 - \eta, \qquad \theta,\ph=\cst
\end{equation}
satisfy the null geodesic equations for the static metric~\eqref{eq:FL_metric_static}, and are therefore null geodesics for $\vect{g}$ as well. They form the lightcone of $O$. Note that although the curves \eqref{eq:null_geodesics_FL} appear as straight lines in terms of the coordinates~$(\chi,\theta,\ph)$, when $K\not= 0$ they are exceptions in the sense that other null geodesics (out of the lightcone of $O$) do not. Any affine parametrisations of the null curves~\eqref{eq:null_geodesics_FL} is, with respect to $\tilde{\vect{g}}$, simply proportional to conformal time $\dd\tilde{v}=\tilde{\omega}^{-1}\dd\eta$, with $\tilde{\omega}=\tilde{\omega}_0=\cst$. In other words, $\tilde{k}^\mu=\cst$. We conclude that, with respect to the original metric~$\vect{g}$,
\begin{equation}
k^\mu = \pa{\frac{a_0}{a}}^2 k^\mu_0,
\end{equation}
where it is understood that $\mu=0$ would refer to a component with respect to $\vect{\partial}_\eta$, not $\vect{\partial}_t$. The frequency measured by a comoving observer at $\eta$ is thus $\omega=[a_0/a(\eta)]\omega_0$, and the redshift between emitted light as $S$ and the observed light at $O$ reads
\begin{empheq}[box=\fbox]{equation}\index{redshift!in a FL spacetime}
1+z = \frac{a_0}{a_S} \geq 1,
\end{empheq}
in agreement with the relation given by Table~\ref{tab:conformal_dictionary}, since $\tilde{z}=0$. By virtue of Eq.~\eqref{eq:null_geodesics_FL}, the redshift is related to the radial coordinate~$\chi$ of $S$ by $\dd z/\dd\chi = a_0 H$, integrated as
\begin{equation}\label{eq:chi_z}\index{conformal!radial distance}
a_0 \chi = \int_0^z \frac{\dd \zeta}{H(\zeta)} 
= \frac{1}{H_0} \int_0^z \dd\zeta\;\pac{ \Omega\e{m0}(1+\zeta)^3+\Omega_{K0}(1+\zeta)^2+\Omega_{\Lambda0} }^{-1/2},
\end{equation}
where we used the first Friedmann equation~\eqref{eq:Friedmann_1} to link $H$ with $a$, i.e. with $z$. In homogeneous cosmology, the redshift can thus be considered a kind of distance measurement. Depending on the cosmological parameters, the integral of Eq.~\eqref{eq:chi_z} can be calculated either analytically or numerically.

The observed redshift of a comoving source (with $\chi=\cst$) generally evolves with time. Consider a light signal observed at $t_0+\dd t_0$; to this reception cosmic time corresponds an emission time $t_S+\dd t_S$, where $\dd t_S=\dd t_0/(1+z)$ by the very definition of the redshift. The associated correction to the redshift reads
\begin{equation}
\dd(1+z) = \frac{\dot{a}_0\dd t_0}{a_S} - a_0\frac{\dot{a}_S\dd t_S}{a_S^2} 
\end{equation}
whence
\begin{equation}
\ddf{z}{t_0} = (1+z) H_0 - H_S .
\end{equation}
This \emph{redshift drift}\index{redshift!drift} was first mentioned by Refs.~\cite{1962ApJ...136..319S,1962ApJ...136..334M}. Its theoretical order of magnitude is $\dd z/\dd t_0\sim 10^{-11}\U{yr^{-1}}$ for $z\sim 1$. Albeit very small, next generation high-resolution spectroscopy experiments, such as the COsmic Dynamics EXperiment (CODEX)~\cite{2007NCimB.122.1165C} proposed for the European Extremely Large Telescope (E-ELT), should be able to measure redshift drifts by the next decades. As forecasted by Ref.~\cite{2012PhR...521...95Q}, such a measurement over $30$ years, applied to $z>2$ quasars, could contribute to observationally distinguish between several models of dark energy. 

\subsection{Light beams}

We now investigate the properties of radial light beams. Like for single light rays, we start with the conformal geometry~$\tilde{\vect{g}}$,  for which calculations are simpler. 

\subsubsection{In the conformal geometry}

The spatial direction of propagation of radial geodesics~\eqref{eq:null_geodesics_FL} is $\vect{\tilde{d}}=-\vect{\partial}_\chi$. The other two spatial basis vectors being orthogonal to $\vect{\tilde{d}}$, it is natural to use them for constructing the Sachs basis as
\begin{equation}
\tilde{\vect{s}}_1 = \frac{-1}{f_K(\chi)} \pd{}{\theta},
\qquad
\tilde{\vect{s}}_2 = \frac{-1}{f_K(\chi)\sin\theta} \pd{}{\ph}.
\end{equation}
It is straightforward to check that these vectors indeed satisfy the transport condition~\eqref{eq:equivalence_transport}. The optical tidal matrix associated with this Sachs basis then reads $\widetilde{\vect{\tidal}}=-\tilde{\omega}_0^2 K\,\identity_2$, where we used that the frequency $\tilde{\omega}$ measured by comoving observers ($\tilde{\vect{u}}=\vect{\partial}_\eta$) is a constant. The full specification of the Sachs basis was actually not necessary to get this result, since the FL geometry is conformally flat, which implies that the optical tidal matrix~$\widetilde{\vect{\tidal}}$ has only a Ricci (pure-trace) part $\widetilde{\Ricfoc}\identity_2$, with $\widetilde{\Ricfoc}=-(1/2)\tilde{R}_{\mu\nu}\tilde{k}^\mu \tilde{k}^\nu=-\tilde{\omega}^2_0 K$. The Jacobi matrix equation
\begin{equation}
\ddf[2]{\widetilde{\vect{\jacobi}}}{\tilde{v}} = -\tilde{\omega}_0^2 K \widetilde{\vect{\jacobi}}
\end{equation}
is then easily solved as
\begin{equation}
\widetilde{\vect{\jacobi}}(S\leftarrow O) = \tilde{\omega}_0^{-1} f_K(\eta_S-\eta_0) \, \identity_2,
\end{equation}
where $\eta_0-\eta_S$ can be replaced by the conformal radial distance~$\chi_\source$ of the source.

\subsubsection{In the original geometry}

Recovering all lensing quantities for the original FL metric~$\vect{g}$ is now easily achieved using the dictionary of Table~\ref{tab:conformal_dictionary}. In particular, the Jacobi matrix reads
\begin{empheq}[box=\fbox]{equation}
\vect{\jacobi}(S\leftarrow O) = -a_S \omega_0^{-1} f_K(\chi_S) \, \identity_2 .
\end{empheq}
Note that the minus sign is here due to our conventional future orientation of the wave four-vector. As expected, light propagation in the FL spacetime exhibits no shear nor rotation: the Jacobi matrix is directly proportional to the angular diameter distance (see \S~\ref{sec:decomposition_Jacobi})
\begin{equation}\label{eq:angular_distance_FL}
D\e{A} = a_S f_K(\chi_S).
\end{equation}

The above result matches the interpretation of $f_K(\chi)\equiv R$ as a conformal \emph{areal} radius. For a given radial coordinate~$\chi$, sources appear larger (closer) as $K$ increases. This, however shall not be interpreted as if spatial curvature had any actual focusing effect. As discussed in \S~\ref{sec:Ricci_Weyl_lensing}, the \emph{physical} source of focusing is the local density of energy and momentum, due to Ricci focusing, which here reads $\Ricfoc=-4\pi G(\rho+p)\omega^2$. Spatial curvature only enters into the game via the dynamics of $a(t)$, related to $\rho$ by the Friedmann equations.

In Table~\ref{tab:distances_FL}, we summarise the expressions of the other observational notions of distance defined in Chap.~\ref{chapter:distances} in a FL spacetime. Note that, contrary to the angular and luminosity distances, the radar and parallax distances are given as the results of an academic exercise, since they cannot be applied to cosmological distances in practice (see \S~\ref{sec:measuring_distances}). As such, they can nevertheless serve to illustrate the difference between the various observational notions of distance in general relativity.

\begin{table}[h!]
\centering
\begin{tabular}{|r|l|}
\hline 
\rowcolor{lightgray}
\textsf{\bfseries Distance} & \multicolumn{1}{c|}{\textsf{\bfseries Expression}} \\
\hline
& \\[-4mm]
radar & $D\e{R} = \dfrac{1}{2}\displaystyle\int_{\eta_0-2\chi}^{\eta_0} a \; \dd\eta$ \\[4mm] 
parallax & $D\e{P} = \pac{ H_0 + \dfrac{f'_K(\chi_S)}{a_0 f_K(\chi_S)} }^{-1}$ \\ 
angular & $D\e{A} = a_S f_K(\chi_S)$ \\ 
luminosity & $D\e{L} = a_0^2 a_S^{-1} f_K(\chi_S)$ \\[1mm]
\hline 
\end{tabular}
\caption{Expressions of the observational distances in a FL geometry, for a source with comoving radial coordinate $\chi$. The radar distance is here defined as a retarded distance.\index{distance!in a FL spacetime}}
\label{tab:distances_FL}
\end{table} 

\section{Optics in perturbation theory}

The presence of perturbations with respect to strict homogeneity and isotropy (see \S~\ref{sec:linear_perturbation_theory}) modifies the propagation of light compared to the previous results. This section reviews the consequent corrections to light's frequency, beam's morphology, etc. For that purpose, we restrict to the Newtonian regime, where the metric reads
\begin{align}
\dd s^2 &= a^2(\eta) \pac{ -(1+2\Phi) \dd\eta^2 + (1-2\Phi) \gamma_{ij} \dd x^i \dd x^j } \\
				&= a^2(\eta) (\bar{g}_{\mu\nu}+\delta\tilde{g}_{\mu\nu}) \dd x^\mu \dd x^\nu
\end{align}
which, as discussed in \S~\ref{sec:Newtonian_regime}, is a good approximation at late times, and provided large-scale perturbations are not concerned. The first assumption is meaningful because gravitational lensing is mostly due to \begin{inparaenum}[(i)]\item collapsed structures, or \item the cosmic web\end{inparaenum}; both are absent in the primordial Universe, and appear during the matter-dominated era. The second assumption is generally satisfied  in gravitational lensing, because the size of the beam dictates the relevant scales.

\subsection{Perturbation of light rays}

We start with the effect of metric perturbation on a single light ray.\footnote{Note of April 2023: I thank Angela Ng for spotting an error in the original version of this paragraph, which is now corrected.} Like in the previous section, we take advantage of the conformal invariance of null geodesic and work with the conformal metric~$\bar{\vect{g}}+\delta\tilde{\vect{g}}$. Note that, in order to be perfectly consistent, the conformal background metric should be denoted $\bar{\tilde{\vect{g}}}$; we here choose to drop the tilde to alleviate notation. In the remainder of this section, a bar thus denotes conformal background quantities, except explicit mention of the contrary.

We consider a background null geodesic $\bar{x}^\mu(\tilde{v})$ and a perturbed null geodesic $x^\mu(\tilde{v})=\bar{x}^\mu(\tilde{v})+\delta x^\mu(\tilde{v})$. Both curves are affinely parameterised by the same $\tilde{v}$, which defines the mapping between them: $\delta x^\mu(\tilde{v})$ is the difference between the positions of the perturbed and background geodesics at the same $\tilde{v}$. We also define the associated wave four-vectors as
$\tilde{k}^\mu\define\dd x^\mu/\dd\tilde{v}$,
$\bar{k}^\mu\define\dd \bar{x}^\mu/\dd\tilde{v}$, and
$\delta\tilde{k}^\mu\define\dd\delta x^\mu/\dd\tilde{v}=\tilde{k}^\mu-\bar{k}^\mu$.

The geodesic equations for the background and perturbed geodesics read
\begin{align}
\ddf{\bar{k}^\mu}{\tilde{v}} + \bar{\Gamma}\indices{^\mu_\nu_\rho} \bar{k}^\nu \bar{k}^\rho &= 0 ,\\
\ddf{\tilde{k}^\mu}{\tilde{v}} + \tilde{\Gamma}\indices{^\mu_\nu_\rho} \tilde{k}^\nu \tilde{k}^\rho &= 0 ,
\end{align}
which, by subtraction, yields at first order
\begin{equation}\label{eq:perturbed_geodesic_equation}
\ddf{\delta\tilde{k}^\mu}{\tilde{v}} + 2\bar{\Gamma}\indices{^\mu_\nu_\rho} \bar{k}^\nu \delta\tilde{k}^\rho
= - \delta\tilde{\Gamma}\indices{^\mu_\nu_\rho} \bar{k}^\nu \bar{k}^\rho ,
\end{equation}
where
$\delta\tilde{\Gamma}\indices{^\mu_\nu_\rho}
\define \tilde{\Gamma}\indices{^\mu_\nu_\rho} - \bar{\Gamma}\indices{^\mu_\nu_\rho}$
denotes the perturbation of the conformal Christoffel symbols.


\subsubsection{Effect on the frequency}

Let us first calculate the effect of perturbations on the observed frequency of the light signal. By definition, we have
\begin{align}
\tilde{\omega} &= -\tilde{u}_\mu \tilde{k}^\mu \\
						&= \bar{\omega} + \delta \tilde{k}^0 + \bar{\omega}(\Phi - \upsilon_i \bar{d}^i) + \mathcal{O}(2).
\end{align}
where we have used the decomposition of the perturbation of the four-velocity introduced in \S~\ref{sec:perturbed_stress-energy_tensor} as $\delta\tilde{\vect{u}}=-\Phi\vect{\partial}_\eta+\upsilon^i\vect{\partial}_i$. We see that only the $0$th component of $\delta\tilde{\vect{k}}$ needs to be determined for computing the perturbed frequency.

Since $\overline{\Gamma}\indices{^0_\nu_\rho}=0$ (see Table~\ref{tab:FL_geometry} with $a=\cst$), the second term on the left-hand side of Eq.~\eqref{eq:perturbed_geodesic_equation} vanishes for $\mu=0$. The right-hand side is given by the perturbations to the Christoffel symbols~\cite{PeterUzan} 
\begin{equation}
\delta\widetilde{\Gamma}\indices{^0_0_\mu} = \partial_\mu\Phi,
\qquad
\delta\widetilde{\Gamma}\indices{^0_i_j} = -(\partial_0\Phi)\,\gamma_{ij},
\end{equation}
so that
\begin{equation}
\delta\widetilde{\Gamma}\indices{^0_\nu_\rho} \bar{k}^\nu \bar{k}^\rho
= 2\bar{\omega} \ddf{\Phi}{\bar{v}} - 2\bar{\omega}^2\partial_0\Phi .
\end{equation}
Therefore, for $\mu=0$, Eq.~\eqref{eq:perturbed_geodesic_equation} is readily integrated as
\begin{equation}
\pac{\delta\tilde{k}^0}^O_S
=
-2\bar{\omega}\pac{\Phi}^O_S 
+ 2\bar{\omega}^2\int_S^O \dd\tilde{v} \; \partial_0\Phi[\eta,\bar{x}^i(\tilde{v})] .
\end{equation}

Finally, we can use the fact that along the background null geodesic, $\dd\tilde{v}=\bar{\omega}^{-1}\dd\eta$, with $\bar{\omega}=\cst$, and obtain
\begin{empheq}[box=\fbox]{equation}\label{eq:perturbed_frequencies}
\frac{\delta\omega_0}{\omega_0} - \frac{\delta\omega_S}{\omega_S}= \frac{-\delta z}{1+z} = -\pac{\Phi + \upsilon_i \bar{d}^i}_S^O + 2\int_{\eta_\source}^{\eta_0} \dd\eta \; \partial_0\Phi[\eta,\bar{x}^i(\eta)],
\end{empheq}
where we have reintroduced the frequencies defined with respect to the full perturbed metric. The quantity $\upsilon_i \bar{d}^i$ can be considered either with respect to the conformal metric (i.e. $\gamma_{ij}\upsilon^i \tilde{d}^j$) or with respect to the full metric (i.e. $g_{ij}\upsilon^i d^j$), because both are equal. Note that the integral term on the right-hand side of Eq.~\eqref{eq:perturbed_frequencies} is not trivially integrated, because the partial derivative~$\partial_\eta$ only hits the first $\eta$; in other words, while the integral is a curvilinear integral along the unperturbed null geodesic $\bar{x}^\mu(\eta)$, the derivative is not performed along this geodesic.

Physically speaking, the bracket term on the right-hand side of Eq.~\eqref{eq:perturbed_frequencies} contains the intuitively expected corrections to the redshift, respectively interpreted as gravitational and Doppler effects with respect to this coordinate system. The gravitational part $[-\Phi]^O_S$ is sometimes referred to as the \emph{Sachs-Wolfe} (SW)\index{Sachs-Wolfe effect} effect, while the quantities $\upsilon^i_O$ and $\upsilon^i_S$ are usually called the observer's and source's \emph{peculiar velocities}\index{peculiar velocity}, they encode the deviation of their motions with respect to the Hubble flow. They are, of course, gauge dependent. The integral term, contrary to the previous ones, depends on the whole light path from $S$ to $O$, and is specific to time-dependent perturbations. The corresponding physical phenomenon is either called \emph{integrated Sachs-Wolfe} (ISW) effect~\cite{1967ApJ...147...73S}\index{integrated Sachs-Wolfe effect} or \emph{Rees-Sciama} (RS)\index{Rees-Sciama effect} effect~\cite{1968Natur.217..511R}, depending on its physical cause.

First note that, in linear perturbation theory during the matter-dominated era ($\Omega\e{m}\approx 1$), since $\delta\propto a$, Eq.~\eqref{eq:Newtonian_1} implies that $\partial_\eta\Phi=0$, so the integrated effect vanishes in this case. There are three possible excursions from this situation:
\begin{enumerate}
\item Before the matter era, i.e. during the radiation era. We then talk about the \emph{early} ISW effect. Note that the above calculations cannot be directly applied to this case, since vector, tensor modes, and anisotropic stress cannot be neglected.
\item After the matter era, when the cosmological constant (or dark energy, or spatial curvature) starts to affect the growth of structures. This is the \emph{late} ISW effect. It has first been detected in 2004 by cross-correlating the CMB map of the Wilkinson Microwave Anisotropy Probe (WMAP) with maps of the large-scale structure~\cite{2004Natur.427...45B}; since then, the late ISW effect has been exploited in a number of studies to put constraints on dark energy (see e.g. Ref.~\cite{2014PTEP.2014fB110N} and references therein). For instance, the recent \textit{Planck} results~\cite{2015arXiv150201595P} exclude $\Lambda=0$ at a $3\sigma$ confidence level, from the ISW effect only.
\item In the matter era, but beyond the linear regime. The time-dependence of the gravitational potential then occurs in the vicinity of virialised structures. In this case we talk about the Rees-Sciama effect. We will see in Chap.~\ref{chapter:SC} an explicit example of this phenomenon in Swiss-cheese cosmological models.
\end{enumerate}

\subsubsection{Effect on the source's position}

We now turn to the spatial part of the perturbed geodesic equation~\eqref{eq:perturbed_geodesic_equation}. The second term on the left-hand side, for $\mu=i$, reads
\begin{equation}
2\bar{\Gamma}\indices{^i_\nu_\rho} \bar{k}^\nu \delta\tilde{k}^\rho
=
2 \tensor[^3]{\Gamma}{^i_j_l} \bar{k}^l \delta\tilde{k}^j
=
- 2\bar{\omega} \, \frac{f'_K(\chi)}{f_K(\chi)} 
\pa{ \delta^i_j - \delta^i_\chi\delta^\chi_j} \delta\tilde{k}^j,
\end{equation}
where we used that $\bar{k}^i=-\bar{\omega} \delta^i_\chi$, and that the nonzero Christoffel symbols~$\tensor[^3]{\Gamma}{^i_j_l}$ of the background spatial metric~$\gamma_{ij}$ such that $l=\chi$ are $\tensor[^3]{\Gamma}{^\theta_\theta_\chi}=\tensor[^3]{\Gamma}{^\ph_\ph_\chi}=f'_K(\chi)/f_K(\chi)$. The right-hand side of Eq.~\eqref{eq:perturbed_geodesic_equation} requires the correction to the Christoffel coefficients
\begin{equation}
\delta\widetilde{\Gamma}\indices{^i_0_0} =  \gamma^{ij} \partial_j\Phi,
\qquad
\delta\widetilde{\Gamma}\indices{^i_j_0} = -\delta^i_j\,\partial_0 \Phi,
\qquad
\delta\widetilde{\Gamma}\indices{^i_j_k} = -2 \delta^i_{(j} \partial_{k)}\Phi + \gamma_{jk} \gamma^{il} \partial_l\Phi,
\end{equation}
so that
\begin{equation}
-\delta\tilde{\Gamma}\indices{^i_\mu_\nu} \bar{k}^\mu \bar{k}^\nu
=
2\ddf{\Phi}{\tilde{v}} \, \bar{k}^i - 2 \bar{\omega}^2 \gamma^{ij}\partial_j\Phi.
\end{equation}
The resulting differential equation is naturally expressed in terms of $\tilde{v}$. We have seen in the previous paragraph that it can be written in terms of $\eta$ since $\dd\tilde{v}=\bar{\omega}^{-1}\dd\eta$; similarly, since at the background level $\dd\eta=-\dd\chi$, we can translate it as
\begin{equation}\label{eq:evolution_perturbation_wave_vector}
\ddf{\delta \tilde{k}^i}{\chi} + 2\frac{f'_K(\chi)}{f_K(\chi)} \pa{ \delta^i_j - \delta^i_\chi\delta^\chi_j} \delta\tilde{k}^j
=
2\ddf{\Phi}{\chi} \, \bar{k}^i + 2 \bar{\omega} \gamma^{ij} \partial_j\Phi.
\end{equation}

\begin{figure}[h!]
\centering
\begin{minipage}{5cm}
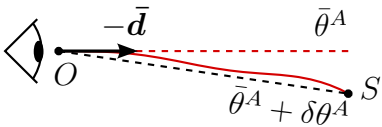
\end{minipage}
\hspace*{1cm}
\begin{minipage}{7cm}
\caption{Perturbation of the angular coordinates $(\theta^A)=(\theta,\ph)$ of the source, given an observation direction~$-\vect{\bar{d}}$.}
\label{fig:perturbed_ray}
\end{minipage}
\end{figure}

Let us focus on the perturbation to the position of the source on the observer's celestial sphere. The problem can be formulated as follows: consider a line of sight~$-\vect{\bar{d}}$, corresponding to the direction $(\bar{\theta},\bar{\ph})\define(\bar{\theta}^A)_{A=1,2}$ towards which the observer looks; the deflection of light by $\Phi$ implies that the light ray deviates from the radial straight line~$\theta^A=\bar{\theta}^A=\cst$, so that the source event~$S$ actually has angular coordinates $\bar{\theta}^A+\delta\theta^A$. This new direction corresponds to the direction in which the observer would see the image if light did follow a radial straight line (see Fig.~\ref{fig:perturbed_ray}).

\noindent The correction~$\delta \theta^A$ is related to the perturbation of the wave four-vector via $\delta\tilde{k}^A=\dd\delta\theta^i/\dd\tilde{v}=-\bar{\omega}\dd\delta\theta^A/\dd\chi$. For $i=A$, the first term on the right-hand side of Eq.~\eqref{eq:evolution_perturbation_wave_vector} vanishes, because $\bar{k}^i\propto \delta^i_\chi$, and the equation is integrated twice as
%
%
\begin{empheq}[box=]{equation}\label{eq:perturbation_source_position}
\delta \theta^A = -2 \int_{0}^{\chi} \dd\chi' \int_{0}^{\chi'} \dd\chi'' \; \pac{\frac{f_K(\chi'')}{f_K(\chi')}}^2 \gamma^{AB}\partial_B \Phi[\eta'',\bar{x}^i(\eta'')],
\end{empheq}
where it is understood that $\eta''=\eta_0-\chi''$, and $\bar{x}^i(\eta'')=(\eta'',\chi'',\bar{\theta},\bar{\ph})$ is the unperturbed ray. The properties of the areal function~$f_K$ allow us to turn the double integral of Eq.~\eqref{eq:perturbation_source_position} into the single integral\footnote{A first step consists in changing the integration scheme as
\begin{equation}
\int_0^\chi \dd\chi' \int_0^{\chi'} \dd\chi'' \; F(\chi',\chi'') = \int_0^\chi \dd\chi'' \int_{\chi''}^{\chi} \dd\chi' \; F(\chi',\chi''),
\end{equation}
and then perform the integration with respect to $\chi'$,
\begin{equation}
\int_{\chi''}^\chi \frac{\dd\chi'}{f^2_K(\chi')} 
= \pac{ -\frac{f_K'}{f_K} }^{\chi''}_\chi 
= \frac{1}{f_K(\chi'') f_K(\chi)} \pac{ f_K(\chi) f'_K(\chi'') - f_K(\chi'') f'_K(\chi) }
= \frac{f_K(\chi-\chi'')}{f_K(\chi'') f_K(\chi)},
\end{equation}
where we used twice the fact that $f_K$ is either $\sin$ or $\sinh$ (or the identity). Renaming $\chi''$ as $\chi'$ finally gives the result of Eq.~\eqref{eq:perturbation_source_position_simplified}
}
\begin{empheq}[box=\fbox]{equation}\label{eq:perturbation_source_position_simplified}
\delta \theta^A = -2 \int_{0}^{\chi} \dd\chi' \; \frac{f_K(\chi')f_K(\chi-\chi')}{f_K(\chi)} \, \gamma^{AB}\partial_B \Phi[\eta',\bar{x}^i(\eta')].
\end{empheq}
This formula agrees with the Newtonian intuition according to which particles are deflected by gravitational fields, i.e. by $-\partial\Phi$, modulo a factor two which distinguishes the deflection of light with the deflection of nonrelativistic particles in GR. Note the form of the kernel
\begin{equation}\label{eq:lensing_efficiency}
\frac{f_K(\chi-\chi')f_K(\chi')}{f_K(\chi)}
\propto \frac{D\e{A}(L \leftarrow O) D\e{A}(S \leftarrow L)}{D\e{A}(S\leftarrow O)},
\end{equation}
sometimes called \emph{lensing efficiency}\index{lensing efficiency}, which peaks for $\chi'=\chi/2$, i.e. when the lens~$L$ is such that $D\e{A}(S \leftarrow L)=D\e{A}(L \leftarrow O)$. This is a generic characteristic of gravitational lensing: the most important contributions to the net deflection are due to gravitational fields lying halfway between the source and the observer.

\subsection{Perturbation of light beams}\label{sec:perturbation_light_beams}

The previous paragraph showed that an inhomogeneous potential~$\Phi$ tends to deflect light with respect to the purely radial geodesics of the background FL spacetime. Considering now a family of neighbouring light rays, we expect from their differential deflection to distort and focus the underlying beam.

\subsubsection{Perturbed optical tidal matrix}

The sources of focusing and distortions of a light beam are encoded in the optical tidal matrix (see Chap.~\ref{chapter:beams}). We keep working in the conformal geometry, and decompose the conformal optical tidal matrix as $\widetilde{\vect{\tidal}}=\overline{\vect{\tidal}}+\delta\widetilde{\vect{\tidal}}$, with $\overline{\vect{\tidal}}=-\bar{\omega}^2 K$. Its perturbation
\begin{equation}
\delta\widetilde{\tidal}_{AB} \define \delta\pa{ R_{\mu\nu\rho\sigma}s_A^\mu k^\nu k^\rho s_B^\sigma }
\end{equation}
contains a priori terms of the form $\delta R \bar{s}\bar{k}\bar{k}\bar{s}\propto \delta(\partial^2g)$, and $\bar{R}\delta s \bar{k}\bar{k}\bar{s},\bar{R}\bar{s}\delta k\bar{k}\bar{s}\propto \delta(\partial g)$. As discussed at the end of the previous chapter, the perturbations of curvature~$\delta(\partial^2 g)$ are generally much larger than the perturbations of lower-order derivatives of the metric. At leading order, we can thus neglect the latters compared to the formers, so that
\begin{align}
\delta\widetilde{\tidal}_{AB} 
&\approx \delta\tilde{R}_{\mu\nu\rho\sigma} \bar{s}^\mu_A \bar{k}^\nu \bar{k}^\rho \bar{s}_B^\sigma 
\qquad \text{neglecting }\bar{R}\delta s \bar{k}\bar{k}\bar{s},\bar{R}\bar{s}\delta k\bar{k}\bar{s} \\
&\approx -2\delta\widetilde{\Gamma}_{\mu\nu[\rho,\sigma]} \bar{s}^\mu_A \bar{k}^\nu \bar{k}^\rho \bar{s}_B^\sigma
\qquad \text{neglecting }\overline{\Gamma}\delta\tilde{\Gamma} \\
&= \frac{1}{2} \pa{ \delta\tilde{g}_{\mu\nu,\rho\sigma} + \delta\tilde{g}_{\rho\sigma,\mu\nu}  } 
	\bar{s}^\mu_A \bar{s}_B^\nu \bar{k}^\rho \bar{k}^\sigma.
\end{align}
Similar considerations allow us to write
\begin{align}
\delta\tilde{g}_{\mu\nu,\rho\sigma} \bar{s}^\mu_A \bar{s}_B^\nu \bar{k}^\rho \bar{k}^\sigma
&\approx \ddf[2]{}{\tilde{v}}\pa{ \delta\tilde{g}_{\mu\nu} \bar{s}_A^\mu \bar{s}_B^\nu }, \\
\delta\tilde{g}_{\rho\sigma,\mu\nu} \bar{s}^\mu_A \bar{s}_B^\nu \bar{k}^\rho \bar{k}^\sigma
&\approx -4\bar{\omega}^2\Phi_{,\mu\nu} \bar{s}_A^\mu \bar{s}_B^\nu,
\end{align}
which finally leads to
\begin{equation}\label{eq:perturbation_optical_tidal_matrix}
\delta\widetilde{\tidal}_{AB} = 
-2 \bar{\omega}^2 \partial^\perp_A\partial^\perp_B\Phi
+ \ddf[2]{}{\tilde{v}}\pa{ \delta\tilde{g}_{\mu\nu} \bar{s}_A^\mu \bar{s}_B^\nu } 
+ \mathcal{O}(\partial\tilde{g}) + \mathcal{O}(2).
\end{equation}
In Eq.~\eqref{eq:perturbation_optical_tidal_matrix}, we have introduced transverse derivatives~$\partial_A^\perp\define \bar{s}_A^\mu\partial_\mu$, such that
\begin{equation}
\partial^\perp_A\partial^\perp_B
\define
\bar{s}_A^\mu \bar{s}_B^\nu \partial_\mu\partial_\nu
=
\frac{1}{f^2_K(\chi)}
\begin{pmatrix}
\pd[2]{}{\theta} & \frac{1}{\sin\theta} \frac{\partial^2}{\partial\theta\partial\ph} \\
\frac{1}{\sin\theta} \frac{\partial^2}{\partial\theta\partial\ph} & \frac{1}{\sin^2\theta}\pd[2]{}{\ph}
\end{pmatrix}.
\end{equation}

\subsubsection{Jacobi matrix at linear order}

Let us now exploit the perturbed optical tidal matrix to determine the correction to the Jacobi matrix. Still in conformal geometry, we decompose it into a background and a perturbation as $\widetilde{\vect{\jacobi}}=\overline{\vect{\jacobi}}+\delta\widetilde{\vect{\jacobi}}$. At linear order in perturbations, the Jacobi matrix equation~\eqref{eq:Jacobi_matrix_equation} reads
\begin{equation}
\ddf[2]{\delta\widetilde{\vect{\jacobi}} }{\tilde{v}}
= \overline{\vect{\tidal}} \, \delta\widetilde{\vect{\jacobi}} 
		+ \delta\widetilde{\vect{\tidal}} \, \overline{\vect{\jacobi}},
\end{equation}
or, in terms of the comoving radial coordinate~$\chi$, and replacing the background quantities by their expressions,
\begin{equation}\label{eq:Jacobi_matrix_equation_perturbed}
\ddf[2]{\delta\widetilde{\vect{\jacobi}} }{\chi}
= - K \, \delta\widetilde{\vect{\jacobi}} - \bar{\omega}^{-3} f_K(\chi) \, \delta\widetilde{\vect{\tidal}}.
\end{equation}
A Green-function method for solving this second-order differential equation then yields
\begin{equation}
\delta\widetilde{\vect{\jacobi}}
= -\int_0^\chi \dd\chi' \;  f_K(\chi') f_K(\chi-\chi') \, \bar{\omega}^{-3}\delta\widetilde{\vect{\tidal}},
\end{equation}
which is easily checked to be a solution of Eq.~\eqref{eq:Jacobi_matrix_equation_perturbed}, with initial conditions $\delta\widetilde{\vect{\jacobi}}_0=(\dd\delta\widetilde{\vect{\jacobi}}/\dd\tilde{v})_0=\vect{0}_2$.

In the expression~\eqref{eq:perturbation_optical_tidal_matrix} of $\delta\widetilde{\vect{\tidal}}$, the total derivative $\dd^2(\delta\tilde{g}_{\mu\nu}\bar{s}^\mu_A\bar{s}^\nu_B)/\dd\tilde{v}^2$ can be integrated by parts, and yields a term on the order of $K\Phi$, which is much smaller than $\partial^2\Phi$ because the gravitational potential varies on distances much smaller than the background spatial curvature radius. Back to the original perturbed geometry, we thus obtain the following formula for the Jacobi matrix, at first order in cosmological perturbations:\index{amplification matrix!due to cosmological perturbations}\index{Jacobi matrix!in cosmological perturbation theory}
\begin{empheq}[box=\fbox]{equation}\label{eq:Jacobi_matrix_perturbed}
\jacobi\indices{_A_B}(S\leftarrow O)
=
-a_S\omega_0^{-1} f_K(\chi) 
\paac{ \delta_{AB}
		 - 2\int_0^\chi \dd\chi' \; \frac{f_K(\chi') f_K(\chi-\chi')}{f_K(\chi)} 
		 	\,\partial^\perp_A \partial^\perp_B \Phi[\eta',\bar{x}^i(\eta')]
		},
\end{empheq}
which is a standard textbook result~\cite{PeterUzan,2006glsw.conf.....M}.

\begin{figure}[h!]
\centering
\begin{minipage}[c]{8cm}
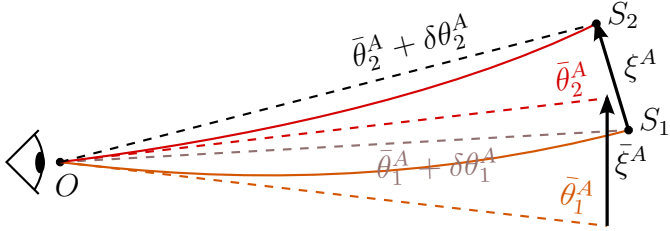
\end{minipage}
\hfill
\begin{minipage}{6cm}
\caption{Position of two light sources~$S_1$, $S_2$, associated with the directions of observation~$\bar{\theta}^A_1$, $\bar{\theta}^A_2$, and their physical separation in screen space for the background ($\bar{\xi}^A$) and perturbed ($\xi^A$) spacetimes.}
\label{fig:amplification_alternative_definition}
\end{minipage}
\end{figure}

Inside the braces, we recognise the amplification matrix~$\vect{\amplification}=\vect{\jacobi}\overline{\vect{\jacobi}}^{-1}$ defined in \S~\ref{sec:amplification_matrix}. Interestingly, this expression of $\vect{\amplification}$ could also have been obtained from Eq.~\eqref{eq:perturbation_source_position}. By definition, this matrix is indeed
\begin{equation}
\amplification\indices{^A_B}
= \frac{\partial \xi^A}{\partial\dot{\xi}^C_0} \frac{\partial\dot{\xi}^C_0}{\partial\xi^B} \bigg|_\text{FL}
= \frac{\partial\xi^A}{\partial\bar{\xi}^B},
\end{equation}
where $\xi^A$ and $\bar{\xi}^A$ are defined as follows and depicted in Fig.~\ref{fig:amplification_alternative_definition}. Consider two (very close) directions of observation defined by the angles $(\bar{\theta}_1^A)$ and $(\bar{\theta}_2^A)$. To these directions of observation correspond, for a given radial coordinate~$\chi$, two sources $S_1$ and $S_2$, whose positions differ whether we consider the perturbed or background spacetime; $\bar{\xi}^A$ (resp. $\xi^A$) represents the physical separation, in screen space, between the sources in the background (resp. perturbed) spacetime.

We assume for simplicity that the system of axes is set so that the directions that we are considering lie in the vicinity of the $\theta=\pi/2$ plane; the physical separation between the points $(\eta,\chi,\theta_1^A)$ and $(\eta,\chi,\theta_2^A)$ is then simply $a(\eta) f_K(\chi) (\theta^A_2-\theta^A_1)$ (at the background level). In particular, it avoids complications due to the $\sin\theta$ term, always present in the geometry of spherical coordinates. We thus have
\begin{equation}
\xi^A = a f_K(\chi) \pac{ \bar\theta_2^A+\delta\theta_2^A - (\bar\theta_1^A+\delta\theta_1^A) }
		= \bar{\xi}^A + \frac{\partial\delta\theta^A}{\partial\bar{\theta}^B} \bar{\xi}^B
\end{equation}
so
\begin{equation}
\mathcal{\amplification}\indices{^A_B} = \delta^A_B + \frac{\partial\delta\theta^A}{\partial\bar{\theta}^B},
\end{equation}
which, using the expression~\eqref{eq:perturbation_source_position_simplified} of $\delta\theta^A$, indeed coincides with Eq.~\eqref{eq:Jacobi_matrix_perturbed}.

Note that the Jacobi and amplification matrices are symmetric here, i.e. at linear order in cosmological perturbations. This agrees with the discussion of \S~\ref{sec:amplification_matrix}, where we have seen that when Weyl lensing is treated as a perturbation, rotation is a second-order quantity contrary to convergence~$\kappa$ and shear~$\gamma$, and we can write
\begin{equation}
\vect{\amplification}
=
\begin{pmatrix}
1-\kappa-\gamma_1 & \gamma_2 \\
\gamma_2 & 1-\kappa+\gamma_1
\end{pmatrix}.
\end{equation}

\subsubsection{Cosmic convergence}\index{convergence!cosmic}\index{cosmic!convergence}

The net convergence~$\kappa$ of a light beam due to cosmological perturbations is called \emph{cosmic convergence}. It represents the order-one correction to the observed angular distance of a given light source, compared to the background FL case, and is extracted from the amplification matrix as
\begin{equation}
\kappa \define \frac{\bar{D}\e{A}-D\e{A}}{\bar{D}\e{A}} = 1-\frac{\tr\vect{\amplification}}{2}.
\end{equation}
Taking the trace of Eq.~\eqref{eq:Jacobi_matrix_perturbed} yields a transverse Laplacian of the gravitational potential
\begin{equation}
\delta^{AB}\partial^\perp_A\partial^\perp_B \Phi
= \bar{S}^{\mu\nu} \partial_\mu\partial_\nu \Phi 
= \Delta\Phi - \partial^2_\chi\Phi,
\end{equation}
where we identified the coordinate Laplacian~$\gamma^{ij}\partial_i\partial_i$ with the background covariant Laplacian $\Delta=\gamma^{ij}\bar{\Dd}_i \bar{\Dd}_i$, because their difference is on the order of $\overline{\Gamma}\partial\Phi$ which is negligible compared to $\partial^2\Phi$. By virtue of the Poisson equation~\eqref{eq:Newtonian_1}, $\Delta\Phi$ can be replaced by $4\pi G a^2\bar{\rho}\delta$, where $\delta$ is matter's density contrast. Besides, the longitudinal derivative~$\partial^2_\chi\Phi$ can also be neglected because of the integration\footnote{Contrary to what is sometimes claimed~\cite{PeterUzan,2006glsw.conf.....M}, the contribution of $\partial^2_\chi\Phi$ does not trivially vanish after integrating by parts. First note that it is only a \emph{partial} derivative, which cannot be directly integrated by parts, for the same reason that we could not do so for the ISW term in Eq.~\eqref{eq:perturbed_frequencies}. However, since the time evolution of $\Phi$ is much slower than its spatial evolution, we can consider $\dd\Phi/\dd\chi=-\bar{\omega}^{-1}\dd\Phi/\dd\tilde{v}=\partial_\chi\Phi-\partial_\eta\Phi\approx \partial_\chi\Phi$. A double integration by parts then gives
\begin{equation}
\int_0^\chi \dd\chi'\; f_K(\chi')f_K(\chi-\chi') \ddf[2]{\Phi}{\chi'}
=
f_K(\chi) \pac{ \Phi(\chi)-\Phi(0) } - 2\int_0^\chi \dd\chi' \; f'_K(\chi')f'_K(\chi-\chi') \Phi(\chi'),
\end{equation}
which does not exactly vanish, but can be neglected as it does not contain second-order derivative of $\Phi$ any more.}
with respect to $\chi'$, and the convergence finally reads
\begin{empheq}[box=\fbox]{equation}\label{eq:cosmic_convergence}
\kappa = \frac{3}{2} H_0^2\Omega\e{m0} 
\int_0^\chi \dd\chi' \; \frac{f_K(\chi')f_K(\chi-\chi')}{f_K(\chi)} \frac{\delta[\eta',\bar{x}^i(\eta')]}{a(\eta')},
\end{empheq}
where we assumed that matter is well modelled by a pressureless dust ($w=0$) in order to write $\bar{\rho}=(a_0/a)^3\bar{\rho}_0$, and introduced the cosmological parameter~$\Omega\e{m0}=8\pi G \bar{\rho}_0/(3H_0^2)$. This expression of $\kappa$ shows that overdensities ($\delta>0$) or underdensities ($\delta<0$) respectively tend to focus and defocus light beams with respect to their background behaviour, in agreement with the effect of \emph{Ricci} focusing~$\Ricfoc$ discussed in Chap.~\ref{chapter:beams}. Weyl lensing~$\Weylfoc$ is indeed absent from Eq.~\eqref{eq:cosmic_convergence}, which could have been derived e.g. from Eq.~\eqref{eq:focusing_theorem} by taking $\sigma=0$. This no longer true for lensing at second order in perturbations, see e.g.~Refs.~\cite{2005PhRvD..71f3537B,2012JCAP...11..045B,2014JCAP...06..023A,2014CQGra..31t2001U,2014CQGra..31t5001U}.

Besides second-order effects, several first-order contributions have also been neglected to obtain the simple result~\eqref{eq:cosmic_convergence}. In particular, by identifying the amplification matrix~$\vect{\amplification}$ with the expression between braces in Eq.~\eqref{eq:Jacobi_matrix_perturbed}, we did not take into account the fact that the observed frequency $\omega_0$ is also affected by cosmological perturbations, leading to a Doppler contribution to the amplification matrix. Physically speaking, this a contribution to the convergence corresponds to the aberration effects discussed in \S~\ref{sec:aberration}. It can actually be large~\cite{2013PhRvL.110b1302B}, and generically dominates over deflection effects on short distances~\cite{2013arXiv1309.6542N,2006PhRvD..73l3526H}. For a more careful derivation of the correction to the convergence, taking into account all the first order contributions, see e.g. in Ref.~\cite{2006PhRvD..73b3523B}, see also Ref.~\cite{2012PhRvD..86b3510D} for calculations which include the effect of vector and tensor modes.

Averaging the expression~\eqref{eq:cosmic_convergence} over a large number of sources yields the effective cosmic convergence which only depends on the line of sight
\begin{equation}
\kappa\e{eff} = \frac{3}{2} H_0^2\Omega\e{m0} 
\int_0^\infty \dd\chi \; g(\chi) f_K(\chi) \frac{\delta[\eta',\bar{x}^i(\eta')]}{a(\eta')},
\end{equation}
where
\begin{equation}
g(\chi) \define \int_\chi^\infty \dd\chi' \; p(\chi') \, \frac{f_K(\chi'-\chi)}{f_K(\chi')}
\end{equation}
is the integrated lensing efficiency, $p(\chi)\dd\chi$ being the probability of finding a source within the interval $[\chi,\chi+\dd\chi]$. The resulting power spectrum~$P_\kappa$ for $\kappa\e{eff}$, in the so-called flat-sky approximation\footnote{This approximation consists in computing the Fourier transform of $\kappa\e{eff}(\theta^A)$ as if the angular variables $\theta^A$ were Cartesian coordinates, hence neglecting the curvature of the celestial sphere. It is justified by the fact that the correlations between lines of sight with large angular separations are small.}, together with an analogue of Limber's approximation~\cite{1992ApJ...388..272K}, is related to the matter density power spectrum $P_\delta$ as
\begin{equation}\label{eq:power_spectrum_cosmic_convergence}
P_\kappa(k)
	\approx \pa{ \frac{3}{2}\, H_0^2\Omega\e{m0} }^2 
	\int_0^\infty \dd\chi \; \frac{g^2(\chi)}{a^2(\eta)} \, P_\delta\pac{\eta,\frac{k}{f_K(\chi)}},
\end{equation}
where it is understood that $\eta=\eta(\chi)=\eta_0-\chi$.

\subsubsection{Cosmic shear}\index{cosmic!shear}\index{shear!cosmic}

Lensing does not only affect the apparent size (or distance) of light sources, but also their shape. Observing the shape of lensed galaxies provides a measurement of this \emph{cosmic shear} effect, which encodes key information on the matter distribution in the Universe. Suppose that we observe remote galaxies, whose intrinsic shape and observed shape are well described by ellipses. The properties of any ellipse~$\mathscr{E}$ can be quantified by a complex number $\eps\define\eps_1+\i\eps_2=|\eps|\ex{-2\i\vartheta}$, which defines the transformation which must be applied to a circle~$\mathscr{C}$ to obtain it, according to
\begin{equation}\label{eq:definition_ellipse}
\mathscr{E} = \exp
\begin{pmatrix}
-\eps_1 & \eps_2 \\
\eps_2 & \eps_1
\end{pmatrix}
\mathscr{C}.
\end{equation}
For a circle~$\mathscr{C}$ with unit radius, the ellipse~$\mathscr{E}$ has semi-minor axis~$b=\ex{-|\eps|}$ along the direction $\vartheta$, and semi-major axis~$a=\ex{|\eps|}$ along the orthogonal direction, so that the ellipticity of $\mathscr{E}$ is given by $(a-b)/(a+b)=\tan|\eps|$. Note that the present $\eps$ is \emph{not} the complex ellipticity usually defined in this context~\cite{2001PhR...340..291B} (though they agree for $|\eps|\ll1$). Our choice, however, together with the general decomposition of the Jacobi matrix introduced in \S~\ref{sec:decomposition_Jacobi}, turns out make the following discussion clearer.

By definition (see \S~\ref{sec:Jacobi_matrix_definition}) the source~$\mathscr{S}$ is related to the image~$\mathscr{I}$ by the Jacobi matrix as $\mathscr{S}=-\omega_O\vect{\jacobi}(S\leftarrow O) \mathscr{I}$. Normalising $\mathscr{S}$ (resp. $\mathscr{I}$) by its physical size (resp. angular size) results into a unity-area ellipse $\mathscr{E}\e{intr}$ (resp. $\mathscr{E}\e{obs}$) characterising the intrinsic shape of the source (resp. observed shape of the image). These two ellipses are thus related by $\mathscr{E}\e{intr}=(-\vect{\jacobi}/\sqrt{\det\vect{\jacobi}}) \mathscr{E}\e{obs}$, which, introducing the decomposition~\eqref{eq:Jacobi_decomposition} of $\vect{\jacobi}$, yields
\begin{equation}
\mathscr{E}\e{obs} = \exp
\begin{pmatrix}
\gamma_1 & -\gamma_2 \\
-\gamma_2 & -\gamma_1
\end{pmatrix}
\begin{pmatrix}
\cos\psi & \sin\psi \\
-\sin\psi & \cos\psi
\end{pmatrix}
\mathscr{E}\e{intr}.
\end{equation}
Now suppose that many galaxies are observed in a small region of the sky, across which $\vect{\jacobi}$ (hence $\gamma,\psi$) can be considered constant. Since the galaxies have in principle random shapes and orientations\footnote{This hypothesis can however be spoiled by the potential trend of galaxies to align each others, or with the cosmic web, due to gravitational interactions~\cite{2015MNRAS.448.3391C,2015arXiv150707843C}.}, we have $\ev{\mathscr{E}\e{intr}}=\mathscr{C}$, and the ensemble average of the observed ellipses reads
\begin{equation}
\ev{\mathscr{E}\e{obs}} =
\exp \begin{pmatrix}
\gamma_1 & -\gamma_2 \\
-\gamma_2 & -\gamma_1
\end{pmatrix}
\mathscr{C}.
\end{equation}
Comparing with Eq.~\eqref{eq:definition_ellipse}, we conclude that this average ellipse is characterised by $\bar{\eps}\e{obs}=-\gamma$. The same reasonings apply to the two-point correlation function of the observed galaxy shapes, which thus coincides with shear correlation function.

While the above considerations are fully general, they can be further exploited in the context of cosmological perturbation theory. In the expression~\eqref{eq:Jacobi_matrix_perturbed} of the Jacobi matrix, we see that the convergence (trace part) and the shear (trace-free part) both come from derivatives of the same function~$\Phi$, so they are not independent from each other. For example, it can be shown that $\kappa$ and $\gamma$ have the same angular power spectrum, $P_\kappa=P_\gamma=P_\eps$. In Ref.~\cite{1993ApJ...404..441K}, Kaiser and Squires exploited this property to design an algorithm for reconstructing the convergence, which is not directly observable, from the shear. Observing the shape of galaxies therefore allows one to infer the properties of the density contrast, via the reconstructed convergence.

\section{Some observations and their interpretation}

In the previous sections we have derived the theoretical optical properties of the standard cosmological model. This provide a framework to interpret cosmological observations, i.e., to validate or falsify the model, and to measure its free parameters such as the $\Omega$s. In this section, we briefly review the main current cosmological probes, namely the Hubble diagram of SNe (\S~\ref{sec:Hubble_diagram}), the CMB (\S~\ref{sec:CMB}), BAO (\S~\ref{sec:BAO}), and mention some other observations in \S~\ref{sec:other_observations}. For all of them, we will emphasize the crucial character of the relation between distances and redshift for their correct interpretation.

\subsection{Hubble diagram}\label{sec:Hubble_diagram}

The Hubble diagram\index{Hubble!diagram} is, conceptually, the simplest observation to interpret. It consists in plotting the luminosity distance~$D\e{L}$, or the distance modulus~$\mu\e{L}$, of objects with known intrinsic luminosity---the so-called standard-(isable) candles\index{standard!-isable candles}---as a function of their redshift. In cosmology, as discussed in \S~\ref{sec:luminosity_distance}, type Ia supernovae (SNeIa) are the best candidates.

\begin{figure}[h!]
\centering
\includegraphics[width=0.7\textwidth]{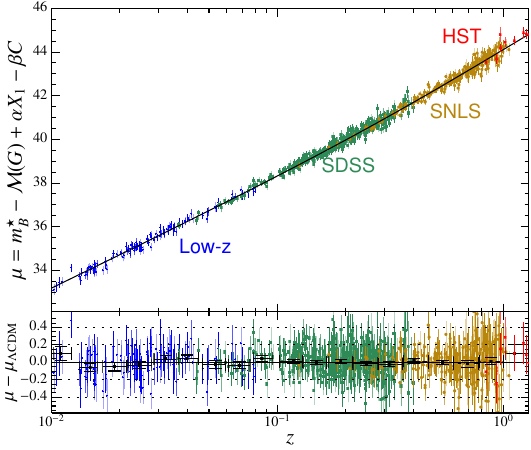}
\caption{A Hubble diagram obtained by the joint lightcurve analysis of 740 SNeIa from four different samples: Low-$z$, SDSS-II, SNLS3, and HST. The top panel depicts the Hubble diagram itself with the $\Lambda$CDM best fit (black line); the bottom panel shows the residuals. From Ref.~\cite{2014A&A...568A..22B}}
\label{fig:Hubble_diagram_JLA}
\end{figure}

In practice, SNeIa are standardised by the measurement of their lightcurve, that is the evolution of the luminosity of the event with time (which typically lasts from a few weeks to a few months). Most current analyses are based on the assumption that the absolute magnitudes of all SNeIa are comparable, and that their variations can be captured by two parameters $X_1$ and $C$, characterising respectively the duration and the colour of the explosion~\cite{1998A&A...331..815T}. Together with the observed peak B-band magnitude~$m^\star\e{B}$ of the SNIa, they allow one to determine its distance modulus according to~\cite{2014A&A...568A..22B}
\begin{equation}
\mu\e{L} = 5\log_{10}\pa{ \frac{D\e{L}}{10\U{pc}} } = m^\star\e{B} - (M\e{B}-\alpha X_1 + \beta C),
\end{equation}
where $M\e{B}$, $\alpha$, and $\beta$ are three nuisance parameters which are fitted simultaneously with the cosmological parameters. The redshift is determined besides by spectroscopic measurements. Figure~\ref{fig:Hubble_diagram_JLA} shows the most recent Hubble diagram~\cite{2014A&A...568A..22B}, obtained from the joint lightcurve analysis (JLA) of 740 SNeIa belonging to four different samples: Low-$z$ survey~\cite{2009ApJ...700..331H}, the SDSS-II supernova survey~\cite{2014arXiv1401.3317S}, the 3-year data release of the SuperNova Legacy Survey (SNLS)~\cite{2010A&A...523A...7G}, and a few high redshift SNe detected with the Hubble Space Telescope (HST)~\cite{2007ApJ...659...98R}.

\begin{figure}[t]
\centering
\includegraphics[width=0.7\textwidth]{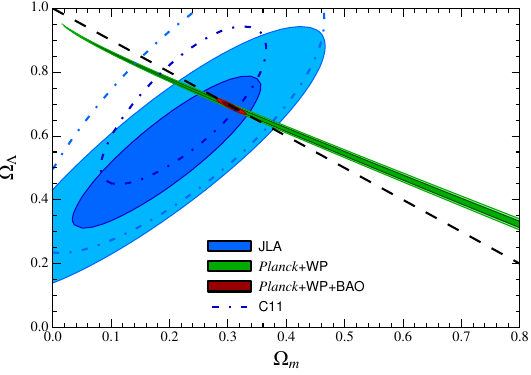}
\caption{Constraints on the cosmological parameters~$\Omega\e{m0}, \Omega_{\Lambda 0}$ obtained from the Hubble diagram of Fig.~\ref{fig:Hubble_diagram_JLA} (blue contours), together with other observations: CMB (green), and CMB+BAO (red). The dot-dashed contours corresponds to the constraints from earlier SN data~\cite{2011ApJS..192....1C}. The dashed line indicates $\Omega\e{m0}+\Omega_{\Lambda0}=1$, i.e. $K=0$. From Ref.~\cite{2014A&A...568A..22B}}
\label{fig:constraints_JLA}
\end{figure}

In all current analyses of the Hubble diagram, including Ref.~\cite{2014A&A...568A..22B}, the data are interpreted \emph{assuming that light propagates through a perfectly homogeneous and isotropic Universe}, so that the theoretical relation between luminosity distance and redshift to which observations are confronted is
\begin{equation}\label{eq:DL_z_FL}
D\e{L}(z) 
= a_0 (1+z) f_K\pa{ \frac{1}{a_0H_0} \int_0^z \dd\zeta\;\pac{ \Omega\e{m0}(1+\zeta)^3+\Omega_{K0}(1+\zeta)^2+\Omega_{\Lambda0} }^{-1/2} },
\end{equation}
which is derived from $D\e{L}=(1+z)^2 D\e{A}$, and with the results of \S~\ref{sec:optics_FL}. We can see in Fig.~\ref{fig:Hubble_diagram_JLA} that this model provides an excellent fit to the data. The resulting constraints on the free parameters~$\Omega\e{m0}, \Omega_{\Lambda0}$ are displayed in Fig.~\ref{fig:constraints_JLA}, which shows in particular that the absence of dark energy is excluded at a 3$\sigma$ confidence level. The Hubble diagram of SNeIa is indeed particularly adapted to investigating the existence and properties of dark energy, as it probes the Universe at low redshift, i.e. at late times. This is the reason why SNeIa provided the first evidence of the acceleration of cosmic expansion\index{accelerated expansion!observation} in the late 1990s~\cite{1998AJ....116.1009R,1999ApJ...517..565P}, a discovery rewarded by the 2011 Nobel Prize.

\begin{figure}[h!]
\centering
\begin{minipage}{0.5\linewidth}
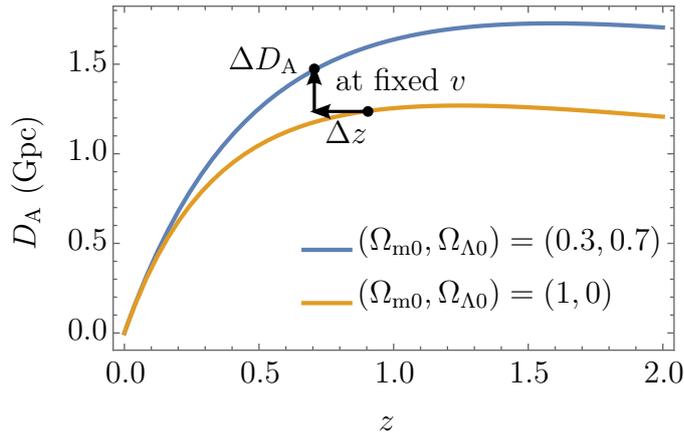
\end{minipage}
\hfill
\begin{minipage}{0.4\linewidth}
\caption{For a given observed redshift $z$, a SN appears farther (smaller and dimmer) through a Universe with dark energy, due to both the acceleration of cosmic expansion and the reduction of Ricci focusing.}
\label{fig:DA_with_without_Lambda}
\end{minipage}
\end{figure}

Physically speaking, the cosmological constant affects SNIa observations in two complementary ways. Consider a source at a given affine-parameter distance~$v$ from us. On the one hand, $\Lambda$ reduces its redshift due to the acceleration of cosmic expansion: if the expansion accelerates, then it was slower in the past, so the recession velocity of a distant object is smaller. In other words, $z_{\Lambda\not=0}(v)<z_{\Lambda=0}(v)$, as confirmed by the fact that, in a FL model,
\begin{equation}
-\omega_0 v = \int_0^z \frac{\dd \zeta}{(1+\zeta)^2 H(\zeta)},
\end{equation}
which can be derived from $\dd z/\dd v=\dd(a_0/a)/\dd v$, using that $\dd/\dd v= k^t \dd/\dd t=\omega_0(1+z)\dd/\dd t$. On the other hand, for a given expansion rate today~$H_0$, the presence of $\Lambda$ reduces the Universe's matter density (it reduces $\Omega\e{m0}$), so it reduces the actual Ricci focusing experienced by light beams, therefore enhancing the observed angular distance: $D\e{A}^{\Lambda=0}(v)>D\e{A}^{\Lambda\not =0}(v)$. As illustrated in Fig.~\ref{fig:DA_with_without_Lambda}, these two effects combine so that $D\e{A}^{\Lambda\not =0}(z)>D\e{A}^{\Lambda=0}(z)$, hence $D\e{L}^{\Lambda\not=0}(z)>D\e{L}^{\Lambda=0}(z)$ as well. We conclude that SNe with a given observed $z$ appear dimmer in a Universe with dark energy than without. Of course, it could also be attributed to a negative spatial curvature~$K<0$, which acts similarly to $\Lambda$. This is the reason why the constraints of Fig.~\ref{fig:constraints_JLA} are degenerate in the direction orthogonal to $K=\cst$.

In the context of linear perturbation theory, the use of the background distance-redshift relation \eqref{eq:DL_z_FL} for modelling the Hubble diagram can be justified by the fact that the corrections are negligible once averaged over many sources. Indeed, regarding the correction to the redshift~\eqref{eq:perturbed_frequencies}, the Doppler contribution vanishes if we suppose that the SNe have random peculiar velocities (see however Ref.~\cite{2015arXiv150201762K}), while the SW and ISW/RS effects are anyway very small. As for the cosmic convergence~$\kappa$, since by definition~$\ev{\delta}=0$, we deduce that $\ev{\kappa}=0$ after averaging over the sky.\footnote{We here identified three notions of averaging with the notation~$\ev{\cdots}$: sky averaging, ensemble averaging, and source averaging. Such an assumption is valid as far as only first-order perturbations are at stake, but it breaks down at second order, as discussed in Refs.~\cite{2005ApJ...632..718K,2015arXiv150308506K,2015JCAP...07..040B}.}
The cosmological perturbations thus do not significantly bias the distance-redshift relation at linear order. However, they are expected to contribute to the scattering of the Hubble diagram: at low redshift because some SNe are more or less redshifted due to their peculiar velocity; at high redshifts because some are magnified and others demagnified. In the analysis of SN data, these effects are taken into account by adding two terms diagonal terms $\sigma\e{pec}^2$ and $\sigma\e{lens}^2$ to the covariance matrix of the $\chi^2$. In Ref.~\cite{2014A&A...568A..22B}, for instance, $\sigma\e{pec}=(5\times 150\U{km/s})/(c z \ln 10)$ and~$\sigma\e{lens}=0.055 z$. For comparison, the intrinsic scatter of SN magnitudes is $\sigma\e{int}\sim 0.1$~\cite{2011ApJS..192....1C}.

If gravitational lensing is worked out at second-order in cosmological perturbations, then the apparent luminosity of the SNe is biased with respect to the background case. This bias remains however very small~\cite{2013JCAP...06..002B,2013PhRvL.110b1301B,2013arXiv1309.6542N,2015arXiv150308506K}. This conclusion does not necessarily hold in nonperturbative approaches, in particular when the fluid description of matter in the Universe is abandoned, as we shall see in Part~\ref{part:Ricci-Weyl}.

Let us finally mention that, although SNIa observations are often presented as the most model-independent cosmological probes, the accuracy of this claim actually depends on the lightcurve fitter used for processing the data~\cite{2014PhLB..733..258B}. The results presented here have been obtained with the Spectral Adaptive Lightcurve Template (SALT2) method~\cite{2007A&A...466...11G}, where the phenomenological parameters $\alpha,\beta$ are fitted simultaneously with the cosmological parameters. So in this approach, the SN distance moduli themselves are measured by assuming a homogeneous FL model. This method is therefore not completely model-independent, in the sense that alternative cosmological models cannot be consistently tested with these data. An alternative method is the Multicolour Light Curve Shape (MLCS) fitter~\cite{2007ApJ...659..122J}, whose calibration is performed using only low-redshift SNe, where only the linear Hubble law is required. Though more model independent, MLCS has the disadvantage of producing results with larger error bars than SALT2.

\subsection{Cosmic microwave background}\label{sec:CMB}

A dissertation on cosmology cannot be without mentioning the observation of the CMB\index{cosmic!microwave background}\index{CMB|see{cosmic microwave background}}, which is certainly the archetype of high-precision cosmological experiments. Its origin, as originally understood by Refs.~\cite{1948Natur.162..680G,1948Natur.162..774A} in 1948, goes back to the early Universe, when the primordial plasma cooled enough for the atomic nuclei to recombine\index{recombination} with electrons, forming (mostly) neutral hydrogen atoms. Light thus suddenly stopped being scattered by charged particles, and started propagating freely, following null geodesics. According to the cosmological principle, this happened everywhere at the same (cosmic) time, so that whatever the direction we look at today, we receive such photons coming from some remote place of the Universe where they were released 13.8 billion years ago.

The first experimental evidence for the CMB was accidentally found in 1964~\cite{1965ApJ...142..419P}, and rewarded by the 1978 Nobel Prize. Since then, considerable efforts were carried out to measure and analyse this primordial radiation with an increasing precision, both with Earth-based and space experiments, such as the COsmic Background Explorer (COBE)~\cite{COBE}, the Wilkinson Microwave Anisotropy Probe (WMAP)~\cite{WMAP}, and lately the \textit{Planck} satellite~\cite{Planck}. The CMB appears today as an almost perfect black body, very well-described by a Planck spectrum of temperature $T\e{CMB}=2.72548\pm0.00057\U{K}$~\cite{2009ApJ...707..916F}, whose peak wavelength is\footnote{According to Wien's displacement law, $\lambda\e{peak} T = b$, with~\cite{NIST} $b=2.897 7729\times 10^{-3}\U{m\cdot K}$.} $\lambda\e{CMB}\sim\text{mm}$ (microwave/radio domain). This observed signal corresponds to an emitted Planck spectrum of temperature $k\e{B}T_*\sim 0.3\U{eV}$ ($\lambda_*\sim \text{\micro m}$) redshifted by $z_*=1089.90 \pm 0.23$~\cite{2015arXiv150201589P} due to cosmic expansion.

\begin{figure}[h!]
\centering
\includegraphics[width=0.8\textwidth]{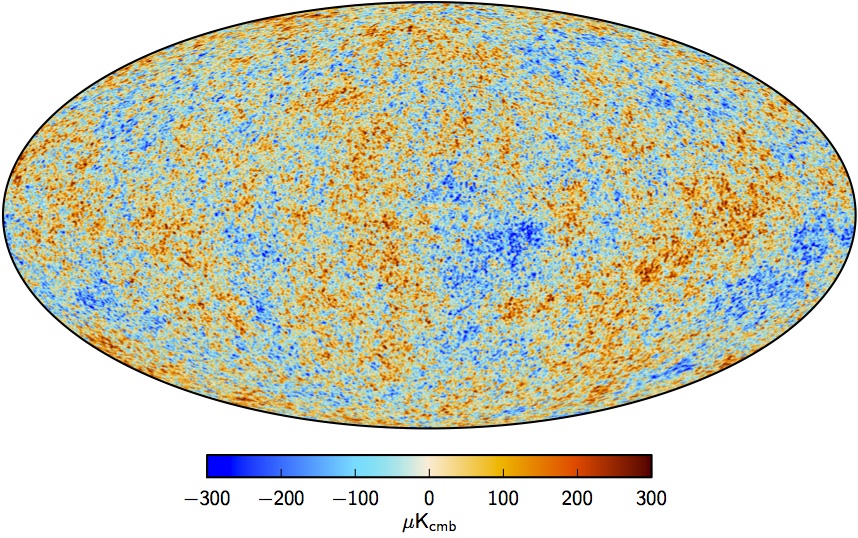}
\caption{Temperature anisotropy of the CMB, as observed by the \textit{Planck} satellite. This map of the whole celestial sphere is obtained by a Mollweide projection. From Ref.~\cite{2015arXiv150201582P}}
\label{fig:CMB_map}
\end{figure}

The CMB is however not perfectly isotropic, and the fluctuations of the observed temperature~$\Theta\define \delta T/\bar{T}\sim 10\U{\micro K}$ (see Fig.~\ref{fig:CMB_map}) actually contain a lot of information about the Universe. The origin of the temperature anisotropies can be separated in two categories: \begin{inparaenum}[(i)] \item primary anisotropies, generated before recombination, and thus related to the physics of inflation and of the primordial plasma; \item secondary anisotropies, due to what happens to the CMB photons after their release and before their observation (gravitational lensing, SZ effect in galaxy clusters, etc.) \end{inparaenum}
We refer the reader to textbooks~\cite{PeterUzan,2008cmb..book.....D} for details about the physics and the analysis of the CMB. In the perspective of the present thesis, we choose to restrict to a single important feature of the CMB anisotropies: the acoustic horizon scale.

\begin{figure}[h!]
\centering
\includegraphics[width=0.9\textwidth]{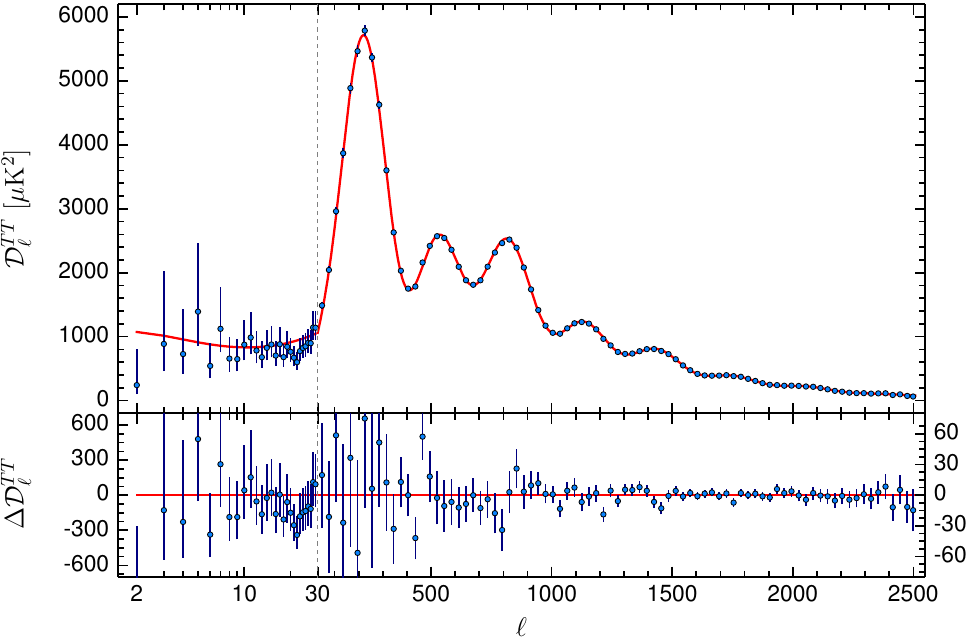}
\caption{Angular power spectrum of the temperature anisotropies of the CMB, as measured by the \textit{Planck} mission. From Ref.~\cite{2015arXiv150201582P}}
\label{fig:power_spectrum_Planck}
\end{figure}

Assuming statistical isotropy, the covariance of the temperature fluctuation~$\Theta$ as observed in two directions $\vect{e}_1$, $\vect{e}_2$ can be decomposed over Legendre polynomials~$P_\ell$ as
\begin{equation}
\ev{ \Theta(\vect{e}_1) \Theta(\vect{e}_2) } 
= \sum_{\ell=0}^\infty \frac{2\ell +1}{4\pi} \, C_\ell P_\ell(\vect{e}_1\cdot\vect{e}_2),
\end{equation}
where $\vect{e}_1\cdot\vect{e}_2=\cos\theta$ denotes the Euclidean scalar product between the unit spatial vectors $\vect{e}_1$, $\vect{e}_2$, and $\theta$ is the angle between them. Physically speaking, $C_\ell$ quantifies the correlation between the temperature of two points in the sky separated by an angle $\theta\sim\pi/\ell$. It thus corresponds to an angular power spectrum. Figure~\ref{fig:power_spectrum_Planck} shows the $C_\ell$, or more precisely the $\mathcal{D}_\ell\define \ell(\ell+1)C_\ell/2\pi$, as measured by the \textit{Planck} mission.

Among the features of this plot, note the oscillation of $\mathcal{D}_\ell$, with a period of $\Delta\ell_*\approx 300$, indicating a particular correlation between points separated by $\theta_*\approx 0.5\deg$. The origin of this correlation lies in the presence of sound waves propagating within the photon/plasma fluid before recombination, sustained by radiation pressure. From its birth at the end of inflation to its disappearance at recombination, such a wave propagates over a distance~$r\e{s}$ called \emph{sound horizon}\index{sound!horizon}. At recombination, two overdensities (or underdensities) are thus more likely to be separated by $r\e{s}$. This implies, on the CMB temperature map, that two hot (or cold) points are more likely to be separated by an angle $\theta_*=r\e{s}/D\e{A}$, where $D\e{A}$ is the angular diameter distance from us to the last scattering surface.

\begin{figure}[h!]
\centering
\includegraphics[width=0.7\textwidth]{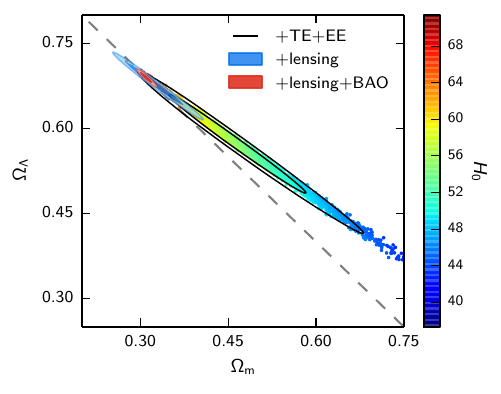}
\caption{Constraints on the cosmological parameters $\Omega\e{m0},\Omega_{\Lambda 0}$, and on the Hubble expansion rate~$H_0$, obtained from the analysis of the CMB. From Ref.~\cite{2015arXiv150201589P}}
\label{fig:constraints_Planck}
\end{figure}

While $r\e{s}$ depends on the physics of the primordial plasma, in particular through the density of baryonic matter~$\Omega\e{b0}$, it requires a model for the angular distance-redshift relation~$D\e{A}(z)$ to be connected with the observable quantity~$\theta_*$. The situation is similar to the analysis of the Hubble diagram, where a model for $D\e{L}(z)$ is required, and once again the standard choice is to use the distance-redshift relation of a FL model. Because this relation involves $\Omega\e{m0}$ and $\Omega_{\Lambda0}$, the analysis of the CMB provides constraints on these parameters, as shown in Fig.~\ref{fig:constraints_Planck}. Its degeneracy direction is kindly orthogonal to the one of SNIa constraints, making the combination of both a powerful and accurate measurement of $\Omega\e{m0},\Omega_{\Lambda0}$. These parameters are not the only ones to be constrained by the CMB, from which can also be extracted crucial information on the amplitude of matter density fluctuations~\cite{2015arXiv150201589P}, cosmic topology~\cite{2015arXiv150201593P}, inflation~\cite{2015arXiv150202114P}, reionisation, etc.

In the standard CMB analyses, gravitational lensing is considered to act essentially as a remapping of anisotropies of the temperature field~\cite{2006PhR...429....1L,2015arXiv150201591P}, according to the first-order formula~\eqref{eq:perturbation_source_position}. This brings an additional contribution to the angular power spectrum~$C_\ell$, but it does not affect the mean angular distance~$D\e{A}(z_*)$ mentioned above, and thus does not lead to any shift in the position of the acoustic peaks. This approach has been questioned recently in Ref.~\cite{2014JCAP...11..036C}, where second-order lensing corrections seemed to affect the average distance to the last-scattering surface by a few percents, a correction which would significantly impact the analysis of the CMB. Almost one year later, however, it was shown independently by Refs.~\cite{2015JCAP...06..050B,2015arXiv150308506K} (one of them being coauthored by the very authors of Ref.~\cite{2014JCAP...11..036C}) that this effect is actually caused by a subtle confusion between source averaging, sky averaging, and ensemble averaging. The distinction between these different ways of averaging physical quantities, and their natural domain of applicability, had been emphasized earlier by Ref.~\cite{2005ApJ...632..718K}; it now seems to be fully understood~\cite{2015JCAP...07..040B}. In the end, this debate validated the standard treatment of CMB lensing.

\subsection{Baryon acoustic oscillation}\label{sec:BAO}\index{baryon acoustic oscillations}\index{BAO|see{baryon acoustic oscillations}}

The acoustic feature present in the CMB corresponds to a rather large scale, which is weakly affected by the gravitational evolution of the Universe between the epoch of recombination and today---contrary to small-scale inhomogeneities which tend to collapse and lose information about their initial conditions. As a consequence, this correlation has survived within the distribution of baryonic matter in the Universe. In this case, it is referred to as the \emph{Baryon Acoustic Oscillation} signal. Because it only grows with cosmic expansion, the BAO scale (or its comoving counterpart~$\chi\e{BAO}$) can be considered a cosmic \emph{standard ruler}\index{standard!ruler}, by analogy with the notion of standard candle.

\begin{figure}[h!]
\centering
\begin{minipage}{0.48\textwidth}
\includegraphics[width=\linewidth]{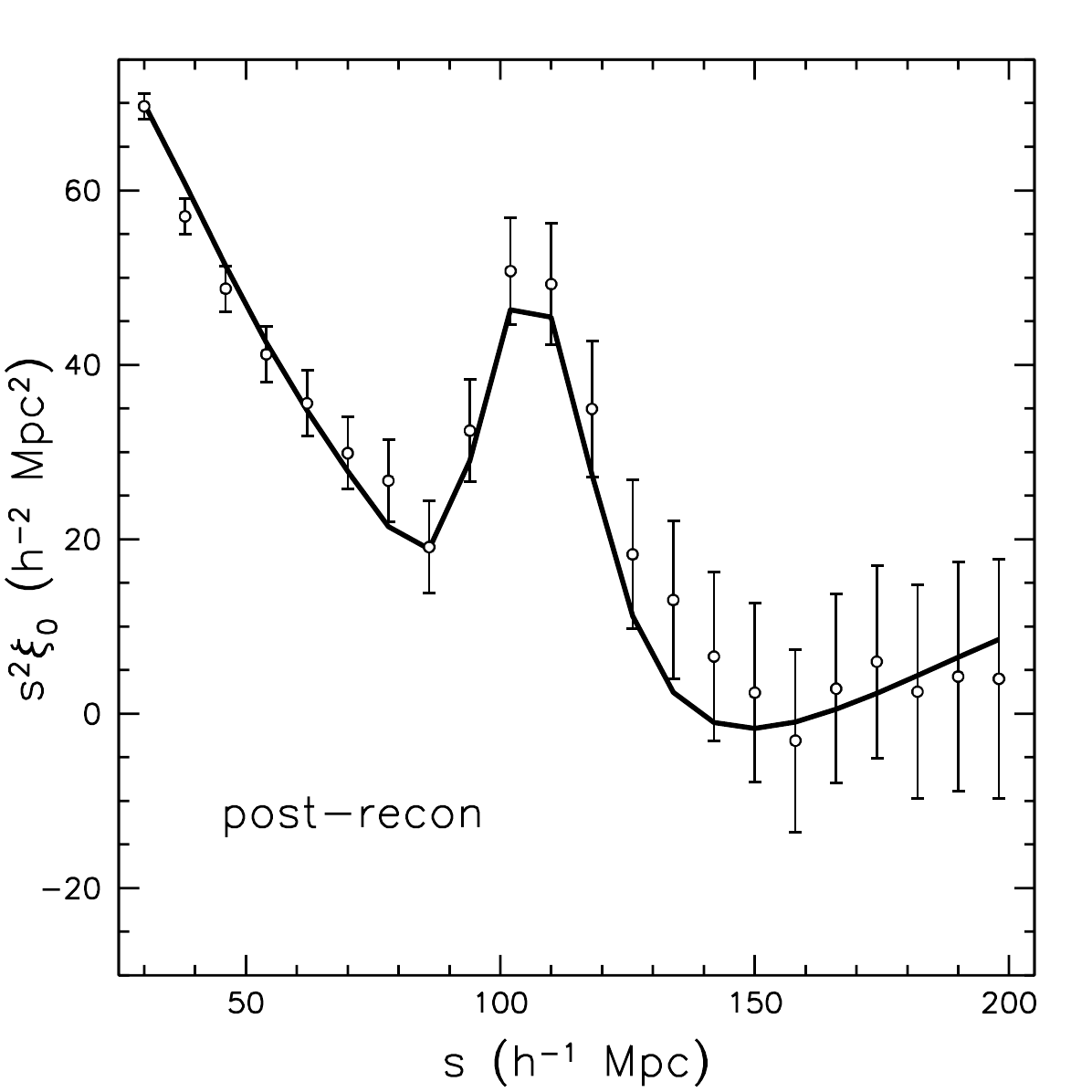}\\
\includegraphics[width=\linewidth]{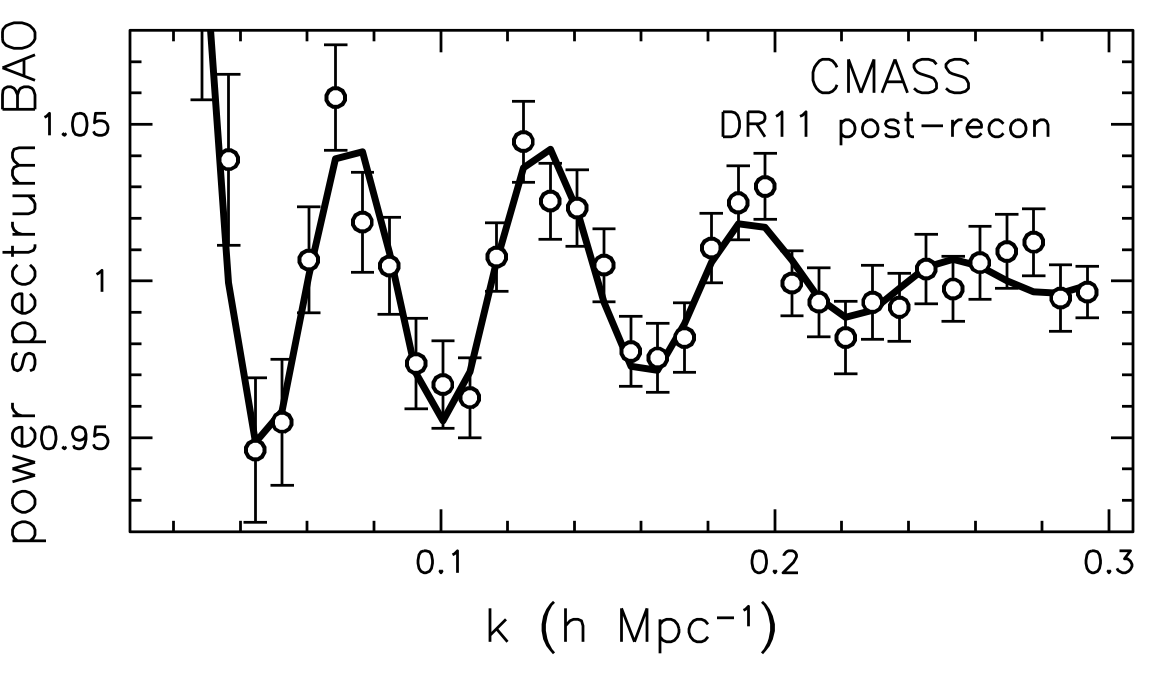}
\end{minipage}
\begin{minipage}{0.5\textwidth}
\includegraphics[width=\linewidth]{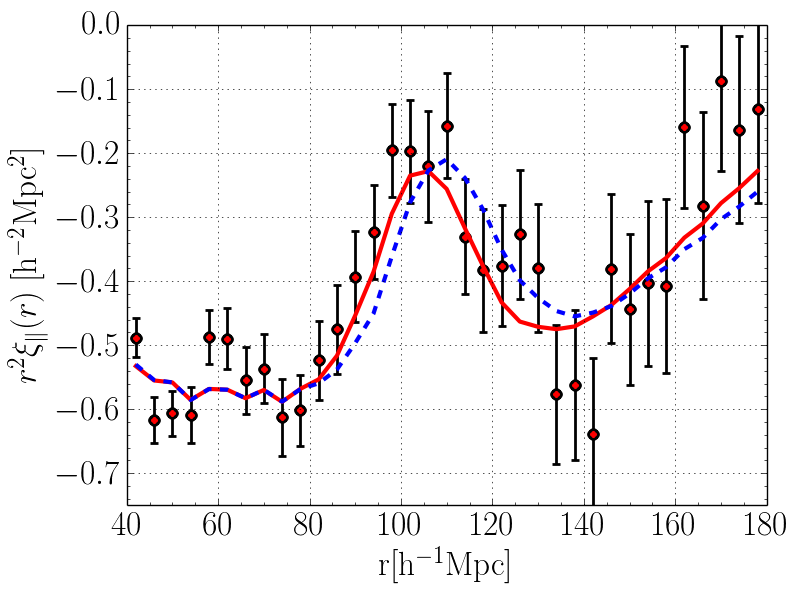}\\
\includegraphics[width=\linewidth]{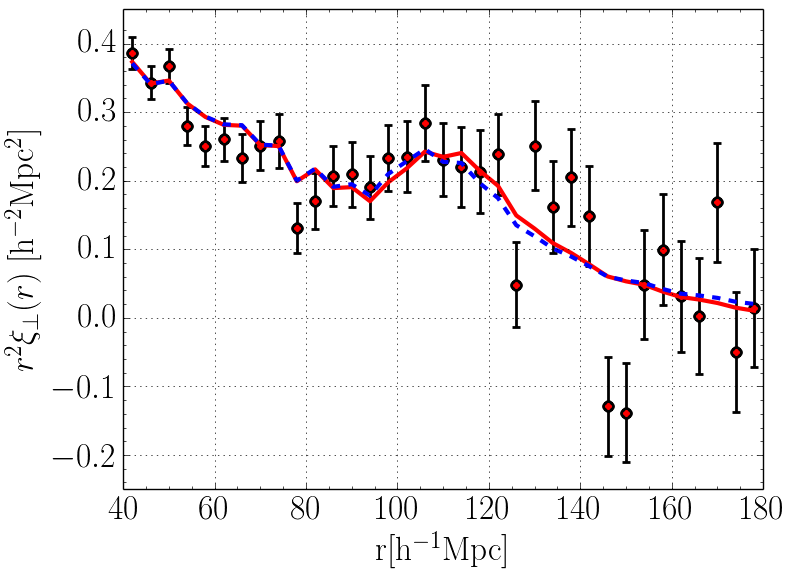}
\end{minipage}
\caption{\textit{Left panel.} Two-point correlation function (top) and power spectrum (bottom) of the distribution of the BOSS CMASS galaxy sample ($0.2 \leq z \leq 0.7$), as a function of comoving distance~$s$ and comoving wavenumber~$k$, respectively. This BAO signal is effectively measured at $z=0.57$. From Ref.~\cite{2014MNRAS.441...24A}. \textit{Right panel}. Two-point correlation function for objects aligned with the line of sight (top), or orthogonal to the line of sight (bottom), measured with the BOSS quasars ($2.1\leq z \leq 3.5$) and the intergalactic medium traced by their Lyman-$\alpha$ forest, as a function of comoving distance~$r$. The effective redshift is $z=2.34$ here. From Ref.~\cite{2015A&A...574A..59D}.}
\label{fig:xi_ps_BOSS}
\end{figure}

From the above reasoning, it is easy to estimate the BAO scale today~$r\e{d}$ as
\begin{equation}
r\e{d} = \frac{a_0}{a_*} \, r\e{s} = (1+z_*) D\e{A}(z_*)\theta_* \approx 150\U{Mpc}\approx 100 h^{-1}\U{Mpc},
\end{equation}
where we used the FL expression of $D\e{A}(z)$. The first detection of the BAO signal in today's distribution of matter has been obtained by the Sloan Digital Sky Survey (SDSS), in the two-point correlation function of low-redshift LRGs~\cite{2005ApJ...633..560E}. To date, the most precise measurement has been realised by the Baryon Oscillation Spectroscopic Survey (BOSS) of SDSS-III~\cite{SDSSIII}, with two complementary experiments: \begin{inparaenum}[(i)] \item a large survey of 1 million galaxies between $0.2 \leq z \leq 0.7$~\cite{2014MNRAS.441...24A}; \item a survey of distant quasars ($2.1\leq z\leq 3.5$), and of the intergalactic medium traced by the Lyman-$\alpha$ forest in their spectrum~\cite{2015A&A...574A..59D}. \end{inparaenum}
The corresponding correlation functions are shown in Fig.~\ref{fig:xi_ps_BOSS}, where the BAO signal is evident ($7\sigma$ confidence level), and in agreement with the order of magnitude obtained above. It is remarkable that this property of the matter distribution can be observed at so different epochs of the Universe: $z=1090$ (Fig.~\ref{fig:power_spectrum_Planck}), $z=2.35$, and $z=0.54$ (Fig.~\ref{fig:xi_ps_BOSS}).

\begin{figure}
\centering
\includegraphics[width=0.465\textwidth]{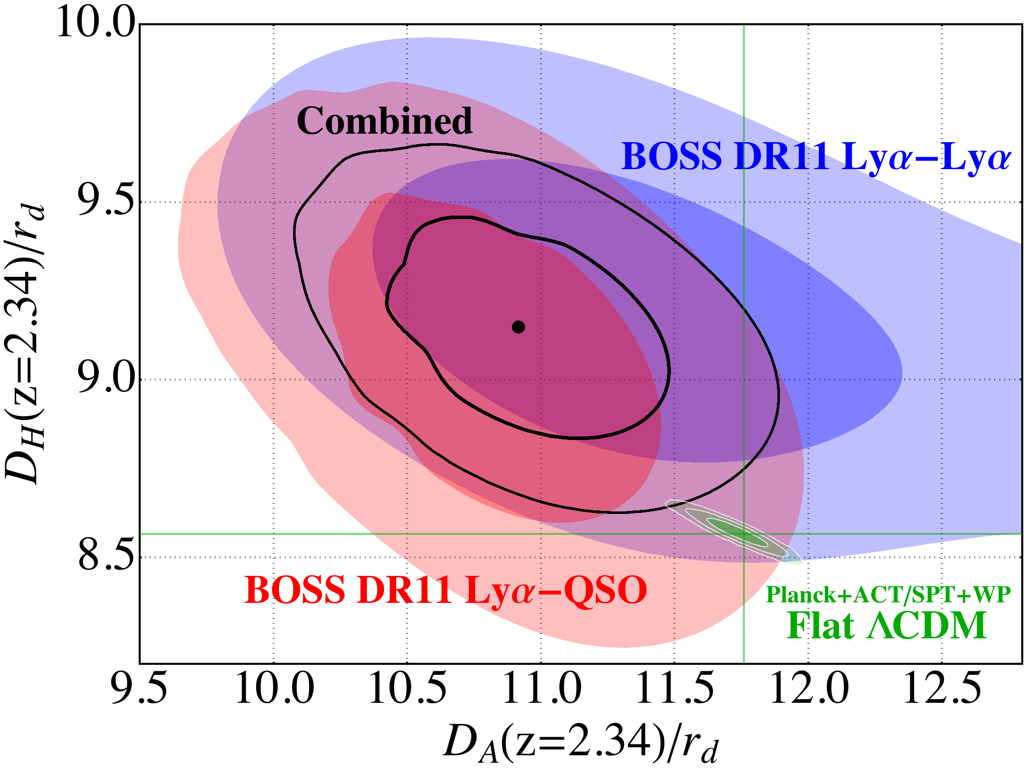}
\includegraphics[width=0.525\textwidth]{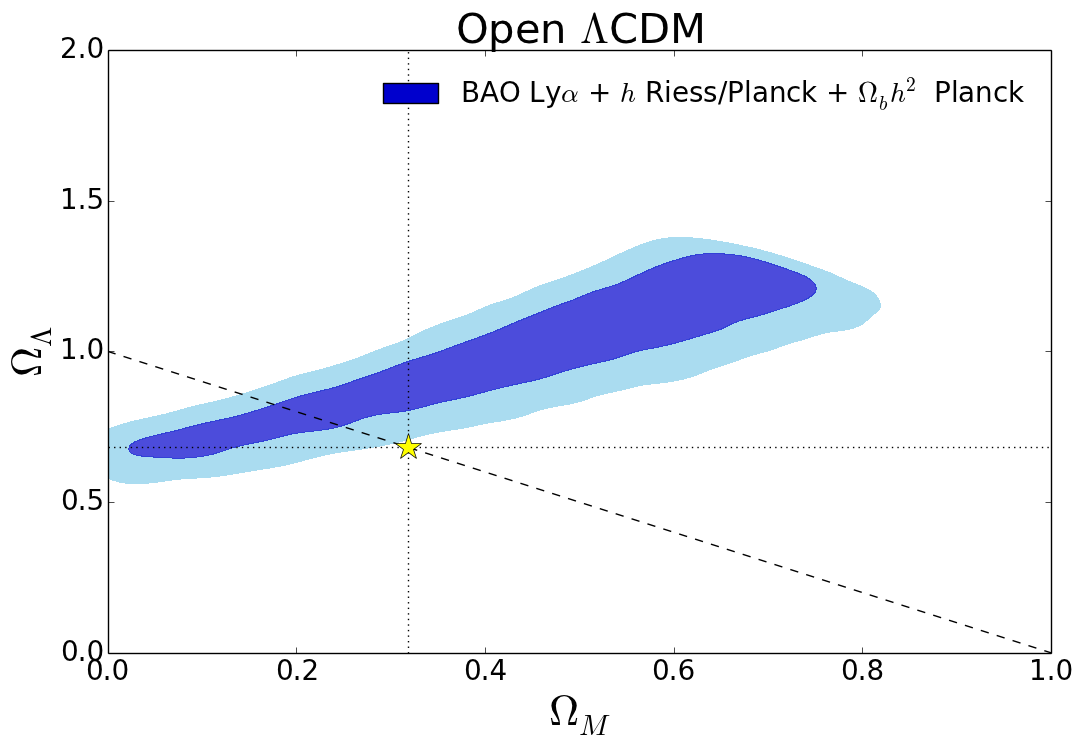}
\caption{\textit{Left panel}. Constraints on the longitudinal and transverse distances $D\e{A}/r\e{d}$, $D_H/r\e{d}$ at $z=2.34$ from BAO measured with quasars and intergalactic medium. The green contours indicate the constraints from the CMB in the case of a flat FL model. \textit{Right panel}. Consequent constraints on the cosmological parameters $\Omega\e{m0},\Omega_{\Lambda 0}$. The yellow star indicates the concordance $\Lambda$CDM model. From Ref.~\cite{2015A&A...574A..59D}.}
\label{fig:constraints_BOSS}
\end{figure}

Experiments such as BOSS have the advantage, with respect to CMB observations, of extracting the BAO signal from a three-dimensional distribution rather than from a two-dimensional map. Hence, additionally to the angular correlation scale~$\theta\e{BAO}$, they yield a redshift correlation scale~$\Delta z\e{BAO}$ associated with BAOs aligned with the line of sight. Assuming a FL model, these quantities are related to the BAO scale today~$r\e{d}=a_0\chi\e{BAO}$ as
\begin{align}
\theta\e{BAO} &= \frac{a}{a_0} \frac{r\e{d}}{D\e{A}(z)} = \frac{r\e{d}}{(1+z) D\e{A}(z)} \\
\Delta z\e{BAO} &= H(z) a_0 \chi\e{BAO} = \frac{r\e{d}}{D_H(z)},
\end{align}
for observations at redshift $z$, with $D_H\define 1/H$, and where we used Eq.~\eqref{eq:chi_z} for writing $a_0 \Delta\chi = \Delta z/H$. Given a set of cosmological parameters, $\theta\e{BAO}$ and $\Delta z\e{BAO}$ are therefore precisely related, so that their comparison allows to test the choice of these cosmological parameters, or the validity of the FL model itself. This procedure is known as the \emph{Alcock-Paczy\'{n}ski test}~\cite{1979Natur.281..358A}\index{Alcock-Paczy\'{n}ski test}. In Fig.~\ref{fig:constraints_BOSS} are represented the observed values of $D\e{A}/r\e{d}$, $D_H/r\e{d}$, and the consequent constraints on $\Omega\e{m0},\Omega_{\Lambda 0}$ obtained by Ref.~\cite{2015A&A...574A..59D}. Note that, here again, the standard interpretation of the observed data relies on the FL $D\e{A}(z)$ relation.

\subsection{Other observations}\label{sec:other_observations}

Let us finally mention a few other cosmological probes which, though less emblematic, have become more and more precise over the last years and are now efficient complements to SNIa, CMB, and BAO observations.

\subsubsection{Baryon and gas fraction in galaxy clusters}

The potential of galaxy clusters as cosmological probes was revealed in the early 1990s, when Ref.~\cite{1993Natur.366..429W} seriously challenged the standard paradigm of that time, according to which $\Omega\e{m0}=1$. By measuring the gas and stellar masses $M\e{gas}$, $M\e{gal}$ of the Coma cluster, respectively from its X-ray and B-band luminosities, the authors of Ref.~\cite{1993Natur.366..429W} estimated the total baryonic mass~$M\e{b}=M\e{gal}+M\e{gas}$ of this cluster and compared it with its total (dynamical) mass~$M\e{tot}$. Assuming a ratio $M\e{b}/M\e{tot}$ equal to the mean cosmological baryonic fraction, 
\begin{equation}
\frac{M\e{b}}{M\e{tot}} = \frac{\Omega\e{b0}}{\Omega\e{m0}},
\end{equation}
and using the value of $\Omega\e{b0}$ obtained by analyses of the Big Bang Nucleosynthesis (BBN), they concluded that $\Omega\e{m0}\approx 0.28$ (for $h=0.7$), a result surprisingly close to the current admitted value. It is worth emphasizing that this discovery occurred \emph{before} the first analyses of the Hubble diagram, which really opened the era of dark energy cosmology.

Three years later, Ref.~\cite{1996PASJ...48L.119S} proposed a more subtle method for constraining cosmological parameters with galaxy clusters. The X-ray data of galaxy clusters are indeed interpreted in such a way that the gas mass fraction extracted from them reads
\begin{equation}
f\e{gas}\define \frac{M\e{gas}}{M\e{tot}} = B(z) D\e{A}^{3/2}(z),
\end{equation}
where $B$ is independent from the cosmological parameters, so that $D\e{A}(z)$ contains all the cosmological dependence. Assuming that $f\e{gas}$ does not depend (on average) on the redshift of the cluster, we conclude that a plot representing the $f\e{gas}$ of several clusters as a function of their redshift must be flat. However, it is not the case if a wrong cosmology---or more generally a wrong distance-redshift relation---is assumed for the data analysis. This idea provides a consistency test of the cosmological model, similarly to the comparison between the longitudinal and transverse BAO signal. Though limited by the intrinsic scatter of $f\e{gas}$, estimated to be $(7.4\pm 2.3)\%$ by Ref.~\cite{2014MNRAS.440.2077M}, this method provides today constraints on $(\Omega\e{m0}, \Omega_{\Lambda0})$ which are competitive with BAO's or SNeIa's (see Fig.~\ref{fig:constraints_fgas}). We keep emphasizing that, like all the other cosmological probes reviewed so far, the $f\e{gas}$ method relies on a particular model for $D\e{A}(z)$, taken to be the FL one.

\subsubsection{Weak lensing tomography}\index{weak lensing tomography}

\begin{figure}[t]
\centering
	\begin{subfigure}[b]{0.47\linewidth}
	\centering
	\includegraphics[width=\linewidth]{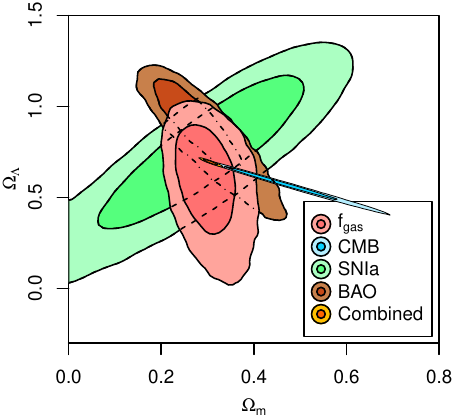}
	\caption{Constraints on $\Omega\e{m0}, \Omega_{\Lambda0}$ by comparing the gas fraction~$f\e{gas}$ within galaxy clusters at different redshifts. From Ref.~\cite{2014MNRAS.440.2077M}}
	\label{fig:constraints_fgas}
	\end{subfigure}
\hfill
	\begin{subfigure}[b]{0.5\linewidth}
	\centering
	\includegraphics[width=\linewidth]{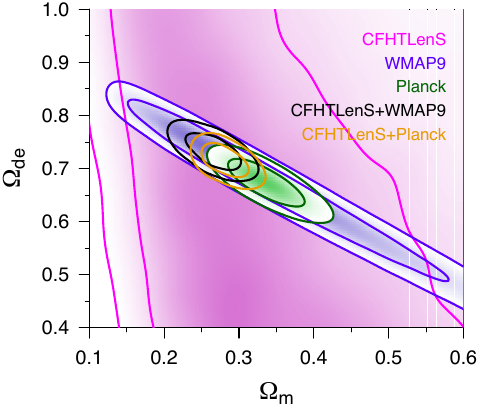}
	\caption{Constraints on $\Omega\e{m0}, \Omega_{\Lambda0}$ imposed by the measured two- and three-point correlation functions of cosmic shear. From Ref.~\cite{2014MNRAS.441.2725F}}
	\label{fig:constraints_CFHTLenS}
	\end{subfigure}
\caption{Some cosmological constraints from $f\e{gas}$ measurements in galaxy clusters (left panel) or weak-lensing tomography (right panel).}
\label{fig:constraints_others}
\end{figure}

We have seen in \S~\ref{sec:perturbation_light_beams} that the observations of the shapes of lensed galaxies gives access to the power spectrum of cosmic convergence, via the one of cosmic shear, as $P_\kappa=P_\gamma=P_\eps$. Besides, the expression~\eqref{eq:power_spectrum_cosmic_convergence} of $P_\kappa$ turns out to involve the cosmological parameters in various ways: directly on $\Omega\e{m0}$ and $H_0$, and indirectly via both the lensing efficiency and the matter power spectrum~$P_\delta$. Measuring the statistics of galaxy ellipticities is therefore a means to constrain the cosmological parameters. As an illustration, Fig.~\ref{fig:constraints_CFHTLenS} shows the constraints on $\Omega\e{m0},\Omega_{\Lambda0}$ obtained from the two- and three-point correlation functions of cosmic shear, as measured by the Canada France Hawaii Telescope Lensing Survey CFHTLenS~\cite{CFHTLenS} in Ref.~\cite{2014MNRAS.441.2725F}. These constraints are rather loose compared to ones obtained with other cosmological probes. Weak lensing (WL) is actually much more efficient at constraining $\sigma_8$\index{$\sigma_8$}, which is the standard deviation of the matter density fluctuations~$\delta$ smoothed over a comoving sphere of radius $R\define 8 h^{-1}\U{Mpc}$,
\begin{equation}
\sigma_8^2
\define \ev[4]{ \pac{\frac{3}{4\pi R^3} \int_{|x^i|\leq R} \dd^3 x \; \delta(x)}^2 }
= \int_0^\infty \frac{\dd k}{k} \pac{\frac{3j_1(kR)}{kR}}^2 P_\delta(k) \sim 1,
\end{equation}
where $j_1$ is the order-one spherical Bessel function~\cite{1994tisp.book.....G}. This can be understood by the fact that $\Omega\e{m0}^2\sigma_8^2$ essentially sets the amplitude of $P_\kappa$, while its dependence in the other cosmological parameters is weaker.

Alternative methods exploiting weak lensing, e.g. from the shear ratio around galaxy clusters~\cite{2003PhRvL..91n1302J}, or using higher-order statistical properties such as shear peak counts~\cite{2010MNRAS.402.1049D,2015arXiv150602192M}, are still under development. They are expected to provide very accurate measurements of the cosmological parameters from future large surveys, such as Euclid~\cite{Euclid}, or the Wide-Field InfraRed Survey Telescope (WFIRST)~\cite{WFIRST}.

\subsubsection{Strong lensing and time delays}

When very massive or compact objects lie between a light source and an observer, the deflection they induce can be large enough for allowing light to have several paths from the one to the other. In this regime of \emph{strong gravitational lensing} (SL)\index{strong lensing}, the observer can thus see multiple images of a single source (gravitational mirages), or images which are so sheared that they appear as luminous arcs around the lens~\cite{1992grle.book.....S}. The size and shape of these arcs, or the Shapiro time delay between two different images, depend on the properties of the lens, but also on the lensing efficiency factor~\eqref{eq:lensing_efficiency}, and therefore on the cosmological parameters. Provided one has an accurate model for the lenses, this property can be exploited to constrain the cosmological parameters from strong lensing measurements~\cite{1964MNRAS.128..307R}. This idea has the advantage, with respect to WL, that it can be applied to single objects~\cite{2014ApJ...788L..35S}, rather than relying on statistics. Its main limitation, however, concerns the uncertainties on the lens properties. Recently, time delays between multiple images observed by the COSmological MOnitoring of GRAvItational Lenses (COSMOGRAIL)~\cite{COSMOGRAIL} were used to measure $h=0.800^{+0.058}_{-0.057}$~\cite{2013ApJ...766...70S}. Besides, strong lensing measurements were recently exploited to constrain spatial curvature via the distance sum rule~$\Omega_{K0}$~\cite{2014arXiv1412.4976R}, or the dark-energy equation of state~\cite{2010MNRAS.406.1055B}.

\subsection{Discussion}

This brief review of the current status of observational cosmology shows that the standard $\Lambda$CDM model consistently fits all the data with an impressively high level of precision, given its simplicity. We emphasized that the interpretation of every observation involves the relation between angular (or luminosity) distance and redshift~$D\e{A}(z)$, from which generally comes its sensitivity with respect to the cosmological parameters. This relation is always assumed to be the one of a FL spacetime, derived in \S~\ref{sec:optics_FL}. In other words, \emph{all the distances necessary to interpret cosmological observations are calculated by assuming that light propagates through a perfectly homogeneous and isotropic Universe}. The success of this strong assumption is particularly striking if we compare the scales involved in all these observations, listed in Table~\ref{tab:scales_observations}. We see that the typical observed angles, i.e. the typical size of the corresponding light beams, span 12 orders of magnitude. To my knowledge, there exists no model in the whole History of physics whose domain of validity is so wide\footnote{One could argue that the validity of quantum electrodynamics has been experimentally verified over more than 11 orders of magnitude, but the latter is more a theory than a model.}.

The issue raised here must be connected to the laws of geometric optics in curved spacetime presented in Part~\ref{part:geometric_optics}. We have seen in Chap.~\ref{chapter:beams} that the evolution of the angular distance is driven by Ricci and Weyl curvatures in quite different ways. Ricci lensing tends to directly reduce $D\e{A}$ by focusing the underlying light beam, while Weyl lensing indirectly reduces it via the beam's shear rate. However, what is considered Ricci or Weyl lensing depends on the size of the beam itself: a distribution of point masses can appear alternatively clumpy or smooth to a beam whose cross-sectional diameter is respectively much smaller or much larger than the typical distance between two point masses. Similarly, our Universe can be considered smooth---Ricci dominated---for the beams involved in BAO observations, but it is certainly very clumpy---Weyl dominated---when SN observations are at stake. This \emph{Ricci-Weyl paradox}\index{Ricci-Weyl paradox} of cosmology is central to the present thesis, whose motivation can be summarised by the following question: why is the FL geometry so efficient at interpreting all the cosmological observations? The next part intends to provide elements of answer.

\bigskip

\begin{table}[h!]
\centering
\begin{tabular}{|c|cc|}
\hline 
\rowcolor{lightgray} {\sf\bfseries observation} & {\sf\bfseries relevant angular scale} & {\sf\bfseries typical value (rad)} \\ 
\hline 
BAO & BAO scale at $z\sim 0.5,2$ & $10^{-1},10^{-2}$ \\ 
CMB & BAO scale at $z\sim 1000$ & $10^{-2}$ \\ 
$f\e{gas}$ & apparent size of a cluster at $z\sim 0.5$ & $10^{-3}$ \\ 
SL & Einstein radius of a galaxy on cosmological scales & $10^{-4}$ \\ 
WL & apparent size of a galaxy at $z\sim 0.5$ & $10^{-5}$ \\ 
SNeIa & apparent size of a SN at $z\sim 0.5 $ & $10^{-13}$ \\ 
\hline 
\end{tabular} 
\caption{Relevant angular size (observed angular aperture) of the light beams involved in various cosmological observations, and their orders of magnitude.}
\label{tab:scales_observations}
\end{table}

%% file: perturbation_ray.pdf_tex
\begingroup%
  \makeatletter%
  \providecommand\color[2][]{%
    \errmessage{(Inkscape) Color is used for the text in Inkscape, but the package 'color.sty' is not loaded}%
    \renewcommand\color[2][]{}%
  }%
  \providecommand\transparent[1]{%
    \errmessage{(Inkscape) Transparency is used (non-zero) for the text in Inkscape, but the package 'transparent.sty' is not loaded}%
    \renewcommand\transparent[1]{}%
  }%
  \providecommand\rotatebox[2]{#2}%
  \ifx\svgwidth\undefined%
    \setlength{\unitlength}{208.83140458bp}%
    \ifx\svgscale\undefined%
      \relax%
    \else%
      \setlength{\unitlength}{\unitlength * \real{\svgscale}}%
    \fi%
  \else%
    \setlength{\unitlength}{\svgwidth}%
  \fi%
  \global\let\svgwidth\undefined%
  \global\let\svgscale\undefined%
  \makeatother%
  \begin{picture}(1,0.26122258)%
    \put(0,0){\includegraphics[width=\unitlength]{perturbation_ray.pdf}}%
    \put(0.08897588,0.13689045){\color[rgb]{0,0,0}\makebox(0,0)[lb]{\smash{$O$}}}%
    \put(0.63640029,0.11311347){\color[rgb]{0,0,0}\makebox(0,0)[lb]{\smash{$S$}}}%
    \put(0.55505278,0.22554413){\color[rgb]{0,0,0}\makebox(0,0)[lb]{\smash{$\bar{\theta}^A$}}}%
    \put(0.17514128,0.22822646){\color[rgb]{0,0,0}\makebox(0,0)[lb]{\smash{$-\vect{\bar{d}}$}}}%
    \put(0.40044839,0.08672864){\color[rgb]{0,0,0}\rotatebox{-7.9661898}{\makebox(0,0)[lb]{\smash{$\bar{\theta}^A+\delta\theta^A$}}}}%
  \end{picture}%
\endgroup%

%% file: amplification_alternative_derivation.pdf_tex
\begingroup%
  \makeatletter%
  \providecommand\color[2][]{%
    \errmessage{(Inkscape) Color is used for the text in Inkscape, but the package 'color.sty' is not loaded}%
    \renewcommand\color[2][]{}%
  }%
  \providecommand\transparent[1]{%
    \errmessage{(Inkscape) Transparency is used (non-zero) for the text in Inkscape, but the package 'transparent.sty' is not loaded}%
    \renewcommand\transparent[1]{}%
  }%
  \providecommand\rotatebox[2]{#2}%
  \ifx\svgwidth\undefined%
    \setlength{\unitlength}{283.30066211bp}%
    \ifx\svgscale\undefined%
      \relax%
    \else%
      \setlength{\unitlength}{\unitlength * \real{\svgscale}}%
    \fi%
  \else%
    \setlength{\unitlength}{\svgwidth}%
  \fi%
  \global\let\svgwidth\undefined%
  \global\let\svgscale\undefined%
  \makeatother%
  \begin{picture}(1,0.37209657)%
    \put(0,0){\includegraphics[width=\unitlength]{amplification_alternative_derivation.pdf}}%
    \put(0.06558741,0.04599411){\color[rgb]{0,0,0}\makebox(0,0)[lb]{\smash{$O$}}}%
    \put(0.83506688,0.13385678){\color[rgb]{0,0,0}\makebox(0,0)[lb]{\smash{$S_1$}}}%
    \put(0.72948849,0.03604172){\color[rgb]{0.83137255,0.33333333,0}\rotatebox{-7.20717314}{\makebox(0,0)[lb]{\smash{$\bar{\theta}_1^A$}}}}%
    \put(0.48968561,0.07035297){\color[rgb]{0.56862745,0.43529412,0.43529412}\rotatebox{2.31506195}{\makebox(0,0)[lb]{\smash{$\bar{\theta}_1^A+\delta\theta_1^A$}}}}%
    \put(0.79270905,0.27504953){\color[rgb]{0,0,0}\makebox(0,0)[lb]{\smash{$S_2$}}}%
    \put(0.72636077,0.18545527){\color[rgb]{0.83137255,0,0}\rotatebox{8.11536319}{\makebox(0,0)[lb]{\smash{$\bar{\theta}_2^A$}}}}%
    \put(0.46056578,0.217199){\color[rgb]{0,0,0}\rotatebox{14.91103264}{\makebox(0,0)[lb]{\smash{$\bar{\theta}_2^A+\delta\theta_2^A$}}}}%
    \put(0.81530696,0.20635112){\color[rgb]{0,0,0}\makebox(0,0)[lb]{\smash{$\xi^A$}}}%
    \put(0.80275077,0.07746999){\color[rgb]{0,0,0}\makebox(0,0)[lb]{\smash{$\bar{\xi}^A$}}}%
  \end{picture}%
\endgroup%

%% file: DA_z_annoted.pdf_tex
\begingroup%
  \makeatletter%
  \providecommand\color[2][]{%
    \errmessage{(Inkscape) Color is used for the text in Inkscape, but the package 'color.sty' is not loaded}%
    \renewcommand\color[2][]{}%
  }%
  \providecommand\transparent[1]{%
    \errmessage{(Inkscape) Transparency is used (non-zero) for the text in Inkscape, but the package 'transparent.sty' is not loaded}%
    \renewcommand\transparent[1]{}%
  }%
  \providecommand\rotatebox[2]{#2}%
  \ifx\svgwidth\undefined%
    \setlength{\unitlength}{325.2140625bp}%
    \ifx\svgscale\undefined%
      \relax%
    \else%
      \setlength{\unitlength}{\unitlength * \real{\svgscale}}%
    \fi%
  \else%
    \setlength{\unitlength}{\svgwidth}%
  \fi%
  \global\let\svgwidth\undefined%
  \global\let\svgscale\undefined%
  \makeatother%
  \begin{picture}(1,0.49366314)%
    \put(0,0){\includegraphics[width=\unitlength]{DA_z_annoted.pdf}}%
    \put(0.10164621,0.0584648){\makebox(0,0)[lb]{\smash{0.0}}}%
    \put(0.25601863,0.0584648){\makebox(0,0)[lb]{\smash{0.5}}}%
    \put(0.41039106,0.0584648){\makebox(0,0)[lb]{\smash{1.0}}}%
    \put(0.56476348,0.0584648){\makebox(0,0)[lb]{\smash{1.5}}}%
    \put(0.71913589,0.0584648){\makebox(0,0)[lb]{\smash{2.0}}}%
    \put(0.05803287,0.10755287){\makebox(0,0)[lb]{\smash{0.0}}}%
    \put(0.05803287,0.21109706){\makebox(0,0)[lb]{\smash{0.5}}}%
    \put(0.05803287,0.31464124){\makebox(0,0)[lb]{\smash{1.0}}}%
    \put(0.05803287,0.41818544){\makebox(0,0)[lb]{\smash{1.5}}}%
    \put(0.41477488,0.00224852){\color[rgb]{0,0,0}\makebox(0,0)[lb]{\smash{$z$}}}%
    \put(0.02118797,0.21134157){\color[rgb]{0,0,0}\rotatebox{90}{\makebox(0,0)[lb]{\smash{$D\e{A}$ (Gpc)}}}}%
    \put(0.3901757,0.21134157){\color[rgb]{0,0,0}\makebox(0,0)[lb]{\smash{$(\Omega\e{m0},\Omega_{\Lambda0})=(0.3,0.7)$}}}%
    \put(0.3901757,0.14984361){\color[rgb]{0,0,0}\makebox(0,0)[lb]{\smash{$(\Omega\e{m0},\Omega_{\Lambda0})=(1,0)$}}}%
    \put(0.35327693,0.33433748){\color[rgb]{0,0,0}\makebox(0,0)[lb]{\smash{$\Delta z$}}}%
    \put(0.36557652,0.39583544){\color[rgb]{0,0,0}\makebox(0,0)[lb]{\smash{at fixed $v$}}}%
    \put(0.24258061,0.42043462){\color[rgb]{0,0,0}\makebox(0,0)[lb]{\smash{$\Delta D\e{A}$}}}%
  \end{picture}%
\endgroup%

%% file: introduction_part_III.tex
\thispagestyle{plain}

\lettrine{T}{he} question of how the clumpiness of the Universe affects the interpretation of cosmological observations has a long history. It was first raised more than 50 years ago, in 1964, by Zel'dovich in Ref.~\cite{1964SvA.....8...13Z}, and by Feynman in a colloquium given at the California Institute of Technology (mentioned e.g. in Ref.~\cite{1967ApJ...150..737G}). The underlying argument is that, on the very small scales probed by, e.g., the light beam coming from a supernova, a fluid description of the surrounding matter shall not hold in principle, as it is rather concentrated in clumps than smoothly distributed. A typical beam is thus expected to encounter less matter than in a strictly homogeneous model. Then followed a series of seminal articles, both on the Soviet side by Dashevskii \& Slysh ~\cite{DashevskiiZeldovich1965,1966SvA.....9..671D,1965AZh....42..863D}, and on the western side by Bertotti~\cite{1966RSPSA.294..195B} and Gunn~\cite{1967ApJ...150..737G,1967ApJ...147...61G}.

While those studies were based on general arguments about geometric optics in an inhomogeneous Universe, Kantowski~\cite{1969ApJ...155...89K}, and later Dyer \& Roeder~\cite{1973ApJ...180L..31D,1973PhDT........17D,1974ApJ...189..167D,1975ApJ...196..671R}, relied on an exact solution of the Einstein equation, namely the Swiss-cheese model. This solution, obtained by gluing together the Schwarzschild and FL spacetime---which makes a `hole' inside the Friedmann-Lema\^itre `cheese'---had been originally proposed by Einstein \& Straus~\cite{1945RvMP...17..120E,1946RvMP...18..148E} as a means to model individual stars within the expanding Universe (see Chap.~\ref{chapter:SC} for more details). These analyses yielded in particular a procedure for determining the effective impact of clumpiness on the angular distance-redshift relation, known as the partially-filled beam approximation, or \emph{Dyer-Roeder approximation}\index{Dyer-Roeder approximation}, the name of Kantowski being usually---but unfairly---omitted in the literature. In agreement with Zel'dovich's original intuition, this approximation predicts that a typical light beam is defocused with respect to the FL behaviour, and therefore bias distance measurements towards larger values. Such a conclusion was criticized by Weinberg in Ref.~\cite{1976ApJ...208L...1W}, who argued on the basis of flux conservation that inhomogeneities should have no mean effect. Although this argument turns out to be inexact, it holds in principle at a very high order of precision (see e.g. Ref.~\cite{2015arXiv150308506K,2015JCAP...06..050B} for recent discussions). In practice, however, Weinberg's approach fails at capturing the consequences of: 
\begin{inparaenum}[(a)]
\item the sparsity of observations---we do not observe an infinity of sources over the whole sky---; and
\item selection effects---some lines of sight can be masked~\cite{Futamase:1989hba}---which were central to the earlier results.
\end{inparaenum}

The whole issue has been then progressively left aside, presumably because no observation managed to arbitrate between the various points of view. It was revived in the 2000s within a new cosmological paradigm, in particular with the perspective of explaining the apparent acceleration of cosmic expansion without the need of dark energy. Most analyses, in this case, focused on the impact on observations of the \emph{large-scale} inhomogeneity, relying either on the standard perturbation theory~\cite{Valageas:1999ch,Cooray:2005yr,Dodelson:2005zt,Bonvin:2005ps,2005PhRvD..71f3537B,Meures:2011gp,2012JCAP...04..036B,2012JCAP...11..045B,2013JCAP...06..002B,2013PhRvL.110b1301B,2013arXiv1309.6542N,2014JCAP...06..023A,2014CQGra..31t2001U,2014CQGra..31t5001U,2014JCAP...11..036C,2015JCAP...08..020F,2015arXiv150308506K}, or on Swiss-cheese models with Lema\^itre-Tolman-Bondi holes~\cite{Marra:2007pm,Brouzakis:2007zi,2007JCAP...02..013B,Biswas:2007gi,2008PhRvD..77b3003M,Clifton:2009nv,Szybka:2010ky,Vanderveld:2008vi,Valkenburg:2009iw,2011JCAP...02..025B,2012PhRvD..85b3510F,Flanagan:2012yv,2013PhRvL.110b1302B,2013JCAP...12..051L,2015arXiv150706590L} or Szekeres~\cite{2009GReGr..41.1737B,2010PhRvD..82j3510B,2014PhRvD..90l3536P,2014JCAP...03..040T} holes, which typically aim at modelling superclusters or cosmic voids (see also Refs.~\cite{Adamek:2014qja,Bolejko:2012ue}). Particular efforts were made in Refs.~\cite{1998astro.ph..1122L,Rasanen:2008be,Rasanen:2009uw,Rasanen:2011bm,2011JCAP...07..008G,2014JCAP...10..073B} in order to connect observables with the backreaction and cosmic averaging problems. In contrast, less attention was paid to the specific issue of clumpiness---with the notable exceptions of Refs.~\cite{Holz:1997ic,Okamura:2009zf,Bruneton:2012ru,Clifton:2009jw,2009JCAP...10..026C,2012PhRvD..85b3502C}, where inhomogeneities are treated similarly to the historical Einstein-Straus Swiss-cheese model. It was exhumed by Clarkson et al.~\cite{Clarkson:2011br}, who reviewed past and present approaches, emphasizing that no definite answer had been formulated so far.

This latter article motivated the present part of my thesis, which represents roughly three quarters of it---the last quarter concerns anisotropic cosmologies and is the subject of Part~\ref{part:anisotropic_cosmologies}. It is divided into two chapters, which present two different approaches to the initial question raised by Zel'dovich and Feynman. In Chap.~\ref{chapter:SC}, I revisit light propagation in Einstein-Straus Swiss-cheese models with the eyes of modern cosmology. In Chap.~\ref{chapter:SL}, I propose a completely new framework for dealing with lensing on small scales, based on the theory of stochastic processes.

%% file: chapter_6.tex
\lettrine{T}{his} chapter is devoted to the analysis of light propagation in the Einstein-Straus Swiss-cheese model. Because it does not rely on a fluid description of matter, this model is indeed particularly adapted to modelling the small-scale inhomogeneity of the Universe, and evaluating its consequences on the interpretation of the Hubble diagram. It consists of three articles, whose main results are summarised in \S~\ref{sec:summary_SC}. The first two articles, given in \S~\ref{hsec:FDU13a} and \ref{hsec:FDU13b}, were done in collaboration with H\'{e}l\`{e}ne Dupuy and Jean-Philippe Uzan; they cover theoretical calculations, cosmological interpretations, and data analysis. The third article, given in \S~\ref{hsec:F14}, contains important theoretical complements on the relation between Swiss-cheese models and the so-called Dyer-Roeder approximation.

\bigskip

\minitoc
\newpage

\section{Summary}\label{sec:summary_SC}

The Einstein-Straus Swiss-cheese\index{Einstein-Straus model}\index{Swiss-cheese model} (hereafter SC) model is constructed from a homogeneous Universe by introducing spherical `holes', with a point mass at their centres, within the FL `cheese'. Inside a hole, spacetime geometry is described by the Schwarzschild metric, or the Kottler---also called Schwarzschild-de Sitter---metric in the presence of a nonzero cosmological constant. The boundary of the hole, where the junction between both geometries is performed, is a sphere of constant comoving radius~$R\e{h}=f_K(\chi\e{h})$. For this junction to be smooth, the central mass~$M$ must be related to the hole radius via
\begin{equation}
M=\frac{4\pi}{3} \rho (a R\e{h})^3,
\end{equation}
where $\rho$ is the average cosmic matter density, and $a$ the scale factor. This result can be proven by two methods. In Art.~\ref{hsec:FDU13a}, we followed the traditional derivation based on the Darmois-Israel junction conditions~\cite{Darmois,1966NCimB..44....1I,1967NCimB..48..463I}, according to which the induced metric and the extrinsic curvature of the junction hypersurface $R=R\e{h}$ must be identical as seen from the interior or from the exterior. In Art.~\ref{hsec:F14}, I proposed a novel approach where the Kottler metric is first written in terms of free-fall coordinates, similar to the coordinates used by Lema\^itre for demonstrating the absence of singularity at the horizon of the Schwarzschild geometry~\cite{1933ASSB...53...51L,1993agns.book..353E}. In terms of those coordinates, the Kottler metric takes a form very similar to the FL metric, and the junction is then performed more naturally, because the coordinate system is actually valid both inside and outside the hole.

The analysis of the propagation of single light rays through a SC model reveals that the presence of holes only marginally affects the relation between redshift~$z$ and affine parameter~$v$. Corrections are due to a subtle mix between the Shapiro time delay caused by the central mass, and the Rees-Sciama\index{Rees-Sciama effect} due the fact that, in comoving coordinates, the gravitational potential inside the hole changes with time\footnote{Alternatively, if one uses the standard Droste coordinates with respect to which the Schwarzschild spacetime is explicitly static, then the radius of the hole grows with cosmic expansion. The gravitational potential experienced by an entering photon is thus lower than when the same photon exits from the hole. The latter thus gains a slight blueshift.}. In Art.~\ref{hsec:F14}, I have rigorously proved that the corresponding fractional correction to $1+z$ is on the order of $r\e{S}/R\e{h}\ll 1$, where $r\e{S}\define 2GM$ is the Schwarzschild radius associated with the central mass.

Regarding light \emph{beams}, I introduced in Art.~\ref{hsec:FDU13a} a technique based on the lensing Wronski matrix, in order to facilitate both the analytical and numerical treatments of the problem. I carefully rederived in Art.~\ref{hsec:F14} the earlier results by Kantowski~\cite{1969ApJ...155...89K} and Dyer \& Roeder~\cite{1974ApJ...189..167D}, and reached the same conclusions: if the masses inside the holes are opaque, with physical radius $r\e{phys}\gg r\e{S}$, then Weyl lensing is essentially negligible, while Ricci focusing is reduced by a factor $f\in[0,1]$, corresponding to the fraction of volume occupied by the FL regions, with respect to the homogeneous case. Note that this \emph{smoothness parameter}\index{smoothness parameter} is denoted~$\alpha$ in Art.~\ref{hsec:F14}. This tends to bias the distance-redshift relation towards larger distances. I then checked those results numerically. This step required to design a numerical ray-tracing code in SC models. I wrote two different versions of it: the first one, exploited in Art.~\ref{hsec:FDU13a}, has its holes arranged on a regular compact hexagonal lattice; the second one, exploited in Art.~\ref{hsec:F14}, has a random distribution of holes, where randomness was implemented by the method of Ref.~\cite{Holz:1997ic}, so that ``each ray creates its own universe''.

The cosmological consequences of these results were analysed in Arts.~\ref{hsec:FDU13a}, \ref{hsec:FDU13b} in two complementary ways, detailed below.

%
\paragraph{Fitting Mock Hubble diagrams.} By randomly throwing rays in a SC model, whose FL regions are characterised by a set of cosmological parameters~$\{\Omega\}$, I generated mock catalogues of SNe. The potential error in the interpretation of SN data caused by inhomogeneity was then quantified by fitting the associated Hubble diagrams with the FL distance-redshift relation, i.e. by wrongly assuming that the light of SNe propagated through a homogeneous Universe. The best-fit \emph{apparent} cosmological parameters~$\{\bar{\Omega}\}$ turned out to significantly differ from the input ones~$\{\Omega\}$. For instance, a SC model constructed from the Einstein-de Sitter universe, i.e. $(\Omega\e{m0},\Omega_{K0},\Omega_{\Lambda 0})=(1,0,0)$, with $f=0.26$, would be observed as $(\bar\Omega\e{m0},\bar\Omega_{K0},\bar\Omega_{\Lambda 0})=(0.5,0.8,-0.3)$, or $(0.15,0,0.85)$ if spatial curvature is forced to vanish (see Fig.~19 of Art.~\ref{hsec:FDU13a}). In other words, the light defocusing effect in SC models tends to mimic the effect of a negative spatial curvature, or a cosmological constant. The effect is however too small to explain SN observations without the need of dark energy. Importantly, the discrepancy~$\bar{\Omega}-\Omega$ between the inferred and actual cosmological parameters drastically reduces as $\Lambda$ increases. This can be understood as follows: the cosmological constant being homogeneous, if it dominates the geometry of spacetime then the difference between a SC and the underlying FL universe is not dramatic.

\paragraph{Re-analysing actual SN data.} The natural questions arising from the above are: How should we interpret SN data in order to account for the effect of small-scale inhomogeneity? What are the values of the cosmological parameters inferred in this case? To the first question, the natural answer provided by Swiss-cheese models is to use the Kantowski-Dyer-Roeder distance-redshift relation, instead of the FL one, in order to fit the Hubble diagram. Note that this option was already considered by Perlmutter et al. in Ref.~\cite{1999ApJ...517..565P}, in order to check whether inhomogeneity could be the origin of the apparent accelerating expansion. It cannot. We repeated this analysis on a more recent data set, namely the SNLS3 catalogue~\cite{2011ApJS..192....1C}, and found that the smoothness parameter~$f$ is not constrained by SN observations. However, fixing a smaller value for $f$, i.e. increasing the clumpiness of the SC, increases the inferred value of $\Omega\e{m0}$ from $0.25$ ($f=1$) to $0.3$ ($f=0$). See Fig.~25 of Art.~\ref{hsec:FDU13a}. That answers the second question. We used this effect in Art.~\ref{hsec:FDU13b} to reconcile the constraints on $\Omega\e{m0}$ obtained by SNLS3 (best-fit value of $0.2$) with the one of \textit{Planck} ($0.31$). Note however that, on the experimental side, recalibrations of the SDSS-II and SNLS lightcurves posterior to our work also managed to reduce this tension, attributed to systematics (see e.g. \S~6.6 of Ref.~\cite{2014A&A...568A..22B}). This conclusion is only partially convincing, since as emphasized in Ref.~\cite{2014PhLB..733..258B}, the calibration of SN lightcurves with the SALT2 method has a degree of model dependence which might force SN data to agree with the FL model.
%

\bigskip

From this series of works, we shall conclude that nature has somehow been kind with us by making a Universe dominated by the cosmological constant today. Indeed, if on the contrary it were dominated by matter, then the net effect of clumpiness on SN data would be larger, leading to a clear discrepancy between the cosmological parameters measured from the Hubble diagram and the ones measured from other probes, such as CMB or BAO experiments. Nevertheless, Art.~\ref{hsec:FDU13b} revealed that, in the era of precision cosmology, such effects may start to become non-negligible.


\cleardoublepage
\hiddensection{Interpretation of the Hubble diagram in a nonhomogeneous universe}
\label{hsec:FDU13a}
\includepdf[pages=1,scale=1,pagecommand={\thispagestyle{plain}}]{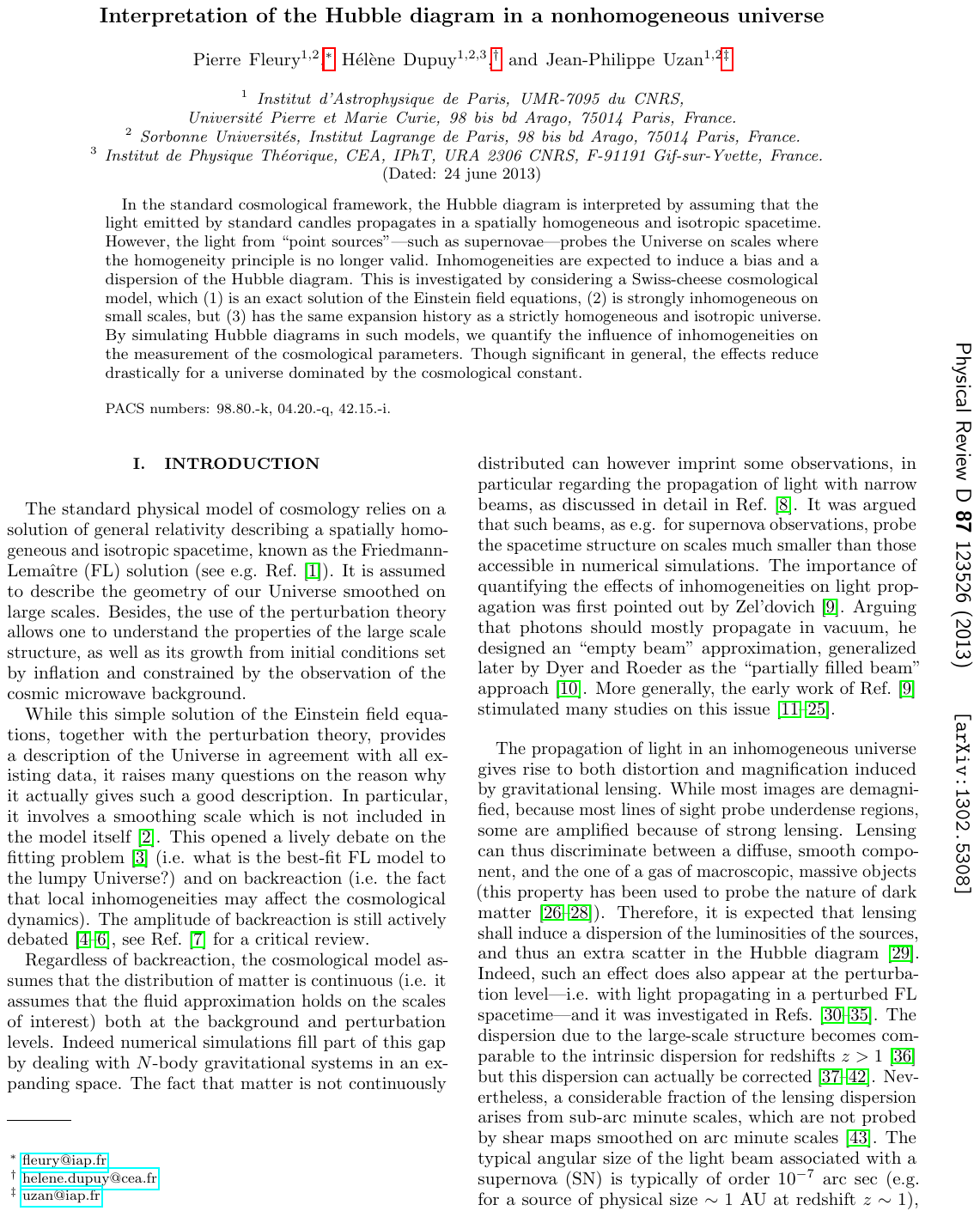} 
\includepdf[pages=2-,scale=1,pagecommand={\thispagestyle{fancy}}]{article_Hubble.pdf}


\cleardoublepage
\hiddensection{Can all cosmological observations be interpreted with a unique geometry?}
\label{hsec:FDU13b}
\includepdf[pages=1,scale=1,pagecommand={\thispagestyle{plain}}]{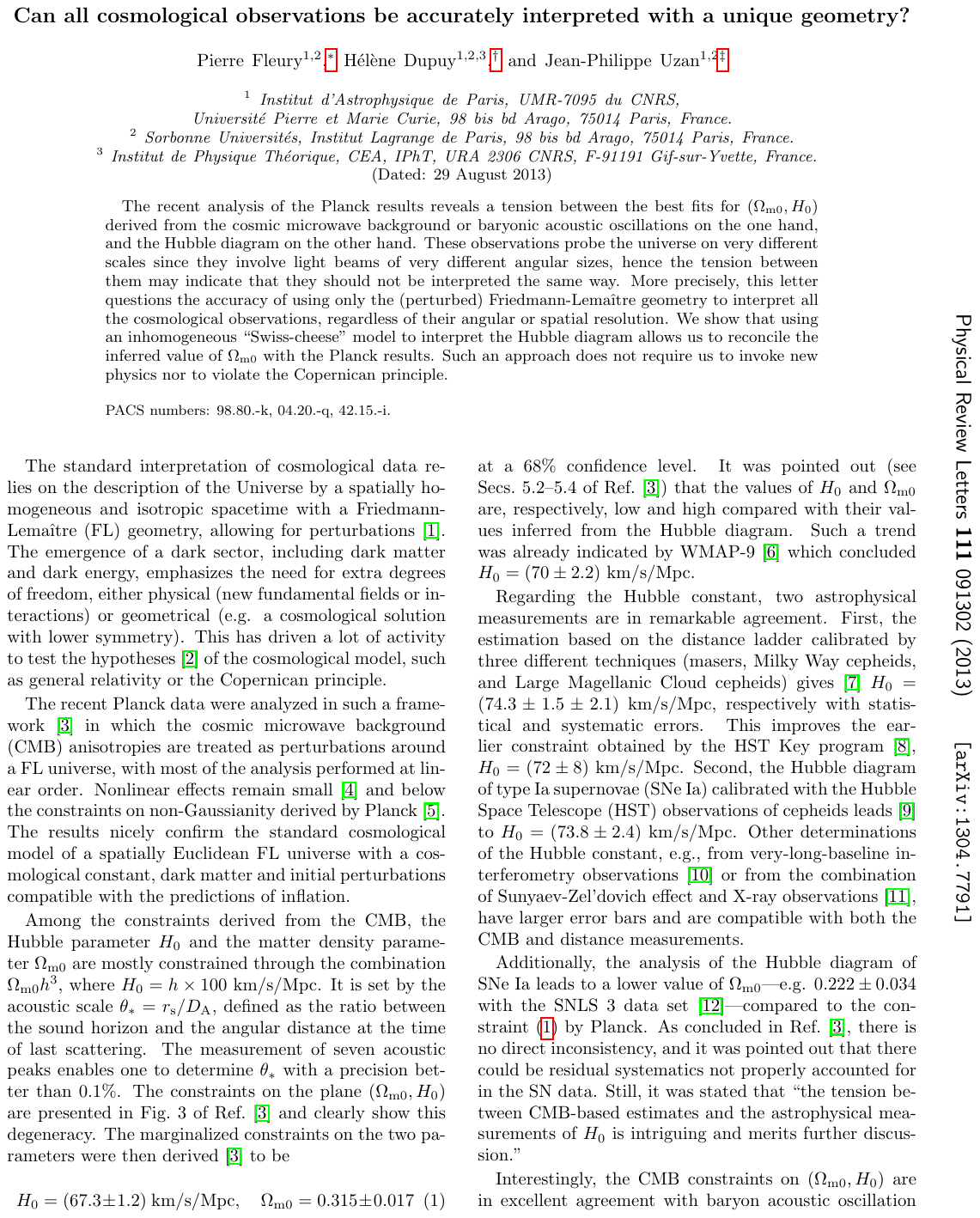}
\includepdf[pages=2-,scale=1,pagecommand={\thispagestyle{fancy}}]{article_Planck.pdf}


\cleardoublepage
\hiddensection{Swiss-cheese models and the Dyer-Roeder approximation}
\label{hsec:F14}
\includepdf[pages=1,scale=1,pagecommand={\thispagestyle{plain}}]{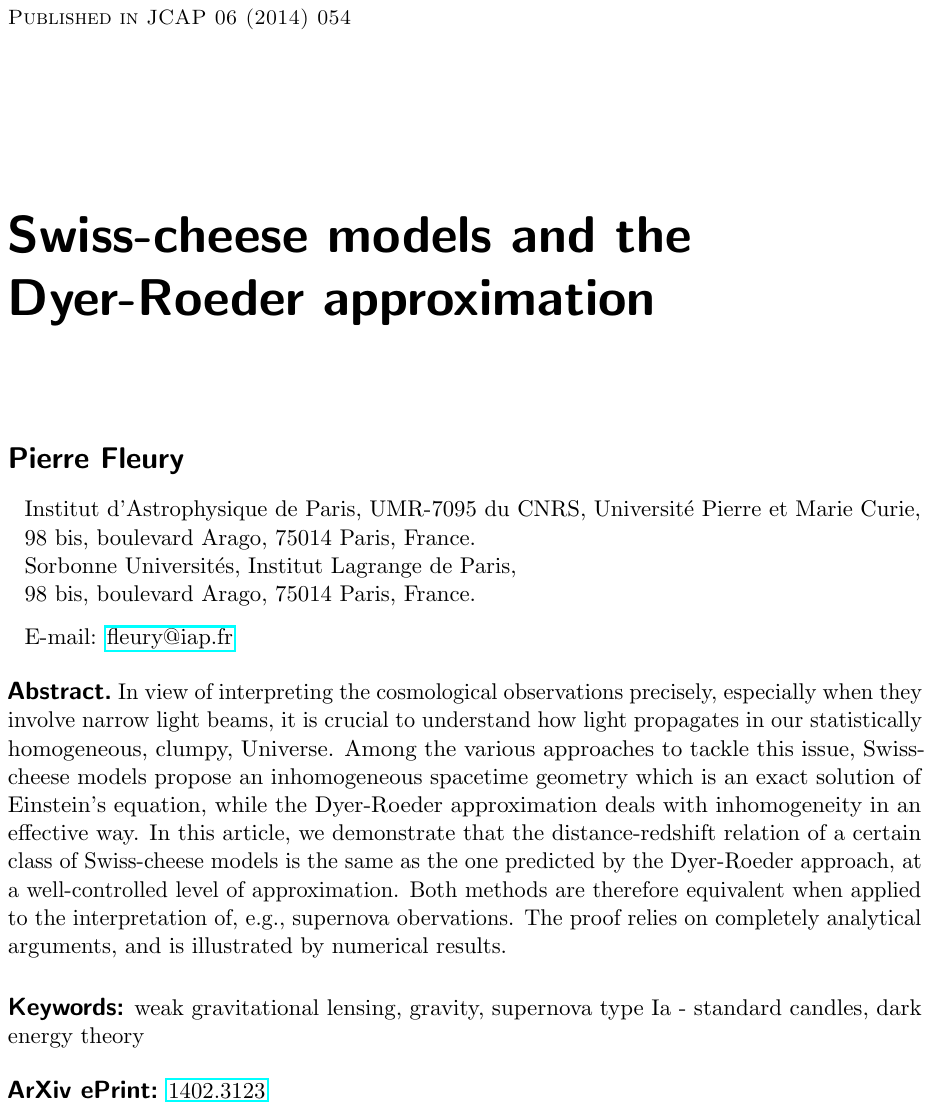}
\includepdf[pages=2-,scale=1,pagecommand={\thispagestyle{fancy}}]{article_DR.pdf}

%% file: chapter_7.tex
\lettrine{A}{lthough} Swiss-cheese models seem to capture some essential features of the small-scale inhomogeneity of the Universe, they remain toy models which suffer from a number of intrinsic limitations. First, they are unable to model at the same time small-scale and large-scale inhomogeneities, such as a cluster or a filament, \emph{with} their substructure. One has to choose between a large-scale description---using for instance LTB or Szekeres holes---or a small-scale description---using Schwarzschild holes---where clumps are then homogeneously distributed, the junction conditions preventing from any over- or underdensity (a larger hole implies a more massive clump inside). Second, even though the distance-redshift relation in a SC model is well approximated by the Kantowski-Dyer-Roeder approximation, the latter only characterizes its mean behaviour, so it tells us nothing about the dispersion, or any higher-order moment of the statistics of gravitational lensing. Determining such statistical quantities in a SC model requires to perform computationally expensive and time-consuming ray-tracing simulations.

Yet extracting this information could be very useful for constraining the cosmological parameters from SN lensing, as emphasized by Marra, Quartin, and Amendola~\cite{2013PhRvD..88f3004M,2014PhRvD..89b3009Q,2015MNRAS.449.2845A}. Those works exploited an efficient weak-lensing code by Kainulainen \& Marra~\cite{Kainulainen:2009dw,Kainulainen:2010at,2011PhRvD..84f3004K}, where inhomogeneities such as dark matter halos of filaments are randomly placed on the line of sight, according to a statistic dictated by the cosmological parameters. This chapter presents a complementary approach, which
\begin{inparaenum}[(i)]
\item focuses on smaller scales;
\item is purely analytical; and
\item does not rely on the weak-lensing approximation.
\end{inparaenum}
It consists of an article written in collaboration with Julien Larena and Jean-Philippe Uzan. Our goal was to design an efficient framework for investigating small-scale lensing, which would be at the same time more practical and flexible than model-based approaches. In particular, it is aimed at eventually being combined with large-scale cosmic lensing.

\index{stochastic lensing}
Small-scale structures are expected to manifest in the lensing equations as a very rapidly fluctuating contribution to the source terms~$\Ricfoc,\Weylfoc$. This is reminiscent of the problem of Brownian motion, e.g., for a dust particle suspended in water due to the myriad of collisions with the molecules forming the liquid. This phenomenon cannot be explained by relying on a purely fluid description of water; one usually adopts a semi-microscopic approach in which collisions are encoded in a stochastic force, mathematically modelled by a white noise. We here apply the same idea to lensing, splitting its sources into an average, slowly varying contribution and a stochastic contribution as
\begin{align}
\Ricfoc &= \ev{\Ricfoc} + \delta\Ricfoc, \\
\Weylfoc &= \ev{\Weylfoc} + \delta\Weylfoc,
\end{align}
where the $\ev{X}$ terms stand for the lensing sources due to the mean universe and the large-scale structure, while the $\delta X$ terms encode the effect of small scales. The present chapter only deals with the latter, i.e. to the \emph{diffusive} behaviour of lensing, its combination with the large-scale behaviour being left for future studies. Summarising the results obtained so far, we derived the Fokker-Planck-Kolmogorov equations governing the evolution of the probability density functions of the lensing observables, and used them to deduce general results on the mean and standard deviation of the angular distance. We then tested the validity of our formalism by applying it to Swiss-cheese models. This allowed us in particular to derive a post-Kantowski-Dyer-Roeder approximation, which turns out to be in excellent agreement with ray-tracing simulations. Regarding the dispersion of the angular distance, however, there can appear discrepancies between the predictions of our stochastic lensing formalism and the output of ray tracing. We found out that those discrepancies stemmed from the non-Gaussianity of the lensing sources, which therefore seem to constitute the main limitation of the present approach. Despite this weakness, the stochastic lensing framework opens a new window towards a precise and consistent treatment of very small scales.


\cleardoublepage
\markright{The theory of stochastic cosmological lensing}
\includepdf[pages=1,scale=1,pagecommand={\thispagestyle{plain}}]{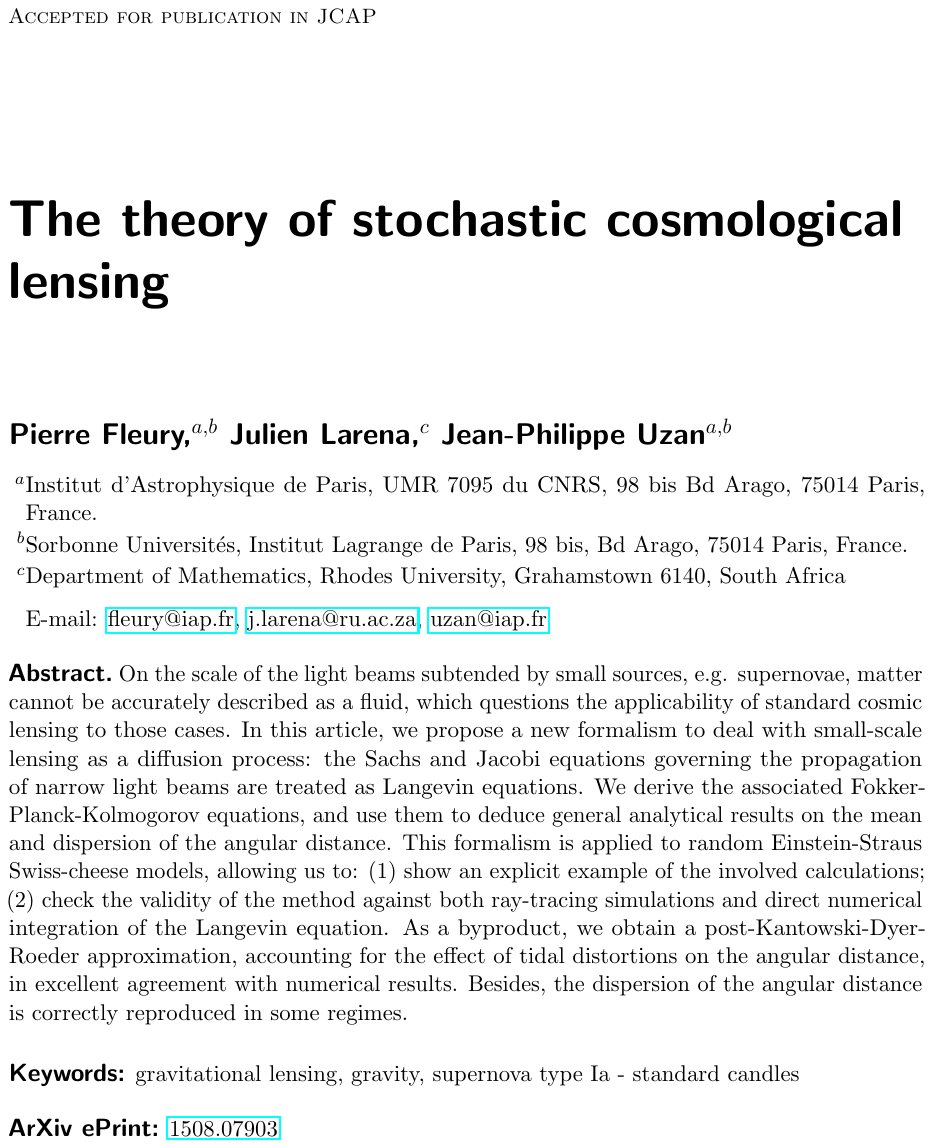}
\includepdf[pages=2-,scale=1,pagecommand={\thispagestyle{fancy}}]{article_stochastic_lensing.pdf}

%% file: introduction_part_IV.tex
\thispagestyle{plain}

\lettrine{A}{part} from homogeneity, the cosmological principle stipulates that the Universe is \emph{isotropic}. Both hypotheses are not redundant, in particular homogeneity does not imply isotropy; for example, the Universe can be the same everywhere while having an expansion faster in some directions than in others. Just like cosmic homogeneity, cosmic isotropy is therefore an assumption to be tested in order to validate or falsify the standard model. For that purpose, it is natural to start by investigating cosmological models where isotropy only is relaxed. The corresponding spacetimes follow Bianchi's classification of three-dimensional homogeneous spaces~\cite{LuigiBianchi}, and were therefore baptised the \emph{Bianchi spacetimes}\index{Bianchi!spacetimes}. For details about their construction and classification, see Ref.~\cite{1969CMaPh..12..108E}. In these models, anisotropy can show up in both the intrinsic and extrinsic curvatures of the homogeneity hypersurfaces. The simplest case is Bianchi~I, whose metric reads
\begin{equation*}
\dd s^2 = -\dd t^2 + a_x^2(t) \dd x^2 + a_y^2(t) \dd y^2 + a_z^2(t) \dd z^2,
\end{equation*}
where $t$ is a cosmic time with the same meaning as in the FL case: it is the proper time measured by fundamental observers\index{fundamental observers} following $x^i=\cst$ worldlines. In this particular example, the homogeneity hypersurfaces ($t=\cst$) are intrinsically Euclidean, hence anisotropy is only present in their extrinsic curvature, i.e. in the difference between the time evolution of the three scale factors~$a_x, a_y, a_z$. As a total, there are nine types of Bianchi models, two of which have two subcases~(VI$_0$, VI$_h$, VII$_0$, VII$_h$). Only types I, V, VII, and IX enjoy an isotropic (FL) limit~\cite{2011CQGra..28r5007P}.

Of course, since we observe a very isotropic CMB, less attention was dedicated to testing cosmic isotropy than homogeneity. However, since an anisotropy of cosmic expansion---such as in the Bianchi~I case---only sources the low multipoles of the CMB temperature map, the precision of any constraint is strongly affected by cosmic variance, which limits the ever achievable level of precision of these constraints. Subtler signatures of anisotropy can also be found in the CMB polarisation, as shown by Ref.~\cite{2007MNRAS.380.1387P} in the case of a type-VII$_h$ Bianchi model. To date no evidence of such signatures have been found~\cite{2015arXiv150201593P}, but the CMB is not the only possible probe of cosmic anisotropy. In particular, it is not particularly adapted to investigating any \emph{late-time anisotropy}, which could be sourced by anisotropic dark energy~\cite{2008JCAP...06..018K}, or emerge from backreaction mechanisms~\cite{2012PhRvD..86f3528M} or if gravity actually follows bimetric theories~\cite{2002PhRvD..66j4025D}. Late-time anisotropy is usually tested from SN data by dividing the sky in two and fitting the Hubble diagram independently in each hemisphere. This allows to look for both local anisotropy (the ``bulk flow'')~\cite{2013A&A...560A..90F,2014JCAP...03..007A,2015ApJ...801...76A} and global anisotropy~\cite{2001MNRAS.323..859K,2007A&A...474..717S,2008JCAP...06..018K,2010JCAP...10..018B,2010JCAP...12..012A,2011MNRAS.414..264C,2013A&A...553A..56K,2013PhRvD..87l3522C,2014MNRAS.444.2820S,2015ApJ...810...47J}. The very sparse sky coverage of current SN data is however a strong limitation to this approach, so that no definite conclusion could be drawn so far~\cite{Jimenez:2014jma}. A very different method has been proposed recently in Refs.~\cite{2013PhRvD..87d3003P,2015PhRvD..92b3501P,2015arXiv150301127P}, where anisotropy is traced from weak-lensing $B$-modes, i.e. curl patterns in the distribution of galaxy ellipticities across the sky.

This last part of the present dissertation is a contribution to the cosmic anisotropy issue. Chapter~\ref{chapter:optics_Bianchi_I} analyses in great details how light propagates in Bianchi~I spacetimes. Chapter~\ref{chapter:sources_anisotropy} is devoted to scalar-vector field theories, which are potential candidates as sources of anisotropy, and puts fundamental constraints on their action on the basis of stability and causality requirements.

%% file: chapter_8.tex
\lettrine{B}{esides} the quest for detecting any large-scale anisotropy in the Universe arises the question of how such a discovery would affect the way we interpret cosmological observations. In other words: how does light propagate through an anisotropic universe? Elements of answer have been provided in a seminal article by Ellis and MacCallum~\cite{1970CMaPh..19...31M}, for general Bianchi models. More specifically, a remarkably simple expression for the angular distance has been derived by Saunders~\cite{saunders_observations_1968,saunders_observations_1969} in the Bianchi~I case. The present chapter is a complement to those earlier studies. It consists of an article, written in collaboration with Cyril Pitrou and Jean-Philippe Uzan, which proposes a comprehensive analysis of geometric optics in the Bianchi~I spacetime. In particular, in addition to Saunders' result, I solved explicitly the Jacobi matrix equation, and therefore determined the impact of Weyl curvature---caused by the anisotropy of the expansion flow---not only on the angular distance, but also on the shear and rotation of images.




\cleardoublepage
\includepdf[pages={1},pagecommand={\thispagestyle{plain}}]{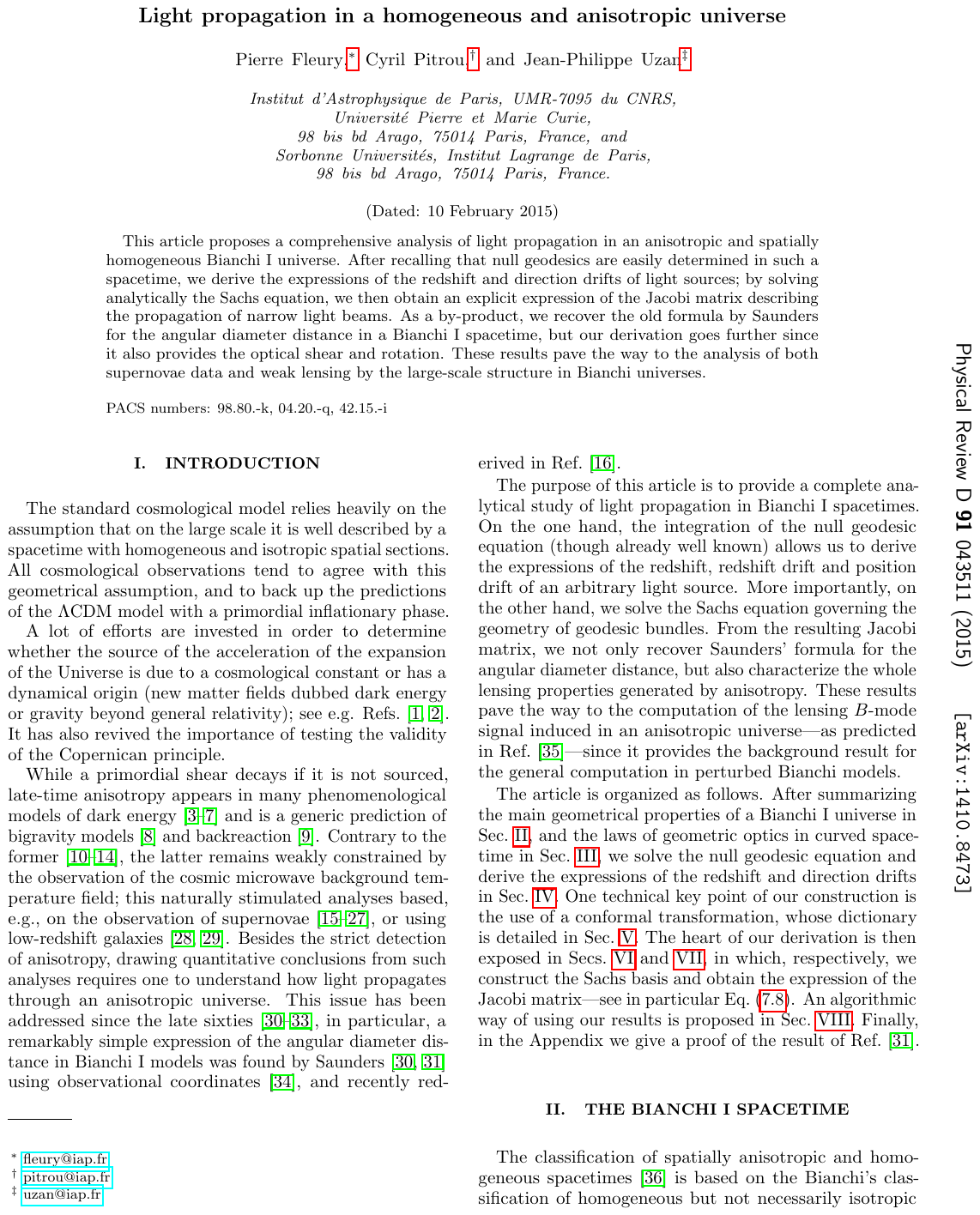}
\includepdf[pages={2-},pagecommand={\thispagestyle{fancy}}]{article_Bianchi.pdf}

%
%

%% file: chapter_9.tex
\lettrine{A}{ny} anisotropy tends to decay if it is not sourced by matter or anisotropic spatial curvature. In the previous chapter, we have seen in the Bianchi~I case that the shear rate tensor~$\sigma^i_j$ of the Hubble flow reads
\begin{equation}
(\sigma^i_j)' + 2\Hc \sigma^i_j = 8\pi G a^2 \pi^i_j.
\end{equation}
If matter has vanishing anisotropic stress ($\pi^i_j=0$), then $\sigma^i_j\propto a^{-2}$ rapidly decreases as the Universe expands; such a component is therefore necessary for anisotropy to persist. The simplest possible sources of anisotropy are vector fields, because they naturally possess a preferred direction. Let us illustrate their effect with the example of a single cosmic electromagnetic field~$\vect{A}$, described by Maxwell's theory, in a Bianchi~I universe. For homogeneity to be respected, the field must have vanishing Lie derivatives along any vector~$\vect{V}$ tangent to the homogeneity hypersurfaces, $\Lie_{\vect{V}}\vect{A}=\vect{0}$. In comoving coordinates, it implies simply $\partial_i A_\mu=0$, so that the field strength of $\vect{A}$ has no magnetic component (with respect to this coordinate system). Using the expression~\eqref{eq:stress-energy_EM_electric_magnetic} of the stress-energy tensor of the electromagnetic field, we conclude that its conformal anisotropic stress reads
\begin{equation}
\pi^i_j = \frac{1}{4\pi a^2} \pa{ \frac{E^2}{3} \, \delta^i_j - E^i E_j }
\end{equation}
with $E_i = -\partial_t A_i$, and $E^i \define \gamma^{ij} E_j$. The gravitational effect of a homogeneous electric field thus consists in decelerating cosmic expansion in the direction of $\vect{E}$, and accelerating it in the orthogonal directions.

Research on potential sources of anisotropy was recently stimulated by low-multipole anomalies in the CMB temperature map, first reported by WMAP~\cite{2004ApJ...605...14E,2005PhRvL..95g1301L,2007ApJS..170..288H} and confirmed by \textit{Planck}~\cite{2014A&A...571A..23P,2015arXiv150607135P}. The independence of both experiments suggests that those anomalies are not due to systematics, but rather are genuine physical features of the CMB. Some of them were interpreted as hints of statistical anisotropy, which could arise either from an anisotropic inflation era~\cite{2008JCAP...04..004P}, or from anisotropic dark energy~\cite{2008JCAP...06..018K}. Such properties can be obtained (among many other possibilities) by modifying the usual scalar-field paradigm with the addition of a vector field, forming the class of \emph{scalar-vector theories}\index{scalar-vector models}. The most emblematic example couples the scalar~$\phi$ with the kinetic term of Maxwell's Lagrangian as $g(\phi)F^{\mu\nu}F_{\mu\nu}$, where $g$ is a positive function. It was highlighted in Ref.~\cite{Watanabe:2009ct} as a counterexample to the so-called cosmic no-hair theorem. In a quite different context, similar models have been proposed as mechanisms for generating magnetic fields during inflation, giving a primordial origin to the large-scale intergalactic magnetic fields that we observe today~\cite{2014JCAP...10..056C}. There is however an infinity of ways to couple a scalar with a vector, leading to a huge variety of models which cannot be investigated one by one. The aim of the article presented in this chapter was to reduce this variety, by excluding the models which do not fulfil the fundamental physical requirements of \emph{stability} (Hamiltonian bounded by below) and \emph{causality} (hyperbolic equations of motion). The work reported here has been performed in collaboration with Juan Pablo Beltr\'{a}n Almeida, Cyril Pitrou, and Jean-Philippe Uzan.

Its conclusion can be summarised under the form of the following {\sf\bfseries theorem}. Consider a field theory with one scalar~$\phi$ and one vector~$\vect{A}$. If
\begin{inparaenum}[(a)] 
\item the fields are minimally coupled to spacetime geometry, and if the associated action
\item contains at most order-one derivatives of $\phi,\vect{A}$, \item is gauge invariant, and
\item is at most quadratic in $\vect{A}$,
\end{inparaenum}
then the most general form of the action leading to a stable and causal theory is
\begin{equation}
S[\phi,\vect{A};\vect{g}] = \int \dd^4 x \sqrt{-g} \pac{ -\frac12 f_0(\phi,K) 
													- \frac14 f_1(\phi) F^{\mu\nu} F_{\mu\nu}
													- \frac14 f_2(\phi) F^{\mu\nu}\tilde{F}_{\mu\nu}
												},
\end{equation}
with $K\define(\partial\phi)^2$, $\tilde{F}_{\mu\nu}\define \eps_{\mu\nu\rho\sigma} F^{\rho\sigma}/2$, and where the coupling functions $f_{0,1,2}$ obey:
\begin{itemize}
\item $\partial f_0/\partial K \geq 0$;
\item $\phi\mapsto f_0(\phi,K\geq 0)$ is bounded by below;
\item $f_1 \geq 0$; 
\item $\partial f_0/\partial K + 2 K\partial^2f_0/\partial K^2 \geq 0$.
\end{itemize}  

\cleardoublepage
\markright{On the stability and causality of scalar-vector theories}
\includepdf[pages={1},pagecommand={\thispagestyle{plain}}]{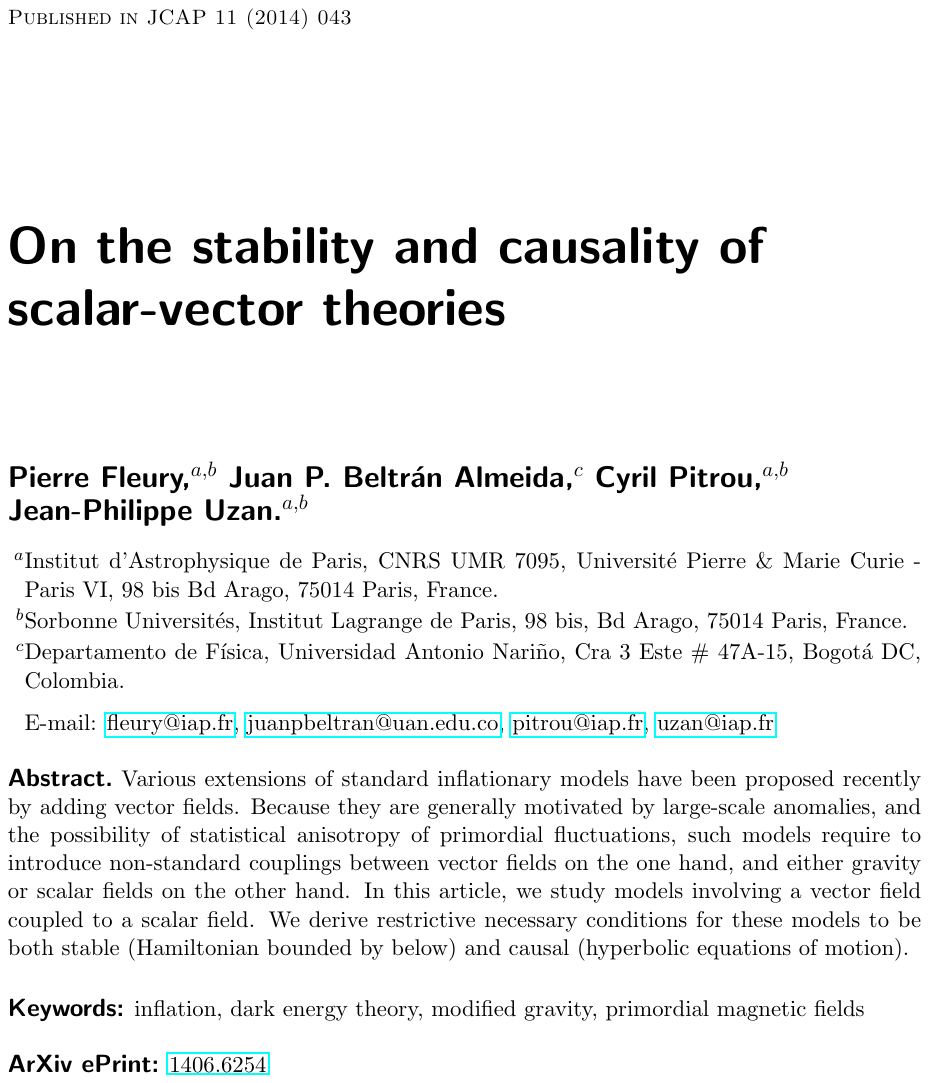}
\includepdf[pages={2-},pagecommand={\thispagestyle{fancy}}]{article_scalar-vector.pdf}

%% file: conclusion.tex
\lettrine{T}{his} thesis aimed at providing an extensive picture of light propagation in cosmology, focusing on the potential effects of the small-scale inhomogeneity and large-scale anisotropy. The first issue was motivated by general arguments, such as the Ricci-Weyl paradox, according to which observations involving very narrow beams---such as SNe---are affected by the inhomogeneity of the Universe in a very different way than observations involving much wider beams. Albeit adapted to the latter, the Friedmann-Lema\^{i}tre geometry should be unable, in principle, to provide an accurate description of the former. In fact, we have deduced from the analysis of Swiss-cheese models in Chap.~\ref{chapter:SC} that the expected effect of small-scale lensing on the interpretation of the Hubble diagram drastically reduces as the cosmological constant increases. In other words, the surprising efficiency of the standard $\Lambda$CDM model at consistently fitting all the cosmological data may be due to $\Lambda$. If this conclusion turns out to be correct, then it would represent a strong argument in favour of the cosmological constant (or any form of homogeneous dark energy) as driving the recent acceleration of cosmic expansion, and against alternative mechanisms such as backreaction. In the era of precision cosmology, however, $\Lambda$ shall no longer suffice to ensure the agreement between, e.g., SN and CMB observations. Hints of such a discrepancy are already present in current data, and we have seen that taking the clumpiness of the Universe into account is capable of reducing the resulting tension. In this context, the stochastic lensing approach developed in Chap.~\ref{chapter:SL} arises as a promising framework for dealing with small-scale lensing, in order to interpret SN observations with the accuracy that future surveys will require.

The possibility of a large-scale anisotropy in the Universe has motivations both from the theoretical and observational points of view. May such a scenario be confirmed, the questions of its physical origin and its consequences on light propagation would naturally follow. The present thesis contributed to both sides. In Chap.~\ref{chapter:optics_Bianchi_I}, I solved all the equations of geometric optics in the Bianchi~I spacetime, which provides a set of theoretical tools to contrain any late time anisotropy, from the analysis of the Hubble diagram of SNe, or from weak gravitational lensing. Besides, by studying in Chap.~\ref{chapter:sources_anisotropy} the properties of stability and causality of a large class of scalar-vector models, I reduced the landscape of physically viable theories for anisotropic dark energy or inflation.

\bigskip

As always in scientific research, this dissertation raises more questions than it provided answers. Most of the work reported here indeed calls for follow-ups, especially Part.~\ref{part:Ricci-Weyl}. Let me mention two of them. First, in Chap.~\ref{chapter:SC}, when fitting the Hubble diagram using Swiss-cheese models (or the Kantowski-Dyer-Roeder approximation), all our ignorance about the actual distribution of matter on small scales was hidden in the smoothness parameter~$f$ (or $\alpha$). This parameter is unconstrained by SN data, and was therefore arbitrarily fixed for practical uses. A more satisfactory approach would be to measure it, either directly from numerical simulations, or indirectly via independent observations. Second, although Chap.~\ref{chapter:SL} established the fundamentals of stochastic lensing, a lot remains to be done: addressing the issue of non-Gaussianity, applying it to the perturbation theory, including the effect of the large-scale structure, etc.

Finally, a few other projects marginally related to the above have called my attention during the last three years. Concerning the fundamentals of gravitational lensing, it seems to me that the range of validity of the infinitesimal beam approximation is not fully understood yet: in which situations the propagation of a light beam can really be described by the geodesic deviation equation? The answer to this question may lead to a better understanding of the transition between the weak and strong lensing regimes, which are currently dealt with using slightly different formalisms. Besides, I am intrigued by the issue of cosmic backreaction. In particular, I am surprised that most of the research activity on this subject has been, so far, dedicated to averaging inhomogeneous cosmologies, while much less was done about the physical consequence of having matter clumps decoupled from cosmic expansion. I intend to address these questions in a near future.

%% file: appendix_GR.tex
\lettrine{T}{his} dissertation extensively exploited the general theory of relativity and the language of differential geometry. The present appendix aims at gathering the main associated definitions and equations. Its title was chosen in the honour of the French mathematician Georges Darmois,\footnote{who turns out to be my great great great grandfather, in the academic sense~\cite{phdtree}} referring to his treatise on GR~\cite{Darmois}.

\bigskip

\minitoc

\newpage

\section{Differential geometry}

The general theory of relativity is naturally formulated in the language of differential geometry. Mathematically speaking, spacetime is described by a four-dimensional differential manifold~$\mathcal{M}$, i.e. a topological space which is locally homeomorphic to $\mathbb{R}^4$. A homeomorphism $\mathcal{M}\rightarrow\mathbb{R}^4$ is called a map, or coordinate system; it associates to each event $E\in\mathcal{M}$ a set of four coordinates~$\{x^\mu\}_{\mu=0\ldots 3}$.

\subsection{Vectors, forms, and tensors}

\subsubsection{Vector fields}

Consider a function~$f:\mathcal{M}\rightarrow\mathbb{R}$. The limit of $[f(x^\mu+\eps u^\mu)-f(x^\mu)]/\eps$ when $\eps$ goes to zero defines the derivative of $f$, at $x^\mu$, in the direction fixed by the four numbers~$\{u^\mu\}$. The linear map that associates to any function~$f$ such a derivative defines the vector~$\vect{u}:f\mapsto\vect{u}(f)$. The action of a vector on a product of function satisfies the \emph{Leibnitz rule}\index{Leibnitz rule}
\begin{equation}
\vect{u}(fg)=\vect{u}(f)g+f\vect{u}(g).
\end{equation}

The set of all vectors, i.e. all derivatives, at an event $E$ of $\mathcal{M}$ is called the \emph{tangent space}\index{tangent!space} of $\mathcal{M}$ at $E$; it is denoted $\mathrm{T}_E\mathcal{M}$. The set of all tangent spaces, $\mathrm{T}\mathcal{M}\define\cup_{E\in\mathcal{M}}\mathrm{T}_E\mathcal{M}$, form the \emph{tangent bundle}\index{tangent!bundle} of $\mathcal{M}$. A section of $\mathrm{T}\mathcal{M}$, i.e. a map $E\in\mathcal{M}\mapsto \vect{u}\in\mathrm{T}_E\mathcal{M}$, defines a \emph{vector field}\index{vector!field}. The set of all vector fields along~$\mathcal{M}$ is a vector space itself, denoted~$\Gamma(\mathcal{M})$.

Any coordinate system~$\{x^\mu\}$ naturally defines a basis for $\Gamma(\mathcal{M})$, namely the four partial derivatives~$\{\partial/\partial x^\mu\}$, usually denoted $\{\vect{\partial}_\mu\}$ for short. Any vector field~$\vect{u}$ is then decomposed over this basis as
\begin{equation}
\vect{u} = u^\mu \vect{\partial}_\mu.
\end{equation}
Of course, the functions $\{u^\mu\}$ depend on the basis at hand. In particular, if another coordinate system~$\{y^\alpha\}$ is chosen, then the components of $\vect{u}$ over $\{\vect{\partial}_\alpha\}$ are easily shown to be $u^\alpha=(\partial y^\alpha/\partial x^\mu) u^\mu$.

\subsubsection{Lie brackets}\index{Lie!brackets}

Any tangent space~$\mathrm{T}_E \mathcal{M}$ enjoys the structure of a Lie algebra, the Lie bracket being simply the commutator between two derivatives. It can indeed be shown that $[\vect{u},\vect{v}]\define \vect{u}\vect{v}-\vect{v}\vect{u}$ is a derivative on $\mathcal{M}$, in the sense that it satisfies the Leibnitz rule. By definition, the Lie bracket also satisfies the Jacobi identity
\begin{equation}
\forall\vect{u},\vect{v},\vect{w}\in \Gamma(\mathcal{M})
\qquad
[\vect{u},[\vect{v},\vect{w}]]
+
[\vect{v},[\vect{w},\vect{u}]]
+
[\vect{w},[\vect{u},\vect{v}]]
=
\vect{0}.
\label{eq:Jacobi_identity}
\end{equation}

Because of the Schwartz theorem, the commutator of two partial derivatives vanishes $[\vect{\partial}_\mu,\vect{\partial}_\nu]=\vect{0}$. The Lie bracket of any two vector fields~$\vect{u},\vect{v}$ thus reads
\begin{equation}
[\vect{u},\vect{v}] 
= [u^\mu\vect{\partial}_\mu,v^\nu\vect{\partial}_\nu]
= \pa{u^\mu \partial_\mu v^\nu - v^\mu\partial_\mu u^\nu}\vect{\partial}_\nu. 
\end{equation}
Now consider an arbitrary basis~$\{\vect{e}_\mu\}$ of $\Gamma(\mathcal{M})$. Because they are not necessarily associated with a coordinate system, the elements of this basis do not commute with each other in general. Their noncommutativity is quantified by \emph{structure functions}~$C\indices{^\rho_\mu_\nu}$\index{structure functions}\index{anholonomy|see{structure functions}}, according to
\begin{equation}
[\vect{e}_\mu,\vect{e}_\nu] = C\indices{^\rho_\mu_\nu} \vect{e}_\rho.
\end{equation}
If all the structure functions of a basis vanish, the basis is called \emph{holonomous}\index{holonomous basis}, and there exists a coordinate system $\{x^\mu\}$ such that $\vect{e}_\mu=\vect{\partial}_\mu$. In the opposite case, the basis is said to be anholonomous. For this reason, the structure functions are also called \emph{anholonomies}.

\subsubsection{Differential forms}\index{differential forms}\index{one-forms|see{differential forms}}

Consider a tangent space~$\mathrm{T}_E\mathcal{M}$ of the spacetime manifold. As in any vector space, one can define linear forms, i.e. linear maps $\mathrm{T}_E\mathcal{M}\rightarrow\mathcal{R}$. The set of all such linear forms defines the \emph{cotangent space}~$\mathrm{T}^*_E\mathcal{M}$\index{cotangent!space} of $\mathcal{M}$ at $E$, and the set of all cotangent space along $\mathcal{M}$ form its \emph{cotangent bundle}~$\mathrm{T}^*\mathcal{M}\define\cup_{E\in\mathcal{M}}\mathrm{T}_E\mathcal{M}$\index{cotangent!bundle}. Just like a vector field is a section of $\mathrm{T}\mathcal{M}$, a differential (one-)form is a section of $\mathrm{T}^*\mathcal{M}$, i.e. a map which associate to any event~$E\in\mathcal{M}$ a linear form $\vect{\omega}\in\mathrm{T}_E\mathcal{M}$. The set of all differential one-forms over $\mathcal{M}$, denoted $\Omega^1(\mathcal{M})$, is a vector space.

To a given coordinate system~$\{x^\mu\}$ is naturally associated a basis~$\{\vect{\dd}x^\mu\}$ of $\Omega^1(\mathcal{M})$. It is defined by the following duality relation
\begin{equation}
\vect{\dd} x^\mu (\vect{\partial}_\nu) = \delta^\mu_\nu.
\end{equation}
Any one-form~$\vect{\omega}$ is then decomposed as
\begin{equation}
\vect{\omega} = \omega_\mu \vect{\dd}x^\mu,
\qquad\text{with}\qquad
\omega_\mu\define\vect{\omega}(\vect{\partial}_\mu).
\end{equation}
Similarly to the vector case, the functions $\omega_\mu$ depend on the coordinate system that subtends the basis. If one picks another coordinate system~$\{y^\alpha\}$, then the components of~$\vect{\omega}$ over $\{\vect{\dd} y^\alpha\}$ are $\omega_\alpha=(\partial x^\mu/\partial y^\alpha)\omega_\mu$. 

\subsubsection{Tensors}\index{tensor!definition}

A means to put together vectors and forms is provided by the \emph{tensor product}~$\otimes$\index{tensor!product}. It is a bilinear and associative combination law, such as, for instance
\begin{equation}
\otimes :
\begin{aligned}
\Gamma(\mathcal{M})\times\Gamma(\mathcal{M}) &\rightarrow \Gamma(\mathcal{M})\otimes\Gamma(\mathcal{M}) \\
(\vect{u},\vect{v}) &\mapsto \vect{u}\otimes\vect{v}
\end{aligned}
\end{equation}
where $\Gamma(\mathcal{M})\otimes\Gamma(\mathcal{M})$ is a vector space. The above example defines a particular $(2,0)$-tensor. More generally, a $(m,n)$-tensor is an element of
\begin{equation}
\mathcal{T}_{m,n}(\mathcal{M})
=
\underbrace{\Gamma(\mathcal{M})\otimes\ldots\otimes\Gamma(\mathcal{M})}_{m\text{ times}}
\otimes
\underbrace{\Omega^1(\mathcal{M})\otimes\ldots\otimes\Omega^1(\mathcal{M})}_{n\text{ times}},
\end{equation}
which is the set of all linear combinations of objects of the form $\vect{u}_1\otimes\ldots\otimes\vect{u}_m\otimes\vect{\omega}_1\otimes\ldots\otimes\vect{\omega}_n$, where each $\vect{u}_i$ is a vector field and each $\vect{\omega}_i$ is a differential one-form on $\mathcal{M}$. The resulting vector space is therefore of dimension~$4^{m+n}$.

A natural basis for $\mathcal{T}_{m,n}(\mathcal{M})$ is obtained by combining the coordinate bases of $\Gamma(\mathcal{M})$ and $\Omega^1(\mathcal{M})$. Any $(m,n)$-tensor~$\vect{X}$ is then decomposed as
\begin{equation}
\vect{X} = X\indices{^{\mu_1\ldots\mu_m}_{\nu_1\ldots\nu_n}} \vect{\partial}_{\mu_1}\otimes\ldots\otimes\vect{\partial}_{\mu_m}\otimes\vect{\dd}x^{\nu_1}\otimes\ldots\otimes\vect{\dd}x^{\nu_n}.
\end{equation}
If one decides to change the coordinate system into $\{y^\alpha\}$, then the components of $\vect{X}$ over the associated basis change as
\begin{equation}
X\indices{^{\alpha_1\ldots\alpha_m}_{\beta_1\ldots\beta_n}}
=
\frac{\partial y^{\alpha_1}}{\partial x^{\mu_1}}\ldots\frac{\partial y^{\alpha_m}}{\partial x^{\mu_m}}\,
X \indices{^{\mu_1\ldots\mu_m}_{\nu_1\ldots\nu_n}}\,
\frac{\partial x^{\nu_1}}{\partial y^{\beta_1}}\ldots\frac{\partial x^{\nu_n}}{\partial y^{\beta_n}}.
\label{eq:coord_change_tensor}
\end{equation}

\subsubsection{Lie derivative}\index{Lie!derivative}

We have seen that vector fields can be considered directional derivatives of functions $\mathcal{M}\rightarrow\mathbb{R}$. The Lie derivative is an extension of this construction, allowing one to define directional derivatives of vectors, forms, and tensors. Let $\vect{u}$ be a vector field, the Lie derivative along $\vect{u}$ is denoted $\Lie_{\vect{u}}$. Its action on a function~$f$ coincides with the action of~$\vect{u}$, $\Lie_{\vect{u}}f\define \vect{u}(f)$. On a vector~$\vect{v}\in\Gamma(\mathcal{M})$, $\Lie_{\vect{u}}$ acts as
\begin{align}
\Lie_{\vect{u}} \vect{v} &\define [\vect{u},\vect{v}],
\end{align}
it is thus linear with respect to $\vect{u}$, and satisfies the Leibnitz rule
\begin{equation}
\Lie_{\vect{u}}(f\vect{v}) = (\Lie_{\vect{u}}f) \vect{v} + f(\Lie_{\vect{u}}\vect{v}).
\end{equation}
The Lie derivative of a one-form $\vect{\omega}\in\Omega^1(\mathcal{M})$ is another one-form~$\Lie_{\vect{u}}\vect{\omega}$ such that
\begin{equation}
\forall\vect{v}\in\Gamma(\mathcal{M})\qquad
\Lie_{\vect{u}} [\vect{\omega}(\vect{v})]
= (\Lie_{\vect{u}}\vect{\omega})\vect{v} + \vect{\omega}[\Lie_{\vect{u}}(\vect{v})],
\end{equation}
which is a kind of Leibnitz rule applied to the contraction of forms and vectors. Its generalisation to tensors is then achieved using the following Leibnitz rule for the tensor product,
\begin{equation}
\Lie_{\vect{u}} (\vect{X}\otimes\vect{Y}) \define (\Lie_{\vect{u}}\vect{X})\otimes\vect{Y} + \vect{X}\otimes(\Lie_{\vect{u}}\vect{Y}),
\end{equation}
for any two tensors~$\vect{X},\vect{Y}$. The above relation implies that the Lie derivative of a $(m,n)$-tensor is a $(m,n)$ tensor.

In terms of components over an arbitrary coordinate basis, we therefore have, for any vector fields~$\vect{u},\vect{v}$, one-form~$\vect{\omega}$, and $(n,m)$-tensor~$\vect{X}$,
\begin{align}
(\Lie_{\vect{u}} \vect{v})^\mu &= u^\nu \partial_\nu v^\mu
															- v^\nu \partial_\nu u^\mu \\
(\Lie_{\vect{u}} \vect{\omega})_\mu &= u^\nu \partial_\nu \omega_\mu
															+ \omega_\nu \partial_\mu u^\nu\\
(\Lie_{\vect{u}}\vect{X})\indices{^{\mu_1\ldots\mu_m}_{\nu_1\ldots\nu_n}} &= u^\rho \partial_\rho X\indices{^{\mu_1\ldots\mu_m}_{\nu_1\ldots\nu_n}} - X\indices{^{\rho\ldots\mu_m}_{\nu_1\ldots\nu_n}}\partial_\rho u^{\mu_1}	- \ldots - X\indices{^{\mu_1\ldots\rho}_{\nu_1\ldots\nu_n}}\partial_\rho u^{\mu_n} \nonumber\\
& \hspace{3.6cm} + X\indices{^{\mu_1\ldots\mu_m}_{\rho\ldots\nu_n}}\partial_{\nu_1} u^\rho + \ldots + X\indices{^{\mu_1\ldots\mu_m}_{\nu_1\ldots\rho}}\partial_{\nu_n} u^\rho.
\end{align}
Note that the parentheses on the left-hand sides of the above relations are often omitted in the literature, so that the components of e.g. $\Lie_{\vect{u}}\vect{v}$ are denoted $\Lie_{\vect{u}}v^\mu$.

\subsection{Linear connections}

\subsubsection{Covariant derivative}

As any vector fibre bundle, $\mathrm{T}\mathcal{M}$ can be equipped with a \emph{linear connection}\index{connection!definition}~$\nabla$, which allows vectors of the fibres to be transported and derived along the manifold. Here, since the fibres of $\mathrm{T}\mathcal{M}$ are nothing but the tangent spaces of $\mathcal{M}$, a connection provides a way to take directional derivatives of vectors, forms, and tensors, which differs from the Lie derivative in general.

Let $\vect{u}$ be a vector field, $\nabla_{\vect{u}}$ is called the \emph{covariant derivative}\index{covariant derivative} along $\vect{u}$ associated with the connection~$\nabla$. Its action on any function~$f$ is the same as the Lie derivative, i.e. $\nabla_{\vect{u}}f\define \vect{u}(f)$, while its effect on vectors is defined by the algebraic properties:
\begin{align}
\nabla_{\vect{u}+f\vect{v}}\vect{w} &= \nabla_{\vect{u}}\vect{w} + f\nabla_{\vect{v}}\vect{w}, \\
\nabla_{\vect{u}}(\vect{v}+\vect{w}) &= \nabla_{\vect{u}}\vect{v}+\nabla_{\vect{u}}\vect{w}, \\
\nabla_{\vect{u}}(f\vect{v}) &= \vect{u}(f) + \nabla_{\vect{u}}\vect{v}.\label{eq:connection_non_linear}
\end{align}
Note that, so far, $\nabla_{\vect{u}}$ has the same properties as $\Lie_{\vect{u}}$, but while the Lie derivative of $\vect{\partial}_\nu$ along $\vect{\partial}_\nu$ vanishes by definition, since $\Lie_{\vect{\partial_\mu}}\vect{\partial}_\nu\define[\vect{\partial}_\mu,\vect{\partial}_\nu]=0$, its covariant counterpart does not in general. In fact, a linear connection is characterised by its specific action on a basis of $\Gamma(\mathcal{M})$, as
\begin{equation}
\nabla_{\vect{\partial}_\mu}\vect{\partial}_\nu
\define\nabla_\mu \vect{\partial}_\nu 
= \Gamma\indices{^\rho_\nu_\mu}\vect{\partial}_\rho,
\end{equation}
where $\Gamma\indices{^\rho_\nu_\mu}$ are called the \emph{connection coefficients}\index{connection!coefficients}. There exist as many different connections as there are such coefficients. Hence the Lie derivative can be considered a special case of covariant derivative, associated with the trivial connection whose coefficients vanish. Note also that~$\nabla$, if considered a map $\Gamma(\mathcal{M})^2\rightarrow\Gamma(\mathcal{M})$, is not a tensor, because is is not perfectly linear with respect to its second argument, as shown by Eq.~\eqref{eq:connection_non_linear}. As a consequence, the $\Gamma\indices{^\rho_\nu_\mu}$ do not change according to Eq.~\eqref{eq:coord_change_tensor} under coordinate transformations.

The covariant derivative of any vector~$\vect{u}$ along a basis vector~$\vect{\partial}_\mu$ can be written in terms of the connection coefficients as
\begin{equation}
\nabla_\mu \vect{u} = \nabla_\mu (u^\nu\vect{\partial}_\nu)
								= (\partial_\mu u^\nu)\vect{\partial}_\nu + \Gamma\indices{^\rho_\nu_\mu} u^\nu\vect{\partial}_\rho
\end{equation}
whose components, with the short-hand notation~$\nabla_\mu u^\nu\define (\nabla_\mu\vect{u})^\nu$ thus read
\begin{equation}
\nabla_\mu u^\nu = \partial_\mu u^\nu + \Gamma\indices{^\nu_\rho_\mu} u^\rho.
\end{equation}

\subsubsection{Extension to forms and tensors}

Covariant derivatives can be extended to act on forms an tensors according to the same rules as the Lie derivatives, that is, assuming a Leibnitz-like rule with respect to both the contraction of forms with vectors, and the tensor product: for any vectors $\vect{u},\vect{v}$, one-form~$\vect{\omega}$, and tensors~$\vect{X},\vect{Y}$, we consider
\begin{align}
\nabla_{\vect{u}}\pac{\vect{\omega}(\vect{v})} &= (\nabla_{\vect{u}}\vect{\omega})(\vect{v}) + \vect{\omega}(\nabla_{\vect{u}}\vect{v}), \label{eq:covariant_derivative_form}\\
\nabla_{\vect{u}} (\vect{X}\otimes\vect{Y}) &= (\nabla_{\vect{u}}\vect{X})\otimes\vect{Y} + \vect{X}\otimes(\nabla_{\vect{u}}\vect{Y}).\label{eq:covariant_derivative_tensor_product}
\end{align}
The covariant derivative of a $(m,n)$-tensor is thus also $(m,n)$-tensor. In terms of components, Eq.~\eqref{eq:covariant_derivative_form} implies
\begin{equation}
\nabla_\mu\omega_\nu \define (\nabla_\mu\vect{\omega})_\nu
										= \partial_\mu \omega_\nu - \Gamma\indices{^\rho_\nu_\mu}\omega_\rho,
\end{equation}
and Eq.~\eqref{eq:covariant_derivative_tensor_product} leads to
\begin{align}
\nabla_\rho X\indices{^{\mu_1\ldots\mu_n}_{\nu_1\ldots\nu_m}}
&\define
(\nabla_\rho \vect{X})\indices{^{\mu_1\ldots\mu_n}_{\nu_1\ldots\nu_m}}\\
&=
\partial_\rho X\indices{^{\mu_1\ldots\mu_n}_{\nu_1\ldots\nu_m}}
+ \Gamma\indices{^{\mu_1}_{\sigma\rho}} X\indices{^{\sigma\ldots\mu_n}_{\nu_1\ldots\nu_m}}
+\ldots
+ \Gamma\indices{^{\mu_n}_{\sigma\rho}} X\indices{^{\mu_1\ldots\sigma}_{\nu_1\ldots\nu_m}}\nonumber\\
&\hspace{3.2cm}
- \Gamma\indices{^\sigma_{\nu_1\rho}} X\indices{^{\mu_1\ldots\mu_n}_{\sigma\ldots\nu_m}}
-\ldots
- \Gamma\indices{^\sigma_{\nu_m\rho}} X\indices{^{\mu_1\ldots\mu_n}_{\nu_1\ldots\sigma}}.
\end{align}

\subsubsection{Parallel transport and geodesics}

Let $\gamma$ be a curve traced on $\mathcal{M}$. We parametrise $\gamma$ with $\lambda\in[0,1]$, so that $\vect{t}\define\dd/\dd\lambda=(\dd x^\mu/\dd\lambda)\vect{\partial}_\mu$ is a \emph{tangent vector} of $\gamma$. A tensor $\vect{X}$ is said to be \emph{parallely transported}\index{parallel transport} along $\gamma$ iff its covariant derivative with respect to $\vect{t}$ vanishes all along $\gamma$,
\begin{equation}
\nabla_{\vect{t}}\vect{X} = \vect{0}.
\end{equation}
The curve~$\gamma$ is a \emph{geodesic}\index{geodesic!definition} iff its tangent vector is parallely transported along itself, $\nabla_{\vect{t}}\vect{t}=\vect{0}$. In terms of components, this requirement is equivalent to\index{geodesic!equation}
\begin{equation}
\ddf[2]{x^\rho}{\lambda} + \Gamma\indices{^\rho_\mu_\nu} \ddf{x^\mu}{\lambda}\ddf{x^\nu}{\lambda} = 0,
\end{equation}
which is called the geodesic equation.

\subsubsection{Torsion}

Any linear connection~$\nabla$ defines two fundamental tensors which characterise its geometrical properties. The first one, called \emph{torsion}\index{torsion}~$\vect{T}$, encodes the tendency of parallelograms constructed by parallely transporting vectors not to close (see Ref.~\cite{Masson} for more details). It is an antisymmetric $(1,2)$-tensor defined as
\begin{equation}
\vect{T}(\vect{u},\vect{v}) \define \nabla_{\vect{u}}\vect{v} - \nabla_{\vect{v}} \vect{u} - [\vect{u},\vect{v}].
\end{equation}
Its components over a coordinate basis are defined as $\vect{T}(\vect{\partial}_\mu,\vect{\partial}_\nu)=T\indices{^\rho_\mu_\nu}\vect{\partial}_\rho$ and read
\begin{equation}
T\indices{^\rho_\mu_\nu} = 2 \Gamma\indices{^\rho_{[\mu\nu]}}.
\end{equation}
Hence the torsion of a connection represents the antisymmetric part of its coefficients. Note that, as a consequence, torsion has no effect on geodesics.

\subsubsection{Curvature}

The second fundamental tensor associated with a connection is its \emph{curvature}\index{curvature}~$\vect{R}$, which quantifies the tendency of vectors to rotate after being parallely transported along a loop. It is a $(1,3)$-tensor defined as
\begin{equation}
\vect{R}(\vect{u},\vect{v}) \vect{w} 
\equiv \nabla_{\vect{u}} \nabla_{\vect{v}} \vect{w} 
- \nabla_{\vect{v}} \nabla_{\vect{u}} \vect{w} 
- \nabla_{[\vect{u},\vect{v}]} \vect{w} .
\label{eq:curvature_definition}
\end{equation}
Its components over a coordinate basis are defined as $\vect{R}(\vect{\partial}_\mu,\vect{\partial}_\nu)\vect{\partial}_\rho=R\indices{^\sigma_\rho_\mu_\nu}\vect{\partial}_\sigma$ and read
\begin{equation}
R\indices{^\sigma_\rho_\mu_\nu} 
=
\partial_\mu \Gamma\indices{^\sigma_\rho_\nu} - \partial_\nu \Gamma\indices{^\sigma_\rho_\mu}
+ \Gamma\indices{^\sigma_\tau_\mu} \Gamma\indices{^\tau_\rho_\nu}
- \Gamma\indices{^\sigma_\tau_\nu} \Gamma\indices{^\tau_\rho_\mu}
\label{eq:curvature_components}
\end{equation}

\subsubsection{Bianchi identity}\index{Bianchi!identity}

The Jacobi identity for the commutator has been given in Eq.~\eqref{eq:Jacobi_identity} for three vector fields, but it is actually valid for any three objects which can be `multiplied' together, including when the underlying product is noncommutative. In particular, it can be applied to the covariant derivatives along three arbitrary vector fields:
\begin{equation}
[\nabla_{\vect{u}},[\nabla_{\vect{v}},\nabla_{\vect{w}}]]
+ [\nabla_{\vect{v}},[\nabla_{\vect{w}},\nabla_{\vect{u}}]]
+ [\nabla_{\vect{w}},[\nabla_{\vect{u}},\nabla_{\vect{v}}]]
=
\vect{0}.
\end{equation}
When applied to a coordinate basis, the above equation yields
\begin{equation}
[\nabla_\mu,\vect{R}(\vect{\partial}_\nu,\vect{\partial}_\rho)]
+ [\nabla_\nu,\vect{R}(\vect{\partial}_\rho,\vect{\partial}_\sigma)]
+ [\nabla_\rho,\vect{R}(\vect{\partial}_\mu,\vect{\partial}_\nu)]
=
\vect{0},
\label{eq:Bianchi_identity}
\end{equation}
which will turn out to be useful in general relativity.

\subsection{Pseudo-Riemannian geometry}

\subsubsection{Metric}\index{metric!definition}

Let us now introduce an additional structure on~$\mathcal{M}$, namely the \emph{metric}~$\vect{g}$. It is defined as a $(0,2)$-tensor which provides a notion of scalar product between two vectors,
\begin{equation}
\vect{g}:
\begin{aligned}
\Gamma(\mathcal{M})^2 &\rightarrow \mathbb{R}\\
(\vect{u},\vect{v}) &\mapsto \vect{g}(\vect{u},\vect{v})=\vect{u}\cdot\vect{v}
\end{aligned},
\end{equation}
which is clearly symmetric. In Riemannian geometry, $\vect{g}$ is positive-definite; in pseudo-Riemannian geometry, this latter assumption is relaxed: there exist nonzero vectors~$\vect{u}$ whose scalar square~$\vect{g}(\vect{u},\vect{u})$ can be zero or even negative. The \emph{signature}\index{signature of the metric}\index{metric!signature} of the metric is defined as the signs of the eigenvalues of $[g_{\mu\nu}]$ seen as a matrix.

In GR, there is one eigendirection in spacetime whose associated eigenvalue has a sign opposite to the others, defining the direction of time. In this thesis, we have adopted the conventional signature $(-+++)$, which means that time is associated with a negative eigenvalue of the metric. A nonzero vector~$\vect{u}$ so that $\vect{g}(\vect{u},\vect{u})<0,=0,>0$ is then said to be respectively timelike, null, or spacelike.

\subsubsection{Dualities and inverse metric}

The metric provides a natural duality between vector fields and one-forms. Indeed, given a vector field~$\vect{u}$ there exists a unique one-form~$\vect{\eta}_{\vect{u}}\define\vect{g}(\vect{u},\cdot)$, which is simply ``do the scalar product with $\vect{u}$''. Because this relation is one-to-one and onto between $\Gamma(\mathcal{M})$ and $\Gamma^*(\mathcal{M})$, to any form $\vect{\omega}$ is conversely associated a vector field $\vect{e}^{\vect{\omega}}$, according to $\vect{\omega}=\vect{g}(\vect{e}^{\vect{\omega}},\cdot)$.

The component of $\vect{\eta}_{\vect{u}}$ over $\vect{\dd}x^\mu$ is by definition $\vect{\eta}_{\vect{u}}(\vect{\partial}_\mu)=g_{\mu\nu}u^\nu$. This quantity is usually denoted $u_\mu$, in order to emphasize the duality, so that $\vect{\eta}_{\vect{u}}=u_\mu\vect{\dd}x^\mu$. In terms of such notations, the metric tensor can be seen as a machine to lower indices. Now, because the duality procedure also allows us to turn forms into vectors, there should be a quantity which on the contrary raises indices, i.e. such that $(\vect{e}^{\vect{\omega}})^\mu\define\omega^\mu=g^{\mu\nu}\omega_\nu$. The coefficients~$g^{\mu\nu}$ can be expressed as functions of $g_{\mu\nu}$ by imposing that the `vectorisation' of a one-form is the inverse of the `one-formisation' of a vector; in other words
\begin{equation}
\forall\vect{u}\in\Gamma(\mathcal{M})\quad\vect{e}^{\vect{\eta}_{\vect{u}}}=\vect{u},
\qquad \text{which implies} \qquad
g_{\mu\rho} \, g^{\rho\nu} = \delta^\nu_\mu .
\end{equation}
If $g_{\mu\nu}$ and $g^{\mu\nu}$ are seen as the coefficients of two matrices, then the above means that those matrices are inverse to each other. For that reason, the $(2,0)$-tensor $g^{\mu\nu}\vect{\partial}_\mu\otimes\vect{\partial}_\nu$ is known as the \emph{inverse metric}.

It is interesting to note that this notion of duality provided by the metric is different from the duality between bases $\{\vect{\partial}_\mu\}$ and $\{\vect{\dd}x^\mu\}$ mentioned previously. One can indeed define four one-forms $\vect{\eta}_\mu\define\vect{\eta}_{\vect{\partial}_\mu}=g_{\mu\nu}\vect{\dd}x^\nu\not=\vect{\dd} x^\mu$. The set $\{\vect{\eta}^\mu\}$ is called metric-dual to $\{\vect{\partial}_\mu\}$, while $\{\vect{\dd}x^\mu\}$ is basis-dual to $\{\vect{\partial}_\mu\}$. Similarly, from the inverse metric we can generate the vector fields $\vect{e}^\mu\define\vect{e}^{\vect{\dd} x^\mu}=g^{\mu\nu}\vect{\partial}_\nu \not= \vect{\partial}_\mu$ which are metric-dual to $\{\vect{\dd}x^\mu\}$. The definition of the inverse metric implies that the sets $\{\vect{\eta}_\mu\}$ and $\{\vect{e}^\mu\}$ are basis-dual to each other,
\begin{equation}
\vect{\eta}_\mu(\vect{e}^\nu) 
= g_{\mu\rho} \vect{\dd} x^\rho \pa{g^{\nu\sigma}\vect{\partial}_\sigma}
= g_{\mu\rho} g^{\nu\sigma} \vect{\dd} x^\rho(\vect{\partial}_\sigma)
= g_{\mu\rho}g^{\nu\rho}
= \delta^\nu_\mu.
\end{equation}
This duality scheme can be summarised on the following diagram:
\begin{center}
\begin{tabular}{ccc}
$\{\vect{\partial}_\mu\}$ & $\xleftrightarrow[]{\text{basis duality}}$ & $\{\vect{\dd}x^\mu\}$ \\[3mm]
\rotatebox{90}{$\xleftrightarrow[]{\text{metric duality}}$} & & \rotatebox{90}{$\xleftrightarrow[]{\text{metric duality}}$} \\
$\{\vect{\eta}_\mu\}$ & $\xleftrightarrow[]{\text{basis duality}}$ & $\{\vect{e}^\mu\}$
\end{tabular}
\end{center}

\subsubsection{Levi-Civita connection}\index{Levi-Civita!connection}

Once a manifold is equipped with both a metric and a connection, one can impose a condition of compatibility between those two structures, that is, roughly speaking, a Leibnitz rule for covariant derivatives with respect to scalar products,
\begin{equation}
\nabla_{\vect{w}}\pac{\vect{g}(\vect{u},\vect{v})}
= \vect{g}(\nabla_{\vect{w}}\vect{u},\vect{v}) + \vect{g}(\vect{u},\nabla_{\vect{w}}\vect{v})
\end{equation}
for any three vector fields $\vect{u},\vect{v},\vect{w}$. Note that the above condition is equivalent to
\begin{equation}
\nabla_{\vect{u}}\vect{g} = \vect{0}.
\end{equation}
Thus, a connection which is compatible with the metric is also said to be \emph{metric preserving}\index{connection!metric preserving}\index{metric!-preserving connection}.

Metric-preserving connections have the interesting property that they are entirely and uniquely determined by their torsion~\cite{2013tegr.book.....A}. A particular case is the one of zero torsion, which defines the \emph{Levi-Civita connection}. It is the connection of GR. Its coefficients are known as the \emph{Christoffel symbols}\index{Christoffel symbols!definition}, and can be expressed as functions of the metric as
\begin{equation}
\Gamma\indices{^\rho_\mu_\nu} \define g^{\rho\sigma} \Gamma_{\sigma\mu\nu},
\qquad \text{with} \qquad
\Gamma_{\sigma\mu\nu}
\define
\frac{1}{2} \pa{ \partial_\mu g_{\nu\sigma} + \partial_\nu g_{\mu\sigma} - \partial_\sigma g_{\mu\nu}}.
\label{eq:Christoffel_symbols}
\end{equation}

\subsubsection{Riemann tensor}\index{Riemann tensor}

The curvature of the Levi-Civita connection is called the \emph{Riemann tensor}. Its components can be expressed in terms of the metric as
\begin{align}
R_{\sigma\rho\mu\nu}
&\define \vect{g}\pac{\vect{\partial}_\sigma,\vect{R}(\vect{\partial}_\mu,\vect{\partial}_\nu)\vect{\partial}_\rho} \\
&= \vect{g}\pac{\vect{\partial}_\sigma,2\nabla_{[\mu}\nabla_{\nu]}\vect{\partial}_\rho}
	\qquad\text{by definition~\eqref{eq:curvature_definition} of $\vect{R}$}\\
&= 2\vect{\partial}_{[\mu} \vect{g}(\vect{\partial}_\sigma,\nabla_{\nu]}\vect{\partial}_\rho)
		-2\vect{g}(\nabla_{[\mu}\vect{\partial}_\sigma,\nabla_{\nu]}\vect{\partial}_\rho)
		\qquad\text{by metric preservation}\\
&= 2\Gamma_{\sigma\rho[\nu,\mu]} - 2\Gamma_{\tau\sigma[\mu}\Gamma\indices{^\tau_\nu_]_\rho}
	\qquad\text{with }\Gamma_{\sigma\rho\nu,\mu}\define\partial_\mu\Gamma_{\sigma\rho\nu}\\
&= \frac{1}{2} \pa{ g_{\sigma\nu,\mu\rho} - g_{\rho\nu,\mu\sigma} }
- \frac{1}{4} g^{\tau\omega} \pa{ g_{\tau\sigma,\mu} + g_{\tau\mu,\sigma} - g_{\sigma\mu,\tau} }
												\pa{ g_{\omega\nu,\rho} + g_{\omega\rho,\nu} - g_{\nu\rho,\omega} }\nonumber \\
&\qquad - (\mu\leftrightarrow\nu) \qquad \text{from Eq.~\eqref{eq:Christoffel_symbols}.}
\end{align}
The components of the Riemann tensor enjoy a number of symmetries summarised below:
\begin{align}
R_{\mu\nu\rho\sigma} &= -R_{\mu\nu\sigma\rho}\label{eq:Riemann_antisymmetry_1} \\
R_{\mu\nu\rho\sigma} &= -R_{\nu\mu\rho\sigma},\label{eq:Riemann_antisymmetry_2}\\
R_{\mu[\nu\rho\sigma]} &= 0, \label{eq:Riemann_Bianchi_1}\\
R_{\mu\nu\rho\sigma} &= R_{\rho\sigma\mu\nu}, \label{eq:Riemann_pair_symmetry}\\
R_{\mu\nu[\rho\sigma;\tau]} &=0 \label{eq:Riemann_Bianchi_2},
\end{align}
where $R_{\mu\nu\rho\sigma;\tau}\define\nabla_\tau R_{\mu\nu\rho\sigma}$.
While the antisymmetry of the second pair of indices~\eqref{eq:Riemann_antisymmetry_1} is a direct consequence of the definition of curvature, the antisymmetry the first pair of indices~\eqref{eq:Riemann_antisymmetry_2} is due to the fact that the underlying connection~$\nabla$ is both metric compatible and torsion free. Equation~\eqref{eq:Riemann_Bianchi_1}, sometimes called first Bianchi identity\index{Bianchi!first identity}, is also due to the fact that $\nabla$ is torsion free. The symmetry~\eqref{eq:Riemann_pair_symmetry} under exchange of the two pairs of indices is a consequence of Eqs.~\eqref{eq:Riemann_antisymmetry_2}, \eqref{eq:Riemann_Bianchi_1}. Those symmetries imply that only $20$ among the $256$ components of the Riemann tensor are independent. Finally, Eq.~\eqref{eq:Riemann_Bianchi_2}, sometimes called second Bianchi identity, is a consequence of \eqref{eq:Bianchi_identity}\index{Bianchi!second identity} and, again, of the fact that $\nabla$ is torsion free.

\subsubsection{Ricci and Einstein tensors}

The \emph{Ricci tensor}\index{Ricci!tensor} is obtained by contracting the first and third indices of the components of the Riemann tensor,
\begin{equation}
\textbf{Ric}(\vect{\partial}_\mu,\vect{\partial}_\nu) \define R_{\mu\nu} \define R\indices{^\rho_\mu_\rho_\nu}.
\end{equation}
Note that the Ricci tensor does not require the metric structure to be defined, it therefore exists also in non-Riemannian geometry. In (pseudo-)Riemannian geometry the Ricci tensor is symmetric, due to Eq.~\eqref{eq:Riemann_pair_symmetry}. Moreover, the second Bianchi identity~\eqref{eq:Riemann_Bianchi_2} implies the interesting relation
\begin{equation}
\nabla_\mu R\indices{^\mu_\nu_\rho_\sigma} = 2 \nabla_{[\rho} R_{\sigma]\nu}.
\end{equation}
By contracting indices $\nu$ and $\sigma$ in the above, we obtain
\begin{equation}
\nabla_\nu R^\nu_\mu = \frac{1}{2}\,\nabla_\mu R,\label{eq:divergence_Ricci}
\end{equation}
where $R\define R^\mu_\mu$ is called the \emph{Ricci scalar}\index{Ricci!scalar}.

If we define the \emph{Einstein tensor}~$\vect{E}$\index{Einstein!tensor} as
\begin{equation}
\vect{E}(\vect{\partial}_\mu,\vect{\partial}_\nu) \define E_{\mu\nu} \define R_{\mu\nu} - \frac{1}{2} R \, g_{\mu\nu},
\end{equation}
then, by virtue of Eq.~\eqref{eq:divergence_Ricci} this tensor is divergence free, $\nabla_\nu E^\nu_\mu=0$.

\section{Gravitation}

In the general theory of relativity, gravitation is encoded in the pseudo-Riemannian geometry of spacetime, equipped with the Levi-Civita connection. As a field theory, the fundamental quantity is therefore the metric~$\vect{g}$ of the spacetime manifold\footnote{In the Palatini formulation of general relativity, the metric and the connection are considered independent dynamical quantities: $\nabla$ is not taken to be the Levi-Civita connection right from the beginning. However, this property emerges from the action of general relativity without the need of any modification~\cite{Baez:1995sj}.}.

\subsection{Geometrodynamics}

\subsubsection{Einstein field equations}

The metric of spacetime is affected by the presence of matter via its stress-energy tensor~$\vect{T}$, according to the \emph{Einstein field equation}\index{Einstein!field equation}
\begin{empheq}[box=\fbox]{equation}
\vect{E} + \Lambda\,\vect{g} = 8\pi G\,\vect{T},
\label{eq:Einstein_equation}
\end{empheq}
where $\Lambda$ is the cosmological constant and $G$ is Newton's constant. In other words, any form of energy or momentum locally generates Ricci curvature. Because the Einstein tensor~$\vect{E}$ is divergence free, the Einstein equation imposes the conservation of energy-momentum
\begin{equation}
\nabla_\nu T^\nu_\mu = 0,
\end{equation}
just like the Maxwell equation imposes the conservation of electric charge.

\subsubsection{Action formulation}

The Einstein field equation~\eqref{eq:Einstein_equation} can be derived from an action principle, with $S=S\e{m}+S\e{EH}+S_\Lambda$. The first term~$S\e{m}$ denotes the action of matter fields, from which derives their stress-energy tensor as\index{stress-energy tensor!definition}
\begin{equation}
T^{\mu\nu} \define \frac{2}{\sqrt{-g}} \frac{\delta S\e{m}}{\delta g_{\mu\nu}} ,
\end{equation}
where $\delta/\delta g_{\mu\nu}$ denotes the standard functional derivative, here with respect to $g_{\mu\nu}$, while $g$ is the metric determinant defined in Eq.~\eqref{eq:metric_determinant}. In other words, $\vect{T}$ is such that, for a small variation~$\delta \vect{g}$ of the metric,
\begin{equation}
S\e{m}[\vect{g}+\delta\vect{g}] - S\e{m}[\vect{g}]
=
\frac{1}{2} \int \dd^4 x\sqrt{-g} \; \delta g_{\mu\nu}  T^{\mu\nu} + \mathcal{O}(|\delta\vect{g}|^2).
\label{eq:stress-energy_tensor_definition}
\end{equation}
Example of such matter actions will be given in \S~\ref{sec:matter}. Note that, since $g^{\mu\rho}g_{\rho\nu}=\delta^\mu_\nu$, the variations of the metric and its inverse are related by $\delta g^{\mu\nu}=-g^{\mu\rho}g^{\nu\sigma}\delta g_{\rho\sigma}$, so that the lowered components~$T_{\mu\nu}$ are given by
\begin{equation}
T_{\mu\nu} = \frac{-2}{\sqrt{-g}} \frac{\delta S\e{m}}{\delta g^{\mu\nu}},
\end{equation}
which is very similar to Eq.~\eqref{eq:stress-energy_tensor_definition}, \emph{except for the minus sign}.

The other two terms of $S$ are respectively the Einstein-Hilbert action\index{Einstein!-Hilbert action}~$S\e{EH}$ and a cosmological constant term, they read
\begin{align}
S\e{EH} &\define \frac{1}{16\pi G}\int \dd^4 x\sqrt{-g} \; R  + \text{boundary term}, \\
S_\Lambda &\define \frac{-1}{8\pi G} \int\dd^4 x\sqrt{-g} \, \Lambda,
\end{align}
where the expression and origin of the boundary term can be found in Ref.~\cite{2004rtmb.book.....P}. It can indeed be shown that
\begin{align}
\frac{1}{\sqrt{-g}}\frac{\delta S\e{EH}}{\delta g^{\mu\nu}} 
&= \frac{1}{16\pi G} \, E_{\mu\nu}, \\
\frac{1}{\sqrt{-g}}\frac{\delta S_\Lambda}{\delta g^{\mu\nu}} &= \frac{\Lambda}{16\pi G} \, g_{\mu\nu}.
\end{align}
The variation of the cosmological constant term is easily performed: thanks to the formula relating the derivative of the determinant of a matrix to its trace we obtain
\begin{equation}
\delta\sqrt{-g} = \frac{1}{2} \, \sqrt{-g} \, g^{\mu\nu} \delta g_{\mu\nu} = -\frac{1}{2} \, \sqrt{-g} \, g_{\mu\nu} \delta g^{\mu\nu}.
\end{equation}
The variation of the Einstein-Hilbert term is more involved, see e.g. Ref.~\cite{2004rtmb.book.....P} for a clear and detailed derivation. In the end, the extremalisation of the complete action~$S$ leads to
\begin{equation}
0 = \frac{\delta S}{\delta g^{\mu\nu}} = \frac{\sqrt{-g}}{16\pi G}\pa{ E_{\mu\nu} + \Lambda\,g_{\mu\nu} - 8\pi G\,T_{\mu\nu} },
\end{equation}
which is the Einstein field equation.

\subsection{Matter}\label{sec:matter}
\index{matter!in general relativity}

\subsubsection{Point particle}\index{point particle}

Consider a spinless point particle with rest mass~$m$. Its movement within the spacetime manifold is a curve~$x\e{p}^\mu(\lambda)$, where $\lambda$ is an arbitrary parameter. The proper time~$\tau$ of this particle is defined as the time measured in its rest frame, i.e. such that $\vect{u}\e{p}=\dd/\dd\tau$, where $\vect{u}\e{p}$ is the four-velocity of the particle, or $\dd\tau^2=-g_{\mu\nu} \dd x\e{p}^\mu \dd x\e{p}^\nu$. In absence of any external force, the action of this particle is\index{action!of a point particle}
\begin{equation}
S\e{p} \define -m\int \dd\tau =  -m \int \dd\lambda \; \sqrt{- g_{\mu\nu} \ddf{x^\mu\e{p}}{\lambda} \ddf{x^\nu\e{p}}{\lambda} }.
\label{eq:action_point_particle}
\end{equation}
If $S\e{p}$ is considered a functional of the trajectory~$x\e{p}^\mu(\lambda)$, then it is straightforward to check that the stationarity of $S\e{p}$ is equivalent to the geodesic equation,
\begin{equation}
\frac{\delta S\e{p}}{\delta x\e{p}^\mu} = 0
\Longleftrightarrow
\ddf[2]{x\e{p}^\rho}{\tau} + \Gamma\indices{^\rho_\mu_\nu} \ddf{x\e{p}^\mu}{\tau} \ddf{x\e{p}^\nu}{\tau} = 0.
\end{equation}
Free-falling particles therefore follow geodesics of the spacetime manifold.

When written as in Eq.~\eqref{eq:action_point_particle}, $S\e{p}$ cannot be used to derive the expression of the stress-energy tensor associated with the point particle, because it is not an integral over a four-dimensional region of spacetime. This issue can be fixed by introducing a Dirac distribution~$\delta\e{D}$ as
\begin{equation}
S\e{p} = -m \int \dd^4x \int \dd\lambda \; \delta^{(4)}[x^\rho - x\e{p}^\rho(\lambda)] 
					\sqrt{- g_{\mu\nu} \ddf{x^\mu\e{p}}{\lambda} \ddf{x^\nu\e{p}}{\lambda} }.
\end{equation}
Applying the definition~\eqref{eq:stress-energy_tensor_definition} of the stress-energy tensor then yields\index{stress-energy tensor!of a point particle}
\begin{align}
T^{\mu\nu}\e{p} &= m \int \dd\tau \; \frac{\delta\e{D}[x^\rho-x\e{p}^\rho(\tau)]}{\sqrt{-g}}\,
															\ddf{x\e{p}^\mu}{\tau} \, \ddf{x\e{p}^\nu}{\tau} \\
							&= \frac{\delta\e{D}[x^i-x\e{p}^i(t)]}{\sqrt{-g}} \, \gamma m \, \ddf{x\e{p}^\mu}{t}\,\ddf{x\e{p}^\nu}{t},
							\label{eq:stress-energy_tensor_particle}
\end{align}
where, in the second expression, we have used the (arbitrary) time coordinate~$t=x^0$ as parameter~$\lambda$ of the particle's trajectory, and introduced the Lorentz factor $\gamma\define\dd t/\dd\tau$.

\subsubsection{Perfect fluid}\index{perfect fluid}

Consider now an ensemble of noninteracting point particles. Because they do not interact, the action of the system is just the sum of the actions of the individual particles, so the resulting stress-energy tensor is
\begin{equation}
\vect{T} \define \sum_{\text{p}} \vect{T}\e{p} .
\end{equation}
Let~$\mathcal{D}$ be a domain of $\mathcal{M}$ centred around an event~$E$, and whose dimensions are small compared to the typical spacetime curvature radius. Within $\mathcal{D}$, all vectors can be considered to approximately belong to the same tangent space~$\mathrm{T}_E\mathcal{M}$, in particular the four-momenta~$\vect{p}\e{p}\define m\e{p}\vect{u}\e{p}$ of the particles in $\mathcal{D}$, which implies that they can be summed. Let~$\bar{\vect{p}}=M \bar{\vect{u}}$ be their ensemble average, where $M$ is such that $\bar{u}^\mu \bar{u}_\mu=-1$. The four-velocity $\bar{\vect{u}}$ defines a preferred frame, the particles' barycentric frame, with respect to which we can pick Fermi normal coordinates. In this coordinate system, the total stress-energy tensor reads
\begin{align}
T^{00} &\coordequal \sum_{\text{p}\in\mathcal{D}} \delta\e{D}[x^k-x\e{p}^k(t)] \, \gamma\e{p} m\e{p},\\
T^{0i} &\coordequal \sum_{\text{p}\in\mathcal{D}} \delta\e{D}[x^k-x\e{p}^k(t)] \, \gamma\e{p} m\e{p} v^i\e{p},\\
T^{ij} &\coordequal \sum_{\text{p}\in\mathcal{D}} \delta\e{D}[x^k-x\e{p}^k(t)] \, \gamma\e{p} m\e{p} v\e{p}^i v\e{p}^j,
\end{align}
where $v^i\e{p}\define \dd x^i\e{p}/\dd\bar{\tau}$ is the velocity of particle p in the barycentric frame.

Now if the domain $\mathcal{D}$ contains a large number of particles, and if we coarse-grain $\vect{T}$ on the scale of $\mathcal{D}$, then
\begin{equation}
T^{00}\coordequal \rho,
\qquad
T^{0i}\coordequal 0,
\qquad
T^{ij}\coordequal p \, \delta^{ij},
\end{equation}
where $\rho$ and $p$ respectively define the energy density and the kinetic pressure of the fluid,
\begin{equation}
\rho \define \frac{1}{V_{\mathcal{D}}} \sum_{\text{p}} \gamma\e{p} m\e{p},
\qquad
p \define \frac{1}{V_{\mathcal{D}}} \sum_{\text{p}} \frac{1}{3} \gamma\e{p} m\e{p} \, \delta_{ij} v^i\e{p} v^j\e{p}.
\end{equation}
When written in a fully covariant form, the stress-energy tensor of the system of particles therefore reads\index{stress-energy tensor!of a perfect fluid}
\begin{equation}
T^{\mu\nu} = (\rho + p) \, \bar{u}^\mu \bar{u}^\nu + p\,g^{\mu\nu}.
\end{equation}

\subsubsection{Scalar field}\index{action!of a scalar field}\index{scalar!field}

The case of matter fields is more easily handled, because their action directly takes the form of the integral of a Lagrangian density over spacetime. In the case of a scalar field~$\phi$, minimally coupled to spacetime geometry, the standard action is
\begin{equation}
S = \int \dd^4 x \; \sqrt{-g} \pac{ -\frac12 \partial_\mu \phi \, \partial^\mu \phi - V(\phi) },
\end{equation}
composed of a kinetic term~$(\partial\phi)^2$ and a potential term~$V(\phi)$. The associated stress-energy tensor is then easily shown to be\index{stress-energy tensor!of a scalar field}
\begin{equation}
T_{\mu\nu} = \partial_\mu \phi \, \partial_\nu \phi - \frac12 (\partial^\rho\phi \,\partial_\rho\phi)\,g_{\mu\nu} - V(\phi) \,g_{\mu\nu}.
\end{equation}

\subsubsection{Vector field}\index{action!of a vector field}

The standard action of a minimally coupled vector field~$\vect{A}$ reads
\begin{equation}
S = \int_\mathcal{M} \dd^4 x \sqrt{-g} \pac{-\frac{1}{4} F^{\mu\nu} F_{\mu\nu} - V(A^2)},
\end{equation}
where $F_{\mu\nu}\define \partial_\mu A_\nu - \partial_\nu A_\mu$ is the field strength of $\vect{A}$, and $A^2\define A^\mu A_\mu$. The associated stress-energy tensor is\index{stress-energy tensor!of a vector field}
\begin{equation}
T_{\mu\nu} =  F_{\mu\rho}F\indices{_\nu^\rho} - \frac{1}{4} (F^{\rho\sigma}F_{\rho\sigma})\,g_{\mu\nu} - V(A^2)\,g_{\mu\nu} .
\end{equation}

%% file: compte_rendu_francais.tex
\minitoc

\newpage

\section*{Introduction}
\phantomsection
\addcontentsline{toc}{section}{Introduction}

Depuis sa naissance, il y a près d'un siècle, à aujourd'hui, le domaine de la cosmologie physique a levé le voile sur un grand nombre de questions fondamentales quant à la nature et l'origine de l'Univers. La fin du vingtième siècle, en particulier, a vu cette discipline muer en une science de haute précision, grâce à des expériences d'une remarquable qualité, qu'il s'agisse de l'observation du fond diffus cosmologique, des supernov\ae, des grands relevés de galaxies, ou encore du lentillage gravitationnel.

La précision atteinte en cosmologie observationnelle contraste toutefois avec la remarquable simplicité du cadre théorique au sein duquel les observations sont interprétées. En particulier, dans le modèle cosmologique standard, la relation entre le décalage spectral~$z$ de sources lumineuses lointaines, dû à leur récession, et leur distance angulaire~$D\e{A}$, relation indispensable à l'analyse de quasiment toutes les observations cosmologiques, est systématiquement calculée \emph{en supposant que la lumière se propage à travers un univers parfaitement homogène et isotrope}. Quoique plausible pour de très larges faisceaux lumineux, cette hypothèse paraît néanmoins très grossière à petite échelle. Or les échelles mises en jeu dans les observations cosmologiques actuelles sont d'une extrême variété (voir table~\ref{tab:echelles_observations}), s'étalant sur 12 ordres de grandeur. Malgré cela, toutes les observations s'avèrent être cohérentes les unes avec les autres lorsqu'interprétées dans le cadre du modèle standard. L'objectif principal de cette thèse a été de comprendre les raisons d'un succès si surprenant.

\begin{table}[h!]
\centering
\begin{tabular}{|c|cc|}
\hline 
\rowcolor{lightgray} {\sf\bfseries observation} & {\sf\bfseries échelle angulaire pertinente} & {\sf\bfseries valeur typique (rad)} \\ 
\hline 
OAB & échelle OAB à $z\sim 0.5,2$ & $10^{-1},10^{-2}$ \\ 
FDC & échelle OAB à $z\sim 1000$ & $10^{-2}$ \\ 
$f\e{gaz}$ & taille apparente d'un amas $z\sim 0.5$ & $10^{-3}$ \\ 
LF & rayon d'Einstein à des distances cosmologiques & $10^{-4}$ \\ 
Lf & taille apparente d'une galaxie à $z\sim 0.5$ & $10^{-5}$ \\ 
SNeIa & taille apparente d'une supernova à $z\sim 0.5 $ & $10^{-13}$ \\ 
\hline 
\end{tabular} 
\caption{Ouvertures angulaires typiques des faisceaux lumineux impliqués dans différentes observations cosmologiques : oscillation acoustique de baryons (OAB) observée dans les grands relevés de galaxies, ou dans les anisotropies du fond diffus cosmologique (FDC) ; fraction de gaz dans les amas de galaxies ($f\e{gaz}$) ; lentillage gravitationnel fort (LF) ou faible (Lf) ; et enfin supernov\ae\xspace de type Ia (SNeIa).}
\label{tab:echelles_observations}
\end{table}

\section{Optique géométrique en espace-temps courbe}

Modéliser la propagation de la lumière à travers le cosmos requiert une compréhension profonde des lois de l'optique géométrique en présence de gravitation, c'est-à-dire en espace-temps courbe.

\subsection{Rayons lumineux}

Les lois de l'électrodynamique classique montrent que, dans le régime eikonal, les ondes électromagnétiques constituant la lumière se propagent en suivant des géodésiques de genre lumière à travers l'espace-temps. Le vecteur tangent à une telle courbe, $k^\mu \define \dd x^\mu/\dd v$, où $v$ est un paramètre affine, n'est autre que le quadrivecteur d'onde ; par définition, il satisfait aux équations
\begin{equation}
k^\mu \nabla_\mu k^\nu = 0, \qquad k^\mu k_\mu = 0.
\label{eq:equation_geodesique}
\end{equation}
Le décalage spectral~$z$ d'une source lumineuse est défini comme étant la différence relative entre la fréquence~$\omega_S$ émise par la source et celle reçue par l'observateur, $\omega_O$, selon
\begin{equation}
z \define \frac{\omega_S-\omega_O}{\omega_O}.
\end{equation}
Cette quantité constitue une mesure de la vitesse de récession de la source (effet Doppler) ainsi que des effets gravitationnels de dilatation du temps (effet Einstein). Soit un référentiel matérialisé par la quadrivitesse~$\vect{u}$, la fréquence d'un signal de quadrivecteur d'onde~$\vect{k}$ est la composante temporelle de celui-ci, c'est-à-dire $\omega=u^\mu k_\mu$. Par conséquent $1+z=(u^\mu k_\mu)_S/(u^\mu k_\mu)_O$. Le décalage spectral d'une source est donc obtenu en résolvant~\eqref{eq:equation_geodesique}.

\subsection{Distances en cosmologie}

En cosmologie, la distance d'une source lumineuse peut être mesurée de deux façons différentes : la première consiste en la comparaison de l'aire de cette source~$A_S$ avec sa taille angulaire apparente $\Omega_O$. La quantité
\begin{equation}
D\e{A} \define \sqrt{\frac{A_S}{\Omega_S}}
\end{equation}
définit alors une notion de distance, connue sous le nom de distance angulaire. Il s'agit de la notion de distance naturellement exploitée dans l'interprétation du FDC, de l'OAB, ou encore en lentillage gravitationnel.

La seconde possibilité revient à comparer la luminosité intrinsèque~$L_S$ de la source à l'intensité lumineuse~$I_O$ observée,
\begin{equation}
D\e{L} \define \sqrt{\frac{L_S}{4\pi I_O}}
\end{equation}
définit alors la notion de distance de luminosité. Il s'agit de la notion de distance la plus commune en astronomie, son rôle en cosmologie étant incarné par l'observation des SNe. Pour une même source, les deux distances $D\e{A},D\e{L}$ sont a priori différentes, mais elles ne sont pas indépendantes ; en fait, il est possible de montrer que si le nombre de photon est conservé entre la source et l'observateur, alors $D\e{L}=(1+z)^2 D\e{A}$.

\subsection{Faisceaux lumineux}

Puisqu'elles font intervenir les notions d'aire et d'intensité lumineuse, les distances définies ci-dessus ne peuvent pas être calculées à partir de la trajectoire d'un simple rayon lumineux : elles nécessitent de considérer un ensemble de rayons, i.e. un faisceau, connectant l'ensemble des points de la source à l'observateur. En optique gravitationnelle, la géométrie d'un faisceau lumineux est commodément décrite par la matrice $2\times 2$ de Jacobi~$\vect{\jacobi}$, reliant la séparation angulaire~$\vect{\theta}_O$ observée entre deux rayons à leur séparation physique~$\vect{\xi}(v)$ en tout autre point du faisceau,
\begin{equation}
\vect{\xi}(v) = \vect{\jacobi}(v) \, \vect{\theta}_O.
\end{equation}
Il est alors clair que la distance angulaire d'une petite source lumineuse est reliée au déterminant de la matrice de Jacobi, selon
\begin{equation}
D\e{A} = \frac{\dd^2\vect{\xi}_S}{\dd^2\vect{\theta}_O} = \sqrt{\det \vect{\jacobi}(v_S)}.
\end{equation}

Si les deux rayons considérés sont très proches l'un de l'autre, alors l'évolution de leur séparation~$\vect{\xi}(v)$ au cours de la propagation est régie par l'équation de déviation géodésique. Il s'ensuit que l'évolution de la matrice de Jacobi avec $v$ satisfait à l'équation de Sachs,
\begin{equation}
\ddf[2]{\vect{\jacobi}}{v} = \vect{\tidal} \vect{\jacobi},
\label{eq:Sachs_FR}
\end{equation}
où $\vect{\tidal}$ est une certaine projection du tenseur de courbure de Riemann, appelée matrice optique de marée. Celle-ci peut être décomposée en une composante associée au tenseur de Ricci~$R_{\mu\nu}$ et une composante associée au tenseur de Weyl~$C_{\mu\nu\rho\sigma}$, selon
\begin{equation}
\vect{\tidal} =
\underbrace{
\begin{pmatrix}
\Ricfoc & 0 \\ 0 & \Ricfoc
\end{pmatrix}
}_{\text{focalisation de Ricci}}
+
\underbrace{
\begin{pmatrix}
- {\rm Re}\,\Weylfoc & {\rm Im}\,\Weylfoc \\
  {\rm Im}\,\Weylfoc & {\rm Re}\,\Weylfoc
\end{pmatrix}
}_{\text{distortions de Weyl}},
\qquad \text{avec} \qquad
\begin{system}
\Ricfoc &\propto R_{\mu\nu},\\
\Weylfoc &\propto C_{\mu\nu\rho\sigma}.
\end{system}
\end{equation}
Cette séparation est à la fois géométriquement et physiquement très sensée. La contribution de Ricci, d'une part, provoque une évolution homothétique de $\vect{\jacobi}$ : elle modifie ses valeurs propres en préservant leur rapport et la direction des axes associés. Il s'agit donc d'une focalisation isotrope du faisceau. La contribution de Weyl, d'autre part, tend au contraire à modifier le rapport des valeurs propres de $\vect{\jacobi}$ et à faire tourner leurs axes, elle provoque donc un cisaillement et une rotation du faisceau. Il est intéressant de remarquer que ces deux contributions à $\vect{\tidal}$ sont liées à des propriétés différentes de la distribution de matière rencontrée par la lumière au cours de sa propagation. En vertu des équations d'Einstein, le tenseur de Ricci est directement lié à la densité locale d'énergie-impulsion, $\Ricfoc$ est ainsi généré par la présence de matière diffuse (gaz, matière noire) traversée par le faisceau. Au contraire, la courbure de Weyl est générée de façon non-locale, à l'extérieur de corps massifs ; $\Weylfoc$ est donc due à la matière située à l'extérieur du faisceau.

\section{Cosmologie au-delà de l'hypothèse d'homogénéité}

La discussion ci-dessus est la raison principale du questionnement quant à l'efficacité du modèle cosmologique standard à interpréter toutes les observations cosmologiques avec précision. En effet, la lumière provenant d'une très petite source, par exemple une supernova, se propage essentiellement à travers le vide intergalactique, où la courbure de l'espace-temps est dominée par sa composante de Weyl. Or l'espace-temps du modèle standard, décrit par la géométrie de Friedmann-Lemaître (FL), possède au contraire une courbure de nature purement Ricci. Cette incompatibilité apparente entre réalité et modèle a été soulevée simultanément par Y. Zel'dovich et R. Feynman en 1964.

\subsection{Observations dans un \og grunivers \fg}

\begin{figure}[h!]
\centering
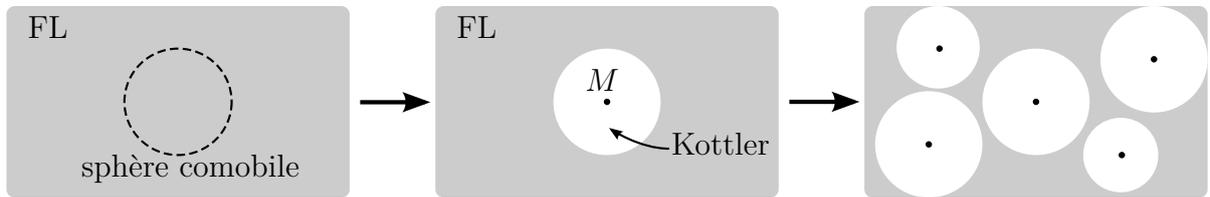
\caption{Construction d'un grunivers à partir d'un modèle homogène et isotrope (FL).}
\label{fig:grunivers}
\end{figure}

Une première façon d'évaluer l'impact de l'inhomogénéité à petite échelle de l'Univers sur la propagation de la lumière consiste en l'utilisation de modèles alternatifs. \`A ce titre, les modèles \og en gruyère\fg (Swiss-cheese models en anglais), que l'on baptisera \og grunivers \fg dans la suite, sont des candidats naturels. Leur construction est la suivante (voir Fig.~\ref{fig:grunivers}) : à partir d'un univers homogène, choisir une sphère comobile, puis concentrer la matière qu'elle contient en son centre. Ceci forme un \og trou\fg au sein du \og fromage \fg homogène initial. \`A l'intérieur du trou, la géométrie spatio-temporelle est décrite par la métrique de Schwarzschild (ou Kottler si $\Lambda\not=0$), tandis qu'à l'extérieur la métrique de FL reste valable. La procédure décrite ci-dessus assure que ces deux géométries se raccordent parfaitement à la frontière du trou, formant un espace-temps bien défini. Physiquement parlant, l'intérieur du trou peut être vu comme représentant le voisinage d'un objet gravitationnellement lié, tel qu'une galaxie ou un amas de galaxie. La masse centrale est donc choisie de l'ordre de $M\sim 10^{11} M_\odot$ (galaxie) ou $M\sim 10^{15}M_\odot$ (amas). Le rayon comobile du trou correspondant, $R\e{h}=(3M/4\pi \rho_0)^{1/3}$, où $\rho_0$ est la masse volumique moyenne de l'Univers aujourd'hui, vaut alors $R\e{h}\sim 1\U{Mpc}$ pour une galaxie, $R\e{h}\sim 20\U{Mpc}$ pour un amas.

L'opération peut ensuite être répétée pour d'autres sphères, toutes disjointes les unes des autres, d'où l'aspect \og en gruyère \fg du résultat. La quantité de trous introduits dans le modèle est quantifié par le paramètre d'homogénéité
\begin{equation}
f \define \lim{V}{\infty}\frac{V\e{FL}}{V},
\end{equation}
où $V$ représente le volume d'une région du modèle, et $V\e{FL}$ la portion de ce volume occupé par des régions homogènes. Les cas $f=0$ ou $f=1$ représentent donc respectivement un univers rempli de masses ponctuelles ou un univers parfaitement homogène. L'avantage principal de cette construction est qu'elle génère un modèle potentiellement très inhomogène, sans toutefois affecter sa dynamique d'expansion ; elle est par conséquent très adaptée à l'étude de la question qui nous intéresse ici.

L'analyse de la propagation de la lumière à travers un grunivers, décrite en détails dans les \S~\ref{hsec:FDU13a} et \S~\ref{hsec:F14}, peut être résumée comme suit :
\begin{enumerate}
\item La relation entre paramètre affine~$v$ et décalage spectral~$z$ est très peu affectée par la présence des inhomogénéités. La correction relative due à un trou est ainsi $(z-z\e{FL})/z\e{FL}= \mathcal{O}(r\e{S}/R\e{h})$, où $r\e{S}\define 2GM$ est le rayon de Schwarzschild de la masse centrale. Cette correction est donc de l'ordre de $10^{-8}$ pour un trou galactique et $10^{-6}$ pour un trou contenant un amas.
\item Les effets de distorsions de Weyl à l'intérieur des trous sont négligeables en première approximation.
\item La focalisation de Ricci est, de manière effective, réduite par le facteur $f$. Tout se passe donc, pour le calcul de la matrice de Jacobi et donc des distances, comme si la lumière se propageait dans un univers homogène de densité réduite $\rho \rightarrow f\rho$.
\end{enumerate}
Les propriétés décrite ci-dessus correspondent à l'approche effective de Kantowski-Dyer-Roeder (KDR). Je les ai démontrées analytiquement, puis vérifiées numériquement à l'aide de deux simulations de propagation de lumière réalisées au cours de cette thèse, l'une présentant une distribution régulière de trous, la seconde une distribution aléatoire.

\begin{figure}[h!]
\centering
\begin{subfigure}[t]{0.55\columnwidth}
\includegraphics[width=\columnwidth]{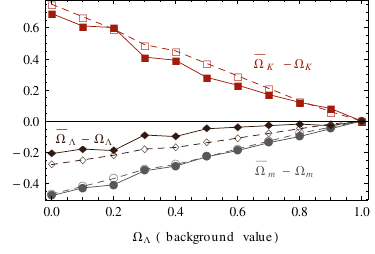}
\caption{}
\label{fig:parametres_vrais_apparents}
\end{subfigure}
\hfill
\begin{subfigure}[t]{0.44\columnwidth}
\includegraphics[width=\columnwidth]{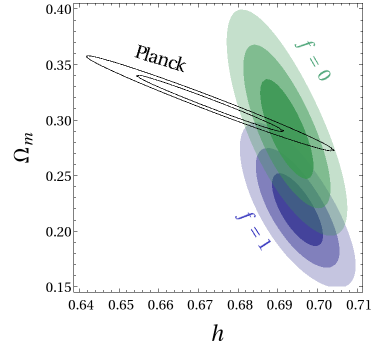}
\caption{}
\label{fig:contraintes}
\end{subfigure}
\caption{\emph{\`A gauche} : différence entre les paramètres cosmologiques apparents~$\{\bar\Omega\}$ et réels~$\{\Omega\}$ dans des grunivers tels que $\Omega_K=0$, $f=0.26$ et $M=10^{11}M_\odot$ (traits et symboles pleins) ou $M=10^{15}M_\odot$ (traits pointillés et symboles vides) dans les trous. \emph{\`A droite} : contraintes observationnelles sur $(h,\Omega\e{m})$ issues de l'analyse du FDC (\textit{Planck}) d'une part, et du diagramme de Hubble (SNLS) d'autre part, en supposant l'Univers soit homogène ($f=1$) soit très inhomogène ($f=0$).}
\end{figure}

Du point de vue de la cosmologie, j'ai dans un premier temps évalué l'erreur potentielle sur les mesures de paramètres cosmologiques que nous commettons en supposant que l'Univers est parfaitement homogène. Pour ce faire, j'ai simulé des catalogues d'observations de SNe, c'est-à-dire un ensemble de doublets $(z,D\e{L})$, dans un grunivers dont la dynamique d'expansion est régie par des paramètres \og vrais \fg~$\{\Omega\}$. J'ai ensuite déterminé les paramètres cosmologiques \og apparents \fg~$\{\bar{\Omega}\}$ obtenus en ajustant ces observations simulées par la courbe théorique $D\e{L}\h{FL}(z|\{\bar{\Omega}\})$, autrement dit, en supposant de façon erronée que la lumière provenant de ces SNe s'est propagée à travers un univers homogène. La différence entre les jeux de paramètres $\{\Omega\}$ et $\{\bar{\Omega}\}$ est représentées sur la Fig.~\ref{fig:parametres_vrais_apparents}, pour un grunivers dont les régions FL présentent des sections spatiales plates ($\Omega_K=0$), avec un paramètre d'homogénéité $f=0.26$. La différence est significative en général, mais diminue fortement avec $\Omega_\Lambda$, ce qui est somme toute très naturel : la constante cosmologique générant une forme de courbure strictement homogène, si celle-ci domine alors la géométrie spatio-temporelle du grunivers ne diffère que très peu du modèle FL dont il est issu.

Dans un second temps, j'ai comparé deux analyses différentes de données réelles de SNe (issues des trois premières années de la campagne SNLS) : d'abord en supposant que leur lumière s'est propagée à travers un univers homogène (approche standard, $f=1$) ; puis en remplaçant le modèle FL sous-jacent par un grunivers très inhomogène ($f=0$), c'est-à-dire en exploitant la relation $D\e{L}(z|\{\Omega\})$ dictée par l'approche de KDR qui, on l'a dit, constitue une bonne approximation des propriété optiques de tels modèles. Les contraintes observationnelles associées, dans le plan $(h,\Omega\e{m})$ où $h\define H_0/(100\U{km/s/Mpc})$, sont représentées sur la Fig.~\ref{fig:contraintes}, ainsi que les contraintes indépendantes établies par l'observation du FDC par la mission \textit{Planck}. Il résulte que la prise en compte de l'inhomogénéité à petite échelle de l'Univers, même à travers un modèle très grossier comme le grunivers, semble améliorer l'accord entre ces différentes observations cosmologiques.

L'analyse résumée ici répond en partie à la problématique principale de cette thèse. Il semble ainsi que la prépondérance de la constante cosmologique dans l'Univers tardif soit un ingrédient crucial du succès du modèle cosmologique standard. En effet, si nous vivions dans un Univers tel que $\Omega_\Lambda\ll \Omega\e{m}$ aujourd'hui, alors les observations impliquant de très fins faisceaux lumineux (par exemple les SNe) seraient bien davantage affectées par l'inhomogénéité de l'Univers, et paraîtraient alors en profond désaccord avec les observations impliquant de plus larges faisceaux (comme pour le FDC). Néanmoins, même dans notre propre Univers, pourtant dominé par $\Lambda$, la Fig.~\ref{fig:contraintes} indique que la précision atteinte par les observations actuelle est désormais suffisante pour que de tels effets soient révélés, et nécessitent par conséquent d'être pris en compte précisément.

\subsection{Lentillage gravitationnel stochastique}

La section précédente a mis en évidence la nécessité de modéliser avec précision l'effet de l'inhomogénéité à petite échelle de l'Univers sur ses propriétés optiques. Bien que révélé par l'étude du grunivers, ce modèle s'avère être trop peu flexible pour permettre une étude réaliste du problème. Une approche effective, qui pourrait prendre en compte les propriétés complexes de l'Univers de façon efficace, serait préférable.

On notera que la situation physique abordée ici, à savoir l'effet d'une multitude de faibles interactions entre un corps (en l'occurrence la lumière) et son environnement, n'est pas sans rappeler le mouvement Brownien. Ce processus désordonné que l'on peut observer, par exemple, pour une poussière micrométrique en suspension sur l'eau, résulte des chocs entre cette poussière et les molécules formant le liquide. Un tel phénomène ne peut donc pas être expliqué en modélisant l'eau comme un fluide, car il n'est pas dû à des courants macroscopiques en son sein. Il n'est cependant pas nécessaire de suivre la dynamique de chaque molécule indépendamment pour décrire le mouvement Brownien : en pratique, leur effet donne lieu à une force stochastique, modélisée par un bruit blanc.

Cette analogie m'a amené à modéliser le lentillage gravitationnel dû aux petites structures de l'Univers par un terme stochastique dans la matrice de marée optique, de sorte que l'équation de Sachs~\eqref{eq:Sachs_FR} prend la forme d'une équation de Langevin,
\begin{equation}
\ddf[2]{\vect{\jacobi}}{v} = (\ev{\vect{\tidal}}+\delta\vect{\tidal})\,\vect{\jacobi},
\label{eq:Sachs-Langevin_FR}
\end{equation}
avec $\ev{\delta\vect{\tidal}}=\vect{0}$. On notera que, malgré une notation suggestive, le terme de fluctuation~$\delta\vect{\tidal}$ n'est pas nécessairement petit par rapport au terme déterministe~$\ev{\vect{\tidal}}$. Les hypothèses d'homogénéité et d'isotropie statistiques de l'Univers impliquent les propriétés suivantes:
\begin{inparaenum}[]
\item $\ev{\Weylfoc} = 0$ ;
\item $\ev{\delta\Ricfoc(v) \Weylfoc(w)} = 0$ ;
\item $\ev{\Re\Weylfoc(v) \,\Im\Weylfoc(w)} = 0$.
\end{inparaenum}
Par ailleurs, le fait que l'on cherche à modéliser de très petites échelles permet de traiter le terme de fluctuation comme un bruit blanc, donc $\delta$-corrélé :
\begin{equation}
\ev{ \delta\Ricfoc(v) \delta\Ricfoc(w) } = C_\Ricfoc(v) \delta(v-w),
\qquad
\ev{ \Weylfoc^\star(v) \Weylfoc(w) } = 2C_\Weylfoc(v) \delta(v-w),
\end{equation}
où l'étoile indique une conjugaison complexe. Les fonctions~$C_X$ sont analogues à des coefficients de diffusion, et sont de l'ordre de $\delta X^2\times \Delta v\e{coh}$, où $\delta X$ indique l'amplitude typique des fluctuations de $X$, et $\Delta v\e{coh}$ l'échelle typique de paramètre affine pendant lequel $X$ reste cohérent.

De l'équation de Sachs-Langevin~\eqref{eq:Sachs-Langevin_FR} peut être déduite l'équation de Fokker-Planck-Kolmogorov régissant l'évolution, avec $v$, de la densité de probabilité~$p(v;\vect{\jacobi},\dot{\vect{\jacobi}})$ de la matrice de Jacobi. Cette équation prend ici la forme
\begin{empheq}[box=]{multline}
\pd{p}{v} = - \dot{\jacobi}_{AB} \pd{p}{\jacobi_{AB}}
					- \ev{\Ricfoc} \jacobi_{AB} \frac{\partial p}{\partial \dot{\jacobi}_{AB}} \\
				+ \frac{1}{2} \pac{ C_\Ricfoc\,\delta_{AE} \delta_{CF} + C_\Weylfoc (\delta_{AC} \delta_{EF} - \eps_{AC}\eps_{EF})} 
					 	\jacobi_{EB}\jacobi_{FD}\,\frac{\partial^2 p}{\partial\dot{\jacobi}_{AB}\partial\dot{\jacobi}_{CD}}
\label{eq:FPK_FR}
\end{empheq}
où les indices $A,B,\ldots$ défilent entre 1 et 2, et $\eps_{AB}$ est antisymétrique, avec $\eps_{12}=1$. Bien que difficilement soluble, cette équation aux dérivées partielles permet toutefois de déterminer les équations différentielles ordinaires régissant les moments de $p$, desquelles on peut ensuite déduire les moments de la distribution de la distance angulaire~$D\e{A}$. Pour les deux premiers, c'est-à-dire la moyenne~$\ev{D\e{A}}$ et la variance~$\sigma_{D\e{A}}$, on trouve en particulier
\begin{gather}
\delta_{D\e{A}}\define\frac{\ev{D\e{A}}-D_0}{D_0}
=
- 2 \int_0^v \frac{\dd v_1}{D^2_0(v_1)}
		\int_0^{v_1} \frac{\dd v_2}{D^2_0(v_2)} 
			\int_0^{v_2}\dd v_3 \; D_0^4(v_3) C_\Weylfoc(v_3) + \mathcal{O}(C_\Weylfoc^2), \\
\ddf[3]{}{x}\pac{ \frac{\sigma^2_{D\e{A}}}{D_0^2} } 
+ 2 D_0^6(2C_\Weylfoc - C_\Ricfoc) \frac{\sigma^2_{D\e{A}}}{D_0^2}
=
2 C_\Ricfoc D_0^6 
+ 6 \int_\obs^x \dd x' \pac{\ddf[2]{\delta_{D\e{A}}}{x}}^2
+ \mathcal{O}(C_\Weylfoc^3),
\label{eq:evolution_variance_DA}
\end{gather}
$D_0$ étant la distance angulaire en l'absence de fluctuations, c'est-à-dire telle que $\dd^2D_0/\dd v^2=\ev{\Ricfoc} D_0$, et $x$ est une variable abstraite telle que $\dd x=\dd v/D_0^2(v)$.

\begin{figure}[h!]
\centering
\includegraphics[width=0.49\columnwidth]{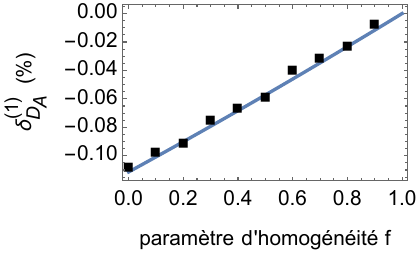}
\hfill
\includegraphics[width=0.49\columnwidth]{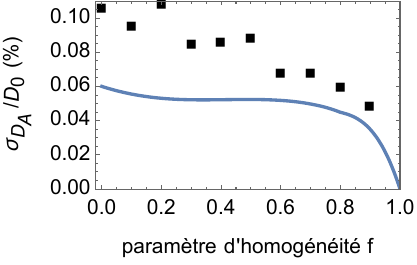}
\caption{Moyenne (à gauche) et écart-type (à droite) de la distance angulaire~$D\e{A}$ de sources de décalage spectral $z=1$ à travers un grunivers. La ligne bleue représente les prédictions du formalisme de lentillage stochastique, tandis que les carrés noir indiquent les résultats de simulations numériques.}
\label{fig:stochlens_numerique}
\end{figure}

Dans le but de tester ce formalisme, je l'ai appliqué au grunivers étudié dans la section précédente, en comparant les prédictions théoriques aux résultats numériques (voir Fig.~\ref{fig:stochlens_numerique}). On voit que la moyenne de $D\e{A}$ est extrêmement bien prédite par le formalisme de lentillage stochastique ; ce n'est pas le cas pour son écart-type. Une étude approfondie a révélé que l'origine du problème réside dans l'hypothèse de gaussianité des fluctuations de $\Weylfoc$, sous-entendue dès lors que l'équation de Fokker-Planck-Kolmogorov~\eqref{eq:FPK_FR} est invoquée.

Malgré cette faiblesse, une telle approche stochastique du lentillage gravitationnel aux petites échelle reste autant prometteuse que novatrice. Elle ouvre la voie vers des méthodes efficaces et flexibles pour estimer, par exemple, le biais et la dispersion du diagramme de Hubble des SNe dus aux petites structures de notre Univers.

\section{Cosmologie au-delà de l'hypothèse d'isotropie}

Le modèle cosmologique standard est fondé sur les hypothèses d'homogénéité et d'isotropie de l'Univers à grande échelle. Il est important de noter que ces deux hypothèses ne sont pas redondantes : l'Univers peut être homogène (identique en tout point) mais anisotrope ; par exemple, son expansion pourrait être plus rapide dans certaines directions que dans d'autres. Comme toute hypothèse fondamentale, l'isotropie cosmique se doit donc d'être testée. Deux aspects de la question peuvent alors être distingués. Du point de vue observationnel, d'abord : comment l'anisotropie modifie-t-elle nos observations et leur interprétation ? Du point de vue théorique ensuite : quelles pourraient être les sources d'une telle anisotropie ?

\subsection{Optique dans un univers anisotrope}

Notre capacité à détecter toute forme d'anisotropie cosmique requiert de comprendre la propagation de la lumière à travers un univers anisotrope. Pour ce faire, il est commode de ne relaxer que l'hypothèse d'isotropie, tout en conservant l'homogénéité. Les modèles correspondants suivent alors la classification de Bianchi des espace tridimensionnels homogènes, et ont donc été baptisés espace-temps de Bianchi. Le plus simple d'entre eux, dit Bianchi~I, présente des sections spatiales euclidiennes, et sa métrique s'écrit
\begin{equation}
\dd s^2 = -\dd t^2 + a^2(t) \sum_{i=1}^3 \pac{\ex{\beta_i(t)} \dd x^i}^2,
\end{equation}
où la somme des $\beta_i$ est nulle. Ce modèle peut être vu comme une extension du modèle isotrope de FL admettant trois facteurs d'échelle $a \ex{\beta_i}$ au lieu d'un seul, et présentant donc une expansion différente le long de chaque direction.

Les propriétés optiques d'un tel modèle peuvent être résumées ainsi :
\begin{itemize}
\item La dérive temporelle du décalage spectral n'est pas isotrope. Si l'on observe un ensemble de sources comobiles ayant toutes le même décalage spectral~$z$ à un instant donné, à un instant ultérieur ces sources n'auront plus le même décalage spectral les unes par rapport aux autres.
\item Les objets dérivent sur la sphère céleste. Si, à un instant donné, l'on pointe un télescope dans une certaine direction pour observer une source lumineuse comobile, alors, à un instant ultérieur la source n'apparaîtra plus dans la ligne de mire. Il s'agit ici d'un effet purement cosmologique, indépendant de l'effet de la rotation de la Terre que l'on suppose avoir corrigé.
\item La relation entre distance angulaire et décalage spectral~$D\e{A}(z)$ dépend de la direction dans laquelle elle est évaluée.
\item Les images sont déformées par rapport à leurs propriétés intrinsèques. L'anisotropie de l'expansion, à cause du lentillage de Weyl qu'elle engendre, tend à cisailler et faire tourner les faisceaux lumineux. Cet effet évolue avec le temps, en général.
\end{itemize}
J'ai volontairement choisi d'exprimer ces propriétés avec des mots plutôt qu'avec des formules mathématiques. Celles-ci sont exposées en détail au Chap.~\ref{chapter:optics_Bianchi_I}. Je précise toutefois que chacun des effets mentionné ci-dessus a été démontré analytiquement. En particulier, j'ai établi une solution exacte de l'équation de Sachs, obtenant ainsi pour la première fois l'expression de la matrice de Jacobi dans un univers anisotrope.

L'étude rapportée ici ouvre la voie vers de nouvelles stratégies observationnelles visant à contraindre l'anisotropie de l'expansion cosmique, en particulier à l'aide du diagramme de Hubble, ou via l'analyse du lentillage gravitationnel faible des galaxies.

\subsection{Modèles scalaire-vecteur}

L'intérêt pour les modèles cosmologiques anisotropes a récemment été ravivé par l'observation d'anomalies dans le spectre de puissance du FDC aux faibles multipôles, attribuables à une anisotropie cosmique. Le cas échéant, la question de son origine physique devrait alors être abordée. \`A ce titre, le candidat le plus naturel serait un champ vectoriel de matière, tel qu'un champ électromagnétique à l'échelle cosmologique. Il est également envisageable qu'un tel champ ait été responsable de la phase primordiale d'inflation (inflaton vectoriel), ou encore de l'accélération actuelle de l'expansion cosmique (quintessence vectorielle). Il s'avère toutefois difficile de générer de tels phénomènes à l'aide d'un champ vectoriel seul (en particulier si celui-ci n'a pas de masse ainsi qu'imposé par l'invariance de jauge) sans provoquer une trop grande anisotropie. Une manière de contourner ce problème consiste à coupler le champ vectoriel avec un champ scalaire qui serait, lui, responsable de la majeure partie de l'inflation ou de l'accélération tardive de l'expansion de l'Univers.

C'est dans ce contexte général que s'inscrit la dernière partie de cette thèse. On y analyse les propriétés de stabilité et de causalité d'une large classe de modèles scalaire-vecteur, afin de déterminer s'ils sont fondamentalement viables, sans avoir besoin de recourir à une confrontation fastidieuse avec les observations. La condition de stabilité se traduit mathématiquement par le fait que le hamiltonien de la théorie est minoré par une certaine valeur. En outre, la causalité d'une théorie est assurée si ses équations du mouvement sont des équations aux dérivées partielles du second ordre, de genre hyperbolique. Cette condition implique en effet que le problème de Cauchy associé à la donnée de conditions initiales, c'est-à-dire de l'état des champs et de leur dérivée temporelle le long d'une hypersurface de genre espace, admet une unique solution.

Les résultats de cette étude peuvent être résumés de la façon suivante. Soit une théorie des champs contenant un scalaire~$\phi$ et un vecteur~$\vect{A}$. Si cette théorie dérive d'un principe de moindre action tel que
\begin{inparaenum}[(i)]
\item $\phi$ et $\vect{A}$ sont minimalement couplés à la gravitation ;
\item l'action ne contient que des dérivées d'ordre inférieur à 1 en ces champs ; et
\item est invariante sous les transformations de jauge de $\vect{A}$ ;
\end{inparaenum}
alors la forme la plus générale de cette action est
\begin{equation}
S[\phi,\vect{A;\vect{g}}]
=
\int \dd^4 x \sqrt{-g} \; \mathcal{L}(\phi,K,X,Y,Z),
\end{equation}
où les scalaires $K,X,Y,Z$ sont définis par
\begin{equation}
K \define \partial_\mu\phi \partial^\mu \phi,
\quad
X \define F^{\mu\nu} F_{\mu\nu},
\quad
Y \define F^{\mu\nu} \tilde{F}_{\mu\nu},
\quad
Z \define (\partial_\mu\phi \tilde{F}\indices{^\mu^\rho}) (\partial_\nu\phi  \tilde{F}\indices{^\nu_\rho}),
\end{equation}
avec $\tilde{F}_{\mu\nu}\define\eps_{\mu\nu\rho\sigma} F^{\rho\sigma}/2$ le dual de Hodge de la deux-forme de courbure associée à $\vect{A}$, elle-même définie par $F_{\mu\nu}\define\partial_\mu A_\nu - \partial_\nu A_\mu$. Si l'on ajoute à cela l'hypothèse que (iv) $\mathcal{L}$ est, au plus, quadratique en $\vect{A}$, alors les seuls théories physiquement viables vérifient
\begin{equation}
\mathcal{L} = \frac{1}{2} \, f_0(\phi,K) - \frac{1}{4} \, f_1(\phi) \, X - \frac{1}{4} \, f_2(\phi) \, Y,
\end{equation}
la fonction $f_1$ étant positive, et la fonction $f_0$ devant vérifier les conditions suivantes :
\begin{inparaenum}[]
\item $f_0(\phi,K\geq 0)$ est minorée ;
\item $\partial f_0/\partial K \geq 0$ ; et
\item $\partial f_0/\partial K + 2 K\partial^2f_0/\partial K^2 \geq 0$.
\end{inparaenum}
Cela réduit de façon drastique l'espace des possibilités pour cette classe de théories.

\section*{Conclusion}
\phantomsection
\addcontentsline{toc}{section}{Conclusion}

Cette thèse avait pour objectif l'étude approfondie de la propagation de la lumière à travers l'Univers, en particulier lorsque les hypothèses d'homogénéité et d'isotropie sont relaxées. Son premier volet a permis de comprendre que la domination de la constante cosmologique dans la dynamique de l'expansion cosmique aujourd'hui n'est pas étrangère au succès du modèle standard. Il semble néanmoins que la précision croissante des observations cosmologiques ne permettra bientôt plus de négliger l'impact de l'inhomogénéité à petite échelle de notre Univers. Le formalisme de lentillage gravitationnel stochastique proposé dans cette thèse constitue alors une méthode prometteuse pour modéliser ces effets de façon précise et efficace. Le second volet de cette thèse, consacré à la possibilité d'une anisotropie à grande échelle de l'Univers, a contribué à la fois à une meilleure compréhension des propriétés optiques de modèles cosmologiques anisotropes, et aux causes potentielles d'une telle anisotropie. Le travail qui vient d'être résumé appelle naturellement de nombreux compléments, tant du point de vue fondamental qu'observationnel, que je souhaite aborder dans un futur proche.

%% file: grunivers.pdf_tex
\begingroup%
  \makeatletter%
  \providecommand\color[2][]{%
    \errmessage{(Inkscape) Color is used for the text in Inkscape, but the package 'color.sty' is not loaded}%
    \renewcommand\color[2][]{}%
  }%
  \providecommand\transparent[1]{%
    \errmessage{(Inkscape) Transparency is used (non-zero) for the text in Inkscape, but the package 'transparent.sty' is not loaded}%
    \renewcommand\transparent[1]{}%
  }%
  \providecommand\rotatebox[2]{#2}%
  \ifx\svgwidth\undefined%
    \setlength{\unitlength}{448bp}%
    \ifx\svgscale\undefined%
      \relax%
    \else%
      \setlength{\unitlength}{\unitlength * \real{\svgscale}}%
    \fi%
  \else%
    \setlength{\unitlength}{\svgwidth}%
  \fi%
  \global\let\svgwidth\undefined%
  \global\let\svgscale\undefined%
  \makeatother%
  \begin{picture}(1,0.16071429)%
    \put(0,0){\includegraphics[width=\unitlength]{grunivers.pdf}}%
    \put(0.0625,0.01785714){\color[rgb]{0,0,0}\makebox(0,0)[lb]{\smash{sphère comobile
}}}%
    \put(0.55357143,0.03571429){\color[rgb]{0,0,0}\makebox(0,0)[lb]{\smash{Kottler}}}%
    \put(0.01785714,0.13392857){\color[rgb]{0,0,0}\makebox(0,0)[lb]{\smash{FL}}}%
    \put(0.48214286,0.08928571){\color[rgb]{0,0,0}\makebox(0,0)[lb]{\smash{$M$}}}%
    \put(0.375,0.13392857){\color[rgb]{0,0,0}\makebox(0,0)[lb]{\smash{FL}}}%
  \end{picture}%
\endgroup%

%% file: backcover.tex
\cleartoleftpage
\thispagestyle{empty}
\renewcommand{\lettrine}[2]{\noindent #1#2}

\begin{otherlanguage}{frenchb}
\vspace*{-0.5cm}
\hrule


\begin{center}
{\sf\bfseries Propagation de la lumière dans des univers inhomogènes ou anisotropes}
\end{center}


{\small
\input{resume.tex}

\medskip

\noindent{\sf\bfseries Mots clés} : cosmologie, lumière, inhomogénéité, modèles \og Swiss cheese \fg, anisotropie, Bianchi~I.
}

\end{otherlanguage}

\bigskip

\hrule



\begin{center}
{\sf\bfseries Light propagation in inhomogeneous and anisotropic cosmologies}
\end{center}


{\small
\input{abstract.tex}

\medskip

\noindent{\sf\bfseries Keywords}: cosmology, light, inhomogeneity, Swiss-cheese models, anisotropy, Bianchi~I.
}

\bigskip

\hrule

%% file: resume.tex
\lettrine{L}{e} modèle standard de la cosmologie est fondé sur les hypothèses d'homogénéité et d'isotropie de l'Univers. Lors de l'interprétation de la plupart des observations, ces deux hypothèses sont appliquées de façon stricte, au sens où l'on suppose que la lumière émise par des sources lointaines se propage jusqu'à nous à travers un espace-temps de Friedmann-Lemaître. L'objectif principal de cette thèse a été d'évaluer la pertinence de ces hypothèses, en particulier lorsque de très petites échelles sont mises en jeu. Après une revue détaillée des lois de l'optique géométrique en espace-temps courbe, on propose une analyse exhaustive de la propagation de la lumière à travers des modèles cosmologiques \og en gruyère\fg, conçus pour modéliser le caractère grumeleux de l'Univers à petite échelle. L'impact sur l'interprétation du diagramme de Hubble est ensuite évalué quantitativement, et s'avère être plutôt faible, en particulier grâce à la constante cosmologique. Lorsqu'appliquées aux données actuelles issues de l'observation de supernovae, les corrections associées tendent toutefois à améliorer l'accord entre les paramètre cosmologiques mesurés à partir du diagramme de Hubble d'une part, et à partir du fond diffus cosmologique d'autre part. Ceci suggère que la précision des observations cosmologiques atteinte aujourd'hui ne permet plus de négliger l'effet des petites structures sur la propagation de lumière à travers le cosmos. Un tel constat a motivé le développement d'un nouveau cadre théorique, inspiré de la physique statistique, visant à décrire ces effets avec précision. Quant à l'hypothèse d'isotropie, cette thèse aborde d'une part les conséquences potentielles d'une anisotropie à grande échelle de l'univers sur la propagation de la lumière, en résolvant de façon explicite toutes les équation de l'optique géométrique dans l'espace-temps de Bianchi~I. D'autre part, on y analyse une classe de sources d'anisotropie, à savoir les modèles scalaire-vecteur pour l'inflation ou l'énergie sombre. La plupart d'entre eux ne sont pas physiquement viables.

%% file: abstract.tex
\lettrine{T}{he} standard model of cosmology is based on the hypothesis that the Universe is spatially homogeneous and isotropic. When interpreting most observations, this cosmological principle is applied stricto sensu: the light emitted by distant sources is assumed to propagate through a Friedmann-Lema\^itre spacetime. The main goal of the present thesis was to evaluate how reliable this assumption is, especially when small scales are at stake. After having reviewed the laws of geometric optics in curved spacetime, and the standard interpretation of cosmological observables, the dissertation reports a comprehensive analysis of light propagation in Swiss-cheese models, designed to capture the clumpy character of the Universe. The resulting impact on the interpretation of the Hubble diagram is quantified, and shown to be relatively small, thanks to the cosmological constant. When applied to current supernova data, the associated corrections tend however to improve the agreement between the cosmological parameters inferred from the Hubble diagram and from the cosmic microwave background. This is a hint that the effect of small-scale structures on light propagation may become non-negligible in the era of precision cosmology. This motivated the development of a new theoretical framework, based on stochastic processes, which aims at describing small-scale gravitational lensing with a better accuracy. Regarding the isotropy side of the cosmological principle, this dissertation addresses, on the one hand, the potential effect of a large-scale anisotropy on light propagation, by solving all the equations of geometric optics in the Bianchi I spacetime. On the other hand, possible sources of such an anisotropy, namely scalar-vector models for inflation or dark energy, are analysed. Most of them turn out to be excluded as physically viable theories.